\documentclass[12pt]{article}
\usepackage{graphicx} 
\begin{document}

\begin{titlepage}
\begin{flushright}
CERN-PH-TH/2008-252
\end{flushright}
\vspace*{1cm}

\begin{center}
{\LARGE {\bf Stochastic backgrounds of 
relic gravitons:}\\
\vskip 0.5cm
{\bf a theoretical appraisal}}
\vskip2.cm
\large{Massimo Giovannini \footnote{e-mail address: massimo.giovannini@cern.ch}}
\vskip 1.cm
{\it Department of Physics, Theory Division, CERN, 1211 Geneva 23, Switzerland}
\vskip 0.5cm
{\it  INFN, Section of Milan-Bicocca, 20126 Milan, Italy }

\end{center}
\begin{abstract}
Stochastic backgrounds or relic gravitons, if ever detected, will constitute a prima facie evidence of physical processes taking place during the earliest stages of the evolution of the plasma. The essentials of the stochastic backgrounds of relic gravitons are hereby introduced and reviewed.  The pivotal observables customarily employed to infer the properties of the relic gravitons  are discussed both in the framework of the $\Lambda$CDM paradigm as well as in neighboring contexts. The complementarity between experiments measuring the polarization of the Cosmic Microwave Background (such as, for instance,  WMAP, Capmap, Quad, Cbi, just to mention a few) and wide band interferometers (e.g. Virgo, Ligo, Geo, Tama) 
is emphasized.  While the analysis of the microwave sky strongly constrains the low-frequency tail of the relic graviton spectrum, wide-band detectors are sensitive to much higher frequencies where the spectral energy density depends chiefly upon the (poorly known) rate of post-inflationary expansion.
\end{abstract}
\end{titlepage}

\pagenumbering{arabic}

\tableofcontents

\newpage
\renewcommand{\theequation}{1.\arabic{equation}}
\setcounter{equation}{0}
\section{The spectrum of the relic gravitons}
\label{sec1}
\subsection{The frequencies and wavelengths of relic gravitons}
Terrestrial and satellite observations, scrutinizing the properties of the electromagnetic spectrum, are unable  to test directly the evolution of the background geometry prior to photon decoupling. 
The redshift probed by Cosmic Microwave Background (CMB in what follows) observations is of the order 
of $z_{\mathrm{dec}} \simeq 1087$ and it roughly corresponds to the peak of  the visibility function, i.e. when most of the CMB photons last scattered free electrons (and protons).  After decoupling the ionization fraction drops; the photons follow null geodesics whose slight inhomogeneities can be directly connected with the fluctuations of the spatial curvature present before matter-radiation decoupling.

The temperature of CMB photons is, today, of the order of $2.725$ K. The same temperature at photon decoupling \footnote{Natural units $\hbar=c=k_{\mathrm{B}}= 1$ will be  adopted. In this system,  
$\mathrm{K} = 8.617\times 10^{-5} \, \mathrm{eV}$.}
must have been of the order of about $2962$ K, i.e.  $0.25$ eV. The CMB temperature increases linearly with the redshift: this fact may be tested empirically by observing at high redshifts clouds of chemical compounds (like CN) whose excited levels may be populated thanks to the higher value of the CMB temperature \cite{smoot,songaila}.

The initial conditions for the processes
leading to formation of CMB anisotropies are  set well before matter radiation equality and right after neutrino decoupling (taking place for temperatures of the order of the MeV) whose 
associated redshift is around $10^{10}$. The present knowledge of particle interactions up to  energy scales of the order of $200$ GeV certainly provides important (but still indirect) clues on the composition of the plasma. 

If ever detected, relic gravitons might provide direct informations on the evolution of the Hubble rate for much higher redshifts. In a rudimentary realization 
of the $\Lambda$CDM paradigm\footnote{In the acronym $\Lambda$CDM, $\Lambda$ 
qualifies the dark-energy component while CDM qualifies the dark matter component.}, the inflationary phase can be modeled in terms of the expanding branch of de-Sitter space. Assuming that, right after inflation, the Universe evolves adiabatically  and is dominated by radiation, the redshift  associated with the end of inflation can be approximately computed as 
\begin{equation}
z + 1 \simeq 3 \times 10^{28} \, \biggl(\frac{T_{\gamma0}}{2.725\, \mathrm{K}}\biggr)^{-1} \biggl(\frac{H}{10^{-5}\, M_{\mathrm{P}}}\biggr)^{1/2} \, \biggl(\frac{g_{\rho}}{106.75}\biggr)^{-1/4},
\label{zinf}
\end{equation}
where $g_{\rho}$ denotes the weighted \footnote{Fermions and bosons 
contribute with different factors to $g_{\rho}$. By assuming that all the species 
of the standard model are in local thermodynamic equilibrium (for instance for 
temperature higher than the top quark mass), $g_{\rho}$  
will be given by $g_{\rho} = 28 + (7/8) 90 =106.75$ where 28 and $90$ count, respectively, the 
bosonic and the fermionic contributions.} number of relativistic degrees of freedom at the onset of the 
radiation dominated evolution and $106.75$ corresponds to the value of the standard model 
of particle interactions.  In Eq. (\ref{zinf}) it has been also assumed, quite generally, that $H \simeq 10^{-5}\, M_{\mathrm{P}}$ as implied by the CMB observations. For more accurate estimates the quasi-de Sitter 
nature of the inflationary expansion must be taken into account. 
In the $\Lambda$CDM paradigm, the basic mechanism responsible for the production of relic gravitons 
is the parametric excitation of the (tensor) modes 
of the background geometry and it is controlled by the rate of variation of space-time curvature.

In the present article the $\Lambda$CDM paradigm will always be assumed as a starting point 
for any supplementary considerations. The reasons for this choice are also practical since 
the experimental results must always be stated and presented in terms of a given 
reference model. Having said this, most of the considerations presented here can also 
be translated (with the appropriate computational effort) to different models.

Given a specific scenario for the evolution of the Universe (like the $\Lambda$CDM model), the relic graviton spectra can be computed.  The amplitude of the relic graviton spectrum over different frequencies depends 
upon the specific evolution of the Hubble rate. 
The theoretical error  on the amplitude increases with the frequency: it is more uncertain (even within a specified scenario) at high frequencies rather than at small frequencies. 

The experimental data, at the moment, do not allow either to rule in or to rule out the presence of a primordial spectrum of relic gravitons compatible with the $\Lambda$CDM scenario.  The typical frequency probed by CMB experiments is of the order of \footnote{We are here enforcing the 
usual terminology stemming  from the prefixes of the International System of units: aHz (for atto Hz i.e. $10^{-18}$ Hz), fHz (for femto Hz, i.e.   $10^{-15}$ Hz) and so on.} $\nu_{\mathrm{p}} =  k_{\mathrm{p}}/(2\pi) \simeq 
10^{-18}\, \mathrm{Hz} = 1 \, \mathrm{aHz}$  
where  $k_{\mathrm{p}}$ is the pivot frequency at which the tensor power spectra are assigned. 
CMB experiments will presumably  
set stronger bounds on the putative presence of a tensor 
background for  frequencies ${\mathcal O}(\mathrm{aHz})$. This 
bound will be significant also for higher frequencies only if
the whole post-inflationary thermal history is assumed to be known and specified. 
\begin{figure}[!ht]
\centering
\includegraphics[height=6.7cm]{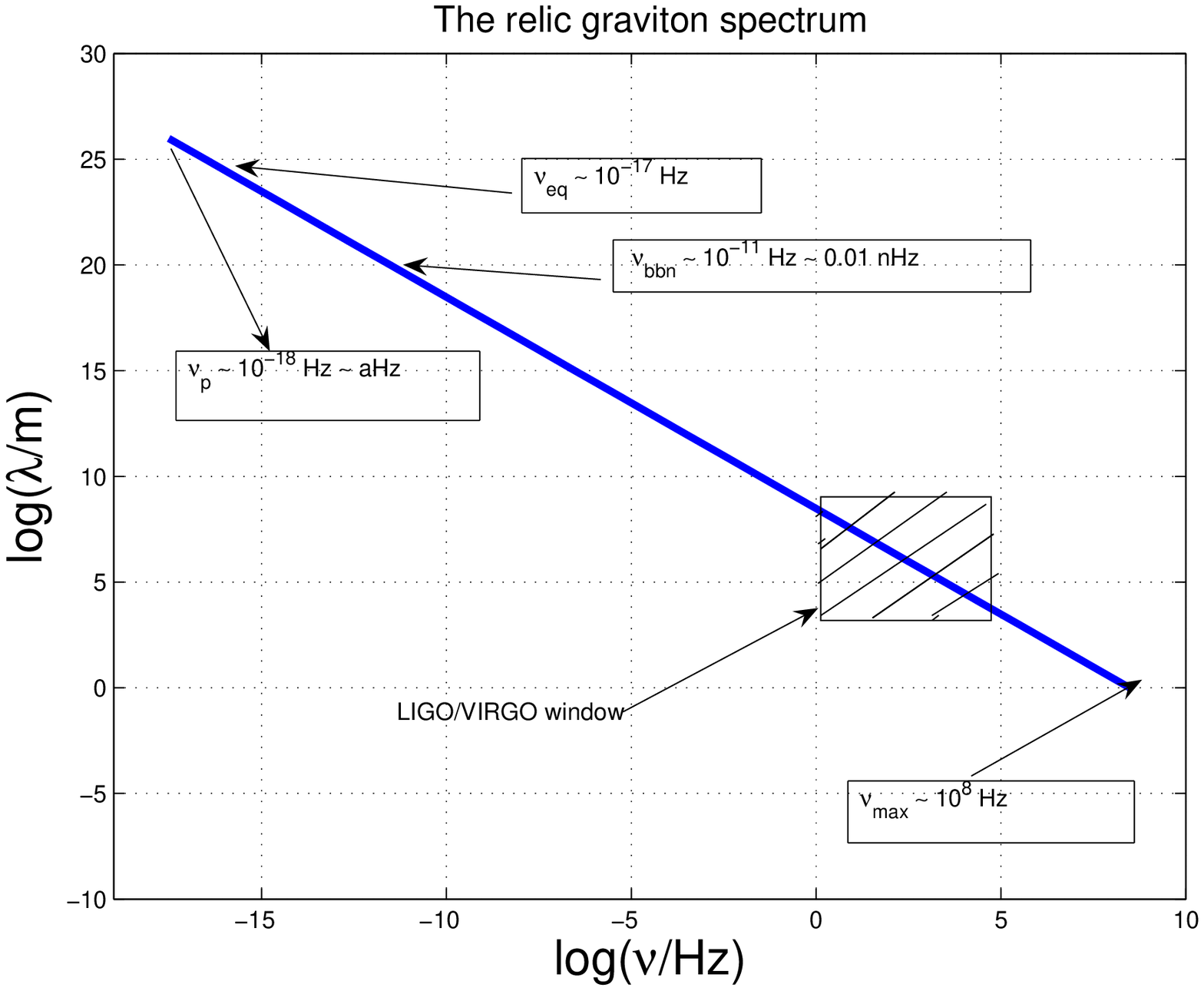}
\includegraphics[height=6.7cm]{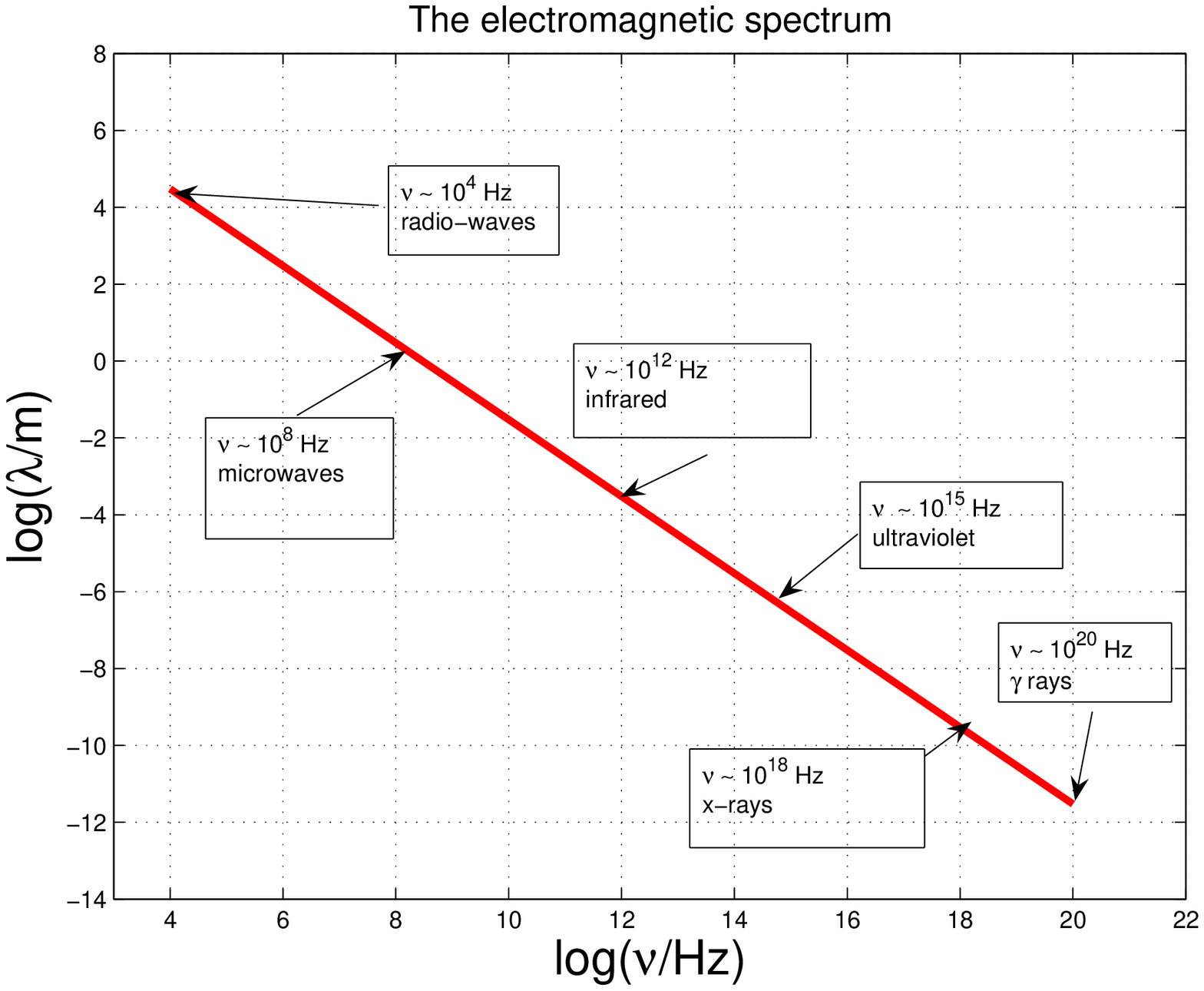}
\caption[a]{A schematic view of the frequency range of the relic graviton spectrum (plot at the left) and of the 
electromagnetic spectrum (plot at the right). In both plots the common logarithms of the (comoving) 
frequency and of the (comoving) wavelengths are reported, respectively, on the horizontal and on the 
vertical axis.}
\label{FigureSP}      
\end{figure}
The typical frequency window of wide-band interferometers (such as Ligo and Virgo)
 is located between few Hz and $10$ kHz, i.e. roughly  speaking, $20$ orders of magnitude 
larger than the frequency probed by CMB experiments.  The frequency range of wide-band 
interferometers will be conventionally denoted by $\nu_{\mathrm{LV}}$.
To compute the relic graviton spectrum over the latter range of frequencies, 
the evolution of our Universe should be known over a broad range of redshifts. 
We do have some plausible guesses on the evolution 
 of the plasma from the epoch of neutrino decoupling down to the epoch of photon 
 decoupling. The latter range of redshifts corresponds to an interval of comoving frequencies 
 going from $\nu_{\mathrm{p}} \simeq 10^{-18}$ Hz up to 
 $\nu_{\mathrm{bbn}} \simeq 0.01\,\mathrm{nHz} \sim 10^{-11}\,$ Hz (at most). 

In Fig. \ref{FigureSP} (plot at the left) the frequency range of the relic graviton spectrum is illustrated,
starting from the (comoving) frequency $\nu_{\mathrm{p}}$ whose associated (comoving) wavelength is of the order of  of $10^{26}$ m, i.e. roughly comparable with  the Hubble radius at the present time. The low-frequency branch of the spectrum can be conventionally defined between $\nu_{\mathrm{p}}$ and $\nu_{\mathrm{bbn}}$. The largest 
frequency of the relic graviton spectrum (i.e. $\nu_{\mathrm{max}}$) is of the 
order of $0.1$ GHz in the $\Lambda$CDM scenario. Thus the high-frequency branch 
of the graviton spectrum can be conventionally defined for $\nu_{\mathrm{bbn}} < \nu < \nu_{\mathrm{max}}$. 
In summary we can therefore say that
\begin{itemize}
\item{} the range $\nu_{\mathrm{p}} < \nu < \nu_{\mathrm{bbn}}$ will be generically referred to as 
the low-frequency domain; in this range the spectrum of relic gravitons basically follows from 
the minimal $\Lambda$CDM paradigm;
\item{} the range $\nu_{\mathrm{bbn}} < \nu < \nu_{\mathrm{max}}$ will be generically referred to as 
the high-frequency domain; in this range the spectrum of relic gravitons is more uncertain.
\end{itemize}
The high-frequency branch of the relic 
 graviton spectrum, overlapping with the frequency 
 window of wide-band detectors (see shaded box in the left plot of Fig. \ref{FigureSP}),  is rather sensitive to the thermodynamic history of the plasma after inflation as well as, for instance, to
the specific features of the underlying gravity theory at small scales. This is why we said that 
 the theoretical error in the  calculation of the relevant observables  increases, so to speak,  with the frequency.

In Fig. \ref{FigureSP} (plot at the right) the electromagnetic spectrum is reported in its 
salient features. It seems instructive  to draw a simple minded parallel  between the electromagnetic spectrum and the spectrum of relic gravitons. Consider first the spectrum of relic gravitons (see Fig. \ref{FigureSP}, plot at the left): between $10^{-18}$ Hz (corresponding to $\nu_{\mathrm{p}}$) and $10$ kHz (corresponding to $\nu_{\mathrm{LV}}$) there are, roughly, 22 decades in frequency. 
A similar frequency gap (see Fig. \ref{FigureSP}, plot at the right), if applied to the well known electromagnetic spectrum,  would drive us from low-frequency radio waves up  to 
x-rays or $\gamma$-rays. As the physics explored by radio waves is very 
different from the physics probed by $\gamma$ rays,
it can be argued that the informations carried by low and high 
frequency gravitons originate from two very different physical regimes of the theory.
 Low frequency gravitons are sensitive to the large scale features of the given cosmological model and of the underlying theory of gravity. High frequency gravitons are sensitive to the small scale features of a given cosmological model and of the underlying theory of gravity. 

The interplay between long wavelength gravitons and CMB experiments will be specifically 
discussed in the subsection \ref{sec12}. The main message will be that, according to current CMB experiments, 
long wavelength gravitons have not been observed yet.  The latter occurrence imposes a very important 
constraint on the low frequency branch of the relic graviton spectrum of the $\Lambda$CDM scenario
whose salient predictions will be introduced in subsection \ref{sec13}. According to the minimal 
$\Lambda$CDM paradigm a very peculiar conclusion seems to pop up: the CMB constraints 
on the low-frequency tail of the graviton spectrum jeopardize the possibility 
of any detectable signal for frequencies comparable with the window 
explored by wide band interferometers (see subsection \ref{sec14}). 
The natural question arising at this point is rather simple: is it possible to have a quasi-flat 
low-frequency branch of the relic graviton spectrum and a sharply increasing spectral 
energy density at high-frequencies? This kind of signal is typical of a class of completions 
of the $\Lambda$CDM paradigm which have been recently dubbed T$\Lambda$CDM 
(for tensor -$\Lambda$CDM). The main predictions of these models will be introduced 
in subsection \ref{sec15}. We shall conclude this introductory section with a discussion 
of two relevant constraints which should be applied to relic graviton backgrounds in general, i.e. 
the millisecond pulsar and the big-bang nucleosynthesis constraint (see subsection \ref{sec16}).

\subsection{Long wavelength gravitons and  CMB experiments}
\label{sec12}
The bounds on the backgrounds of relic gravitons stemming from CMB experiments are phrased in terms of 
$r_{\mathrm{T}}$ which is the ratio between the tensor and the scalar power spectra
at the same conventional scale (often called pivot scale).
While the use of $r_{\mathrm{T}}$ is practical (see 
e. g. the 5-year WMAP data \cite{WMAP51,WMAP52,WMAP53,WMAP54,WMAP55}),
it assumes the $\Lambda$CDM scenario insofar as the  curvature perturbations are adiabatic. 
Within the $\Lambda$CDM model, the tensor and scalar power spectra can be parametrized as
\begin{eqnarray}
&&{\mathcal P}_{\mathrm{T}}(k) = {\mathcal A}_{\mathrm{T}} \biggl(\frac{k}{k_{\mathrm{p}}}\biggr)^{n_{\mathrm{T}}}, \qquad 
{\mathcal A}_{\mathrm{T}} = r_{\mathrm{T}} {\mathcal A}_{{\mathcal R}},
\label{int1}\\
&& {\mathcal P}_{{\mathcal R}}(k) = {\mathcal A}_{{\mathcal R}}   \biggl(\frac{k}{k_{\mathrm{p}}}\biggr)^{n_{\mathrm{s}}-1}, \qquad 
{\mathcal A}_{{\mathcal R}} = (2.41 \pm 0.11) \times 10^{-9}, \qquad k_{\mathrm{p}} = 0.002 \,\, \mathrm{Mpc}^{-1}.
\label{int2}
\end{eqnarray}
where $k_{\mathrm{p}}$ is the  pivot wave-number and ${\mathcal A}_{{\mathcal R}}$ 
is the amplitude of the power spectrum of curvature perturbations \footnote{The perturbations 
of the spatial curvature, conventionally denoted by ${\mathcal R}$ are customarily employed to characterize the scalar fluctuations of the geometry since  ${\mathcal R}$ is approximately constant (in time) across the radiation-matter transition.} computed at $k_{\mathrm{p}}$; $n_{\mathrm{s}}$ and $n_{\mathrm{T}}$ are, respectively, the scalar and the tensor spectral indices \footnote{As it is clear 
from Eqs. (\ref{int1}) and (\ref{int2}) there is a difference in the way the scalar and the tensor 
spectral indices are assigned: while the scale-invariant limit corresponds to $n_{\mathrm{s}} \to 1$ 
for the curvature perturbations, the  scale invariant limit for the long wavelength gravitons 
corresponds to $n_{\mathrm{T}}\to 0$.}. 
The value of $k_{\mathrm{p}}$ is conventional and it corresponds to an effective harmonic 
$\ell_{\mathrm{eff}} \simeq 30$. The figure for ${\mathcal A}_{\mathcal R}$ quoted in Eq. (\ref{int2}) 
corresponds to the value inferred from the WMAP 5-year data \cite{WMAP51,WMAP52,WMAP53,WMAP54,WMAP55} in combination with 
the minimal $\Lambda$CDM model\footnote{See also \cite{WMAPfirst1,WMAPfirst2,WMAPfirst3,WMAPthird1,WMAPthird2} for earlier WMAP data releases.}.
In the $\Lambda$CDM model the origin of 
${\mathcal A}_{{\mathcal R}}$ stems from adiabatic 
curvature perturbations which are present after neutrino decoupling but before matter radiation equality (taking place at a redshift $z_{\mathrm{eq}} = 3176_{-150}^{+151}$ according to the WMAP 5-yr data  \cite{WMAP51,WMAP52,WMAP53}). The dominant component of curvature perturbations is adiabatic meaning that, over large scales, the fluctuations in the specific entropy are vanishing, at least in the minimal version of the model. 
The adiabatic nature of the fluctuations induces a simple relation between the first acoustic peak of the TT power spectra and the first anticorrelation peak of the TE power spectra  \cite{WMAPfirst2}: this is, to date, the best 
evidence that  curvature perturbations are, predominantly, adiabatic.
It is useful to translate the comoving wave number $k_{\mathrm{p}}$ into a comoving frequency  
\begin{equation}
\nu_{\mathrm{p}} = \frac{k_{\mathrm{p}}}{2\pi}  = 3.092\times 10^{-18} \,\, \mathrm{Hz} \equiv 3.092 \,\,\mathrm{aHz},
\label{int3}
\end{equation}
so, as anticipated, $\nu_{\mathrm{p}}$ is of the order of the aHz.
 The amplitude at the pivot scale\footnote{In the first release of the WMAP data 
the scalar and tensor pivot scales were chosen to be different and, in particular, $k_{\mathrm{p}}=0.05\,\, \mathrm{Mpc}^{-1}$
for the scalar modes. In the subsequent releases of data the two pivot scales have been taken to coincide.} 
is controlled exactly by $r_{\mathrm{T}}$.
The combined analysis of the CMB data, of the 
large-scale (LSS) structure data \cite{LSS1,LSS2} and of the supernova (SN) data \cite{SN1,SN2} 
can lead to quantitative upper limits on $r_{\mathrm{T}}$ which are 
illustrated in Tabs. \ref{TABLE1} and \ref{TABLE2}  as they are emerge from the combined analyses of different data sets.
\begin{table}[!ht]
\begin{center}
\begin{tabular}{||l|c|c|c|c|c|c||}
\hline
Data & $r_{\mathrm{T}}$ &$n_{\mathrm{s}}$ & $\alpha_{\mathrm{S}}$&$\Omega_{\Lambda}$&$\Omega_{\mathrm{M}0}$&$k_{\mathrm{eq}} \mathrm{Mpc}$\\
\hline
 WMAP5 alone& $<0.58$ &$ 1.087_{-0.073}^{0.072}$&$-0.050\pm 0.034$ & $0.722_{-0.057}^{+0.054}$&$0.278_{-0.054}^{0.057}$ & $0.00993$ \\
WMAP5 + Acbar& $<0.54 $& $1.083_{-0.062}^{0.063}$&$-0.048\pm 0.027$&$0.719_{-0.047}^{+0.048}$&$0.281_{-0.048}^{0.047}$ &$0.01004$\\
WMAP5+ LSS + SN &$<0.54$ &$ 1.093_{-0.069}^{0.068}$&$-0.055_{-0.028}^{0.027}$&$0.714_{-0.016}^{+0.017}$&$0.286_{-0.017}^{0.016}$&$0.01008$\\
WMAP5+ CMB data & $<0.64$ & $1.127_{-0.071}^{0.075}$&$ -0.072_{-0.030}^{+0.031}$&$0.704_{-0.054}^{0.055}$&$0.296_{-0.055}^{0.054}$&$0.01013$\\
\hline
\end{tabular}
\caption{The change in determination of the parameters of the tensor background for three different choices 
of cosmological data sets.}
\label{TABLE1}
\end{center}
\end{table}
The inferred  values of the scalar spectral index (i.e. $n_{\mathrm{s}}$), of the dark energy and dark matter fractions (i.e., respectively, $\Omega_{\Lambda}$ and $\Omega_{\mathrm{M}0}$), and of the typical wavenumber of equality $k_{\mathrm{eq}}$ are reported in the remaining  columns.  While different analyses can be performed, it is clear, by looking at Tabs. \ref{TABLE1} and \ref{TABLE2}, that the typical upper bounds on $r_{\mathrm{T}}(k_{\mathrm{p}})$ range between, say, $0.2$ and $0.4$.
More stringent limits can be obtained by adding supplementary assumptions. 
\begin{table}[!ht]
\begin{center}
\begin{tabular}{||l|c|c|c|c|c|c||}
\hline
Data & $r_{\mathrm{T}}$ &$n_{\mathrm{s}}$ & $\alpha_{\mathrm{S}}$&$\Omega_{\Lambda}$&$\Omega_{\mathrm{M}0}$&$k_{\mathrm{eq}} \mathrm{Mpc}$\\
\hline
 WMAP5 alone& $<0.43$ &$0.986 \pm 0.22$&$ 0 $ & $0.770_{-0.032}^{+0.033}$&$0.230_{-0.033}^{0.032}$ & $0.00936$ \\
WMAP5 + Acbar& $< 0.40  $& $0.985_{-0.020}^{0.019}$&$0 $&$0.767 \pm 0.032$&$0.233\pm 0.032$ &$0.00944$\\
WMAP5+ LSS + SN &$<0.20$ &$0.968 \pm 0.015$&$0 $&$0.725 \pm 0.015$&$0.275 \pm 0.015$&$0.00999$\\
WMAP5+ CMB data & $<0.36$ & $0.979\pm 0.020$&$0$&$0.775\pm 0.032$&$0.225\pm 0.032$&$0.00922$\\
\hline
\end{tabular}
\caption{Same as in Tab. \ref{TABLE1} but assuming no running in the (scalar) spectral index (i.e. $\alpha_{\mathrm{S}} =0$).}
\label{TABLE2}
\end{center}
\end{table}
In Tab. \ref{TABLE1}  the quantity 
$\alpha_{\mathrm{S}}$ determines the frequency dependence of the scalar spectral index. In the simplest 
case $\alpha_{\mathrm{S}} =0$ and the spectral index 
is frequency-independent (i.e. $n_{\mathrm{s}}$ does not run with the frequency). It can also happen, however, that $\alpha_{\mathrm{S}} \neq 0$ which implies 
an effective frequency dependence of the spectral index.
If the inflationary phase is driven by a single scalar degree of freedom 
(as contemplated in the minimal version of the $\Lambda$CDM scenario) 
and if the radiation dominance kicks in almost suddenly after inflation, 
the whole tensor contribution can be solely parametrized  in terms of $r_{\mathrm{T}}$. 
The rationale for the latter statement is that  $r_{\mathrm{T}}$ not only determines the tensor amplitude but also, thanks 
to the algebra obeyed by the slow-roll parameters, the  slope of the tensor power spectrum, customarily denoted by 
$n_{\mathrm{T}}$. To lowest order in the slow-roll 
expansion, therefore, the tensor spectral index is slightly red and it is related to $r_{\mathrm{T}}$ (and to the slow-roll parameter) as 
\begin{equation}
n_{\mathrm{T}} \simeq - \frac{r_{\mathrm{T}}}{8} \simeq  - 2 \epsilon, \qquad 
\epsilon = -\frac{ \dot{H}}{H^2},
\end{equation}
where $\epsilon$ measures the rate of decrease of the Hubble 
parameter during the inflationary epoch \footnote{The overdot will denote
throughout the paper a derivation with respect to the cosmic 
time coordinate $t$ while the prime will denote a derivation with respect 
to the conformal time coordinate $\tau$.}. Within the established set of conventions 
the scalar spectral index $n_{\mathrm{s}}$ 
is given by $n_{\mathrm{s}} = (1 - 6 \epsilon + 2 \overline{\eta})$ and it depends not only upon $\epsilon$ 
but also upon the second slow-roll parameter 
$\overline{\eta} = \overline{M}_{\mathrm{P}}^2 V_{,\varphi\varphi}/V$ (where 
$V$ is the inflaton potential, $V_{,\varphi\varphi}$ denotes the second derivative  of the potential with respect to the inflaton field and $\overline{M}_{\mathrm{P}} = 1/\sqrt{8\pi G}$).
It is sometimes assumed that also $n_{\mathrm{T}}$ is not constant but it is rather a function of the 
wavenumber, i.e. 
\begin{equation}
n_{\mathrm{T}}(k) =  - 2 \epsilon + \frac{\alpha_{\mathrm{T}}}{2} \ln{(k/k_{\mathrm{p}})}, \qquad \alpha_{\mathrm{T}} = \frac{r_{\mathrm{T}}}{8}\biggl[(n_{\mathrm{s}} -1) + \frac{r_{\mathrm{T}}}{8}\biggr],
\label{int4}
\end{equation}
where $\alpha_{\mathrm{T}}$ now measures the running of the tensor spectral index.

As already mentioned, among the  CMB experiments  a central 
role is played by  WMAP   
\cite{WMAP51,WMAP52,WMAP53,WMAP54,WMAP55}  (see also \cite{WMAPfirst1,WMAPfirst2,WMAPfirst3} for first year data release and \cite{WMAPthird1,WMAPthird2} for the third year data release. In connection with  
\cite{WMAP51,WMAP52,WMAP53,WMAP54,WMAP55},  the WMAP 5-year data have been also combined with observations of the  Acbar   \footnote{The Arcminute Cosmology Bolometer Array 
Receiver (ACBAR) operates in three frequencies, i.e. $150$, $219$ and $274$ GHz.} experiment \cite{AC1,AC2,AC3,AC4} .
The TT, TE  and, partially EE angular power spectra\footnote{Following the custom the TT correlations will  simply denote the angular power spectra of the temperature autocorrelations. The TE and the EE power spectra  denote, respectively, the cross power spectrum between temperature and polarization and the polarization autocorrelations.} have been measured 
by the WMAP experiment. Other (i.e. non space-borne) experiments are now measuring polarization observables, in particular there are
\begin{itemize}
\item{} the Dasi (degree angular scale interferometer) experiment \cite{dasi1,dasi2,dasi3} operating at south pole ;
\item{}   the Capmap (cosmic anisotropy polarization mapper) experiment \cite{capmap1,capmap2} ;
\item{} the  Cbi (cosmic background imager) experiment \cite{cbi1,cbi2};
\item{} the  Quad experiment \cite{quad1,quad2,quad3};
\end{itemize}
 as well as various other experiments at different stages of development  \footnote{Other planned 
experiments have, as specific target, the polarization of the CMB. In particular it is worth 
quoting here the recent projects Clover \cite{CLOVER}, Brain \cite{BRAIN}, Quiet \cite{QUIET} 
and Spider \cite{SPIDER} just to mention a few.}.  
In the near future the Planck explorer satellite \cite{planck} might be able to set  more direct limits on $r_{\mathrm{T}}$ by measuring (hopefully) the BB angular power spectra.

\subsection{The relic graviton spectrum in the $\Lambda$CDM model}
\label{sec13}
Having defined the frequency range of the spectrum of relic gravitons, it is 
now appropriate to illustrate the possible signal which is expected within the 
$\Lambda$CDM scenario. 
\begin{figure}[!ht]
\centering
\includegraphics[height=6.7cm]{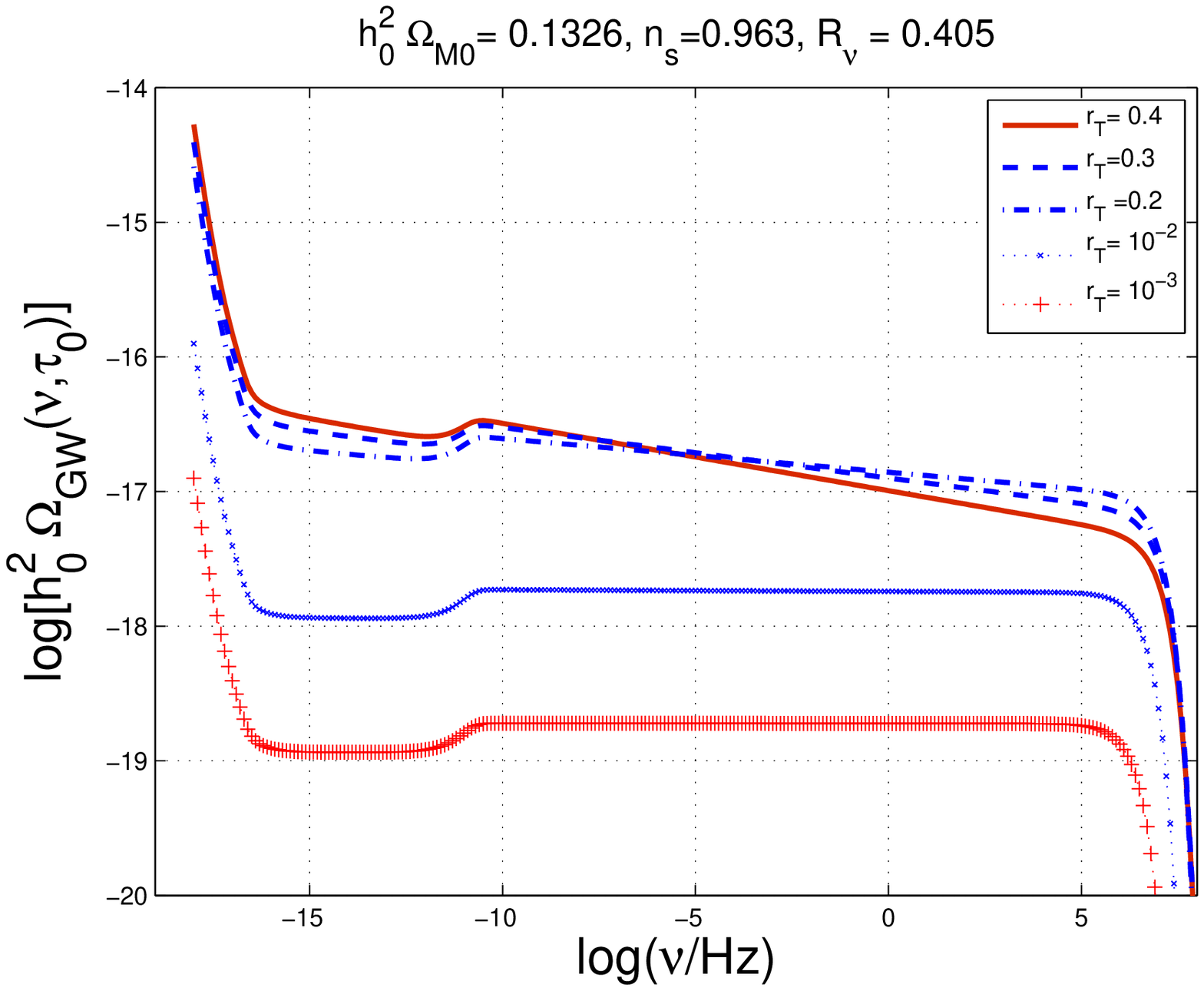}
\includegraphics[height=6.7cm]{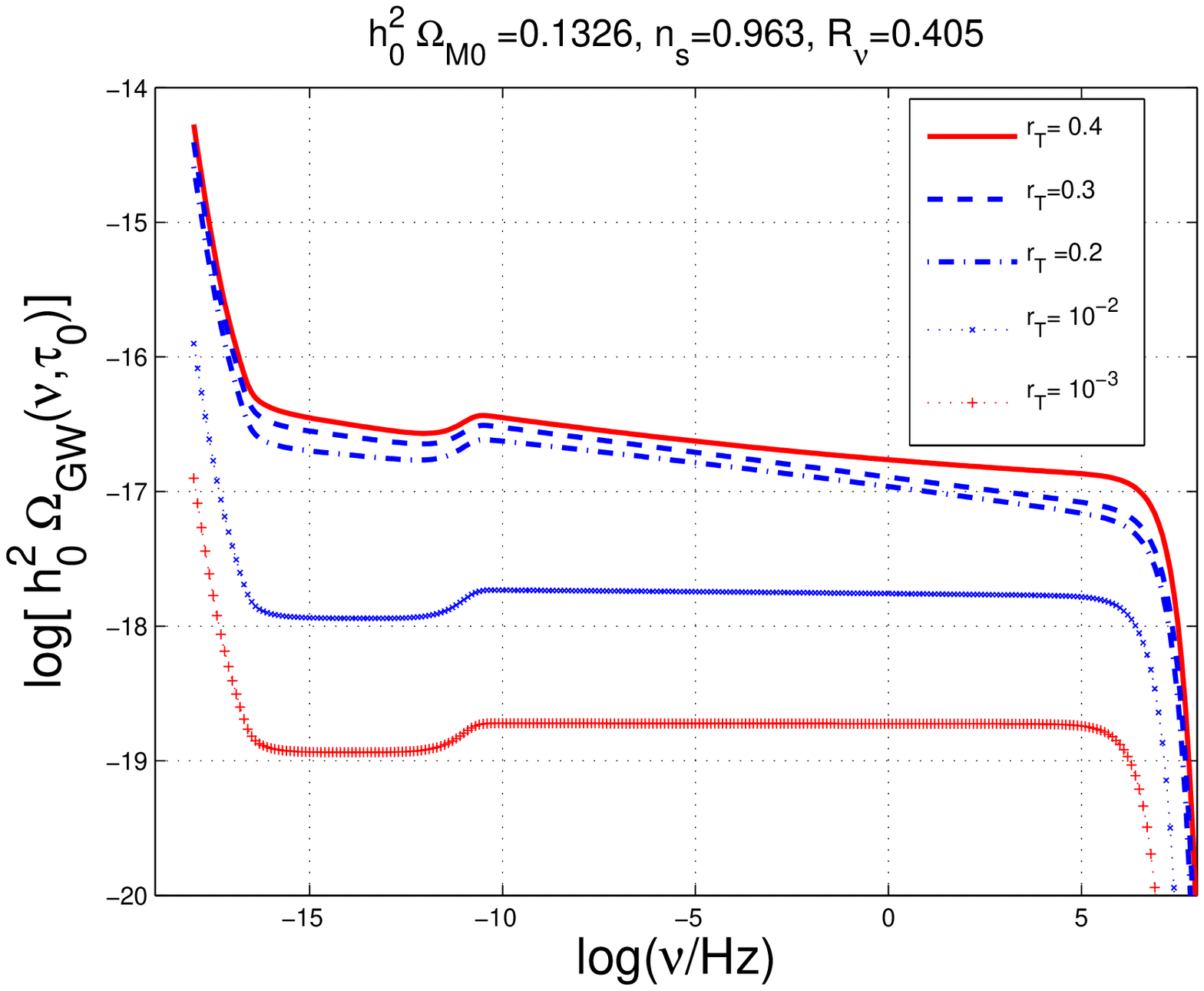}
\caption[a]{The spectral energy density of the relic gravitons is illustrated  the context 
of the $\Lambda$CDM scenario and for different values of $r_{\mathrm{T}}$ (see Eqs. (\ref{int1})--(\ref{int2})). 
In the plot at the right the tensor spectral index $n_{\mathrm{T}}$ does depend upon the frequency ($\alpha_{\mathrm{T}}$ follows from Eq. (\ref{int4}) with the $\Lambda$CDM choice of parameters). The plot at the left corresponds to $\alpha_{\mathrm{T}} =0$.}
\label{Figure1}      
\end{figure}
In Fig. \ref{Figure1} the spectrum of the relic gravitons is reported in the case of the minimal
$\Lambda$CDM scenario for different values of $r_{\mathrm{T}}$. On the horizontal axis 
the common logarithm of the comoving frequency is reported. The 
spectral energy density per logarithmic interval of frequency and in critical units is illustrated on the vertical axis. More quantitatively  $\Omega_{{\rm GW}}(\nu,\tau_0)$ is defined as \footnote{In the present review the $\ln{}$ will denote the natural logarithm while the $\log{}$ will always denote 
the common logarithm.}
\begin{equation}
\Omega_{{\rm GW}}(\nu,\tau_0)\,=\,\frac{1}{\rho_{\mathrm{crit}}}\,
\frac{{\rm d} \rho_{{\rm GW}}}{{\rm d} \ln{\nu}},
\label{int5}
\end{equation}
where $\rho_{\mathrm{crit}} = 3 H_{0}^2 \overline{M}_{\mathrm{P}}^2$ is the critical energy density. 
Since $\rho_{\mathrm{crit}}$ depends upon $H_{0}^2$ (i.e. the present value of the Hubble rate), it is 
practical to plot directly $h_{0}^2\, \Omega_{\mathrm{GW}}(\nu,\tau_{0})$ at the present (conformal) time $\tau_{0}$. The proper definition of $\Omega_{\mathrm{GW}}(\nu,\tau_{0})$ in terms of the energy-momentum 
pseudo-tensor in curved space-time is postponed to section \ref{sec5}. 
The salient features of the relic graviton spectra 
arising in the context of the $\Lambda$CDM scenario 
can be appreciated by looking carefully 
at Fig. \ref{Figure1}. 

The infra-red branch of the relic graviton spectrum 
(see also Fig. \ref{FigureSP}) extends, approximately, 
from $\nu_{\mathrm{p}}$ up to a new frequency scale which can be numerically determined by integrating the evolution equations of the tensor modes and of the background geometry 
across the matter-radiation transition. A semi-analytic estimate of this frequency is given by 
\begin{equation}
\nu_{\mathrm{eq}} = \frac{k_{\mathrm{eq}}}{2 \pi} = 1.281 \times 10^{-17} \biggl(\frac{h_{0}^2 \Omega_{\mathrm{M}0}}{0.1326}\biggr) \biggl(\frac{h_{0}^2 \Omega_{\mathrm{R}0}}{4.15 \times 10^{-5}}\biggr)^{-1/2}\,\, \mathrm{Hz}.
\label{int6}
\end{equation}
At intermediate frequencies Fig. \ref{Figure1} exhibits
a further suppression which is due to the coupling of the tensor modes with the anisotropic stress 
provided by the collisionless species which are present 
prior to matter-radiation equality.  This aspect has been recently 
emphasized in Ref. \cite{wnu1} (see also \cite{wnu2,wnu3,wnu4}).
Figure \ref{Figure1} assumes that the only colisionless species are provided by massless neutrinos, as the $\Lambda$CDM model stipulates and this 
corresponds, as indicated, to $R_{\nu} = 0.405$. The quantity 
$R_{\nu}$ measures the contribution of $N_{\nu}$ families of massless neutrinos to the radiation plasma:
\begin{equation}
R_{\nu} = \frac{r_{\nu}}{r_{\nu} + 1}, \qquad r_{\nu} = 0.681 \biggl(\frac{N_{\nu}}{3}\biggr),\qquad R_{\gamma} + R_{\nu} = 1. 
\label{ANIS4}
\end{equation}
The frequency range of the suppression due to neutrino free-streaming  extends from $\nu_{\mathrm{eq}}$ up to $\nu_{\mathrm{bbn}}$ which is given, approximately, by 
\begin{equation}
\nu_{\mathrm{bbn}}= 
2.252\times 10^{-11} \biggl(\frac{g_{\rho}}{10.75}\biggr)^{1/4} \biggl(\frac{T_{\mathrm{bbn}}}{\,\,\mathrm{MeV}}\biggr) 
\biggl(\frac{h_{0}^2 \Omega_{\mathrm{R}0}}{4.15 \times 10^{-5}}\biggr)^{1/4}\,\,\mathrm{Hz}\simeq 0.01 \,\,\mathrm{nHz}.
\label{int7}
\end{equation}
Both in Eqs. (\ref{int6}) and (\ref{int7}) $\Omega_{\mathrm{M}0}$ and $\Omega_{\mathrm{R}0}$ denote, respectively, the present critical fraction of matter and radiation with typical values drawn from the best fit to the WMAP 5-yr data alone and within 
the $\Lambda$CDM paradigm. In Eq. (\ref{int7}) $g_{\rho}$ denotes the effective number of relativistic degrees of freedom entering the total energy density of the plasma.  While $\nu_{\mathrm{eq}}$  is still close 
to the aHz, $\nu_{\mathrm{bbn}}$ is rather in the nHz range. 
In Fig. \ref{Figure1} (plot at the left) the spectral index 
$n_{\mathrm{T}}$ is frequency independent; in the plot at the 
right, always in Fig. \ref{Figure1}, the spectral index does depend on the wavenumber. These two possibilities correspond, respectively, 
to $\alpha_{\mathrm{T}} =0$ and $\alpha_{\mathrm{T}} \neq 0$ in Eq. 
(\ref{int4}). 
In the regime $\nu < \nu_{\mathrm{eq}}$ a numerical calculation 
of the transfer function is mandatory for a correct evaluation 
of the spectral slope. In the approximation of a sudden  transition between the radiation and matter-dominated regimes the spectral energy density goes, approximately, as $\nu^{-2 + n_{\mathrm{T}}}$. 
The spectra illustrated have been computed within the approach developed in \cite{mgn1,mgn2} and include also other two effects 
which can suppress the amplitude of the quasi-flat plateau. These effects are related to the  contribution of the dark energy and to the evolution of the effective number of relativistic species. From a 
quantitative point of view both effects are, however, less relevant 
than neutrino free streaming. 

Apart from the modification induced by the neutrino free-streaming the slope of the spectral energy density for $\nu > \nu_{\mathrm{eq}}$ is quasi flat and it is determined 
by the wavelengths which reentered the Hubble radius during the radiation-dominated stage of expansion. 
The suggestion that relic gravitons can be produced  in isotropic Friedmann-Robertson-Walker models is due to Ref.  \cite{gr1} (see also \cite{star1}) and was formulated before the inflationary paradigm.
 After the formulation of the inflationary scenario the focus 
has been to compute reliably the low frequency branch of the relic graviton spectrum.  In \cite{flat1,flat2,flat3} the 
low-frequency  branch of the spectrum has been  computed with slightly different analytic approaches  but 
always assuming an exact de Sitter stage of expansion prior to the radiation-dominated phase.
The analytical calculation (whose details will be described in section \ref{sec6}) shows that in the range 
$\nu_{\mathrm{p}} < \nu < \nu_{\mathrm{eq}}$, the spectral energy density 
of the relic gravitons (see Eq. (\ref{int5})) should 
approximately go as $\Omega_{\mathrm{GW}}(\nu,\tau_{0})\simeq \nu^{-2}$. 
Within the same approximation, for $\nu > \nu_{\mathrm{eq}}$ the spectral 
energy density is exactly flat (i.e. $\Omega_{\mathrm{GW}}(\nu,\tau_{0})\simeq \nu^{0}$). This result, obtainable by means of analytic calculations (see also \cite{flat5,flat6,flat7,flat8}), is a bit crude in the light of more recent developments. To assess the accurately spectral energy density
 it is necessary to take into account that the infrared branch is gradually passing from a quasi-flat slope (for $\nu> \nu_{\mathrm{eq}}$) to the slope $\nu^{-2}$ which is the one computed within the sudden approximation \cite{flat5,flat6,flat7,flat8}. It is useful to quote some of the previous reviews which covered, in a more dedicated perspective, the subject of the stochastic backgrounds of relic gravitons.
The review article by Thorne \cite{rev1} does not deal 
solely with relic graviton backgrounds while the 
reviews of Refs. \cite{rev2,rev3,rev4} are more topical. 

The flat plateau of the spectral energy density extends, approximately, between $\nu_{\mathrm{eq}}$ and a certain $\nu_{\mathrm{max}}$. Also the maximal amplified 
frequency can be computed once the model of smooth transition between inflation and radiation is known. 
The smoothness of the transition 
determines specifically the precise amount of exponential suppression for $\nu > \nu_{\mathrm{max}}$. 
A simple estimate of $\nu_{\mathrm{max}}$ is given by 
\begin{equation}
\nu_{\mathrm{max}}  = 0.346 \,\biggl(\frac{\epsilon}{0.01}\biggr)^{1/4} 
\biggl(\frac{{\mathcal A}_{\mathcal R}}{2.41 \times 10^{-9}}\biggr)^{1/4}
\biggl(\frac{h_{0}^2 \Omega_{\mathrm{R}0}}{4.15 \times 10^{-5}}\biggr)^{1/4} \,\mathrm{GHz},
\label{int8}
\end{equation}
where, as in Eqs. (\ref{int1}) and (\ref{int2}), ${\mathcal A}_{\mathcal R}$ denotes the amplitude of the power spectrum 
of curvature perturbations evaluated at the pivot wavenumber $\nu_{\mathrm{p}}$.
It is worth noticing that between $\nu_{\mathrm{bbn}}$ and $\nu_{\mathrm{max}}$ there are approximately 20 orders of magnitude in frequency.
In the $\Lambda$CDM scenario the spectrum has, in this range, always the same slope (i.e. $n_{\mathrm{T}}$ is frequency-independent in Eq. (\ref{int1})). 

Some details of the calculations leading to the spectral energy densities illustrated in Fig. \ref{Figure1} 
can be found in sections \ref{sec5} and \ref{sec6}. Without dwelling on the details it is 
however clear, as anticipated, that the constraints on the long wavelength 
gravitons make it difficult (if not impossible) to have a detectable spectral energy density at the 
scale of wide-band interferometers. The latter statement, valid in the minimal $\Lambda$CDM 
scenario,  will be sharpened in the following subsection.

\subsection{Short wavelength gravitons and wide-band interferometers} 
\label{sec14}
In the $\Lambda$CDM scenario the spectral energy density of the relic gravitons has its larger amplitude 
 in the low-frequency branch. As the frequency 
increases the spectral energy density diminishes so that 
it is plausible to expect a rather small amplitude over the frequencies 
corresponding to wide-band interferometers (see, for instance, Fig. \ref{Figure1} for $\nu \simeq \nu_{\mathrm{LV}} = 100\, \mathrm{Hz}$). 

Wide-band interferometers operate in a window ranging from few Hz up to $10$ kHz (see also Fig. \ref{FigureSP}). The available interferometers 
are Ligo \cite{LIGO}, Virgo \cite{VIRGO}, Tama \cite{TAMA} and Geo
\cite{GEO}.  In loose terms these instruments are 
Michelson interferometers with two important differences: the mirrors 
are suspended and Fabry-P\'erot cavities are used to increase 
the optical path of the photons. It would be too pretentious 
to describe in detail, in the present script, also the experimental apparatus and we therefore suggest Ref. \cite{saulson} where the basics 
of wide-band interferometers are introduced in a self-contined perspective.  

The sensitivity of a given pair of wide-band detectors to a stochastic background of relic gravitons depends upon 
the relative orientation of the instruments. The wideness of the band 
(important for the correlation among different instruments)
is not as large as $10$ kHz  but typically narrower and, in an optimistic perspective,  it could range up to $100$ Hz. The putative frequency of wide-band detectors will therefore be 
indicated as $\nu_{\mathrm{LV}}$, i.e. in loose terms, the Ligo/Virgo frequency. 
 There are daring projects of wide-band detectors in space like the Lisa \cite{LISA}, the Bbo
\cite{BBO} and the Decigo \cite{DECIGO} projects. The common feature of these three projects 
is that they are all space-borne missions and that they are all sensitive to frequencies 
smaller than the mHz  ($1\, \mathrm{mHz} = 10^{-3}\, \mathrm{Hz}$). 
While wide-band interferometers are now operating 
and might even reach their advanced sensitivities during the incoming decade, the wished sensitivities of space-borne interferometers are still on the edge of the achievable technologies. 

\begin{figure}[!ht]
\centering
\includegraphics[height=6.7cm]{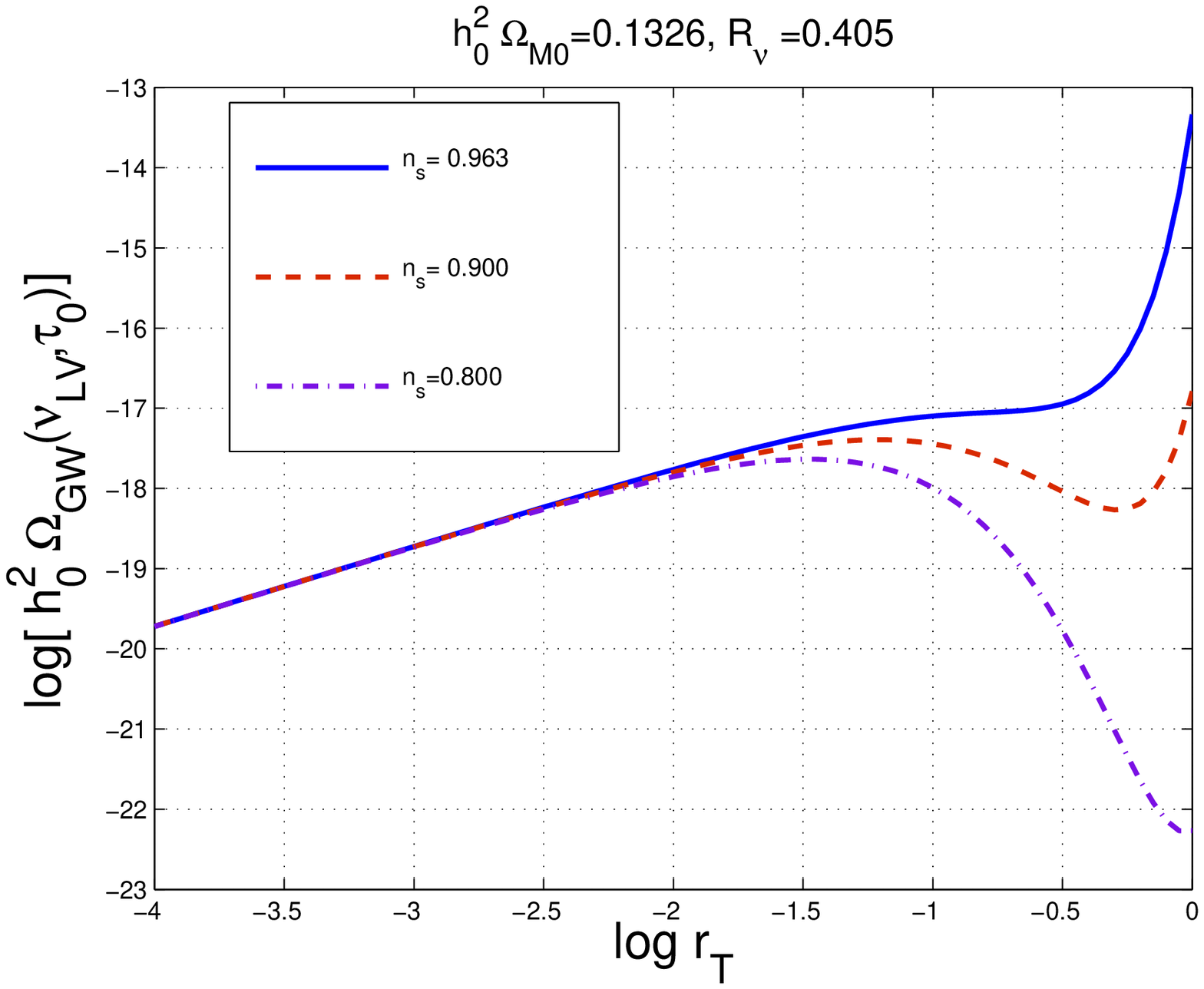}
\includegraphics[height=6.7cm]{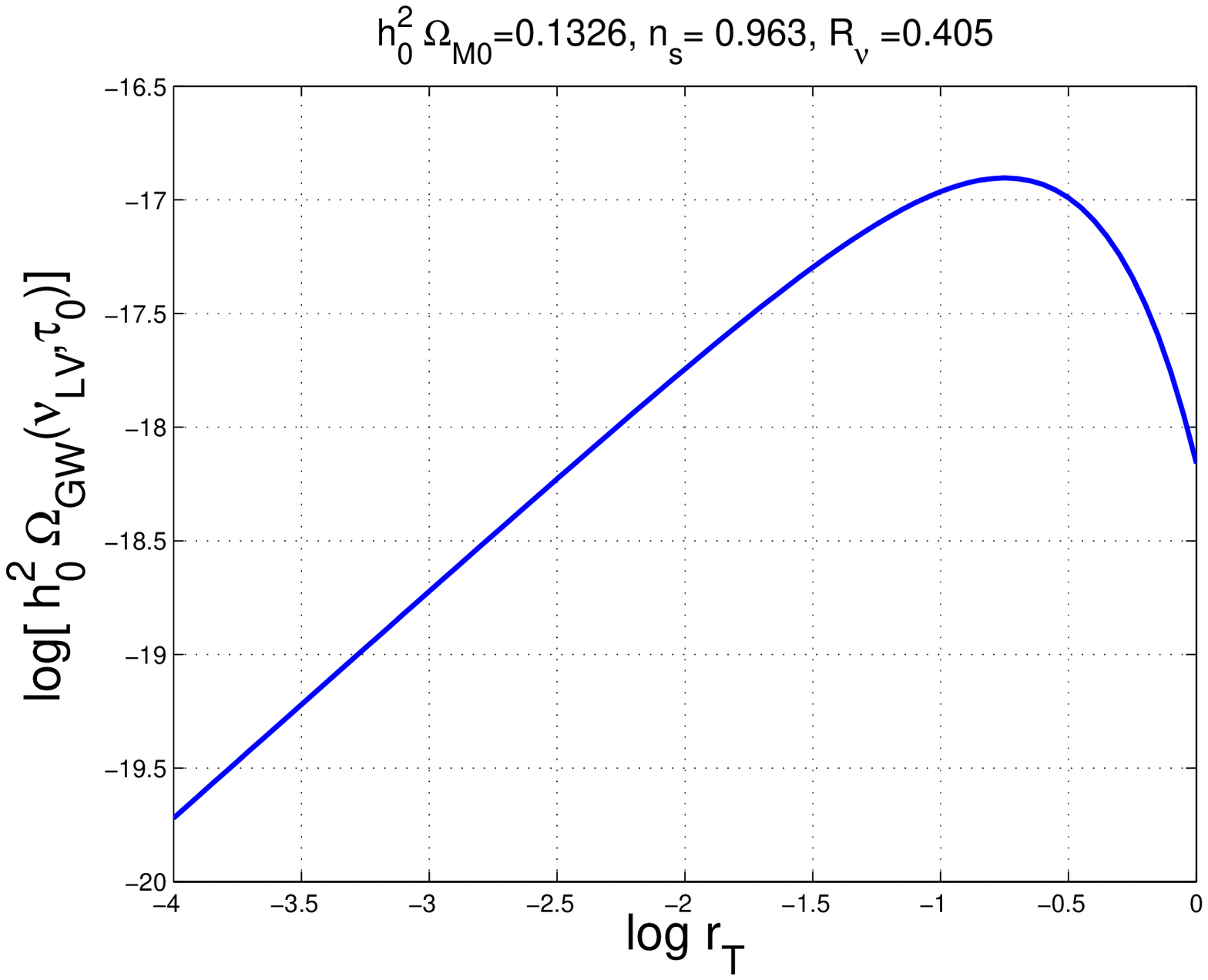}
\caption[a]{The spectral 
energy density of the relic gravitons in the context of the $\Lambda$CDM model evaluated at 
the Ligo/Virgo frequency as a function of the tensor-to-scalar ratio. To guide the intuition consider 
that (advanced) wide-band interferometers will achieve, optimistically, a sensitivity of $10^{-10}$ in $\Omega_{\mathrm{GW}}(\nu,\tau_{0})$. Taking into account the current bounds on $r_{\mathrm{T}}$ (see Tabs. \ref{TABLE1} and \ref{TABLE2}) it is clear that the $\Lambda$CDM scenario is definitively too small to provide 
a serious target for wide-band detectors. }
\label{Figure2}      
\end{figure}

Since $\nu_{\mathrm{bbn}} < \nu_{\mathrm{LV}} < \nu_{\mathrm{max}}$,  wide-band interferometers 
represent an ideal instrument to investigate the relic graviton spectrum at  high-frequencies.
The  spectral energy density of the relic gravitons produced within the $\Lambda$CDM model is quite minute and it is undetectable by interferometers even in their advanced version where the sensitivity is expected to improve by 5 or even 6 orders of magnitude in comparison with the present performances \cite{LIGOS1,LIGOS2,LIGOS3} (see also \cite{virgoligo} and \cite{auriga}). 
In Fig. \ref{Figure2} the spectral energy density 
is reported for $\nu = \nu_{\mathrm{LV}}$ and always in the case of the prediction stemming from the minimal $\Lambda$CDM scenario.
In Fig. \ref{Figure2},  the common logarithm of the spectral energy density is illustrated as a function of the common logarithm of 
$r_{\mathrm{T}}$.

In Ref. \cite{LIGOS2} (see also \cite{LIGOS1,LIGOS3})
 the current limits on the presence of an isotropic 
background of relic gravitons have been assessed. According to the Ligo collaboration 
(see Eq. (19) of Ref. \cite{LIGOS2}) the spectral energy density of a putative 
(isotropic) background of relic gravitons can be parametrized as\footnote{
The variable $\beta$  is used in Eq. (\ref{LIGOpar}) just because this is the notation 
endorsed by the Ligo collaboration and there is no reason to change it. At the same time, in the present review, $\beta$ will be used also with different meanings.
In section \ref{sec6}, $\beta$ quantifies the theoretical error on the maximal frequency of the relic graviton spectrum(see e.g. Eq. (\ref{W5}) and discussion therein). In section \ref{sec7} $\beta$ parametrizes a portion of the azimuthal structure of the Stokes parameters. Since none of these variables appear in the same context, potential 
clashes of conventions are avoided.}:
\begin{equation}
\Omega_{\mathrm{GW}}(\nu,\tau_{0}) = \Omega_{\mathrm{GW},\beta} \biggl(\frac{\nu}{100\,\mathrm{Hz}}\biggr)^{\beta + 3}.
\label{LIGOpar}
\end{equation}
The parametrization of Eq. (\ref{LIGOpar}) fits very well 
with Fig.  \ref{Figure2} where the pivot frequency $\nu_{\mathrm{LV}}=100 
\mathrm{Hz}$ coincides 
with the pivot frequency appearing in the parametrization (\ref{LIGOpar}). For 
the scale-invariant case (i.e. $\beta= -3$ in eq. (\ref{LIGOpar}))
the Ligo collaboration sets a $90 \%$ upper limit of $1.20\times 10^{-4}$ on 
the amplitude appearing in Eq. (\ref{LIGOpar}), i.e. $\Omega_{\mathrm{GW},-3}$.
Using different sets of data (see \cite{LIGOS1,LIGOS3}) the Ligo collaboration 
manages to improve the bound even by a factor $2$ getting down to 
$6.5\times 10^{-5}$. Thus Fig. \ref{Figure2} together 
with the upper limit of Eq. (\ref{LIGOpar}) shows that the current Ligo sensitivity is still too small to detect 
the relic graviton background arising within the $\Lambda$CDM paradigm.

\subsection{ Beyond the $\Lambda$CDM paradigm  and high-frequency gravitons}
\label{sec15}
In the case of an exactly scale invariant spectrum
the correlation of the two (coaligned) LIGO detectors with 
central corner stations in Livingston (Lousiana) and in Hanford 
(Washington) might reach a sensitivity to a flat spectrum 
which is \cite{mg4,mg5,mg6}
 \begin{equation}
h_0^2\,\, \Omega_{\rm GW}(\nu_{\mathrm{LV}},\tau_{0}) \simeq 6.5 \times 10^{-11} \,\, 
\biggl(\frac{1\,\,\mathrm{yr}}{T} \biggr)^{1/2}\,\,\mathrm{SNR}^2, \qquad \nu_{\mathrm{LV}} =0.1 \,\, \mathrm{kHz}
\label{SENS}
\end{equation} 
where $T$ denotes the observation time and $\mathrm{SNR}$ is the signal to noise ratio.  Equation (\ref{SENS}) is in close agreement with the 
sensitivity of the advanced Ligo apparatus \cite{LIGO} to an exactly scale-invariant spectral energy density \cite{int1,int2,int3,int4}.  Equation (\ref{SENS}) together with the 
plots of Fig. \ref{Figure2} suggest that the relic graviton 
background predicted by the $\Lambda$CDM paradigm is not directly 
observable by wide-band interferometers in their advanced version.

CMB observations probe the aHz region of the spectral energy density of Fig. \ref{Figure1}. Wide-band interferometers probe a frequency range 
between few Hz and $10$ kHz. In both ranges, the signal of the $\Lambda$CDM scenario might be too small to be directly detectable. 
\begin{figure}[!ht]
\centering
\includegraphics[height=6.7cm]{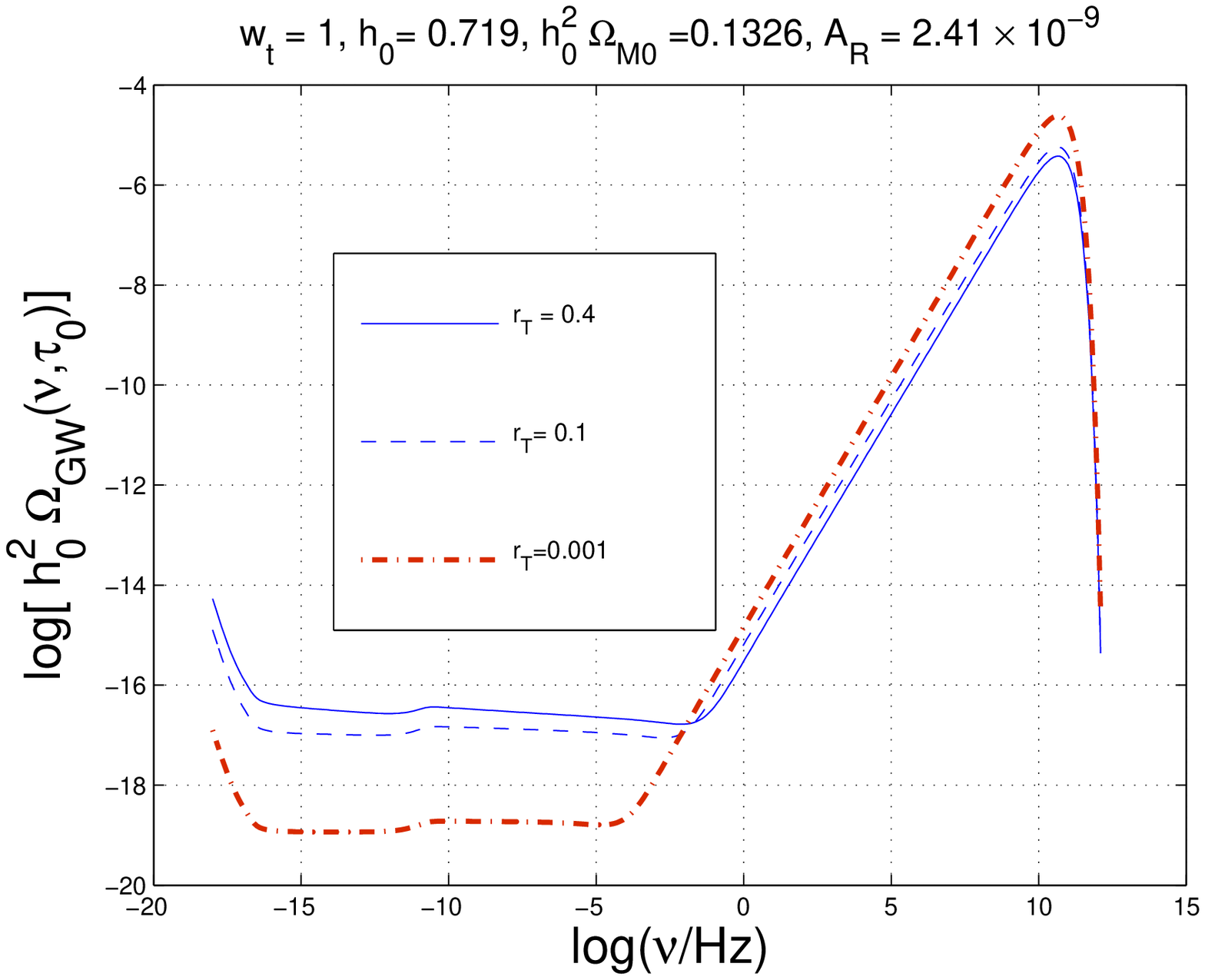}
\includegraphics[height=6.7cm]{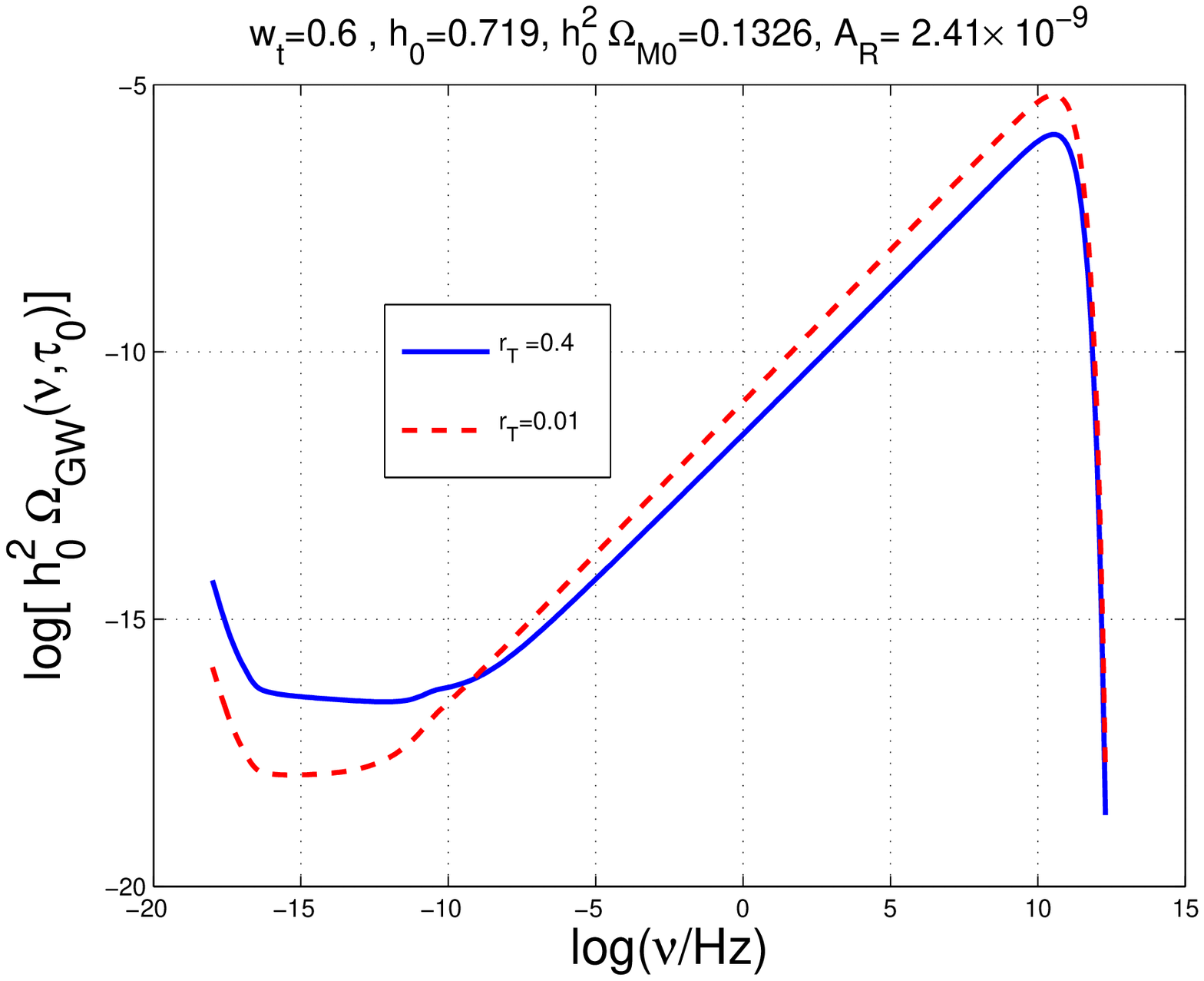}
\caption[a]{The spectral energy density of relic gravitons 
in the minimal  T$\Lambda$CDM scenario where the post-inflationary thermal history 
contemplates a phase where the sound speed of the plasma $c_{\mathrm{st}}$  is close to the speed of light $c$.
In the plot at the right $c_{\mathrm{st}}=c$ (corresponding to a constant barotropic index $w_{\mathrm{t}} =1$); 
 in the plot at the left $c_{\mathrm{st}}=0.77\,c$ (corresponding to a constant barotropic index $w_{\mathrm{t}} =0.6$). In both plots the other cosmological parameters have been chosen to coincide with the best fit to the WMAP 5-yr data alone.}
\label{Figure2a}      
\end{figure}
In Fig. \ref{Figure2a} the spectral energy density 
is computed in an extension of the $\Lambda$CDM 
paradigm which has been dubbed tensor-$\Lambda$CDM (T$\Lambda$CDM for short) \cite{mgn1,mgn2}. 
In the T$\Lambda$CDM scenario the transition 
from the inflationary phase to the radiation-dominated epoch is mediated by a rather long stiff phase. By stiff 
phase we mean a phase where the total sound speed 
of the plasma is larger than the sound speed 
of a radiation-dominated plasma (i.e. $1/\sqrt{3}$ in natural units).  In the simplest realization 
of the scenario the barotropic index $w_{\mathrm{t}}$ is 
constant during the stiff phase. For instance, in Fig. 
\ref{Figure2a} the cases $w_{\mathrm{t}} =1$ and 
$w_{\mathrm{t}} =0.6$ have been illustrated. Since 
$w_{\mathrm{t}}$ denotes the ratio between the total 
pressure and the total energy density, it is rather plausible to demand that $w_{\mathrm{t}} \leq 1$. The 
latter requirement implies that the sound speed is always 
smaller than the speed of light.  As suggested in 
\cite{mg1} (see also \cite{mg2,mg3}) the presence 
of a stiff phase can have the effect of increasing the spectral energy density at high frequencies. The increase takes place for frequencies larger than the mHz
and is typically maximal in the GHz region.
The spectral energy densities illustrated in Fig.  \ref{Figure2a} 
suggest that it is not impossible to imagine situations where the spectral 
energy density of the relic gravitons satisfies 
all the constraints demanded by CMB physics 
but, at the same time, it is sufficiently large
to be observed by wide-band interferometers.
The results reported in Fig. \ref{Figure2a} refer to the minimal T$\Lambda$CDM 
model where only one post-inflationary phase stiffer than radiation is contemplated. 
The barotropic index could have, however, a more complicated dependence. Already 
in the examples of Fig. \ref{Figure2a} the numerical integration implies that the barotropic index 
does depend, effectively, upon the scale factor (see, e. g., discussions in section \ref{sec6} 
on the transfer function for the spectral energy density).

The comparison of Fig. \ref{Figure1} and \ref{Figure2a}  suggests, in short, 
the following subjects of reflection:
\begin{itemize}
\item{} the  theoretical error in the estimate of the spectral energy density increases with the frequency;
\item{} departures from the standard post-inflationary thermal history can be directly imprinted in the primordial spectrum of the relic gravitons;
\item{} in the incoming decade the observations 
of wide-band interferometers could be analyzed in conjunction with 
more standard data sets (i.e. CMB data supplemented by large-scale structure data and by the observations of type Ia supernovae)  to constrain the spectral energy density of the relic gravitons both at small and at high frequencies.
\end{itemize}
The presence of post-inflationary phases stiffer than radiation is, after all, rather 
natural and this was the original spirit of \cite{mg1}. We do not know 
which was the rate of the post-inflationary expansion and since guesses 
cannot substitute experiments it would be productive to use the T$\Lambda$CDM 
paradigm as reference model for a unified analysis of the low-frequency data stemming 
from CMB and of the high-frequency data provided by wide-band interferometers.  
Already in \cite{mg1} (see also \cite{mg2,mg3}) a rather special candidate for a post-inflationary phase 
stiffer than radiation was the case when the sound speed equals the speed of light, i.e. 
the case when the energy density of the sources driving the geometry is dominated by 
the kinetic term of a (minimally coupled) scalar field. This particular case was also prompted by 
various classes of quintessence models.  A specific example of this dynamics was provided in \cite{PV}.

A more detailed account of the techniques leading to Fig. \ref{Figure2a} will be 
swiftly presented in section \ref{sec6} and can be found in \cite{mgn1,mgn2}.
Without going through the details it is however important to stress  that the calculations should be accurate 
enough not only in the high-frequency region but also in the low-frequency part of the spectrum. 
Indeed, as stressed above, one of the purposes of the T$\Lambda$CDM scenario 
is to convey the idea that low-frequency and high-frequency measurements of the relic graviton
background can be analyzed in a single theoretical framework.

\subsection{The millisecond pulsar bound and the nucleosynthesis constraint}
\label{sec16}
 The spectral energy density of the relic gravitons must be compatible not only with the CMB constraints (bounding, from above, the value of $r_{\mathrm{T}}$) but also with 
 the pulsar timing bound\cite{pulsar1,pulsar2} and the big-bang nucleosynthesis constraints \cite{bbn1,bbn2,bbn3}.  The pulsar timing bound demands
\begin{equation}
\Omega(\nu_{\mathrm{pulsar}},\tau_{0}) < 1.9\times10^{-8},\qquad 
\nu_{\mathrm{pulsar}} \simeq \,10\,\mathrm{nHz},
\label{PUL}
\end{equation}
where $\nu_{\mathrm{pulsar}}$ roughly corresponds to the inverse 
of the observation time during which the pulsars timing has been monitored. 
The spectral energy densities illustrated in Figs. \ref{Figure1} and \ref{Figure2a} satisfy the pulsar timing bound. 

The most constraining bound for the high-frequency branch 
of the relic graviton spectrum is represented by big-bang nucleosynthesis.  Gravitons, being relativistic, can potentially increase the expansion rate at the BBN epoch. The increase in the expansion rate will affect, in particular, 
the synthesis of $^{4}\mathrm{He}$. To avoid the overproduction 
of  $^{4}\mathrm{He}$ the expansion rate the number of relativistic species 
must be bounded from above.

The BBN bound is customarily expressed in terms of (equivalent) extra fermionic species. During the radiation-dominated era, the energy density of the plasma 
can be written as $\rho_{\mathrm{t}} = g_{\rho} (\pi^2/30) T^4$ where $T$ denotes here the common (thermodynamic) temperature of the various species. 
An (ultra)relativistic fermion species with
two internal degrees of freedom
and in thermal equilibrium contributes $2\cdot7/8 = 7/4 = 1.75$ to $g_{\rho}$.  Before
neutrino decoupling the contributing relativistic particles are photons,
electrons, positrons, and $N_\nu = 3$ species of neutrinos, giving
\begin{equation}
    g_{\rho} = \frac{11}{2} + \frac{7}{4}N_{\nu} = 10.75.
\end{equation}
The neutrinos have decoupled before electron-positron annihilation so that they
do not contribute to the entropy released in the annihilation.
While they are relativistic, the neutrinos still retain an equilibrium energy
distribution, but after the annihilation
their (kinetic) temperature is lower, $T_\nu = (4/11)^{1/3}T$.  Thus
\begin{equation}
    g_{\rho} = 2 + \frac{7}{4}N_\nu\left(\frac{T_\nu}{T}\right)^4
           = 2 + 0.454N_\nu = 3.36,
\end{equation}
after electron-positron annihilation.
By now assuming that there are some additional relativistic degrees of
freedom, which also have decoupled by the time of electron-positron
annihilation,  or just some additional component $\rho_X$ to the energy
density with a radiation-like equation of state (i.e. $p_{X} = \rho_{X}/3$), the effect on the 
expansion rate will be the same as that of having
some (perhaps a fractional number of)
additional neutrino species.  Thus its contribution can be represented by
replacing $N_\nu$ with $ N_\nu + \Delta N_{\nu}$ in the above.  Before
electron-positron annihilation we have $\rho_X = (7/8)\Delta N_{\nu} \rho_\gamma$
and after electron-positron annihilation we have
$\rho_X = (7/8) (4/11)^{4/3} \,\Delta N_{\nu} \,\rho_{\gamma} \simeq 0.227\,\Delta N_{\nu} \,\rho_\gamma$.
The critical fraction of CMB photons can be directly computed from the 
value of the CMB temperature and it is notoriously given by
$h_{0}^2 \Omega_\gamma \equiv \rho_\gamma/\rho_{\mathrm{crit}} = 2.47\times10^{-5}$.
If the extra energy density component has stayed radiation-like until today,
its ratio to the critical density, $\Omega_X$, is given by
\begin{equation}
h_{0}^2   \Omega_X \equiv h_{0}^2\frac{\rho_X}{\rho_{\mathrm{c}}} = 5.61\times10^{-6}\Delta N_{\nu} 
\biggl(\frac{h_{0}^2 \Omega_{\gamma0}}{2.47 \times 10^{-5}}\biggr).
\end{equation}
If the additional species are relic gravitons, then  \cite{bbn1,bbn2,bbn3}: 
\begin{equation}
h_{0}^2  \int_{\nu_{\mathrm{bbn}}}^{\nu_{\mathrm{max}}}
  \Omega_{{\rm GW}}(\nu,\tau_{0}) d\ln{\nu} = 5.61 \times 10^{-6} \Delta N_{\nu} 
  \biggl(\frac{h_{0}^2 \Omega_{\gamma0}}{2.47 \times 10^{-5}}\biggr),
\label{BBN1}
\end{equation}
where $\nu_{\mathrm{bbn}}$ and $\nu_{\mathrm{max}}$ are given, respectively, by Eqs. (\ref{ANIS5}) and (\ref{STFR4}).
Thus the constraint of Eq. (\ref{BBN1}) arises from the simple consideration 
that new massless particles could eventually increase the expansion rate 
at the epoch of BBN. 
The extra-relativistic species do not have to be, however, fermionic \cite{bbn2} 
and therefore the bounds on $\Delta N_{\nu}$ can be translated  into bounds 
on the energy density of the relic gravitons. 

A review of the constraints on  $\Delta N_{\nu}$ can be found 
in \cite{bbn2}. Depending on the combined data sets (i.e. various light elements abundances and different combinations of CMB observations), the standard BBN scenario implies that the bounds on $\Delta N_{\nu}$ range from $\Delta N_{\nu} \leq 0.2$ 
to $\Delta N_{\nu} \leq 1$. Similar figures, 
depending on the priors of the analysis, have been obtained in a more recent analysis \cite{bbn3}.  All the relativistic species present inside the 
Hubble radius at the BBN contribute to the potential increase in the expansion rate and this  explains why the integral in Eq. (\ref{BBN1}) must be performed from $\nu_{\mathrm{bbn}}$ to $\nu_{\mathrm{max}}$ (see also 
\cite{mg2} where this point was stressed in the framework of a specific model).  
The existence of the exponential suppression for $\nu>\nu_{\mathrm{max}}$  (see Figs. \ref{Figure2a})
guarantees the convergence of the integral also in the case when the integration 
is performed up to $\nu \to \infty$. The constraint of Eq. (\ref{BBN1})  can be relaxed in some 
non-standard nucleosynthesis scenarios \cite{bbn2}, but, in what follows, the 
validity of Eq. (\ref{BBN1}) will be enforced by adopting 
$\Delta N_{\nu} \simeq 1$  which implies, effectively 
\begin{equation}
h_{0}^2  \int_{\nu_{\mathrm{bbn}}}^{\nu_{\mathrm{max}}}
  \Omega_{{\rm GW}}(\nu,\tau_{0}) d\ln{\nu} < 5.61\times 10^{-6} \biggl(\frac{h_{0}^2 \Omega_{\gamma0}}{2.47 \times 10^{-5}}\biggr). 
\label{BBN2}
\end{equation}
The spectral energy densities illustrated in Figs. \ref{Figure1} and \ref{Figure2a} are both compatible with the big-bang nucleosynthesis 
bound. Thus the big-bang nucleosyntheis 
constraint does not forbid a potentially detectable signal in the 
high-frequency branch of the relic graviton spectrum. Potential deviations of the thermal history of the plasma must anyway occur before big-bang nucleosynthesis.

\newpage
\renewcommand{\theequation}{2.\arabic{equation}}
\setcounter{equation}{0}
\section{The polarization of relic gravitons and of relic photons}
\label{sec2}
\subsection{Basic notations}
\label{sec21}
As discussed in the introduction, in the $\Lambda$CDM paradigm the 
background line element can be written 
\begin{equation}
ds^2 = \overline{g}_{\mu\nu} dx^{\mu} dx^{\nu} = a^2(\tau)[ d\tau^2 - \overline{\gamma}_{ij}(\vec{x}) d x^{i} dx^{j}],
\label{SI1}
\end{equation}
where, in the spatially flat case, $\overline{\gamma}_{ij}$ will coincide with $\delta_{ij}$ and the 
Friedmann-Lema\^itre equations can be written as 
\begin{eqnarray}
&& 3 {\mathcal H}^2 = a^2 \ell_{\mathrm{P}}^2 \rho_{\mathrm{t}},
\label{FL1}\\
&& 2({\mathcal H}^2 - {\mathcal H}') =  a^2 \ell_{\mathrm{P}}^2 (\rho_{\mathrm{t}} + p_{\mathrm{t}}),
\label{FL2}\\
&& \rho_{\mathrm{t}}' + 3 {\mathcal H} (\rho_{\mathrm{t}} + p_{\mathrm{t}}) =0,
\label{FL3}
\end{eqnarray}
where ${\mathcal H} = a'/a$; the prime denotes a derivation 
with respect to the conformal time coordinate $\tau$. The Hubble 
rate is customarily defined in the synchronous frame where 
the time coordinate (conventionally denoted by $t$) obeys $dt = a(\tau) 
d\tau$. Denoting with a dot a derivation with respect to the cosmic time $t$,  $H = \dot{a}/a$, and, by definition, $H = a {\mathcal H}$. In Eqs. (\ref{FL1})--(\ref{FL3})  $\rho_{\mathrm{t}}$ and $p_{\mathrm{t}}$ are, respectively, the total energy density and the total pressure of the plasma, i.e. 
\begin{eqnarray}
&&\rho_{\mathrm{t}} = \rho_{\mathrm{b}} + \rho_{\mathrm{c}} + 
\rho_{\gamma} + \rho_{\nu} + \rho_{\Lambda}, \qquad \rho_{\mathrm{M}} = 
\rho_{\mathrm{b}} + \rho_{\mathrm{c}},
\label{rex}\\
&& p_{\mathrm{t}} = \frac{\rho_{\gamma}}{3} + \frac{\rho_{\nu}}{3} - \rho_{\Lambda}.
\label{pex}
\end{eqnarray}
The total matter fraction of the critical energy density, i.e. 
$\Omega_{\mathrm{M}0} = \rho_{\mathrm{M}0}/\rho_{\mathrm{crit}}$ consists of baryons and 
(i.e. $\rho_{\mathrm{b}}$) and  cold dark matter particles (i.e. $\rho_{\mathrm{c}}$). In Tabs. \ref{TABLE1} and \ref{TABLE2}  the values of $\Omega_{\mathrm{M}0}$ are given as they are inferred within 
the $\Lambda$CDM scenario. In similar terms $\Omega_{\Lambda} = \rho_{\Lambda}/\rho_{\mathrm{crit}}$ denotes the critical fraction 
of dark energy. 
 In what follows, if not otherwise stated, 
 the cosmological parameters will be fixed to the best 
fit of the WMAP-5yr data alone, i.e. 
\begin{equation}
(\Omega_{\mathrm{b}0},\, \Omega_{\mathrm{c}0}, \,\Omega_{\mathrm{\Lambda}}, \,h_{0}, \,n_{\mathrm{s}},\,\epsilon)= (0.0441,\, 0.214,\, 0.742,\, 0.719,\, 0.963,\,0.087).
\label{bestfit}
\end{equation}
where $\Omega_{\mathrm{b}0},\, \Omega_{\mathrm{c}0}, \,\Omega_{\mathrm{\Lambda}}$ denote, respectively, the (present) critical fractions of baryons, CDM particles and dark energy; $h_{0}$ fixes the present 
value of the Hubble rate; $n_{\mathrm{s}}$, as already mentioned in section \ref{sec1},  is the spectral index of curvature perturbations and $\epsilon$ is the reionization optical depth.

At the beginning of the previous section we started by stressing 
analogies and differences between relic gravitons and relic photons.
The most important one is that both gravitons and photons 
carry two polarizations. This observation is important 
 for a quantitative understanding of the present endevours aimed at measuring the E-mode and the B-mode 
polarization of the CMB. In the present section the description 
of the polarization of the gravitons will be developed by stressing, when 
possible, the analogy with polarization observables of the electromagnetic field. 

\subsection{Linear and circular tensor polarizations}
\label{sec22}
Recalling that $i,j,k,...$ are  indices defined on the 
three-dimensional  Euclidean sub-manifold, the tensor fluctuations of the geometry are parametrized in terms of the rank-two tensor $h_{ij}$ 
\begin{equation}
\delta_{\mathrm{t}}^{(1)} g_{i j} = -  a^2\,\, h_{ij},\qquad h_{i}^{i} = \nabla_{i} h^{i}_{j} =0,
\label{SI2}
\end{equation}
where $\nabla_{i}$ is the covariant derivative with respect to $\overline{\gamma}_{ij}$; if 
$\overline{\gamma}_{ij} = \delta_{ij}$, $\nabla_{i}= \partial_{i}$.  In Eq. (\ref{SI2}) 
the subscript refers to the tensor nature of the fluctuation while the superscript denotes the perturbative 
order.  The tensor fluctuation $h_{ij}(\vec{x},\tau)$ can be 
decomposed in terms of the two linear polarizations, i.e. 
 \begin{equation}
 h_{ij}(\vec{x},\tau) = \sum_{\lambda} h_{(\lambda)}(\vec{x},\tau) \epsilon_{ij}^{(\lambda)}(\hat{k}),\qquad \epsilon_{ij}^{(\lambda)} \epsilon_{ij}^{(\lambda')} = 2 \delta^{\lambda\lambda'},
 \label{SI4}
 \end{equation}
 where $\lambda = \oplus,\,\, \otimes$ denote the two polarizations and where  
 \begin{eqnarray}
&& \epsilon_{ij}^{\oplus}(\hat{k}) = ( \hat{a}_{i} \hat{a}_{j} - \hat{b}_{i} \hat{b}_{j}), 
\label{SI5}\\
&& \epsilon_{ij}^{\otimes}(\hat{k}) = (\hat{a}_{i} \hat{b}_{j} + \hat{a}_{j} \hat{b}_{i}).
\label{SI6}
\end{eqnarray}
In Eqs. (\ref{SI5}) and (\ref{SI6}), $\hat{a}$, $\hat{b}$ and
$\hat{k} = \vec{k}/|\vec{k}|$ represent a 
triplet of mutually orthogonal unit vectors, i.e.
\begin{equation}
\hat{a}\cdot\hat{b} =0,\qquad \hat{a}\cdot \hat{k} =0,\qquad \hat{b}\cdot\hat{k} =0.
\label{SI7}
\end{equation}
If the direction of propagation coincides with the $\hat{z}$, the unit vectors $\hat{k}$, $\hat{a}$ and $\hat{b}$ can be chosen as:
\begin{equation}
\hat{k} = (0,\, 0,\, 1), \qquad \hat{a} = (1,\, 0,\, 0),\qquad 
\hat{b} = (0,\, 1,\, 0).
\label{cartesian}
\end{equation}
Using Eq. (\ref{cartesian}), Eqs. (\ref{SI5}) and (\ref{SI6}) become 
\begin{eqnarray}
\epsilon_{xx}^{\oplus} = 1, \qquad \epsilon_{yy}^{\oplus} = -1,
\label{SI8}\\
\epsilon_{xy}^{\otimes} = 1, \qquad \epsilon_{yx}^{\otimes} = 1.
\label{SI9}
\end{eqnarray}
If $\hat{k}$ coincides with the radial direction, the unit 
vectors $\hat{k}$, $\hat{a}$ and $\hat{b}$ can be chosen, in spherical coordinates, as:
\begin{eqnarray}
&& \hat{k} = (\sin{\vartheta} \cos{\varphi}, \sin{\vartheta}\sin{\varphi}, \cos{\vartheta}),
\label{SI10}\\
&& \hat{a} = (\cos{\vartheta}\cos{\varphi}, \cos{\vartheta}\sin{\varphi}, - \sin{\vartheta}),
\label{SI11}\\
&& \hat{b} = (\sin{\varphi}, -\cos{\varphi}, 0).
\label{SI12}
\end{eqnarray}
Since $\epsilon_{ij}^{(\lambda)}(\hat{k}) \epsilon_{ij}^{(\lambda')}(\hat{k}) = 2 \delta^{\lambda\lambda'}$, it is straightforward to prove that 
\begin{equation}
h_{ij}(\vec{x},\tau) h^{ij}(\vec{x},\tau) = 2 [ h_{\otimes}(\vec{x},\tau)^2 + h_{\oplus}(\vec{x},\tau)^2].
\label{SI13}
\end{equation}

As in the case of electromagnetic waves, it 
is often desirable to pass from the linear to the circular polarizations: 
\begin{eqnarray}
&& \epsilon_{ij}^{(\mathrm{L})}(\hat{k}) = \frac{1}{\sqrt{2}} [ \epsilon_{ij}^{\oplus}(\hat{k}) + i \epsilon_{ij}^{\otimes}(\hat{k})],
\label{SI14}\\
&&  \epsilon_{ij}^{(\mathrm{R})}(\hat{k})= \frac{1}{\sqrt{2}} [ \epsilon_{ij}^{\oplus}(\hat{k}) - i \epsilon_{ij}^{\otimes}(\hat{k})].
\label{SI15}
\end{eqnarray}
Equations (\ref{SI14}) and (\ref{SI15}) also imply that 
$\epsilon_{ij}^{(\mathrm{L})*}= \epsilon_{ij}^{(\mathrm{R})}$ and 
$\epsilon_{ij}^{(\mathrm{R})*}= \epsilon_{ij}^{(\mathrm{L})}$. A rotation of $\hat{a}$ and $\hat{b}$ in the plane 
orthogonal to $\hat{k} = \vec{k}/|\vec{k}|$
\begin{eqnarray}
&& \hat{a}' = \cos{\alpha}\, \hat{a} + \sin{\alpha}\, \hat{b},
\label{SI16}\\
&& \hat{b}' = - \sin{\alpha}\, \hat{a} + \cos{\alpha}\, \hat{b},
\label{SI17}
\end{eqnarray}
implies, using Eqs. (\ref{SI5}) and (\ref{SI6}), 
\begin{eqnarray}
&& \tilde{\epsilon}_{ij}^{\oplus} = \epsilon_{ij}^{\oplus} \cos{2\alpha} + \epsilon_{ij}^{\otimes} \sin{2\alpha}, 
\label{SI18}\\
&& \tilde{\epsilon}_{ij}^{\otimes} = - \epsilon_{ij}^{\oplus} \sin{2\alpha} + \epsilon_{ij}^{\otimes} \cos{2\alpha},
\label{SI19}
\end{eqnarray}
where the tilde denotes the two transformed (linear) polarizations. Under the transformation
given in Eqs. (\ref{SI16}) and (\ref{SI17})  the two circular polarizations 
defined in Eqs. (\ref{SI14}) and (\ref{SI15})  transform as 
\begin{equation}
\tilde{\epsilon}_{ij}^{(\mathrm{L})}(\hat{k}) = e^{-2i \alpha} \epsilon_{ij}^{(\mathrm{L})}(\hat{k}), \qquad 
\tilde{\epsilon}_{ij}^{(\mathrm{R})}(\hat{k}) = e^{2i \alpha} \epsilon_{ij}^{(\mathrm{R})}(\hat{k}).
\label{SI20}
\end{equation}
The transformation properties of the circular polarization under a 
rotation in the plane orthogonal to the direction of propagation 
are closely analog to the transformation properties, under the same 
rotation, of the polarization of the electromagnetic field. This 
analogy will now be exploited to introduce the E-mode and B-mode 
polarization.

Before proceeding with the discussion it is appropriate to recall a very basic 
aspect of rotations which can have, however, some confusing impact 
of the polarization analysis especially in the case of the tensor modes. Consider, for simplicity, a coordinate 
system characterized by two basis vectors, i.e. $\hat{e}_{x} = \hat{r} \cos{\vartheta}$ and $\hat{e}_{y} = \hat{r} \sin{\vartheta}$. If we now 
perform a clockwise (i.e. right-handed) rotation of the axes $\hat{e}_{x}$ 
and $\hat{e}_{y}$, the rotated basis $(\hat{e}'_{x},\,\hat{e}'_{y})$ will be given  as in Eqs. (\ref{SI16}) 
and (\ref{SI17}) by replacing $(\hat{a}',\,\hat{b}') \to (\hat{e}'_{x},\,\hat{e}'_{y})$ and $(\hat{a},\,\hat{b}) \to (\hat{e}_{x},\,\hat{e}_{y})$. 
Some authors, for different reasons, instead of rotating the coordinate system prefer to rotate the polarization vector. If angles are 
in the right-handed sense for the rotation of the axes, they are in the left-handed sense for the rotation of the vectors.

\subsection{Polarization of the CMB radiation field}
The radiation field can be described by the polarization tensor, i.e.
\begin{equation}
\rho_{ij} = E_{i} E_{j}^{*},
\label{Tr1}
\end{equation}
where $E_{i}$ and $E_{j}$ are the electric components of the radiation field. Assuming, for 
sake of simplicity, that the radiation field propagate along the $\hat{z}$ axis, then 
the various entries  of $\rho_{ij}$ can be written in a matrix form 
\begin{equation}
\rho_{ij} =  \left(\matrix{|E_{x}|^2 
& E_{x} \,E_{y}^{*} &\cr
E_{y} \,E_{x}^{*} & |E_{y}|^2 &\cr}\right), 
\label{Tr2}
\end{equation}
where $x$ and $y$ denote the components of the electric field orthogonal 
to the direction of propagation coinciding, in this set-up, with the third Cartesian axis. A full description of the radiation field can be achieved by 
studying the four Stokes parameters \cite{jackson} conventionally named $I$, $Q$, $U$ and $V$:
\begin{eqnarray}
&& I = |E_{x}|^2 + |E_{y}|^2, \qquad V = 2 \,\mathrm{Im}[ E_{x}^{*}\, E_{y}],
\label{Tr3}\\
&& U = 2 \,\mathrm{Re}[ E_{x}^{*}\, E_{y}] , \qquad 
 Q =  |E_{x}|^2 - |E_{y}|^2. 
\label{Tr4}
\end{eqnarray}
It is immediate, from the definitions (\ref{Tr3}) and (\ref{Tr4}), to write the intensities 
of the radiation field along the different Cartesian axis as a function of the Stokes parameters, i.e. 
\begin{eqnarray}
&& |E_{x}|^2 = \frac{I  + Q}{2}, \qquad |E_{x}|^2 = \frac{I  - Q}{2},
\label{Tr5}\\
&& E_{x} \,E_{y}^{*} = \frac{ U + i V}{2}, \qquad  E_{x}^{*} \,E_{y} = \frac{ U - i V}{2}.
\label{Tr6}
\end{eqnarray}
Equations (\ref{Tr5}) and (\ref{Tr6}) can be inserted back into Eq. (\ref{Tr2}) with the result that 
\begin{equation}
\rho_{ij} = \frac{1}{2}\left(\matrix{ I + Q
& U - i V\cr
U + i V & I - Q \cr}\right)  = \overline{P}_{ij} + P_{ij},
\label{Tr7}
\end{equation}
where 
\begin{equation}
\overline{P}_{ij} = \frac{1}{2}\left(\matrix{ I
& -i V\cr
  i V & I \cr}\right), \qquad P_{ij} =  \frac{1}{2}\left(\matrix{  Q
& U \cr U &  - Q \cr}\right).
\label{Tr8}
\end{equation}
From Eq. (\ref{Tr7}) it also follows that 
\begin{equation}
\rho_{ij} = \frac{1}{2} \left( I\,{\bf 1} +U\, \sigma_{1} + V\, \sigma_2 +Q\, \sigma_3 \right)
\label{Tr8a}
\end{equation}
where ${\bf 1}$ is the $2\times 2$ unit matrix and 
\begin{equation}
\sigma_{1}= \left(\matrix{ 0
& 1\cr
1&0 \cr}\right), \qquad \sigma_{2}= \left(\matrix{ 0
& -i\cr
i& 0\cr}\right), \qquad \sigma_{3} = \left(\matrix{ 1& 0\cr 0& -1\cr}\right),
\label{Tr8b}
\end{equation}
are the Pauli matrices.
Consider now a rotation of an angle $\alpha$ on the plane orthogonal 
to the direction  of propagation of the monochromatic wave. It is 
easy to show that 
$I$ and $V$ are left invariant while $Q$ and $U$ do transform by a rotation of $2\alpha$. By indicating with a tilde the transformed
Stokes parameters the result can be expressed as 
\begin{equation}
\tilde{Q} = \cos{2 \alpha} \,Q + \sin{2 \alpha}\, U, \qquad \tilde{U} = -\sin{2\alpha} \,Q + \cos{2 \alpha} \,U.
\label{Tr9}
\end{equation}
Equations (\ref{SI18})--(\ref{SI19}) and (\ref{Tr9}) express 
the fact that the polarization of the graviton and of the 
radiation field do change for a rotation on the plane orthogonal to the direction 
of propagation of the radiation (either gravitational or electromagnetic).
It is possible to construct polarization observables which are invariant 
for rotations on the plane orthogonal to the direction of propagation of the radiation: because 
of their properties under parity transformations they are called E-and B-modes.

\subsection{E- and B-modes}
The fluctuations of the geometry induce fluctuations of the Stokes parameters 
whose spectral properties are, ultimately, the aim of CMB polarization experiments.
In general terms the fluctuation of each Stokes parameter can  be written as 
\begin{eqnarray}
\Delta_{\mathrm{I}}(\hat{n},\tau)&=& \Delta^{(\mathrm{T})}_{\mathrm{I}}(\hat{n},\tau) + \Delta^{(\mathrm{S})}_{\mathrm{I}}(\hat{n},\tau) + \Delta^{(\mathrm{V})}_{\mathrm{I}}(\hat{n},\tau),
\label{Itot}\\
\Delta_{\mathrm{Q}}(\hat{n},\tau)&=& \Delta^{(\mathrm{T})}_{\mathrm{Q}}(\hat{n},\tau) + \Delta^{(\mathrm{S})}_{\mathrm{Q}}(\hat{n},\tau) + \Delta^{(\mathrm{V})}_{\mathrm{Q}}(\hat{n},\tau),
\label{Qtot}\\
\Delta_{\mathrm{U}}(\hat{n},\tau)&=& \Delta^{(\mathrm{T})}_{\mathrm{U}}(\hat{n},\tau) + \Delta^{(\mathrm{S})}_{\mathrm{U}}(\hat{n},\tau) + \Delta^{(\mathrm{V})}_{\mathrm{U}}(\hat{n},\tau).
\label{Utot}
\end{eqnarray}
In Eqs. (\ref{Itot})--(\ref{Utot})  the superscript reminds that the various 
fluctuations of the Stokes parameters are induced, respectively, by the tensor, scalar and vector modes of the geometry. While some of the results of the present section will be generally valid, the focus, in what 
follows, will be on the tensor contribution.
Defining the two linear combinations 
\begin{equation}
\Delta_{\pm}(\hat{n},\tau) = \Delta_{\mathrm{Q}}(\hat{n},\tau) \pm i \Delta_{\mathrm{U}}(\hat{n},\tau),
\label{Tr9b}
\end{equation}
and denoting with a tilde the transformed quantities,  Eq. (\ref{Tr9}) implies that
$ \Delta_{\pm}(\hat{n},\tau)$ transform as
\begin{equation}
 \tilde{\Delta}_{\pm}(\hat{n},\tau) = e^{\mp 2 i\alpha}\,
\Delta_{\pm}(\hat{n},\tau).
\label{Tr10}
\end{equation}
In more general terms, consider a generic function of $\hat{n}$ (be it $f(\hat{n})$).
Under a rotation of an angle $\alpha$ on the plane 
orthogonal to $\hat{n}$, $f(\hat{n})$ is said to transform as a 
function of (integral) spin-weight $\pm s$ provided 
\begin{equation}
\tilde{f}(\hat{n}) = e^{ \mp i s\alpha} f(\hat{n}).
\label{Tr11}
\end{equation}
In other words  $\Delta_{+}(\hat{n})$ and $\Delta_{-}(\hat{n})$ transform, respectively, 
as functions of spin weight $+2$ and $-2$. The
circular polarizations of the gravitons introduced in
 Eqs. (\ref{SI18})--(\ref{SI19}) transform, respectively,  as functions of spin weight $+2$ and $-2$. 
The brightness perturbations for the intensity of the radiation field  
(i.e. $\Delta_{\mathrm{I}}(\hat{n})$) transform, on the contrary, as quantities of spin weight $0$.
 The fluctuations in the intensity of the radiation field, being a spin-0 quantity, can be expanded in ordinary spherical harmonics as 
\begin{eqnarray}
\Delta_{\mathrm{I}}(\hat{n},\tau) &=& \sum_{\ell\,m} a^{(\mathrm{T})}_{\ell\,m} Y_{\ell\,m}(\hat{n}), 
\nonumber\\
a^{(\mathrm{T})}_{\ell\,m} &=& \int d\hat{n} Y_{\ell\,m}^{*}(\hat{n}) \Delta_{\mathrm{I}}(\hat{n},\tau)
\label{Tr12}
\end{eqnarray}
The spin-s quantity will naturally be expanded in terms of a generalization of the ordinary spherical harmonics 
which are called spin-s spherical harmonics or also spin-weighted spherical harmonics. Owing to this observation,  $\Delta_{\pm}(\hat{n},\tau)$ can be expanded in terms of spin-$\pm2$ spherical harmonics 
$_{\pm 2}Y_{\ell\,m}(\hat{n})$, i.e. 
\begin{equation}
\Delta_{\pm}(\hat{n},\tau) = \sum_{\ell \, m} a_{\pm2,\,\ell\, m} \, _{\pm 2}Y_{\ell\, m}(\hat{n}).
\label{intN2}
\end{equation}

Given a quantity  of spin-weight $s$ it is possible to construct quantities of spin-weight $0$ by the repeated use 
of appropriate differential operators which can either raise or lower the spin-weight of a given function (see 
subsection \ref{sec25} for a specific discussion). Consequently, from $\Delta_{+}(\hat{n},\tau)$  and $\Delta_{-}(\hat{n},\tau)$ it is possible to construct two fluctuations of spin $0$ which can be eventually expanded 
in ordinary spherical harmonics. By demanding that the two fluctuations of spin-weight $0$ are eigenstates 
of parity the E- and B-modes are defined as 
\begin{eqnarray}
\Delta_{\mathrm{E}}(\hat{n},\tau) &=& \sum_{\ell\, m} \overline{N}_{\ell}^{-1} \,  
a^{(\mathrm{E})}_{\ell\, m}  \, Y_{\ell\, m}(\hat{n}),\qquad \overline{N}_{\ell} = \sqrt{\frac{(\ell - 2)!}{(\ell +2)!}},
\nonumber\\
\Delta_{\mathrm{B}}(\hat{n},\tau) &=& \sum_{\ell\, m} \overline{N}_{\ell}^{-1} \,  
a^{(\mathrm{B})}_{\ell\, m}  \, Y_{\ell\, m}(\hat{n}),
\label{intN4}
\end{eqnarray}
where
\begin{equation}
a^{(\mathrm{E})}_{\ell\, m} = - \frac{1}{2}(a_{2,\,\ell m} + a_{-2,\,\ell m}), \qquad  
a^{(\mathrm{B})}_{\ell\, m} =  \frac{i}{2} (a_{2,\,\ell m} - a_{-2,\,\ell m}).
\label{intN3}
\end{equation}

From $a^{(\mathrm{T})}_{\ell\, m}$, $a^{(\mathrm{E})}_{\ell\, m}$ and $a^{(\mathrm{B})}_{\ell\, m}$ 
the angular power spectra can be defined. In particular 
the EE, BB, TT and TE angular power spectra are given by:  
\begin{eqnarray}
C_{\ell}^{(\mathrm{EE})} &=& \frac{1}{2\ell + 1} \sum_{m = -\ell}^{\ell} 
\langle a^{(\mathrm{E})*}_{\ell m}\,a^{(\mathrm{E})}_{\ell m}\rangle,\qquad 
C_{\ell}^{(\mathrm{BB})} = \frac{1}{2\ell + 1} \sum_{m=-\ell}^{\ell} 
\langle a^{(\mathrm{B})*}_{\ell m}\,a^{(\mathrm{B})}_{\ell m}\rangle,
\label{intN5}\\
C_{\ell}^{(\mathrm{TT})} &=& \frac{1}{2\ell + 1} \sum_{m = -\ell}^{\ell} 
\langle a^{(\mathrm{T})*}_{\ell m}\,a^{(\mathrm{T})}_{\ell m}\rangle,\qquad 
C_{\ell}^{(\mathrm{TE})} = \frac{1}{2\ell + 1} \sum_{m=-\ell}^{\ell} 
\langle a^{(\mathrm{T})*}_{\ell m}\,a^{(\mathrm{E})}_{\ell m}\rangle,
\label{intN6}
\end{eqnarray}
where $\langle ...\rangle$ denotes the ensemble average.  Two further power spectra 
can be defined and they are:
\begin{equation}
C_{\ell}^{(\mathrm{EB})} = \frac{1}{2\ell + 1} \sum_{m = -\ell}^{\ell} 
\langle a^{(\mathrm{E})*}_{\ell m}\,a^{(\mathrm{B})}_{\ell m}\rangle,\qquad 
C_{\ell}^{(\mathrm{TB})} = \frac{1}{2\ell + 1} \sum_{m=-\ell}^{\ell} 
\langle a^{(\mathrm{T})*}_{\ell m}\,a^{(\mathrm{B})}_{\ell m}\rangle.
\label{intN7}
\end{equation}
Overall, the existence of linear polarization allows for 6 different  power 
spectra.

In the minimal version of the $\Lambda$CDM paradigm the adiabatic fluctuations of the scalar curvature 
lead to a polarization which is characterized exactly by the condition $a_{2,\,\ell m} = a_{-2,\,\ell m}$, i.e. $a_{\ell m}^{(\mathrm{B})} =0$.  This observation implies that, in the $\Lambda$CDM scenario, 
the non-vanishing angular power spectra are given by the TT, EE and TE correlations.
In the T$\Lambda$CDM scenario  the TT, EE and TE angular power spectra are 
supplemented by a specific prediction for
the B-mode autocorrelation (see section \ref{sec7}).
 
\subsection{Spin-2 spherical harmonics}
\label{sec25}
Spherical harmonics 
of higher spin appear in matrix elements calculations in nuclear physics (see e.g. the classic treatise of
 Blatt and Weisskopf \cite{blatt}, and, in a similar perspective 
the book of Edmonds \cite{edmonds}). 
The comprehensive treatments  of Biedenharn and Louk \cite{bie} and  
of Varshalovich et al. \cite{var} can also be usefully consulted.

The spin-s harmonics have been introduced, in their present form, 
by Newman and Penrose \cite{new} and their group theoretical 
interpretation has been discussed in \cite{sud}. The 
spin-s spherical harmonics have been applied to the discussion 
of CMB polarization induced by relic gravitons in a number 
of papers \cite{B1,tot,B2}.  They are rather crucial in the formulation of the so-called total angular momentum approach. Discussions of the spin-weighted spherical harmonics 
in a cosmological context can also be found in \cite{B3,B4}. 
The spin weighted spherical harmonics will now be introduced by 
following the spirit of Ref. \cite{sud} which has 
been also used, with different conventions, in \cite{B1}. In subsection \ref{sec26}
the (equivalent) approach of \cite{tot,B2} will be more specifically outlined.

Th functions  $_{\pm 2}Y_{\ell\,m}(\hat{n})$  appearing in Eq. (\ref{intN2}) 
are the spin-2 spherical harmonics \cite{sud}. 
Consider the representations
of the operator specifying three-dimensional rotations, i.e. $\hat{R}$; 
this problem  is usually approached within the matrix element, i.e. 
${\cal D}_{m \, m'}^{(j)}(R) = \langle j, \, m'| \hat{R} | j,\,m\rangle$ where 
$j$ denotes the eigenvalue of $J^2$ and $m$ denotes the eigenvalue of $J_{z}$. 
Now, if we replace $m' \to - {\rm s}$, $j\to \ell$, we have the definition 
of spin-s spherical harmonics in terms of the 
${\cal D}_{ -{\rm s},\, m}^{(\ell)}(\alpha,\beta, 0)$, i.e.  
\begin{equation}
_{\rm s}Y_{\ell\, m}(\alpha,\beta) = \sqrt{\frac{2\ell +1}{4\pi}} {\cal D}_{-{\rm s},\, m}^{(\ell)}(\alpha,\beta, 0),
\label{EB4}
\end{equation}
where $\alpha$, $\beta$ and $\gamma$ (set to zero in the above definition) 
are the Euler angles defined as in \cite{sakurai}. If $s=0$, ${\cal D}_{0,\, m}^{(\ell)}(\alpha,\beta, 0) = \sqrt{(2\ell +1)/4\pi} Y_{\ell \, m}(\alpha,\beta)$ where 
$Y_{\ell \, m}(\alpha,\beta)$  are the ordinary spherical harmonics. 
The spin-s spherical harmonics can be obtained from the spin-0 spherical harmonics 
by using repeatedly the differential operators:
\begin{eqnarray}
&& K_{+}^{(\mathrm{s})}(\hat{n})= -(\sin\vartheta)^{s}\,\left[ \partial_\vartheta
    + {i\over \sin\theta} \partial_\varphi \right] \frac{1}{(\sin\vartheta)^{s}},
\label{Kplus}\\
&& K_{-}^{(\mathrm{s})}(\hat{n})= -\frac{1}{(\sin{\vartheta})^s}\,\left[ \partial_\vartheta
    - {i\over \sin\vartheta} \partial_\varphi \right] (\sin\vartheta)^{s}.
 \label{Kminus}
 \end{eqnarray}
The notation spelled out in Eqs. (\ref{Kplus}) and (\ref{Kminus}) (which is not usual) will be employed to emphasize the interpretation of $K_{\pm}^{(\mathrm{s})}$  as ladder operators (see \cite{sud}). 
The operator $ K_{+}^{(\mathrm{s})}$ raises the spin weight of a 
function by one unit. Consider, therefore, the ordinary spherical 
harmonics $Y_{\ell\,m}(\hat{n})$ defined as
\begin{eqnarray}
Y_{\ell\,m}(\hat{n}) &=& N_{\ell}^{m} P_{\ell}^{m}(\mu) e^{i m \varphi}, 
\qquad N_{\ell}^{m} = \sqrt{\frac{2\ell + 1}{4\pi}} \sqrt{\frac{(\ell -m)!}{(\ell + m)!}},\qquad \mu = \cos{\vartheta}
\label{Ydef1}\\
 P_{\ell}^{m}(\mu) &=& (-1)^{m} (1- \mu^2)^{m/2} \frac{d^{m}}{d \mu^{m}} P_{\ell}(\mu),\qquad  
 P_{\ell}^{-m}(\mu) = (-1)^{m} \frac{(\ell - m)!}{(\ell + m)!} P_{\ell}^{m}(\mu),
 \label{Ydef2}
\end{eqnarray}
where $P_{\ell}(\mu)$ are the Legendre polynomials and $P_{\ell}^{m}(\mu)$ the 
associated Legendre functions.  It is appropriate to mention here that the factor $(-1)^{m}$ 
(i.e. Condon-Shortley phase) can either be included in the normalization 
factor $N_{\ell}^{m}$ or (as it has been done) in the definition 
of the associated Legendre functions appearing in Eq. (\ref{Ydef2}). When using 
the recurrence relations of the associated Legendre functions the Condon-Shortley phase introduces 
a sign difference every time $m$ is odd. The conventions expressed 
by Eqs. (\ref{Ydef1}) and (\ref{Ydef2}) will be  followed 
throughout the present discussion and, in particular, in section \ref{sec7} where 
the correlation functions of the E-modes and of the B-modes will be specifically 
computed with different techniques.

According to Eq. (\ref{Tr11}), $Y_{\ell\,m}$ transform 
with $ s =0$, i.e. they have spin weight $0$. By applying once 
$K_{+}^{(0)}(\hat{n})$ to $Y_{\ell\,m}(\hat{n})$ we do get a function 
of spin weight $1$, i.e. 
\begin{equation}
K_{+}^{(0)}(\hat{n})Y_{\ell\,m}(\hat{n}) = - \biggl[ \partial_{\vartheta} + \frac{i}{\sin{\vartheta}}\partial_{\varphi}\biggr] Y_{\ell\, m}(\hat{n}).
\label{Ydef3}
\end{equation} 
We can then apply once more\footnote{This time, in $K_{+}^{(\mathrm{s})}(\hat{n})$, $s = 1$ since 
$K_{+}^{(0)}(\hat{n}) Y_{\ell\,m}(\hat{n})$ is a quantity of spin weight $1$.}
 $K_{+}^{(1)}(\hat{n})$ to $K_{+}^{(0)}(\hat{n}) Y_{\ell\,m}(\hat{n})$.
The result of this simple manipulation will be 
\begin{equation}
K_{+}^{(1)}(\hat{n})\,[K_{+}^{(0)}(\hat{n})Y_{\ell\,m}(\hat{n})] = \biggl[\partial_{\vartheta}^2 -  \cot{\vartheta} \partial_{\vartheta} - \frac{\partial_{\varphi}^2}{\sin^2{\vartheta}} + 
\frac{2 i}{\sin{\vartheta}}(\partial_{\vartheta}\partial_{\varphi} - \cot{\vartheta} \partial_{\varphi})\biggr] Y_{\ell\,m}(\hat{n}).
\label{Ydef4}
\end{equation}
The right hand side of Eq. (\ref{Ydef4}) is $_{+2} Y_{\ell\, m}(\vartheta,\varphi)$ (up to an overall normalization).
Including the appropriate normalization factor, $_{\pm 2} Y_{\ell\, m}(\vartheta,\varphi)$, i.e. 
the spherical harmonics of spin weight $s = \pm 2$ are given by:
\begin{equation}
_{\pm 2} Y_{\ell\, m}(\vartheta,\varphi)=\sqrt{\frac{(\ell -2)!}{(\ell + 2)!}} 
\biggl[\partial_{\vartheta}^2 -  \cot{\vartheta} \partial_{\vartheta} - \frac{\partial_{\varphi}^2}{\sin^2{\vartheta}} \pm
\frac{2 i}{\sin{\vartheta}}(\partial_{\vartheta}\partial_{\varphi} - \cot{\vartheta} \partial_{\varphi})\biggr] Y_{\ell\,m}(\hat{n}).
\label{EB5}
\end{equation}
The spin weights $s = \pm 2$ are both needed since the transformation of 
the polarization involve both spin weights (see  Eq. (\ref{intN2})). In fact,
since $_{\pm 2} Y_{\ell\, m}(\vartheta,\varphi)$ form a complete and orthogonal basis on the sphere, i.e. 
\begin{eqnarray}
&& \int d\hat{n}\, _sY^*_{\ell'\,m'}(\vartheta,\varphi) _sY_{\ell\,m}(\vartheta,\varphi) = \delta_{\ell'\,\ell}
    \,\delta_{m'm},
\label{norm1}\\
&& \sum_{\ell,m}\, _sY_{\ell\,m}^*(\vartheta,\varphi) _sY_{\ell\,m}(\vartheta',\varphi')
    = \delta(\varphi-\varphi')\,\delta(\cos\vartheta-\cos\vartheta'),
\label{norm2}
\end{eqnarray}
$\Delta_{\pm}(\hat{n},\tau)$ can be expanded in terms of $_{\pm 2} Y_{\ell\, m}(\vartheta,\varphi)$ as in Eq. 
(\ref{intN2}). The coefficients off the expansion will be given by 
\begin{eqnarray}
a_{2,\,\ell\,m} &=& \int d\hat{n} \,\, _{2}Y_{\ell\, m}^{*}(\hat{n}) \Delta_{+}(\hat{n},\tau), 
\label{norm3}\\
a_{-2,\,\ell\,m} &=& \int d\hat{n} \,\, _{-2}Y_{\ell\, m}^{*}(\hat{n}) \Delta_{-}(\hat{n},\tau).
\label{norm4}
\end{eqnarray}
Integrating by parts in Eqs. (\ref{norm3}) and (\ref{norm4}) allows for a different form of the expansion
coefficients $a_{\pm 2,\,\ell\,m}$:
\begin{eqnarray}
a_{2,\,\ell\,m} &=& \overline{N}_{\ell} \int d\hat{n} Y_{\ell\, m}^{*} K^{(1)}_{-}(\hat{n})[K^{(2)}_{-}(\hat{n}) \Delta_{+}(\hat{n},\tau)], 
\label{norm6}\\
a_{-2,\,\ell\,m} &=& \overline{N}_{\ell} \int d\hat{n} Y_{\ell\, m}^{*} K^{(-1)}_{-}(\hat{n})[K^{(-2)}_{+}(\hat{n}) \Delta_{-}(\hat{n},\tau)],
\label{norm7}
\end{eqnarray}
where, as already mentioned, $\overline{N}_{\ell} = \sqrt{(\ell -2)!/(\ell +2)!}$. In Eqs. (\ref{norm6}) 
and (\ref{norm7}) there appear only ordinary (i.e. spin-weight $0$) spherical harmonics. This 
occurrence suggests a complementary approach to the problem: instead of expanding 
$\Delta_{\pm}(\hat{n},\tau)$ in terms of spin-2 spherical harmonics, fluctuations of spin-weight 
$0$ can be directly constructed (in real space) from $\Delta_{\pm}(\hat{n},\tau)$ by 
repeated application of the ladder operators defined in Eqs. (\ref{Kplus}) and (\ref{Kminus}).

The E-mode and B-mode polarization in real space  are then, in explicit terms:
\begin{eqnarray}
&&\Delta_{\mathrm{E}}(\hat{n}, \tau)  = - \frac{1}{2} \{ K^{(1)}_{-}(\hat{n})[K^{(2)}_{-}(\hat{n}) \Delta_{+}(\hat{n},\tau)]
+ K^{(-1)}_{-}(\hat{n})[K^{(-2)}_{+}(\hat{n}) \Delta_{-}(\hat{n},\tau)]\},
\label{EX1}\\
&& \Delta_{\mathrm{B}}(\hat{n}, \tau) = \frac{i}{2} \{ K^{(1)}_{-}(\hat{n})[K^{(2)}_{-}(\hat{n}) \Delta_{+}(\hat{n},\tau)]- K^{(-1)}_{-}(\hat{n})[K^{(-2)}_{+}(\hat{n}) \Delta_{-}(\hat{n},\tau)]\}.
\label{EX2}
\end{eqnarray}
The quantities $\Delta_{\mathrm{E}}(\hat{n}, \tau)$ and $\Delta_{\mathrm{B}}(\hat{n}, \tau)$ 
can be expanded in terms of ordinary spherical harmonics, as already suggested in Eq. (\ref{intN4}):
\begin{equation}
\Delta_{\mathrm{E}}(\hat{n},\tau) = \sum_{\ell\, m} \overline{N}_{\ell}^{-1} \,  
a^{(\mathrm{E})}_{\ell\, m}  \, Y_{\ell\, m}(\hat{n}),\qquad 
\Delta_{\mathrm{B}}(\hat{n},\tau) = \sum_{\ell\, m} \overline{N}_{\ell}^{-1} \,  
a^{(\mathrm{B})}_{\ell\, m}  \, Y_{\ell\, m}(\hat{n}),
\label{intN4a}
\end{equation}
The ``electric" and ``magnetic" components of polarization are 
eigenstates of parity and may be defined, from $a_{\pm,\,\,\ell\,m}$ as already mentioned in Eq. (\ref{intN3}):
\begin{equation}
a_{\ell m}^{(\rm E)}=-\frac{1}{2}(a_{2, \ell m}+a_{-2,\ell m}),\qquad
a_{\ell m}^{(\rm B)}= \frac{i}{2}(a_{2,\ell m}-a_{-2,\ell m}).
\label{EB6}
\end{equation}
Under parity  the components appearing in Eqs. (\ref{EB6}) transform
\begin{equation}
 a_{\ell m}^{(\rm E)} \to (-1)^{\ell }\,\,a_{\ell m}^{\rm E},\qquad
 a_{\ell m}^{(\rm B)} \to (-1)^{\ell +1 }\,\,a_{\ell m}^{\rm B}.
\label{EB7}
\end{equation}
Therefore, the E-modes have the same parity of the temperature correlations  
which have, in  turn, the same parity of conventional spherical harmonics, i.e. 
$(-1)^{\ell}$. On the contrary, the B-modes have $(-1)^{\ell + 1}$ parity. The same analysis can be directly performed in real space, i.e. using Eqs. (\ref{EX1}) and (\ref{EX2}). Denoting the radial direction 
with $\hat{n}$ and the tangential directions with $\hat{e}_{1}$ and $\hat{e}_{2}$, the ladder operators 
defined in Eqs. (\ref{Kplus}) and (\ref{Kminus}) are consistent with the following choice of \footnote{As discussed 
at the end of subsection \ref{sec21} the sign of $\varphi$ can be flipped. This possibility is not related 
to a parity transformation and it has to do with the way two-dimensional rotations are introduced. This aspect 
will also be relevant in section \ref{sec7} for explicit derivations.}
$\hat{e}_{1}$ and $\hat{e}_{2}$:
\begin{equation}
\hat{e}_{1} = (\cos{\vartheta}\cos{\varphi}, -\cos{\vartheta} \sin{\varphi}, - \sin{\vartheta}),\qquad 
\hat{e}_{2} = (- \sin{\varphi}, - \cos{\varphi}, 0).
\label{EB7a}
\end{equation}
A parity transformation (i.e. a space inversion) implies, in spherical coordinates, that 
\begin{equation}
r\to r,\qquad \vartheta \to \pi - \vartheta,\qquad \varphi \to \pi + \varphi.
\label{EB7b}
\end{equation}
The transformation (\ref{EB7b}) implies that the two basis vectors defined in Eq. (\ref{EB7a}) 
transform as $\hat{e}_{1} \to \hat{e}_{1}$ and $\hat{e}_{2} \to - \hat{e}_{2}$, i.e. 
while $\hat{e}_{1}$ does not change $\hat{e}_{2}$ flips its sign under space inversion. 
It follows that space-inversion does not flip the sign of $\Delta_{\mathrm{E}}(\hat{n})$ 
but it does flip the sign of $\Delta_{\mathrm{B}}(\hat{n})$, i.e. under the transformation 
(\ref{EB7b}), $\Delta_{\mathrm{E}}(\hat{n})\to \Delta_{\mathrm{E}}(\hat{n})$ while 
$\Delta_{\mathrm{B}}(\hat{n})\to -\Delta_{\mathrm{B}}(\hat{n})$.

 Using Eqs. (\ref{norm6}) and (\ref{norm7}) inside Eq. (\ref{EB6}) a more explicit 
expression for $a_{\ell m}^{(\rm E)}$ and $a_{\ell m}^{(\rm B)}$ can be obtained and it is:
\begin{eqnarray}
&& a_{\ell m}^{(\rm B)} = \frac{ i\, \overline{N}_{\ell}}{2} \int d\hat{n} Y_{\ell\,m}(\hat{n})^{*} \{ K^{(1)}_{-}(\hat{n}) [K^{(2)}_{-}(\hat{n}) \Delta_{+}(\hat{n},\tau)]  - K^{(-1)}_{-}(\hat{n})[K^{(-2)}_{+}(\hat{n}) \Delta_{-}(\hat{n},\tau)]\},
\label{EB8}\\
&& a_{\ell m}^{(\rm E)} = - \frac{ \overline{N}_{\ell}}{2} \int d\hat{n} Y_{\ell\,m}(\hat{n})^{*} \{ K^{(1)}_{-}(\hat{n})[K^{(2)}_{-}(\hat{n}) \Delta_{+}(\hat{n},\tau)]  + K^{(-1)}_{-}(\hat{n})[K^{(-2)}_{+}(\hat{n}) \Delta_{-}(\hat{n})]\}.
\label{EB9}
\end{eqnarray}
The contribution of long wavelength gravitons to Eqs. (\ref{EB8}) and (\ref{EB9}) will be discussed in section \ref{sec7}.
It is often useful to observe that the differential 
operators appearing in the definition of the spin-weighted spherical harmonics 
(see, e.g. Eq. (\ref{EB5})) can be expressed in terms 
of the usual differential operators arising in the theory of the 
orbital angular momentum in non-relativistic quantum mechanics (see, e. g. 
\cite{sakurai}). Indeed, recalling that 
\begin{eqnarray}
&& L_{\pm}= e^{\pm i \varphi}[ \pm \partial_{\vartheta} + i \cot{\vartheta} \, \partial_{\varphi}],\qquad L_{z} = - i \partial_{\varphi}, 
\label{AM1}\\
&& L^2 = \frac{1}{2}(L_{+} \, L_{-} + L_{-}\, L_{+}) + L_{z}^2,
\label{AM2}
\end{eqnarray}
it can be easily deduced that 
\begin{eqnarray}
&& 
\cot{\vartheta} \partial_{\varphi} = - \frac{i}{2} [ e^{-i \varphi} \,L_{+} + e^{i\varphi}\, L_{-}],
\label{AM3}\\
&& 
\partial_{\vartheta} \partial_{\varphi} = \frac{i}{2} [ e^{-i\varphi}\, L_{+} \,L_{z} - e^{i \varphi}\, L_{-} \,L_{z}]
\label{AM4}\\
&& - \cot{\vartheta} \partial_{\vartheta} - \frac{\partial_{\varphi}^2}{\sin^2{\vartheta}} = 
L^2 + \partial_{\vartheta}^2.
\label{AM5}
\end{eqnarray}
Equations (\ref{AM3})--(\ref{AM5}) allow often to express combinations of spin-$2$ spherical harmonics 
in terms of ordinary (i.e. spin-weight $0$) spherical harmonics using the properties 
of the ladder operators associated to the (orbital) angular momentum, i.e. $L_{\pm}$:
\begin{equation}
L_{\pm} Y_{\ell\,m} = \sqrt{\ell (\ell + 1) - m( m \pm 1)} Y_{\ell, m \pm 1}, \qquad  
L_{z} Y_{\ell\,m} = m Y_{\ell\,m},\qquad  L^2 Y_{\ell\,m}= \ell (\ell + 1) Y_{\ell\,m},
\label{AM6}
\end{equation}
where $L_{\pm}$ and $L_{z}$ obey the well known commutation relations 
 $[L_{\pm}, L_{z}]= \mp L_{\pm}$ and $[L_{+}, L_{-}] = 2 L_{z}$. 
 
Looking at Eq. (\ref{AM6}) it is tempting draw a parallel between the 
(orbital) ladder operators and the ladder operators raising (or lowering) 
the spin weight of a given function (see Eqs. (\ref{Kplus}) and (\ref{Kminus}).
This problem has been discussed and solved  in \cite{sud}. It is 
possible to formulate the parallel in terms of a putative $O(4)$ group. Half of the generators 
will be connected with the orbital angular momentum operators, while the other 
half will allow to increase (or decrease) the spin weight of a given function. 
The two sets of generators commute. The operators $K_{\pm}^{(\mathrm{s})}$
are not directly, though, the ladder operators stemming from the second set of generators. This has to do with the fact that in Eq. (\ref{EB4}) the third 
Euler angle (i.e. $\gamma$) has been fixed to zero. The $K_{\pm}^{(\mathrm{s})}$
are ladder operators defined within a putative $O(4)$ group in the case $\gamma \neq 0$. When $\gamma \to 0$ the dependence upon $\gamma$ drops and we are left with Eqs. (\ref{Kplus}) and (\ref{Kminus}). 

\subsection{Polarization on the  2-sphere}
\label{sec26}
In a more geometric perspective, the spin-2 harmonics 
are introduced  by describing the polarization tensor on the 2-sphere which 
represents the microwave sky.  
In Eq. (\ref{Tr8}) the tensor $P_{ij}$ describes  the properties of the 
radiation field and it is symmetric and trace-free (i.e. $P_{ij} = P_{ji}$ and
 $P_{i}^{i} = 0$). Equation (\ref{Tr8}) holds in flat space-time. On the 
 2-sphere the line element can be 
 written as 
 \begin{equation}
 ds^2 = d \vartheta^2 + \sin^2{\vartheta} \varphi^2, \qquad g_{\vartheta\vartheta}= 1,\qquad g_{\varphi\varphi} = \sin^2{\vartheta}.
 \label{SP1}
 \end{equation}
 The polarization matrix $P_{ij}$ will now be generalized as 
\begin{equation} 
P_{ab}(\hat{n})= \frac{1}{2} \left(
\begin{array}{cc}
Q(\hat{n}) & -U(\hat{n})\sin\vartheta\\
-U(\hat{n})\sin\vartheta & -Q(\hat{n})\sin^2\vartheta\\
\end{array}
\right),
\label{SP2}
\end{equation}
satisfying $P_{ab}=P_{ba}$, and $g^{ab}P_{ab}=0$,
where $\hat{n}$ is a unit vector
in the direction $(\vartheta, \varphi)$. The sign of the off-diagonal entries in Eq. 
(\ref{SP2}) is opposite with respect to the one obtained in Eq. (\ref{Tr8}). This 
is just because we want to match with the conventions adopted, for instance, in 
\cite{B2,B3,B4}. To avoid possible confusions, furthermore, the Latin indices 
$a,\,b,\,c,\,d, ....$ run over the two-dimensional space. 

As already mentioned, for scalar functions defined on the 2-sphere, such as the
temperature anisotropies, the
spherical harmonic functions $Y_{\ell m}(\hat{n})$ are the complete
orthonormal basis. For the $2\times 2$ tensors defined on the
2-sphere, such as $P_{ab}$ in Eq.(\ref{SP2}), the 
complete orthonormal set of tensor spherical harmonics can be written 
as \cite{B2,B3,B4}:
\begin{eqnarray}
&&Y_{(\ell m)ab}^{\rm G} = N_{\ell}
     \left( \nabla_{a} \nabla_{b} Y_{(\ell m)} - {1\over2} g_{ab} \nabla_{c}\nabla^{c}Y_{(\ell m)}\right),
\label{SP3}\\
&&   Y_{(\ell m)ab}^{\rm C} = { N_{\ell} \over 2}
     \left(\vphantom{1\over 2}
      \nabla_{a}\nabla_{c} Y_{(\ell m)} \epsilon^c{}_b +\nabla_{b} \nabla_{c}Y_{(\ell m)} \epsilon^c{}_a \right),
\label{SP4}
\end{eqnarray}
where $\nabla$, in this subsection,  denotes a covariant derivation on the 2-sphere, e.g.
\begin{eqnarray}
&&\nabla_{a} \nabla_{b} Y_{(\ell m)} = \partial_{a} \partial_{b} Y_{(\ell m)} - \Gamma_{ab}^{c} 
\partial_{c} Y_{(\ell m)}, \qquad  \nabla_{c} \nabla^{c} Y_{(\ell m)}= \partial_{c} \partial^{c} Y_{(\ell m)} + 
\Gamma_{c a}^{c} \partial^{a} Y_{(\ell m)}.
\label{SP5}\\
&& \Gamma_{a c}^{c} = \frac{1}{2} \,g^{c d} ( - \partial_{d} g_{a b} + \partial_{b} g_{d c} + \partial_{c} g_{b d}).
\label{SP5a}
\end{eqnarray}
Using Eq. (\ref{SP1}) into Eq. (\ref{SP5a}), 
the  Christoffel connections $\Gamma_{ab}^{c}$ on the 
2-sphere are 
\begin{equation}
\Gamma_{\varphi\vartheta}^{\varphi} = \frac{1}{\tan{\vartheta}},\qquad 
\Gamma_{\varphi\varphi}^{\vartheta} = - \sin{\vartheta}\cos{\vartheta}.
\label{SP6}
\end{equation}
In Eq. (\ref{SP4}) the normalization factor is given by\footnote{Notice that $N_{\ell}$ differs from $\overline{N}_{\ell}$ defined in Eqs. (\ref{intN4}) (see also Eqs. (\ref{norm6}) and (\ref{norm7})) by a factor 
$\sqrt{2}$. This difference will be ultimately relevant to relate ($a_{\ell m}^{(\mathrm{G})}$, $a_{\ell m}^{(\mathrm{C})}$) to ($a_{\ell m}^{(\mathrm{E})}$, $a_{\ell m}^{(\mathrm{B})}$).} 
$N_{\ell}\equiv \sqrt{ {2 (\ell-2)! / (\ell+2)!}}$, while 
\begin{equation}  
\epsilon^a\, _b= \left(
\begin{array}{cc}
0 & \sin\theta\\
-1/\sin\theta & 0\\
\end{array}
\right),
\label{SP7}
\end{equation}
is the Levi-Civita symbol on the 2-sphere.   The differential operators acting in
Eqs. (\ref{SP3}) and (\ref{SP4}) are interpreted as a generalized gradient and curl operators, i.e. 
\begin{equation}
\nabla_{ab}^{(\mathrm{G})} =  \nabla_{a} \nabla_{b} - {1\over2} g_{ab} \nabla_{c}\nabla^{c},
\qquad  \nabla_{ab}^{(\mathrm{C})} =   \epsilon^c{}_b \nabla_{a}\nabla_{c} + \epsilon^c{}_a \nabla_{b} \nabla_{c}
 \label{SP7a}
 \end{equation}
The explicit form of the various components of
$Y^{\mathrm{G}}_{(\ell m)ab}$ and $Y^{\mathrm{C}}_{ab}$ can be computed. For instance 
using Eqs. (\ref{SP3}), (\ref{SP5}) and (\ref{SP6}), the explicit components of 
 $Y^{\mathrm{G}}_{(\ell m)ab}$:
\begin{eqnarray}
&& Y_{(\ell m)\vartheta\vartheta}^{(\mathrm{G})}(\vartheta,\varphi) = \frac{N_{\ell}}{2} \biggl[ \partial_{\vartheta}^2Y_{(\ell m)}
- \frac{\partial_{\varphi}^2 Y_{(\ell m)}}{\sin^2{\vartheta}} - \frac{\partial_{\vartheta}Y_{(\ell m)}}{\tan{\vartheta}}\biggr],
\label{SP8}\\
&& Y_{(\ell m)\varphi\varphi}^{(\mathrm{G})}(\vartheta,\varphi) = \frac{N_{\ell}}{2}[ \partial_{\varphi}^2Y_{(\ell m)}
- \sin^2{\vartheta} \partial_{\vartheta}^2\,\,Y_{(\ell m)} + \sin{\vartheta} \cos{\vartheta} \partial_{\vartheta}Y_{(\ell m)}]
\label{SP9}\\
&& Y_{(\ell m)\vartheta\varphi}^{(\mathrm{G})}(\vartheta,\varphi) = 
N_{\ell} \biggl[ \partial_{\vartheta} \partial_{\varphi} Y_{(\ell m)} - \frac{\partial_{\varphi}Y_{(\ell m)}}{\tan{\vartheta}}\biggr].
\label{SP10}
\end{eqnarray}
The expressions 
obtained in Eqs. (\ref{SP8}), (\ref{SP9}) and (\ref{SP10}) can be simplified by recalling 
Eqs. (\ref{AM1}), (\ref{AM2}) and (\ref{AM6}).
Equations (\ref{SP8}), (\ref{SP9}) and (\ref{SP10}) can be simply rewritten as 
\begin{eqnarray}
&& Y_{(\ell m)\vartheta\vartheta}^{(\mathrm{G})}(\vartheta,\varphi) = 
\frac{N_{\ell}}{2} [ 2 \, \partial_{\vartheta}^2\, Y_{(\ell m)}+ \ell(\ell+1) Y_{(\ell m)}],
\label{SP15}\\
&& Y_{(\ell m)\varphi\varphi}^{(\mathrm{G})}(\vartheta,\varphi) = - \frac{N_{\ell}}{2}[ 2 \partial_{\vartheta}^2 Y_{(\ell m)} + \ell (\ell + 1) Y_{(\ell m)}],
\label{SP16}\\
&& Y_{(\ell m)\vartheta\varphi}^{(\mathrm{G})}(\vartheta,\varphi) = 
m i N_{\ell} \biggl[ \partial_{\vartheta} \partial_{\varphi} Y_{(\ell m)} - \frac{Y_{(\ell m)}}{\tan{\vartheta}}\biggr].
\label{SP17}
\end{eqnarray}
The same exercise can be conducted for the various components 
of $Y_{(\ell m)a b}^{(\mathrm{C})}(\vartheta,\varphi)$. 
The $ Y_{(\ell m)ab}^{(\mathrm{G})}$ and 
$Y_{(\ell m)ab}^{(\mathrm{C})}$ can be written in the form of  $2\times 2$ matrices:
\begin{equation}
   Y_{(\ell m)ab}^{\mathrm{G} }(\hat{n})={N_{\ell}\over 2} \left( \begin{array}{cc}
   W_{(\ell m)}(\hat{n}) & X_{(\ell m)}(\hat{n}) \sin\vartheta\\
   \noalign{\vskip6pt}
   X_{(\ell m)}(\hat{n})\sin\vartheta & -W_{(\ell m)}(\hat{n})\sin^2\vartheta \\
   \end{array} \right),
\label{SP18}
\end{equation}
and as
\begin{equation}
   Y_{(\ell m)ab}^{(\mathrm{C})}(\hat{n})={N_{\ell}\over 2} \left( \begin{array}{cc}
   -X_{(\ell m)}(\hat{n}) & W_{(\ell m)}(\hat{n}) \sin\vartheta \\
   \noalign{\vskip6pt}
   W_{(\ell m)}(\hat{n})\sin\theta & X_{(\ell m)}(\hat{n})\sin^2\vartheta \\
   \end{array} \right),
\label{SP19}
\end{equation}
where
\begin{equation}
W_{(\ell m)}(\hat{n}) =  [2\partial^2_{\vartheta} + \ell(\ell+1) ] Y_{(\ell m)}(\hat{n}),\qquad   X_{(\ell m)}(\hat{n}) = {2im \over \sin\vartheta}
     (\partial_{\vartheta} -
     \cot\vartheta ) Y_{(\ell m)}(\hat{n}).
\label{SP21}
\end{equation}
In terms of the spin-2 harmonics ${}_{\pm2}Y_{(lm)}(\hat{n})$
\begin{equation}
     W_{(\ell m)}(\hat n) \pm i X_{(\ell m)}(\hat n) = \sqrt{{ (\ell+2)!
     \over (\ell-2)!}}\,_{\pm2}Y_{(\ell m)}(\hat{n}),
\label{SP22}
\end{equation}
which is Eq. (\ref{EB5}). Using the orthonormality 
of the spherical harmonics $Y_{\ell m}(\hat{n})$ it is easy to prove 
the orthonormality conditions, i.e. 
\begin{eqnarray}
&& \int d\hat{n}\,Y_{(\ell m)ab}^{(\mathrm{G})\,*}(\hat{n})\,Y_{(\ell' m')}^{(\mathrm{G})\,\,ab}(\hat{n})
      =\int d\hat{n}\,Y_{(\ell m)ab}^{(\mathrm{C})\,*}(\hat{n})\,Y_{(\ell' m')}^{(\mathrm{C})\,\,ab}(\hat{n})
      =\delta_{\ell' \ell } \delta_{m m'},
\nonumber\\
&&   \int d \hat{n} \,Y_{(\ell m) ab}^{(\mathrm{G})\,*}(\hat{n})\,
Y_{(\ell' m')}^{(\mathrm{C})\,\,ab}(\hat{n})
     =0,
\label{SP23}
\end{eqnarray}
where $d\hat{n} = \sin{\vartheta} d\vartheta d\varphi$ denotes, as 
usual, the integration over the solid angle. 
Since $Y_{(\ell m) ab}^{\mathrm{G}}$ and  $Y_{(\ell m) ab}^{\mathrm{C}}$
form an appropriate orthonormal basis, the polarization 
can be expanded as 
\begin{equation}
     P_{ab}(\hat{n})= 
     \sum_{\ell=2}^\infty\sum_{m=-\ell}^{\ell}
     \left[ a_{\ell m}^{(\mathrm{G})}
     Y_{(\ell m)ab}^{(\mathrm{G})}(\hat{n}) + a_{\ell m}^{(\mathrm{C})} Y_{(\ell m)ab}^{(\mathrm{C})}
     (\hat{n}) \right],
\label{SP24}
\end{equation}
where expansion coefficients
 $a_{\ell m}^{(\mathrm{G})}$ and $ a_{\ell m}^{(\mathrm{C})}$ represent
the electric and magnetic type components of the polarization,
respectively.
Note that the sum starts from $\ell=2$,
since relic gravitons generate only perturbations of multipoles
from the quadrupoles up. The expansion coefficients are obviously
\begin{equation}
     a_{\ell m}^{(\mathrm{G})}=\int \, d\hat{n}\, {P}_{ab}(\hat{n})
              Y_{(\ell m)}^{(\mathrm{G}) \,ab\, *}(\hat{n}), \quad
     a_{\ell m}^{(\mathrm{C})}=\int d\hat{n}\, {P}_{ab}(\hat{n})
              Y_{(\ell m)}^{{(\mathrm{C})} \, ab\, *}(\hat{n}).
\label{SP25}
\end{equation}
In the notations of \cite{B1} the $a_{\ell m}^{(\mathrm{G})}$
and $a_{\ell m}^{(\mathrm{C})}$ can be related to 
the $a_{\ell m}^{(\mathrm{E})}$ and $a_{\ell m}^{(\mathrm{B})}$ 
already introduced in Eq. (\ref{EB6}). The relation between the two sets 
of expansion coefficients is simply:
\begin{equation}
a_{\ell m}^{(\mathrm{G})}= \frac{\overline{N}_{\ell}}{N_{\ell}} a_{\ell m}^{(\mathrm{E})}= \frac{ a_{\ell m}^{(\mathrm{E})}}{\sqrt{2}}, \qquad 
a_{\ell m}^{(\mathrm{C})}=\frac{\overline{N}_{\ell}}{N_{\ell}} a_{\ell m}^{(\mathrm{B})}= \frac{ a_{\ell m}^{(\mathrm{B})}}{\sqrt{2}},
\label{SP26}
\end{equation}
The two approaches to the spin weighted spherical harmonics 
described in the present section are  equivalent and can be used interchangeably depending 
upon the specific problem.
\newpage
\renewcommand{\theequation}{3.\arabic{equation}}
\setcounter{equation}{0}
\section{The action of the relic gravitons}
\label{sec3}
\subsection{Second-order fluctuations of the Einstein-Hilbert action}
By perturbing the Einstein-Hilbert action, to second-order in the 
amplitude of the tensor fluctuations we have, formally, that:
\begin{equation}
\delta_{\mathrm{t}}^{(2)}S = - \frac{1}{16\pi G} \int d^{4} x [ \delta_{\mathrm{t}}^{(2)} (\sqrt{-g})\, \overline{R} + 
\sqrt{-\overline{g}}\, \delta_{\mathrm{t}}^{(2)} R + \delta_{\mathrm{t}}^{(1)}( \sqrt{-g})\,\,\delta_{\mathrm{t}}^{(1)}R],
\label{SI4a}
\end{equation}
where $\overline{R}$ denotes the background Ricci scalar; $\delta^{(1)} R$ and $\delta^{(2)}R$ denote 
respectively, the first and second-order fluctuations of $R= g^{\mu\nu} R_{\mu\nu}$.
In Eq. (\ref{SI4a}) the possible coupling 
to the anisotropic stress has been neglected. This is customary 
during the early evolution of the geometry since, in the context 
of the $\Lambda$CDM paradigm, during 
the early inflationary phase the sources of anisotropic stress 
can be safely ignored unless the number of effective 
e-folds is close to minimal. 
Later on the anisotropic stress of the fluid plays a role and cannot 
be neglected at least if we aim at a reasonable quantitative 
discussion of the relic graviton spectrum (see also section \ref{sec1} and Fig. \ref{Figure2}).
By introducing the first-order  fluctuations of the background geometry $\overline{g}_{\mu\nu}$ we have that 
\begin{eqnarray}
&& g_{\mu\nu} = \overline{g}_{\mu\nu} + H_{\mu\nu},
\label{G1}\\
&& g^{\mu\nu} = \overline{g}^{\mu\nu} - H^{\mu\nu} + H^{\mu\alpha}H_{\alpha}^{\nu}
\label{G2}\\
&& \sqrt{-g} = \sqrt{-\overline{g}}\biggl[ 1 + \frac{1}{2} \overline{g}^{\rho\alpha} H_{\rho\alpha} 
- \frac{1}{4} \overline{g}^{\rho\alpha} \overline{g}^{\nu\sigma} H_{\alpha\nu} H_{\sigma\rho} +
\frac{1}{8} \overline{g}^{\rho\alpha} \overline{g}^{\nu\sigma} H_{\rho\alpha} H_{\nu\sigma}\biggr].
\label{G3}
\end{eqnarray}
Recalling now Eqs. (\ref{SI1}) and (\ref{SI2}),   Eqs. (\ref{G1})--(\ref{G3}) become
\begin{equation}
\delta_{\mathrm{t}}^{(1)} g^{i j} = \frac{ h^{ij}}{a^2},\qquad 
\delta_{\mathrm{t}}^{(2)} g^{i j} = -  \frac{h^{i}_{k} h^{kj}}{a^2}.
\label{SI3}
\end{equation}
The first- and second-order 
fluctuations of the Christoffel connections are:
\begin{eqnarray}
&& \delta_{\rm t}^{(1)} \Gamma_{i j}^{0} = \frac{1}{2} ( h_{ij}' + 2 {\cal H} h_{ij}),\qquad  \delta_{\rm t}^{(1)} \Gamma_{i 0}^{j} = \frac{1}{2} {h_{i}^{j}}',
\qquad  \delta_{\rm t}^{(1)} \Gamma_{ij}^{k} = 
\frac{1}{2}( \partial_{i} h^{k}_{j} + \partial_{j} h_{i}^{k} - \partial^{k} h_{ij}),
\nonumber\\
&& \delta_{\rm t}^{(2)} \Gamma_{i 0}^{j} = - \frac{1}{2} h^{i k} h_{k j}' ,\qquad \delta_{\rm t}^{(2)} \Gamma_{ij}^{k} = \frac{1}{2} h^{i \ell} [ \partial_{\ell} h_{j k} -
\partial_{k} h_{j\ell} - \partial_{j} h_{k \ell}],
\label{SI5a}
\end{eqnarray}
where the prime denotes a derivation with respect to the conformal time coordinate.
Using the result of Eq. (\ref{SI5a}) the 
 first- and second- order fluctuations of the Ricci tensor can be written in explicit terms:
\begin{eqnarray}
\delta_{\rm t}^{(1)} R_{ij} &=& \frac{1}{2} [h_{ij}'' + 2 {\cal H} h_{ij}' - \nabla^2 h_{ij}] 
+ ({\cal H}' + 2 {\cal H}^2) h_{ij},
\label{rij1}\\
\delta_{\rm t}^{(2)} R_{00} &=& \frac{1}{4} h_{ij}' {h^{ij}}' - \frac{{\cal H}}{2} 
h_{ij} {h^{ij}}' + \frac{1}{2} h^{ij} \nabla^2 h_{ij},
\label{r00}\\
\delta_{\rm t}^{(2)} R_{ij} &=& \frac{1}{2} h^{k \ell} [ \partial_{k} \partial_{\ell} h_{ij} - 
\partial_{k} \partial_{j} h_{\ell i} - \partial_{k} \partial_{i} h_{j \ell}] 
\nonumber\\
&-& \frac{1}{2} \partial_{j} [ h^{k \ell} ( \partial_{\ell} h_{ik} - \partial_{k} h_{\ell i} - 
\partial_{i} h_{k\ell})] - \frac{{\cal H}}{2} h^{k\ell} h_{k \ell}' \delta_{ij}
\nonumber\\
&+& \frac{{\cal H}}{2} h^{\ell}_{j} h_{\ell i}' + \frac{{\cal H}}{2} h^{\ell}_{i} h_{\ell j}' 
- \frac{1}{4} {h^{k}_{j}}' h_{ik}' - \frac{{\cal H}}{2} {h^{k}_{j}}' h_{ik} - 
\frac{1}{4} {h^{k}_{i}}' h_{kj}' - \frac{{\cal H}}{2} {h^{k}_{i}}' h_{kj}
\nonumber\\
&-& \frac{1}{4} [ \partial_{i} h_{k}^{\ell} + \partial_{k} h^{\ell}_{i} - \partial^{\ell}h_{ik}]
[\partial_{\ell} h^{k}_{j} + \partial_{j} h^{k}_{\ell} - \partial^{k}h_{j \ell}].
\label{rij2}
\end{eqnarray}
The Ricci scalar is zero to first order in the tensor fluctuations, i.e. $\delta_{\rm t}^{(1)} R =0$. This is due to the traceless nature of these fluctuations.
To second-order, however, $\delta_{\rm t}^{(2)} R \neq 0$ and its form is:
\begin{eqnarray}
\delta_{\rm t}^{(2)} R &=& \frac{1}{a^2} \biggl\{ \frac{3}{4} h_{k \ell}' {h^{k\ell}}' + 
{\cal H} h_{k\ell}' h^{k\ell} + \frac{1}{2} h^{k\ell} \nabla^2 h_{k\ell} - 
\frac{1}{4}\partial_{i} h^{k\ell}  \partial^{i} h_{k\ell}\biggr\}
\nonumber\\
&+& \frac{1}{a^2} \biggl\{ - \frac{1}{2} \partial_{i}[ h^{k\ell}(\partial_{\ell} h^{i}_{k} 
- \partial_{k} h_{\ell}^{i} - \partial^{i} h_{k\ell})] 
\nonumber\\
&-& \frac{1}{4}[ \partial_{i} h^{\ell}_{k} \partial_{\ell} h^{k}_{i} - \partial_{i} h^{\ell}_{k}
\partial^{k} h_{i\ell} + \partial_{k} h^{\ell i} \partial_{\ell} h^{k}_{i} - \partial^{\ell} h_{ik}  
\partial^{i} h^{k}_{\ell} + \partial^{\ell} h_{ik} \partial^{k} h_{i \ell}]\biggr\}.
\label{r}
\end{eqnarray}
Using the results of Eqs. (\ref{rij1})--(\ref{r}) into Eq. (\ref{SI4a}) the second-order action for the tensor modes 
reads, up total derivatives, 
\begin{equation}
S_{\mathrm{gw}} = \delta_{\mathrm{t}}^{(2)}  S = \frac{1}{8\ell_{\mathrm{P}}^2} \int d^{4} x \sqrt{-\overline{g}} \,\,
\overline{g}^{\mu\nu} \,\,\partial_{\mu} h_{ij} \partial_{\nu} h^{ij}, 
\label{action2}
\end{equation}
where, as already mentioned in section \ref{sec1},
\begin{equation}
\ell_{\mathrm{P}}^2 = 8\pi G = \frac{1}{\overline{M}_{\mathrm{P}}^2} = \frac{8\pi}{M_{\mathrm{P}}^2},\qquad M_{\mathrm{P}} = 1.221\times 10^{19} \, \mathrm{GeV}.
\label{defPL}
\end{equation}

\subsection{Lagrangian densities}
The action (\ref{action2}) 
can be written in various ways which differ by the addition (or subtraction) of a total conformal time derivative.
Recalling the standard notations
\begin{equation}
S_{\mathrm{gw}} = \int d\tau L_{\mathrm{gw}}(\tau), \qquad L_{\mathrm{gw}}(\tau) = \int d^{3} x {\mathcal L}_{\mathrm{gw}}(\vec{x},\tau),
\label{L1}
\end{equation}
the Langrangian density ${\mathcal L}^{(1)}(\vec{x},\tau)$ can be recast in the form 
\begin{equation}
{\mathcal L}_{\mathrm{gw}}^{(1)}(\vec{x},\tau) = \frac{a^{2}(\tau)}{2}[ (\partial_{\tau} h)^2 - \overline{\gamma}^{ij}\nabla_{i} h \nabla_{j} h],
\label{L2}
\end{equation}
where the canonical amplitude $h$ has been introduced
\begin{equation}
h = \frac{h_{\otimes}}{\sqrt{2} \ell_{\mathrm{P}}}= \frac{h_{\oplus}}{\sqrt{2} \ell_{\mathrm{P}}}.
\label{L3}
\end{equation}
By now introducing the canonical amplitude as $a h= \mu$, Eq. (\ref{L1}) can be transformed as 
 \begin{equation}
 {\mathcal L}^{(2)}_{\mathrm{gw}}(\vec{x},\tau) = \frac{1}{2}[(\partial_{\tau} \mu)^2 + {\mathcal H}^2 \mu^2 
 - 2 {\mathcal H} \mu\mu' - \overline{\gamma}^{ij} \nabla_{i} \mu\nabla_{j} \mu].
 \label{L4}
 \end{equation}
 If a total $\tau$ derivative is dropped,  an equivalent form of the Lagrangian density can be obtained 
\begin{equation}
{\mathcal L}_{\mathrm{gw}}^{(3)}(\vec{x},\tau) =  \frac{1}{2} [(\partial_{\tau} \mu)^2 + ({\mathcal H}^2 + 
{\mathcal H}') \mu^2 - \overline{\gamma}^{ij} \nabla_{i}\mu\nabla_{j}\mu].
\label{L5}
\end{equation}
All the three Lagrangian densities of Eqs. (\ref{L2}), (\ref{L4}) and (\ref{L5}) lead to the same Euler-Lagrange equations. 
\subsection{Hamiltonian densities} 
In Eq. (\ref{L2}) the canonical field is $h$ and the canonical momentum is $\Pi = a h'$. 
Conversely, in Eq. (\ref{L4}) the canonical field is $\mu$ and the associated canonical 
momentum is $\tilde{\pi} = \mu' - {\mathcal H} \mu$. Finally, according to Eq. (\ref{L5}) the 
canonical momentum is $\pi = \mu'$ while the canonical field is always $\mu$. 
The three Lagrangian densities of Eqs. (\ref{L2}), (\ref{L4}) and (\ref{L5}) will then lead 
to three corresponding Hamiltonians, i.e. 
\begin{eqnarray}
&& H_{\mathrm{gw}}^{(1)}(\tau) = \frac{1}{2} \int d^{3} x \biggl[ \frac{\Pi^2}{a^2} + a^2 \overline{\gamma}^{ij} 
\nabla_{i} h \nabla_{j} h\biggr],
\label{H1}\\
&& H_{\mathrm{gw}}^{(2)}(\tau) = \frac{1}{2} \int d^{3} x [\tilde{\pi}^2 + 2 {\mathcal H} \mu \tilde{\pi} + 
\overline{\gamma}^{ij} \nabla_{i} \mu \nabla_{j}\mu],
\label{H2}\\
&& H_{\mathrm{gw}}^{(3)}(\tau) =  \frac{1}{2} \int d^{3} x \biggl[ \pi^2 - \frac{a''}{a} \mu^2 + 
\overline{\gamma}^{ij} \nabla_{i} \mu \nabla_{j}\mu\biggr],
\label{H3}
\end{eqnarray}
The Hamiltonians of Eqs. (\ref{H1}), (\ref{H2}) and (\ref{H3}) are related by successive 
canonical transformations. To prove this statement it is enough to show that 
Eq. (\ref{H2}) can be obtained from Eq. (\ref{H1}) by means of an appropriate 
canonical tansformation and that, in turn, Eq. (\ref{H3}) can be obtained from Eq. (\ref{H2})
through another canonical transformation. 
To pass from the Hamiltonian of Eq. (\ref{H1}) to Eq. (\ref{H2}) it is practical to consider 
a generating functional depending upon the new canonical fields (i.e. $\mu$) and upon the old 
canonical momenta (i.e. $\Pi$):
\begin{equation}
{\mathcal G}_{(1)\to (2)}[\mu,\Pi] = - \int\frac{\mu \,\Pi}{a} d^3 x.
\label{1to2}
\end{equation}
By taking the functional derivative of ${\mathcal G}_{(1)\to (2)}[\mu,\Pi] $ with respect to $\Pi$  and 
with respect to $\mu$ we get, up to a sign, the connection between the new and old pivot variables, namely:
\begin{equation}
 h = - \frac{\delta {\mathcal G}_{(1)\to(2)}}{\delta \Pi}  = \frac{\mu}{a}, \qquad \tilde{\pi} = - \frac{\delta {\mathcal G}_{(1)\to(2)}}{\delta \mu}  = \frac{\Pi}{a}.
\label{G12b}
\end{equation}
Since the generating functional depends explicitly upon time, the new Hamiltonian 
will be related to the old one through a partial time derivative of the generating functional, i.e. 
\begin{equation}
H_{\mathrm{gw}}^{(2)}(\tau) = H_{\mathrm{gw}}^{(1)}(\tau)  + \frac{\partial {\mathcal G}_{(1)\to(2)}}{\partial \tau},
\label{H1to2}
\end{equation}
as it can be explicitly verified by using Eqs. (\ref{H1}), (\ref{H2}) and (\ref{1to2}) into Eq. (\ref{H1to2}).
A further canonical transformation allows to go from Eq. (\ref{H2}) to (\ref{H3});  the relevant generating functional  is 
\begin{equation}
{\mathcal G}_{(2)\to (3)}[\mu,\pi] = - \int d^3 x \biggl( \mu \pi - \frac{{\mathcal H}\mu^2}{2}\biggr),
\label{2to3}
\end{equation}
 depending upon the old coordinates (i.e. $\mu$) and upon the new momenta (i.e. 
$\pi$). The relations between the new and old variables are given by
\begin{equation}
\tilde{\pi} = \frac{\delta {\mathcal G}_{(2)\to(3)}}{\delta \mu}  = \pi - {\mathcal H}\mu,
\qquad \mu = \frac{\delta {\mathcal G}_{(2)\to(3)}}{\delta \pi}= \mu,
\label{GEN2}
\end{equation}
stipulating that, in this case, the canonical momentum gets shifted by ${\mathcal H} \mu$  while the canonical field is left 
invariant.  Since the generating functional depends explicitly upon the conformal time coordinate, we will simply have that 
\begin{equation}
H_{\mathrm{gw}}^{(3)}(\tau) = H_{\mathrm{gw}}^{(2)}(\tau) + \frac{\partial {\mathcal G}_{(2)\to(3)}}{\partial \tau}
\label{H2to3}
\end{equation}
as it can be explicitly verified by using Eqs. (\ref{H2}), (\ref{H3}) and (\ref{GEN2}) into Eq. (\ref{H2to3}).

\subsection{Evolution equations in different regimes}
From Eq. (\ref{action2}) the evolution equations of $h_{i}^{j}$ will be given by 
\begin{equation}
{h_{i}^{j}}'' + 2 {\mathcal H} {h_{i}^{j}}' - \nabla^2 h_{i}^{j} =0.
\label{Eq1}
\end{equation}
The canonical field  $h$ (see Eq. (\ref{L3})) will also obey Eq. (\ref{Eq1}). 
The Hamilton equations derived from Eq. (\ref{H1}) read:
\begin{equation}
h' = \frac{\Pi}{a^2},\qquad \Pi' = a^2 \nabla^2 h,
\label{Eq2}
\end{equation}
which has exactly the same content as Eq. (\ref{Eq1}). In similar terms the Hamilton's equations 
can be derived from Eq. (\ref{H2}) and the result is 
\begin{equation}
\mu' = \tilde{\pi} + {\mathcal H} \mu,\qquad \tilde{\pi}' = - {\mathcal H}\tilde{\pi}  + \nabla^2 \mu.
\label{Eq3}
\end{equation}
Bearing in mind that $a h = \mu$, Eqs. (\ref{Eq2}) and (\ref{Eq3}) all reduce to Eq. (\ref{Eq1}) since 
the different Hamiltonians are related by canonical transformations. 
The same conclusion follows by deriving the Hamilton's equations using Eq. (\ref{H3}).  
It is practical, for some applications, to change the time parametrization. For instance, in terms 
of the rescaled time coordinate $\sigma$ we will have that the evolution for the canonical amplitude 
$h$ obeys the simple equation 
\begin{equation}
\frac{\partial^2 h}{\partial \sigma^2} - a^4 \nabla^2 h=0, \qquad a^2(\tau) d\tau = d\sigma.
\end{equation}

Before concluding this section it should be pointed out that Eq. (\ref{Eq1}) is accurate as long as the sources of 
anisotropic stress are totally absent. This approximation is, strictly speaking, unrealistic. Indeed 
we do know that there are sources of anisotropic stress. Typically, after neutrino decoupling, the neutrinos free stream and the effective energy-momentum tensor acquires, to first-order in the amplitude 
of the plasma fluctuations, an anisotropic stress, i.e. 
\begin{equation}
\delta T_{i}^{j} = - \delta p  \delta_{i}^{j} + \Pi_{i}^{j}
\end{equation}
where $\Pi_{i}^{j}$ is the contribution of the anisotropic stress, satisfying $\nabla_{i} \Pi_{j}^{i}=0$ and 
$\Pi_{i}^{i} =0$. The presence of the anisotropic stress clearly affects the evolution of the tensor 
modes. To obtain the wanted equation we perturb the Einstein equations to first-order and we get:
\begin{equation}
{h_{i}^{j}}'' + 2 {\mathcal H} {h_{i}^{j}}' - \nabla^2 h_{i}^{j} = - 16 \pi G a^2 \Pi_{i}^{j}.
\label{Eq1a}
\end{equation}
This form of the evolution equation for the tensor modes is the one 
required to compute the effects related to the finite value 
of the anisotropic stress. 
\newpage
\renewcommand{\theequation}{4.\arabic{equation}}
\setcounter{equation}{0}
\section{Quantization of the tensor modes}
\label{sec4}
There are   analogies between the quantum state of relic gravitons  and the 
quantum treatment of visible light. Quantum effects are not crucial 
 to treat first-order interference of the radiation field (i.e. Young interferometry) \cite{QO1}. 
First-order interference in quantum optics correspond to the calculation of the two-point function 
of the relic gravitons. Quantum effects arise, in optics, from second-order 
interference, i.e. when computing (and measuring) the interference 
between the intensities of the radiation field. Second-order interference 
effects are associated with the possibilities of counting photons and have been 
pioneered by Hanbury-Brown and Twiss in the early fifties \cite{QO1,QO2}.  
Hanbury-Brown-Twiss interferometry is based on photon counting statistics. 

Having said that we are not even close (experimentally) to study graviton counting statistics (as we do it 
with the photons), second order interference effects would allow, in principle, to 
assess the coherence properties of  relic graviton backgrounds.
The quantum state of the relic gravitons can be described in terms 
of a generalized coherent state usually called squeezed state. Squeezed states 
can be described in terms of quadrature operators where one of the 
modes of the radiation field is always broadened by the time evolution, while 
the other one is squeezed. 

\subsection{Heisenberg description}
The quantization of the canonical Hamiltonian of Eq. (\ref{H3})  is 
performed  by promoting the normal modes of the action  to field operators in the Heisenberg 
description and by imposing (canonical) equal-time commutation relations:
\begin{equation}
[ \hat{\mu}(\vec{x},\tau), \hat{\pi}(\vec{y},\tau)] = i \delta^{(3)} (\vec{x} - \vec{y}).
\label{cancomma}
\end{equation}
The operator corresponding to the  Hamiltonian (\ref{H3}) becomes: 
\begin{equation}
\hat{H}(\tau) = \frac{1}{2} \int d^3 x 
\biggl[ \hat{\pi}^2 - \frac{a''}{a} \hat{\mu}^2 
+  (\partial_{i} \hat{\mu})^2\biggr].
\label{ham2a}
\end{equation}
In Fourier space the quantum fields  $\hat{\mu}$ and $\hat{\pi}$ can be expanded as
\begin{eqnarray}
\hat{\mu}(\vec{x},\tau) = \frac{1}{2 (2\pi)^{3/2} } \int d^3 k \biggl[ \hat{\mu}_{\vec{k}} e^{- i \vec{k} \cdot \vec{x} }
+ \hat{\mu}_{\vec{k}}^{\dagger} \, e^{ i \vec{k} \cdot \vec{x} }\biggr],\qquad
\hat{\pi}(\vec{y},\tau) = \frac{1}{2 (2\pi)^{3/2} } \int d^3 p \biggl[ \hat{\pi}_{\vec{p}} e^{- i \vec{p} \cdot \vec{y} }
+ \hat{\pi}_{\vec{p}}^{\dagger} \, e^{ i \vec{p} \cdot \vec{y} }\biggr].
\label{expansiona}
\end{eqnarray}
Demanding the validity of the canonical commutation relations of 
Eq. (\ref{cancomma}),  the Fourier components must obey:
\begin{eqnarray}
&& [ \hat{\mu}_{\vec{k}}(\tau), \hat{\pi}_{\vec{p}}^{\dagger}(\tau) ] = i \delta^{(3)}(\vec{k} - \vec{p}),
\qquad
[ \hat{\mu}_{\vec{k}}^{\dagger}(\tau), \hat{\pi}_{\vec{p}}(\tau) ]= i \delta^{(3)}(\vec{k} - \vec{p}),
\nonumber\\
&& [ \hat{\mu}_{\vec{k}}(\tau), \hat{\pi}_{\vec{p}}(\tau) ]= i \delta^{(3)}(\vec{k} + \vec{p}),
\qquad [ \hat{\mu}_{\vec{k}}^{\dagger}(\tau), \hat{\pi}_{\vec{p}}^{\dagger}(\tau) ]= i \delta^{(3)}(\vec{k} + \vec{p}).
\label{fcomma}
\end{eqnarray}
Inserting now Eq. (\ref{expansiona}) into Eq. (\ref{H3}) the Fourier space representation
of the quantum Hamiltonian \footnote{In order to derive the following equation, 
the relations $\hat{\mu}_{-\vec{k}}^{\dagger} \equiv  \hat{\mu}_{\vec{k}}$ and 
$\hat{\pi}_{-\vec{k}}^{\dagger} \equiv  \hat{\pi}_{\vec{k}}$ should be used .} can be obtained:
\begin{equation}
\hat{H}(\tau) = \frac{1}{4} \int d^3 k \biggl[(\hat{\pi}_{\vec{k}} \hat{\pi}^{\dagger}_{\vec{k}} + 
\hat{\pi}_{\vec{k}}^{\dagger} \hat{\pi}_{\vec{k}}) + \biggl( k^2 - \frac{a''}{a}\biggr) (\hat{\mu}_{\vec{k}} \hat{\mu}^{\dagger}_{\vec{k}} + 
\hat{\mu}_{\vec{k}}^{\dagger} \hat{\mu}_{\vec{k}}) \biggr].
\label{ham3a}
\end{equation} 
The evolution of $\hat{\mu}$ and $\hat{\pi}$  is therefore dictated, in the Heisenberg representation,  by:
\begin{eqnarray}
i \hat{\mu}' = [\hat{\mu},\hat{H}],\qquad
i \hat{\pi}' = [\hat{\pi},\hat{H}],
\label{pieq1}
\end{eqnarray}
where, as usual, units $\hbar=1$ are assumed.
Using now the mode expansion (\ref{expansiona}) and the Hamiltonian in the form (\ref{ham3a})
the evolution for the Fourier components of the operators is 
\begin{equation}
\hat{\mu}_{\vec{k}}'= \hat{\pi}_{\vec{k}},\qquad
 \hat{\pi}_{\vec{k}}' = - \biggl( k^2  - \frac{a''}{a}\biggr) \hat{\mu}_{\vec{k}},
\label{pieq2a}
\end{equation}
implying 
\begin{equation}
\hat{\mu}_{\vec{k}}'' +\biggl[k^2 - \frac{a''}{a}\biggr] \hat{\mu}_{\vec{k}}=0.
\label{thirdform}
\end{equation}
It is not a surprise that the evolution equations of the field operators, in the Heisenberg 
description, reproduces, for $\hat{\mu}_{\vec{k}}$ the classical evolution equation derived 
before.
The general solution of the system is then 
\begin{eqnarray}
&& \hat{\mu}_{\vec{k}}(\tau) = \hat{a}_{\vec{k}}(\tau_0) f_{k}(\tau) + \hat{a}_{-\vec{k}}^{\dagger}(\tau_0) f_{k}^{\ast}(\tau),
\label{solmu}\\
&& \hat{\pi}_{k}(\tau) = \hat{a}_{\vec{k}}(\tau_0) g_{k}(\tau) + \hat{a}_{-\vec{k}}^{\dagger}(\tau_0)g_{k}^{\ast}(\tau),
\label{solpi}
\end{eqnarray}
where the mode functions obey: 
\begin{equation}
f_{k}'= g_{k},\qquad g_{k}' = -  \biggl[ k^2 - \frac{a''}{a}\biggr] f_{k}.
\label{fkeq}
\end{equation}
If the form of the Hamiltonian is different by a time-dependent canonical transformation, also the canonical momenta will differ and, consequently, the relation of $g$ to $f$ may be different. For instance, in the case of the Hamiltonian 
of Eq. (\ref{H2}) we will have, instead, 
\begin{equation}
f_{k}'= g_{k} + {\mathcal H} f_{k}, \qquad g_{k}' = - {\mathcal H} g_{k} - k^2 f_{k}.
\label{fkeq2}
\end{equation}
Consider now the canonical commutation relations expressed by Eq. (\ref{cancomma}).  Using Eqs. (\ref{expansiona})
together with Eqs. (\ref{solmu}) and (\ref{solpi}) into Eq. (\ref{cancomma}), the mode functions 
have to obey the condition:
\begin{equation}
f_{k}(\tau) \,g_{k}^{*}(\tau) - g_{k}(\tau)\, f_{k}^{*}(\tau) = i.
\label{WR}
\end{equation}
Since, by construction, the Hamiltonians of Eqs. (\ref{H2}) and (\ref{H3}) are related 
by canonical transformations, the mode functions of Eqs. (\ref{fkeq}) and (\ref{fkeq2}) 
will have both to obey Eq. (\ref{WR}). In different terms, the commutation relations 
between field operators should be preserved by the time evolution and this is equivalent 
to the Wronskian normalization condition of Eq. (\ref{WR}). 
\subsection{Generalized coherent states of relic gravitons}
Consider the Hamiltonian given in Eq. (\ref{H2}) in the spatially flat case:
\begin{equation}
H_{\mathrm{gw}}(\tau) = \frac{1}{2} \int d^{3} x [\hat{\pi}^2 + {\mathcal H} (\hat{\mu} \hat{\pi} + \hat{\pi} \hat{\mu} )
+ \partial_{i} \hat{\mu} \partial^{i} \hat{\mu}],
\label{heff}
\end{equation}
dropping, for simplicity, the tilde from the momenta. Defining the creation and annihilation operators
\begin{equation}
\hat{a}_{\vec{k}} = \sqrt{\frac{k}{2}} \biggl( \hat{\mu}_{\vec{k}} + \frac{i}{k} \hat{\pi}_{\vec{k}}\biggr), \qquad \hat{a}_{-\vec{k}}^{\dagger}= \sqrt{\frac{k}{2}} \biggl( \hat{\mu}_{\vec{k}} - \frac{i}{k} \hat{\pi}_{\vec{k}}\biggr),
\label{A1}
\end{equation}
and recalling that  $\hat{\mu}_{-\vec{k}}^{\dagger} = \mu_{\vec{k}}$ and $\hat{\pi}_{-\vec{k}}^{\dagger} = \pi_{\vec{k}}$,  
Eq. (\ref{A1}) imply
\begin{equation} 
\hat{\mu}_{\vec{k}} = \frac{\hat{a}_{\vec{k}}+\hat{a}^{\dagger}_{\vec{k}}}{\sqrt{2k}},
\qquad 
\hat{\pi}_{\vec{k}}=-i\sqrt{\frac{k}{2}}\left(\hat{a}_{\vec{k}}-\hat{a}^{\dagger}_{-\vec{k}}\right).
\label{A3}
\end{equation}
Since $\hat{a}_{\vec{k}}(\tau)$ and $\hat{a}_{-\vec{k}}^{\dagger}(\tau)$ obey 
$[ \hat{a}_{\vec{k}}(\tau), \hat{a}_{\vec{p}}^{\dagger}(\tau)] = \delta^{(3)}(\vec{k} - \vec{p})$, 
inserting Eq. (\ref{A3}) into Eq. (\ref{heff}), $\hat{H}_{\mathrm{gw}}$ can be written as
\begin{equation}
\hat{H}_{\mathrm{gw}}(\tau) = \frac{1}{2} \int d^{3} k \{ k[ \hat{a}_{\vec{k}} \hat{a}_{\vec{k}}^{\dagger} +   
\hat{a}_{\vec{k}}^{\dagger} \hat{a}_{\vec{k}}] + i {\mathcal H} (\hat{a}_{\vec{k}}^{\dagger} \hat{a}_{-\vec{k}}^{\dagger} 
- \hat{a}_{\vec{k}} \hat{a}_{-\vec{k}}^{\dagger}|) \}.
\label{A4}
\end{equation}
The evolution of $\hat{a}_{\vec{k}}(\tau)$ and $\hat{a}_{-\vec{k}}^{\dagger}(\tau)$ obeys: 
\begin{equation}
\frac{d \hat{a}_{\vec{k}}}{d\tau} = - i k \hat{a}_{\vec{k}} + {\mathcal H}  \hat{a}_{-\vec{k}}^{\dagger},\qquad \frac{d  \hat{a}_{-\vec{k}}^{\dagger}}{d\tau} =  i k  \hat{a}_{-\vec{k}}^{\dagger} + {\mathcal H} \hat{a}_{\vec{k}}.
\label{A6}
\end{equation}
The solution of Eq. (\ref{A6}) is:
\begin{eqnarray}
\hat{a}_{\vec{k}}(\tau) &=& u_{k}(\tau) \hat{a}_{\vec{k}}(\tau_{0}) + v_{k}^{*}(\tau) \hat{a}_{-\vec{k}}(\tau_{0}), 
\label{A7}\\
\hat{a}_{-\vec{k}}^{\dagger}(\tau) &=& v_{k}^{*}(\tau) \hat{a}_{\vec{k}}(\tau_{0}) + u_{k}^{*}(\tau) \hat{a}^{\dagger}_{-\vec{k}}(\tau_{0}),
\label{A8}
\end{eqnarray} 
where $\tau_{0}$ is the initial integration time. The unitarity of the time evolution demands 
that $|u_{k}(\tau)|^2 - |v_{k}(\tau)|^2 = 1$. A useful parametrization of $u_{k}(\tau)$ and $v_{k}(\tau)$ is given in terms of a real amplitude and two phases as:
\begin{equation}
u_{k}(\tau) = e^{-i \vartheta_{k}(\tau)} \cosh{r_{k}(\tau)}, \qquad v_{k}(\tau) = e^{-i \vartheta_{k}(\tau) - 2i \varphi_{k}(\tau)} 
\sinh{r_{k}(\tau)}. 
\label{A9}
\end{equation}
Equation  (\ref{A6})  determine the evolution equations for $u_{k}(\tau)$ and $v_{k}(\tau)$. Using then Eq. (\ref{A9}) the evolution equations for $r_{k}$, $\varphi_{k}$ and $\vartheta_{k}$ can be obtained:
\begin{eqnarray}
&& r_{k}' = {\mathcal H} \cos{2 \varphi_{k}}, 
\label{A10}\\
&& \varphi_{k}' = - k - {\mathcal H} \frac{\sin{2 \varphi_{k}}}{\tanh{2 r_{k}}}, 
\label{A11}\\
&& \vartheta_{k}' = k + {\mathcal H} \tanh{r_{k}} \sin{2 \varphi_{k}}.
\label{A12}
\end{eqnarray}
Note that $\vartheta_{k}$ does not appear neither in Eq. (\ref{A10}) nor in Eq. (\ref{A11}). 
It is interesting, at this point, to compute the two-point functions connected with the two 
canonically conjugate operators, i.e. $\hat{\mu}(\vec{x},\tau)$ and $\hat{\pi}(\vec{x},\tau)$. In terms of the creations and annihilations operators defined in Eqs. (\ref{A1}) and (\ref{A3})--(\ref{A4}) 
the canonically conjugate operators can be written as 
\begin{eqnarray}
&& \hat{\mu}(\vec{x},\tau) = \frac{1}{(2\pi)^{3/2}} \int \frac{d^{3} k}{\sqrt{ 2 k}} [ \hat{a}_{\vec{k}}(\tau) e^{- i \vec{k}\cdot{\vec{x}}} 
+  \hat{a}_{\vec{k}}^{\dagger}(\tau) e^{- i \vec{k}\cdot{\vec{x}}}],
\label{A13}\\
&& \hat{\pi}(\vec{x},\tau) = -\frac{i}{(2\pi)^{3/2}} \int d^{3} k \sqrt{\frac{k}{2}} [ \hat{a}_{\vec{k}}(\tau) e^{- i \vec{k}\cdot{\vec{x}}} 
-  \hat{a}_{\vec{k}}^{\dagger}(\tau) e^{- i \vec{k}\cdot{\vec{x}}}].
\label{A14}
\end{eqnarray}
There is a slight difference in the normalizations adopted between Eqs. (\ref{solmu})--(\ref{solpi})
 and Eqs. (\ref{A13})--(\ref{A14}). This difference is due to the fact that, in Eqs. (\ref{solmu})--(\ref{solpi})
 the mode functions $f_{k}$ are normalized, asymptotically, in such a way that $f_{k} \to 1/\sqrt{2 k}$. 
 In Eqs. (\ref{A1})--(\ref{A3}) the factors $\sqrt{2 k}$ and $\sqrt{k/2}$ have been included in the definition creation and annihilation operators.
 
After simple calculations the two-point functions of the field operators and of their 
related canonical momenta becomes :
\begin{eqnarray}
&& \langle 0| \hat{\mu}(\vec{x},\tau)  \hat{\mu}(\vec{y},\tau) |0 \rangle = \int_{0}^{\infty} \frac{k^2}{4\pi^2} d\ln{k} \frac{\sin{kr}}{kr} [ 
|u_{k}(\tau)|^2 + |v_{k}(\tau)|^2 + u_{k}(\tau)v_{k}^{*}(\tau) + u_{k}^{*}(\tau) v_{k}(\tau)],
\label{A15}\\
&&  \langle 0| \hat{\pi}(\vec{x},\tau)  \hat{\pi}(\vec{y},\tau) |0 \rangle = \int_{0}^{\infty} \frac{k^4}{4\pi^2} d\ln{k} \frac{\sin{kr}}{kr} [ 
|u_{k}(\tau)|^2 + |v_{k}(\tau)|^2 - u_{k}(\tau)v_{k}^{*}(\tau) - u_{k}^{*}(\tau) v_{k}(\tau)],
\label{A16}
\end{eqnarray}
where $r=|\vec{x} - \vec{y}|$. Again, as already remarked, the non-strandard pre-factors apperaing in the Fourier amplitudes of Eqs. (\ref{A15}) and (\ref{A16}) are a consequence of the normalizations of Eq. (\ref{A1}). 
In the limit $|\vec{x} - \vec{y}|\to 0$ (and making use of the definitions of Eq. (\ref{A9})), Eqs. (\ref{A15}) and (\ref{A16}) lead to 
\begin{eqnarray}
&& \langle 0| [\hat{\mu}(\vec{x},\tau)]^2 |0 \rangle = \int_{0}^{\infty} \frac{k^2}{4\pi^2} d\ln{k} [ 
\cosh{ 2 r_{k}(\tau)} + \sinh{2 r_{k}(\tau)}\, \cos{2 \varphi_{k}(\tau)}],
\label{A17}\\
&&  \langle 0| [\hat{\pi}(\vec{x},\tau)]^2  |0 \rangle = \int_{0}^{\infty} \frac{k^4}{4\pi^2} d\ln{k} [ 
\cosh{ 2 r_{k}(\tau)} - \sinh{2 r_{k}(\tau)}\, \sin{2 \varphi_{k}(\tau)}].
\label{A18}
\end{eqnarray}
Equations (\ref{A17}) and (\ref{A18}) show that the canonical field 
is broadened while the conjugate momentum gets squeezed by keeping constant the product of their respective root mean squares. The latter behaviour is evident as soon as the relevant wavelengths are larger than the Hubble radius. To demonstrate the two previous statements consider, indeed, Eqs. (\ref{A10}) and (\ref{A11}).
Their solution for wavelengths larger than the Hubble radius (i.e. to leading order 
in $k\tau < 1$) is:
\begin{equation}
r_{k}(\tau) \simeq \ln{a} + {\mathcal O}(k\tau),\qquad \varphi_{k}(\tau) \simeq {\mathcal O}(k\tau).
\label{A19}
\end{equation}
Using Eq. (\ref{A19}) into Eqs. (\ref{A17}) and (\ref{A18}) we do get 
\begin{eqnarray}
&& \Delta \hat{\mu}_{k} = \frac{1}{\sqrt{2}}[\cosh{ 2 r_{k}(\tau)} + \sinh{2 r_{k}(\tau)}\, \cos{2 \varphi_{k}(\tau)}] =  \frac{1}{\sqrt{2}}e^{2 r_{k}(\tau)},
\label{A20}\\
&& \Delta \hat{\pi}_{k} =  \frac{1}{\sqrt{2}}[\cosh{ 2 r_{k}(\tau)} - \sinh{2 r_{k}(\tau)}\, \cos{2 \varphi_{k}(\tau)}] =   \frac{1}{\sqrt{2}}e^{- 2 r_{k}(\tau)}.
\label{A21}
\end{eqnarray}
The rationale for the behaviour exhibited by Eq. (\ref{A19}) can be understood also from a slightly different 
perspective. The relation between $u_{k}(\tau)$ and $v_{k}(\tau)$ and the tensor mode functions $f_{k}(\tau)$ 
and $g_{k}(\tau)$ is 
\begin{equation}
f_{k}(\tau) = \frac{u_{k}(\tau) + v_{k}(\tau)}{\sqrt{2 k}}, \qquad g_{k}(\tau) = \sqrt{\frac{k}{2}} [ u_{k}(\tau) - v_{k}(\tau)].
\label{A22}
\end{equation}
where the evolution of $f_{k}(\tau)$ and $g_{k}(\tau)$ is obtained by solving 
Eq. (\ref{fkeq2}). Indeed, by solving Eq. (\ref{fkeq2}) in the limit $k^2 \ll ({\mathcal H}^2 + {\mathcal H}')$  
we get
\begin{equation}
f_{k}(\tau) \simeq A_{1}(k) a(\tau) + A_{2}(k) a(\tau) \int^{\tau} \frac{d\tau'}{a^2(\tau')}, \qquad g_{k}(\tau) = f_{k}'(\tau) - 
{\mathcal H} f_{k}.
\label{A23}
\end{equation}
From Eq. (\ref{A23}) the first derivative of $f_{k}$ with respect to $\tau$ is nothing but 
\begin{equation}
f_{k}'(\tau) = {\mathcal H} f_{k}(\tau) + \frac{A_{2}(k)}{a(\tau)}.
\label{A24}
\end{equation}
By computing $g_{k}(\tau)$ from Eq. (\ref{A23}) it is clear that, in the limit 
$k \tau \ll 1$
\begin{equation}
f_{k}(\tau) \simeq A_{1}(k) a(\tau), \qquad g_{k}(\tau) \simeq \frac{A_{2}(k)}{a(\tau)}, 
\label{A25}
\end{equation}
which has exactly the same physical content of Eqs. (\ref{A20}) and (\ref{A21}).
 When the Universe expands, $g_{k}(\tau)$ decreases 
and that the solution associated with $A_{2}(k)$ becomes progressively subleading. 
However, this observation does not imply that $g_{k}(\tau)$ disappears since 
the evolution must be unitary. This feature of squeezed quantum state 
suggests the possibility of associating an effective entropy to the process of graviton production\cite{QC1,QC2,QC3}. 

In the Schr\"odinger description quantum state of the relic gravitons is closely related 
to the squeezed states of the radiation field \cite{LG} (see also \cite{STO,YU}). 
Defining, for practical reasons, 
\begin{equation}
x_{k} = \hat{\mu}_{\vec{k}}, \qquad \tilde{x}_{k} = \hat{\mu}_{-\vec{k}},\qquad p_{k} = \hat{\pi}_{\vec{k}}, \qquad 
\tilde{p}_{k} = \hat{\pi}_{-\vec{k}},
\label{A26}
\end{equation}
the wavefunction of the ground state will be, for a given mode $k$,
\begin{equation}
\Psi_{k}(x,\tilde{x}) = \langle x\,\tilde{x}| \Sigma_{k}(r_{k},\varphi_{k})|0\, \rangle = \sqrt{\frac{\sigma_{k}}{\pi}} 
e^{ - \frac{\sigma_{k}}{2} 
(x^2 + \tilde{x}^2)},\qquad \sigma_{k} = e^{- 2 r_{k}},
\label{A27}
\end{equation}
where, as previously discussed, $\varphi_{k} \to 0$ has been assumed. In Eq. (\ref{A27}) 
$\Sigma_{k}(r_{k}, \varphi_{k})$ is the so-called two-mode squeezing operator \cite{REVSQ1,REVSQ2} which will now be written 
in its generic form:
\begin{equation}
\Sigma_{k}(r_{k},\varphi_{k}) = \exp{[ r_{k}(\tau)( e^{- 2\,i\,\varphi_{k}} \hat{a}_{\vec{k}} \, \hat{a}_{-\vec{k}} - e^{2\,i\,\varphi_{k}}
 \hat{a}_{\vec{k}}^{\dagger} \, \hat{a}_{-\vec{k}}^{\dagger})]}
\label{A28}
\end{equation}
where the creation and destruction operators are the ones computed in $\tau_{0}$, i.e. $\hat{a}_{\vec{k}}\equiv 
\hat{a}_{\vec{k}}(\tau_{0})$ by definition of Schr\"odinger description. The state $|0\,\rangle$ is 
annihilated both by $\hat{a}_{\vec{k}}$ and by $\hat{a}_{-\vec{k}}$. These two-modes appear simultaneously 
since gravitons are produced from the vacuum whose total momentum vanish. 
The $x$ and $\tilde{x}$ have been dubbed, in the literature, as superfluctuant operators (see, e. g., \cite{QC1,QC2,QC3}).

The statistical properties of squeezed states can be addressed by employing a 
useful analogy with quantum optics.  The idea is to pretend to resolve single gravitons and to study 
the statistical properties of the second-order interference effects. In the case 
of the gravitons this problem is addressed by defining the (normalized) Glauber correlation 
function not for the photons (as customarily done) but for the gravitons.  
For simplicity let us consider a single mode of the field. In this case 
the normalized Glauber intensity correlation can be written as 
\begin{equation}
g^{(2)}(t, t +\Delta t)= \frac{\langle\,:\,\hat{I}(t + \Delta t) \hat{I}(t)\,:\, \rangle}{\langle \,:\,\hat{I}(t)^2\,:\,\rangle},
\label{GL1}
\end{equation}
the colons denote normal ordering and $\hat{I}$ denotes the operator corresponding to the intensity 
of the radiation field. The normal ordering is related to the fact that, in the optical domain, most 
measurements of the electromagnetic field are based on the absorption of photons via the 
photoelectric effect \footnote{Needless to say that there is no analog of 
photoelectric detection for (single) relic gravitons. In this sense the following considerations should be regarded as a conditional predictions based on the analogy 
between squeezed states of photons and squeezed states of gravitons.}. In the case of a single mode of the field Eq. (\ref{GL1}) can also be written as 
\begin{equation}
g^{(2)}(t,t) = g^{(2)}(0) = \frac{\langle \hat{a}^{\dagger} \hat{a}^{\dagger} \hat{a} \hat{a} \rangle}{\langle \hat{a} \hat{a} \rangle^2}.
\label{GL2}
\end{equation}
In Eq. (\ref{GL2})  the normalized two point function is written for coincident spatial points (i.e. $\Delta t=0$).
The Hanbury-Brown and Twiss experiment, in some sense, probes directly the properties 
of $g^{(2)}(0)$. In the case of coherent states, $g^{2}(0)=1$: this is the case when the photoelectric counts 
obey a Poisson statistics. In the case of chaotic  light, the joint detection probability greater than that for two independent events. This can be verified from Eq. (\ref{GL2}) by assuming that the state of the photons/gravitons is given by a 
thermal mixture. Equation (\ref{GL2}) can also be recast in the form 
\begin{equation}
g^{(2)}(0) -1 = \frac{\langle(\Delta \hat{n})^2\rangle - \langle \hat{n} \rangle}{\langle \hat{n}\rangle^2},
\label{GL3}
\end{equation}
where $\langle(\Delta \hat{n})^2\rangle = \langle \hat{n}^2 \rangle - \langle \hat{n}\rangle^2$
 is the photon number variance and $\hat{n} = \hat{a}^{\dagger} \hat{a}$.  From Eq. (\ref{GL3}) 
 it is also customary to define the Mandel's Q-parameter, i.e. 
 \begin{equation}
 Q = \langle \hat{n}\rangle [ g^{(2)}(0) -1],
\label{GL4}  
 \end{equation}  
 which vanishes exactly in the case of a coherent state. The latter statement can be easily appreciated since, for 
 a coherent state, $\hat{a} |\alpha\rangle = \alpha  |\alpha\rangle$. Thus, $\langle(\Delta \hat{n})^2\rangle = |\alpha|^2 
= \langle \hat{n} \rangle$.  In the case of chaotic (thermal) light it turns out that $g^{(2)}(0) = 2$. This result 
can be  easily drived by using the so-called Glauber-Sudarshan $P$-representation of the density matrix, i.e. 
\begin{equation}
\hat{\rho} = \int d^{2} \alpha P(\alpha) |\alpha \rangle \langle \alpha |, \qquad P(\alpha) = \frac{1}{\pi\, \overline{n}} e^{- |\alpha|^2 /\overline{n}},
\label{GL5}
\end{equation}
where $\overline{n}$ is the Bose-Einstein occupation number.  In the $P$-representation 
we have that 
\begin{equation}
g^{(2)}(0)- 1= \frac{\int P(\alpha)[ |\alpha|^2 - \langle |\alpha|^2\rangle]^2 \, d^{2} \alpha}{[\int P(\alpha) |\alpha|^2 \, d^{2} \alpha]^2},
\label{GL6}
\end{equation}
where $\langle |\alpha|^2 \rangle = \int d^{2}\alpha |\alpha|^2\, P(\alpha)$.  By performing the required 
integrations it is easy to show that $g^{(2)}(0) -1 = 1$, i.e. $g^{(2)}(0) = 2$. 
So far it has been shown that while a purely coherent state implies $g^{(2)}=1$ (i.e. Poissonian statistic)
a thermal state implies that $g^{(2)}(0) = 2$. In the case of the squeezed states it  can be shown 
that
\begin{equation}
g^{(2)}(0) = 3 + \frac{1}{\langle \hat{n}\rangle }, \qquad Q = 2 \langle \hat{n} \rangle + 1,
\label{GL7}
\end{equation}
where $ \langle \hat{n} \rangle = \sinh^2{r_{k}}$ is the multiplicity. 
The coherent state leads to a radiation field with Poissonian statistics. Thermal states (as well as squeezed 
states) have a statistics which is, according to the quantum optical terminology, 
superpoissonian. The latter statement is often dubbed by saying that if $g^{(2)}(0) > 1$ photons are 
bunched while, in the opposite case (i.e. $g^{(2)}(0) < 1$) the photons are said to be 
anti-bunched.  The quantum optical language is much more effective for a mathematical description of the semi-classical limit than the usual considerations related to the limit $\hbar\to 0$. Squeezed states
are genuine quantum states with many particles. They are, in some sense, like coherent states with the crucial 
difference that their statistics is super-Poissonian.  The possibility of scrutinizing 
the statistical properties of many-gravitons systems would rely on our ability of resolving 
single gravitons which is not even close to the present technological capabilities.  
\newpage
\renewcommand{\theequation}{5.\arabic{equation}}
\setcounter{equation}{0}
\section{Relic graviton backgrounds: observables}
\label{sec5}
In the literature  relic graviton backgrounds are characterized in terms of different quantities and, in 
particular, the most common ones are\footnote{It is understood that all the mentioned quantities can be expressed either in terms 
of the wave-number or in terms of the frequency since $k = 2 \pi \nu$.}:
\begin{itemize}
\item{} the power spectrum ${\mathcal P}_{\mathrm{T}}(k,\tau)$;
\item{} the spectral energy density of the relic gravitons  $\Omega_{\mathrm{GW}}(k,\tau)$;
\item{} the spectral amplitude ${\mathcal S}_{h}(\nu,\tau)$.
\end{itemize}
The three listed variables can be related in different regimes. For instance the power spectrum 
has a simple relation to the spectral energy density when the relevant wavelengths 
are inside  the Hubble radius. In section \ref{sec6} it will be argued that, for numerical applications, 
the transfer function for the spectral energy density is more practical to compute than the 
transfer function for the power spectrum or for the spectral amplitude itself.
The power spectrum is actually a strongly oscillating function of the conformal 
time coordinate for wavelengths shorter than the Hubble radius (i.e. $k \tau >1$); in the same limit 
the spectral energy density is asymptotically constant.

\subsection{The tensor power spectrum and the spectral amplitude}
The two-point function of the tensor modes of the geometry is defined as:
\begin{equation}
\langle \hat{h}_{ij}(\vec{x}, \tau) \hat{h}_{ij}(\vec{y},\tau) \rangle
\equiv \langle 0| \hat{h}_{ij}(\vec{x}, \tau) \hat{h}_{ij}(\vec{y},\tau) |0
\rangle,
\label{EXPEC1}
\end{equation}
where the state $|0\rangle$ is annihilated by $\hat{a}_{k,\lambda}$
for $\lambda = \otimes,\,\oplus$. Recalling Eqs. (\ref{SI4}) and (\ref{expansiona}), 
$\hat{\mu}_{ij} = a \hat{h}_{ij}$, the expansion of $\hat{h}_{ij}$ will then be: 
\begin{equation}
 \hat{h}_{ij}(\vec{x},\tau)= \frac{\sqrt{2} \ell_{\mathrm{P}}}{(2\pi)^{3/2}} \sum_{\lambda = \oplus,\,\otimes}
  \int d^{3}k\,\, \epsilon_{ij}^{(\lambda)}(\hat{k})\, 
 \biggl[ \hat{a}_{\vec{k},\lambda} \,F_{k,\lambda}(\tau) e^{- i \vec{k} \cdot \vec{x}} +  \hat{a}_{\vec{k},\lambda}^{\dagger}\, F_{k,\lambda}^{*}(\tau) 
 e^{ i \vec{k} \cdot \vec{x}}\biggr],
 \label{EXPEC2}
 \end{equation}
 where  $\epsilon_{ij}(\hat{k})$ has been defined in Eqs. (\ref{SI5})--(\ref{SI6})  and where 
\begin{equation} 
[\hat{a}_{\vec{k},\lambda}, \hat{a}^{\dagger}_{\vec{p},\lambda'}] = \delta^{(3)}(\vec{k} - \vec{p})
\delta_{\lambda\lambda'}.
\label{EXPEC3}
\end{equation}
In Eq. (\ref{EXPEC2}) $F_{k,\lambda}$ (and the associated 
$G_{k,\lambda}$) are the (complex) tensor mode functions obeying
\begin{equation}
F'_{k,\lambda} = G_{k,\lambda}, \qquad G_{k,\lambda}' = - 2 {\mathcal H} G_{k,\lambda} - k^2 F_{k,\lambda}.
\label{EXPEC4}
\end{equation}
It is immediate to realize that 
\begin{equation}
F_{k,\lambda} = \frac{f_{k,\lambda}}{a}, \qquad G_{k,\lambda} = 
\frac{1}{a} (g_{k,\lambda} - {\mathcal H} f_{k,\lambda}),
\label{EXPEC5}
\end{equation}
where $f_{k,\lambda}$ and $g_{k,\lambda}$ have been 
introduced, for a single graviton polarization, in Eqs. (\ref{solpi})--(\ref{fkeq}). 
After computing the expectation value we get:
\begin{equation}
\langle 0| \hat{h}_{i}^{j}(\vec{x},\tau) \hat{h}_{j}^{i}(\vec{y},\tau) |0\rangle =
\frac{8 \ell_{\mathrm{P}}^2}{a^2} \int \frac{d^3 k}{(2\pi)^3} |f_{k}(\tau)|^2 e^{ - i \vec{k}\cdot\vec{r}}.
\label{EXtens2}
\end{equation}
which, thanks to spatial isotropy, can also be written as: 
\begin{equation}
\langle 0| \hat{h}_{i}^{j}(\vec{x},\tau) \hat{h}_{j}^{i}(\vec{y},\tau) |0\rangle = 
\int d\ln{k} {\mathcal P}_{\mathrm{T}}(k,\tau) \frac{\sin{kr}}{kr},
\label{EXPEC7}
\end{equation}
where ${\mathcal P}_{T}(k,\tau)$ is, by definition, the tensor power spectrum:
\begin{equation}
 {\mathcal P}_{\mathrm{T}}(k,\tau) = 4 \ell_{\mathrm{P}}^2 \frac{k^3}{\pi^2} |F_{k}(\tau)|^2,\qquad F_{k}(\tau) 
= \frac{f_{k}(\tau)}{a(\tau)}.
\label{EXtens4}
\end{equation}
In Eq. (\ref{EXPEC7}),  ${\mathcal P}_{\mathrm{T}}(k,\tau) $ is the tensor power spectrum which has been 
implicitly introduced in Eq. (\ref{int1}) when talking about 
the customary parametrization of tensor power spectra in the 
context of the $\Lambda$CDM model.  
It is often useful, for practical applications, to consider 
$h_{ij}(\vec{x}, \tau)$ as a classical field 
characterized by a  Fourier amplitude 
obeying a specific stochastic average. Then we can write 
\begin{equation}
h_{ij}(\vec{x},\tau) = \frac{1}{(2\pi)^{3/2}} \int d^{3} k h_{ij}(\vec{k},\tau) e^{- i \vec{k} \cdot \vec{x}}, \qquad h_{ij}(\vec{k},\tau) = 
h_{ij}^{*}( - \vec{k},\tau),
\label{EXPEC8}
\end{equation}
where the Fourier amplitudes obey:
\begin{equation}
\langle h_{ij}^{*}(\vec{k},\tau) h_{ij}(\vec{p},\tau) \rangle = 
\frac{2 \pi^2}{k^3} {\mathcal P}_{\mathrm{T}}(k,\tau) \delta^{(3)}(\vec{k} - \vec{p}).
\label{EXPEC9}
\end{equation}
Following the suggestions of \cite{B1}, it is useful to introduce, for some applications 
 the stochastic fields
\begin{equation}
e_{1}(\vec{k}) = \frac{e_{\oplus}(\vec{k}) - i e_{\otimes}(\vec{k})}{\sqrt{2}}, \qquad 
e_{2}(\vec{k}) = \frac{e_{\oplus}(\vec{k}) + i e_{\otimes}(\vec{k})}{\sqrt{2}}.
\label{STa1}
\end{equation}
Equation (\ref{EXPEC9}) implies that 
\begin{eqnarray}
&& 2 \langle e^{*}_{1}(\vec{k})\, e_{1}(\vec{p}) \rangle = 2 \langle e^{*}_{2}(\vec{k})\, e_{2}(\vec{p}) \rangle =
\frac{2\pi^2}{k^3} {\mathcal P}_{\mathrm{T}}(k) \delta^{3} (\vec{k} - \vec{p}),
\label{STa2}\\
&& \langle e^{*}_{1}(\vec{k})\, e_{2}(\vec{p}) \rangle = \langle e^{*}_{2}(\vec{k})\, e_{1}(\vec{p}) \rangle =0,
\label{STa3}
\end{eqnarray}
where ${\mathcal P}_{\mathrm{T}}(k)$ denotes the tensor power spectrum and where 
the factor $2$ in front of the averages arises as a consequence of the $\sqrt{2}$ appearing in Eq. (\ref{STa1}).
In Eqs. (\ref{STa1}), (\ref{STa2}) and (\ref{STa3}) the conformal time coordinate is absent. In Eq. (\ref{EXPEC9}) 
the conformal time appears explicitly. Indeed, Eqs.  (\ref{STa1}), (\ref{STa2}) and (\ref{STa3}) 
tacitly assume that $h_{\oplus}(\vec{k},\tau) = e_{\oplus}(\vec{k}) T_{e}(k,\tau)$ and that $h_{\otimes}(\vec{k},\tau) = e_{\otimes}(\vec{k}) T_{e}(k,\tau)$. This factorization is related to the concept 
of transfer function for the amplitude which will be discussed in section \ref{sec6}. 
The decomposition of Eqs.  (\ref{STa1}), (\ref{STa2}) and (\ref{STa3}) 
is useful when all the polarization have to be treated simultaneously typically in problems involving 
long wavelength gravitons (see Eqs.  (\ref{Boltz73})--(\ref{Boltz74}) and discussion therein). Furthermore 
the decomposition of Eqs.  (\ref{STa1}), (\ref{STa2}) and (\ref{STa3}) allows to factorize the dependence upon the 
initial spectrum which is useful for numerical applications. 

It is finally common to characterize the stochastic backgrounds 
of relic gravitons in terms of the so-called spectral amplitude. The definition of the 
spectral amplitude can be read-off from the definition of  ${\mathcal P}_{\mathrm{T}}(k,\tau)$. More precisely, 
taking the limit $\vec{x} \to \vec{y}$ in Eq. (\ref{EXtens2}) we will have that  
\begin{equation}
\langle 0| \hat{h}_{i}^{j}(\vec{x},\tau) \hat{h}_{j}^{i}(\vec{x},\tau) |0\rangle =\int d\ln{k} {\mathcal P}_{\mathrm{T}}(k) \equiv 
4 \int \nu ~{\mathcal S}_{h}(\nu)~  d\ln{\nu} ,
\label{T11}
\end{equation}
where the second equivalence defines the spectral amplitude ${\mathcal S}_{h}(\nu)$ by recalling, once more, that 
the comoving wavenumber is related to the comoving frequency as $k = 2 \pi \nu$. 

\subsection{Energy-momentum tensors for the relic gravitons}
According to Eq. (\ref{action2}), 
each polarization of the graviton obeys the action of a minimally coupled scalar field. 
This observation was discussed, in particular, in \cite{ford1,ford2} by Ford 
and Parker (see also, e.g. \cite{flat6}). Following then Refs. \cite{ford1,ford2} and within our set of notations 
the energy-momentum tensor of the relic gravitons becomes 
\begin{equation}
T_{\mu}^{\nu} = \frac{1}{4 \ell_{\mathrm{P}}^2} \biggl[ \partial_{\mu} h_{ij} \partial^{\nu} h^{ij} - \frac{1}{2} \delta_{\mu}^{\nu} \overline{g}^{\alpha\beta} \partial_{\alpha} h_{ij} \partial_{\beta}h^{ij} \biggr],
\label{ENM0}
\end{equation}
which can be obtained from the action of the gravitons by taking the functional 
derivative with respect to $\overline{g}_{\mu\nu}$. By making explicit the 
sum over the polarizations, Eq. (\ref{ENM0}) becomes:
\begin{equation}
T_{\mu}^{\nu} = \frac{1}{2\ell_{\mathrm{P}}^2} \sum_{\lambda} \biggl[ \partial_{\mu} h_{(\lambda)}
 \partial^{\nu} h^{(\lambda)} - \frac{1}{2} \overline{g}^{\alpha\beta} \partial_{\alpha} h_{(\lambda)}
 \partial_{\beta} h_{(\lambda)} \delta_{\mu}^{\nu}\biggr],
 \label{ENM1}
 \end{equation}
 i.e., even more explicitly, 
 \begin{equation}
T_{\mu}^{\nu} = \frac{1}{2\ell_{\mathrm{P}}^2} \biggl[ \partial_{\mu} h_{\otimes}
 \partial^{\nu} h^{\otimes} - \frac{1}{2} \overline{g}^{\alpha\beta} \partial_{\alpha} h_{\otimes}
 \partial_{\beta} h_{\otimes} \delta_{\mu}^{\nu}\biggr] + 
 \frac{1}{2\ell_{\mathrm{P}}^2} \biggl[ \partial_{\mu} h_{\oplus}
 \partial^{\nu} h^{\oplus} - \frac{1}{2} \overline{g}^{\alpha\beta} \partial_{\alpha} h_{\oplus}
 \partial_{\beta} h_{\oplus} \delta_{\mu}^{\nu}\biggr]. 
 \label{ENM2}
 \end{equation}
 By now using the same rescalings defined in Eq. (\ref{L3}) for $h_{\otimes}$ and 
 $h_{\oplus}$ in terms of the canonical amplitude $h$ we do get, from Eq. 
 (\ref{ENM2}), 
 \begin{equation}
 T_{\mu}^{\nu} = 2 \biggl[ \partial_{\mu} h
 \partial^{\nu} h - \frac{1}{2} \overline{g}^{\alpha\beta} \partial_{\alpha} h
 \partial_{\beta} h \delta_{\mu}^{\nu}\biggr].
 \label{ENM3}
 \end{equation}
 From Eq. (\ref{ENM3}) it is clear that the total energy momentum 
pseudo-tensor summed over the 
two polarizations of the graviton is twice the energy-momentum tensor 
of a minimally coupled scalar degree of freedom provided 
the amplitudes of the two polarizations are defined in terms of $h$ as in Eq. (\ref{L3}).
The $(00)$ and $(ij)$ components 
of the energy-momentum pseudo-tensor of Eq. (\ref{ENM3}) are:
\begin{eqnarray}
&& T_{0}^{0} = \frac{1}{a^2} [ (\partial_{\tau} h)^2 + 
(\partial_{m} h)^2],
\label{ENM4a}\\
&& T_{i}^{j} = - \frac{1}{a^2} [(\partial_{\tau} h)^2 - 
(\partial_{m} h)^2] \delta_{i}^{j} - \frac{2}{a^2} \partial_{i} h\partial^{j} h,
\label{ENM4}
\end{eqnarray}
where $\partial_{\tau}$ denotes a derivation with respect to the conformal time coordinate $\tau$.
Since, by definition, 
\begin{equation}
T_{0}^{0} = \rho_{\mathrm{GW}}^{(1)}, \qquad T_{i}^{j} = - p_{\mathrm{GW}}^{(1)}\delta_{i}^{j} + \overline{\Pi}_{i}^{j},
\label{ENM5}
\end{equation}
the energy density and pressure of the relic gravitons can then be written as 
 \begin{eqnarray}
&& \rho_{\rm gw}^{(1)} = \frac{1}{a^2} [ (\partial_{\tau} h)^2 + (\partial_{m} h)^2],
 \label{ENM6}\\
 && p_{\rm gw}^{(1)} = \frac{1}{a^2} \biggl[ (\partial_{\tau} h)^2 -
 \frac{1}{3} (\partial_{m} h)^2 \biggr],
 \label{ENM7}\\
 &&  \overline{\Pi}_{i}^{j} = - \frac{2}{a^2}
 \biggl[ \partial_{i} h \partial^{j} h - \frac{1}{3} (\partial_{m} h)^2 \delta_{i}^{j} 
 \biggr].
\label{ENM8}
 \end{eqnarray}
The superscript in the energy density and pressure (i.e. $\rho_{\mathrm{GW}}^{(1)}$ and $p_{\mathrm{GW}}^{(1)}$) is convenient since different prescriptions for assigning the energy-momentum 
pseudo-tensor will be compared in a moment.

The other possible definition of the energy-momentum pseudo-tensor stems from the generalization, to curved 
space-times, of the usual flat space-time approach \cite{landau} (see also \cite{isacson1,isacson2}).  
The nonlinear corrections to the Einstein tensor, will consist, to lowest order,  of quadratic combinations of $h_{ij}$ that can be formally expressed as 
\begin{equation}
\ell_{\rm P}^2 {\cal T}_{\mu}^{\nu} = - \delta^{(2)}_{\rm t} {\cal G}_{\mu}^{\nu},
\label{ps1}
\end{equation}
where the superscript at the right hand side denotes the second-order 
fluctuation of the corresponding quantity while 
the subscript refers to the tensor nature of the fluctuations. 
This procedure is essentially the one described in 
\cite{isacson1,isacson2} and has been re-explored, in a 
cosmological context, in  \cite{abramo1,abramo2} (see also \cite{bab}).
Recalling the form of the Einstein tensor,
\begin{equation}
\delta^{(2)}_{\rm t} {\cal G}_{00} = - \ell_{\rm P}^2 {\cal T}_{00}
= \delta_{\rm t}^{(2)} R_{00} - \frac{1}{2} \overline{g}_{00} \delta_{\rm t}^{(2)} R,
\end{equation}
we obtain
\begin{equation}
\ell_{\rm P}^2 {\cal T}_{00} = {\cal H} h_{k\ell}' h^{k\ell} + 
\frac{1}{8} ( h_{k\ell}' {h^{k\ell}}'  + \partial_{i} h_{k\ell} \partial^{i} h^{k\ell}) + 
{\cal D}_{00},
\end{equation} 
where ${\cal D}_{00}$ is a total derivative, i.e. 
\begin{equation}
{\cal D}_{00} = \frac{1}{8} \partial_{\ell}[ \partial_{i} h^{k\ell} h^{i}_{k} - 2 \partial_{k} h^{\ell}_{i} h^{k i}]
\end{equation}
From Eqs. (\ref{rij2}) and (\ref{r}) it is also possible to write:
\begin{eqnarray}
\delta_{\rm t}^{(2)} R_{ij} &=& \frac{1}{4} ( \partial_{k} h^{\ell}_{i} \partial^{k} h_{j\ell} + 
\partial^{\ell} h_{i k} \partial_{\ell} h^{k}_{j}) - 
\frac{1}{4} \partial_{i} h_{k\ell} \partial_{j} h^{k\ell}  
\nonumber\\
&-& \frac{{\cal H}}{2} h^{k\ell} h_{k\ell}' \delta_{ij} + 
\frac{{\cal H}}{2} ( h^{\ell}_{j} h_{\ell i}' + h_{i}^{\ell} h_{\ell j}')
\nonumber\\
&-&\frac{{\cal H}}{2} ( {h^{k}_{j}}' h_{ik} + {h_{i}^{k}}' h_{k j}) - 
\frac{1}{4}({h^{k}_{j}}' h_{ik}' + {h^{k}_{i}}' h_{k j}')
+ {\cal D}_{ij},
\nonumber\\
\delta^{(2)}_{\rm t}R &=& \frac{1}{a^2}\biggl[ \frac{3}{4} h_{k\ell}' {h^{k\ell}}' + 
{\cal H} h_{k\ell}' h^{k\ell} - \frac{3}{4} \partial_{i}h^{k\ell} \partial^{i} h_{k\ell}\biggr]
+ \frac{1}{a^2} {\cal D}_{R}, 
\end{eqnarray}
where ${\cal D}_{ij}$ and ${\cal D}_{R}$ are further total derivative
\begin{eqnarray}
{\cal D}_{ij} &=& \frac{1}{2} \partial_{k} [ h^{k\ell}  ( \partial_{\ell} h_{ij} - 
\partial_{j} h_{i\ell} - \partial_{i} h_{j\ell})]
- \frac{1}{2} \partial_{j} [ h^{k\ell} (\partial_{\ell} h_{ik} - \partial_{k} h_{\ell i} - 
\partial_{i} h_{k\ell})] 
\nonumber\\
&-& \frac{1}{4} \partial_{\ell}[ \partial_{k} h^{\ell}_{i} h^{k}_{j} + 
h_{i k} \partial^{k} h_{j}^{\ell}].
\nonumber\\
{\cal D}_{R} &=& \partial_{i}[ h^{k\ell} \partial^{i} h_{k\ell}] + 
\frac{1}{4}\partial_{\ell} [ \partial_{i} h^{k\ell} h^{i}_{k} 
- 2 \partial_{k} h^{\ell}_{i} h^{k i}],
\end{eqnarray}
Therefore, up to total derivatives, the following result holds:
\begin{equation}
\ell_{\rm P}^2 \,\,{\cal T}_{00} = {\cal H} h_{k\ell}' h^{k\ell} + 
\frac{1}{8} ( h_{k\ell}' {h^{k\ell}}'  + \partial_{i} h_{k\ell} \partial^{i} h^{k\ell}),
\end{equation} 
and
\begin{equation}
\ell_{\rm P}^2\,\, {\cal T}_{ij}  = \frac{3}{8}\delta_{ij}[ \partial_{m} h_{k\ell} \partial^{m} h^{k\ell}
- h_{k\ell}' {h^{k\ell}}'] + \frac{1}{2} {h^{k}_{j}}' h_{ik}' + 
\frac{1}{4} \partial_{i} h_{k\ell} \partial_{j} h^{k\ell} - 
\frac{1}{2} \partial_{k} h_{i}^{\ell} \partial^{k} h_{\ell j}.
\end{equation}
To pass from doubly covariant indices to mixed ones, it is useful to recall that, to second 
order,
\begin{equation}
\delta^{(2)} {\cal G}_{\mu}^{\nu} = \delta^{(2)} [ g^{\nu\alpha} {\cal G}_{\mu\alpha}]
= \delta^{(2)} g^{\nu\alpha} \overline{{\cal G}}_{\mu\alpha} + 
\overline{g}^{\nu\alpha} \delta^{(2)} {\cal G}_{\mu\alpha} + 
\delta^{(1)} g^{\nu\alpha} \delta^{(1)} {\cal G}_{\mu\alpha}.
\end{equation}
By looking at the form of the specific terms arising in the previous equation it is clear that
${\cal T}_{0}^{0} = \overline{g}^{00} {\cal T}_{00}$ and that 
${\cal T}_{i}^{j} = \overline{g}^{jk} {\cal T}_{ki}$.
The expressions for ${\cal T}_{0}^{0}$ and ${\cal T}_{i}^{j}$ are 
\begin{eqnarray}
&&{\cal T}_{0}^{0} =  \frac{1}{a^2 \ell_{\rm P}^2} \biggl[ {\cal H} 
h_{k\ell }' h^{k\ell} + \frac{1}{8} ( \partial_{m} h_{k\ell} \partial^{m} h^{k\ell} + 
h_{k\ell}' {h^{k\ell}}')\biggr],
\label{PSsec1}\\
&& {\cal T}_{i}^{j} = \frac{{\cal T}}{3} \delta_{i}^{j} + \Sigma_{i}^{j}, 
\label{PSsec2}
\end{eqnarray}
where 
\begin{eqnarray}
&& {\cal T} = \frac{1}{a^2 \ell_{\rm P}^2}\biggl[ \frac{5}{8} h_{k\ell}' {h^{k\ell}}' - \frac{7}{8} 
\partial_{m} h_{k\ell} \partial^{m} h^{k\ell} \biggr],
\label{TRsec}\\
&& \Sigma_{i}^{j} = 
\frac{1}{a^2 \ell_{\rm P}^2} \biggl\{ \frac{1}{6} \biggl[ h_{k\ell}' {h^{k\ell}}' - 
\frac{1}{2} \partial_{m} h_{k\ell} \partial^{m} h^{k\ell} \biggr] \delta_{i}^{j}
+ \frac{1}{2} \partial_{m} h_{\ell i} \partial^{m} h^{\ell j} - 
\frac{1}{4} \partial_{i} h_{k\ell} \partial^{j}  h^{k\ell}
\nonumber\\
&& - 
\frac{1}{2} h_{k i}' {h^{k j}}' \biggr\},
\label{ANISsec}
\end{eqnarray}
 with $\Sigma_{i}^{i} =0$.  
 (see also Eqs. (\ref{TRsec}) 
and (\ref{ANISsec})). These expressions coincide with the ones 
obtained, for instance, in \cite{abramo1,abramo2} and are also 
consistent with the ones of \cite{isacson1,isacson2}.
From Eqs. (\ref{PSsec1}) and (\ref{TRsec})
 the components of the energy and pressure density can be easily 
 obtained since, by definition, $\rho_{\rm gw}^{(2)}= {\cal T}_{0}^{0}$ and 
 $p_{\rm gw}^{(2)} = - {\cal T}/3$.  As discussed in Eqs. (\ref{ENM1})--(\ref{ENM7}) 
 it is rather useful to derive the explicit form of $\rho_{\rm gw}^{(2)}$ and 
 $p_{\rm gw}^{(2)}$ in terms of the normalized canonical amplitude defined 
 in Eq. (\ref{L3}).  The result of this calculation is simply:
 \begin{eqnarray}
\rho_{\rm gw}^{(2)} &=& \frac{1}{a^2}\biggl\{ 
8 {\cal H} h \partial_{\tau} h  + [ \partial_{m} h \partial^{m} h 
+ (\partial_{\tau} h)^2]\biggr\},
\label{PS1}\\
p_{\rm gw}^{(2)} &=& - \frac{1}{3 a^2} [ 5 (\partial_{\tau} h)^2 - 7 (\partial_{m} h) (\partial^{m} h)].
\label{PS2}
\end{eqnarray}
By comparing Eqs. (\ref{PS1})--(\ref{PS2}) with Eqs. (\ref{ENM6})--(\ref{ENM7})
we can remark that the first term appearing in Eq. (\ref{PS1}) is absent from 
Eq. (\ref{ENM6}). Moreover, also $p_{\mathrm{GW}}^{(1)}$ and 
$p_{\mathrm{GW}}^{(2)}$ seems to be superficially different. 
As it will be shown in a moment the equivalence of the two approaches 
is clear as soon as the relevant wavelengths are larger than the Hubble radius 
at a given time. 

Before proceeding with this step, it is relevant to remark that
the components of the energy-momentum 
pseudo-tensor given in Eqs. (\ref{PSsec1}) and (\ref{PSsec2}) 
are not covariantly conserved. However,  since the 
Bianchi identity $\nabla_{\mu} {\cal G}_{\nu}^{\mu}=0$
 should be valid to all orders, we will also have that:
\begin{equation}
\delta_{\rm t}^{(2)} ( \nabla_{\mu} {\cal G}^{\mu}_{\nu}) =0,
\label{bianchi1}
\end{equation}
whose explicit form is 
\begin{eqnarray}
&&\partial_{\mu} \delta_{\rm t}^{(2)} {\cal G}^{\mu}_{\nu} + 
\delta_{\rm t}^{(2)} \Gamma_{\mu\alpha}^{\mu} \overline{\cal G}_{\nu}^{\alpha} + 
\overline{\Gamma}_{\mu\alpha}^{\mu} \delta^{(2)}_{\rm t} {\cal G}_{\nu}^{\alpha} +
\delta_{\rm t}^{(1)} \Gamma_{\mu\alpha}^{\mu} \delta^{(1)}_{\rm t} {\cal G}_{\nu}^{\alpha} 
\nonumber\\
&&- \delta_{\rm t}^{(2)} \Gamma_{\nu\alpha}^{\beta} \overline{{\cal G}_{\beta}^{\alpha}} - 
\overline{\Gamma}_{\nu\alpha}^{\beta} \delta_{\rm t}^{(2)} {\cal G}_{\beta}^{\alpha} -
\delta_{\rm t}^{(1)} \Gamma_{\nu\alpha}^{\beta} \delta^{(1)}_{\rm t} {\cal G}_{\beta}^{\alpha}=0.
\end{eqnarray}
Recalling  now the components of the energy-momentum 
pseudo-tensor  and the results for the fluctuations of the Christoffel symbols 
we have 
\begin{equation}
\frac{\partial \rho_{\rm gw}}{\partial\tau} + 3 {\cal H}( \rho_{\rm gw} + p_{\rm gw}) 
- \frac{2 ({\cal H}^2 - {\cal H}')}{a^2 \ell_{\rm P}^2} \delta_{\rm t}^{(2)} \Gamma_{k0}^{k}=0,
\end{equation}
that can also be written as 
\begin{equation} 
\frac{\partial \rho_{\rm gw}}{\partial\tau} + 3 {\cal H}( \rho_{\rm gw} + {\cal P}_{\rm gw}) =0
\label{conteq}
 \end{equation}
 where 
\begin{equation}
 {\cal P}_{\rm gw} =  p_{\rm gw} +  \frac{({\cal H}^2 - {\cal H}')}{ 3 {\cal H} a^2} h_{k\ell}' h^{k\ell}.
 \label{Pmod1}
\end{equation}

\subsection{The energy density of the relic gravitons}
By choosing the prescription of Eq. (\ref{ENM4a}) the energy density of the relic gravitons will be given by:
\begin{equation}
\rho_{\mathrm{gw}}^{(1)}(\tau) = \langle 0| T_{0}^{0} | 0 \rangle =  \frac{1}{a^4}[ \langle 0| \hat{\pi}^2 |0 \rangle 
+ {\mathcal H}^2 \langle 0| \hat{\mu}^2 |0 \rangle 
- {\mathcal H}  \langle 0|(\hat{\mu} \hat{\pi} + \hat{\pi} \hat{\mu})|0 \rangle+
   \langle 0| \partial_{i}\hat{\mu}\partial^{i} \hat{\mu} |0 \rangle].
\label{enmom1}
\end{equation}
The expectation values appearing in Eq. (\ref{enmom1}) can be computed in different ways. 
For instance, from Eqs. (\ref{expansiona}) and (\ref{solmu})--(\ref{solpi}), 
Eq. (\ref{enmom1})
\begin{equation}
\rho_{\mathrm{gw}}^{(1)}(\tau) = \frac{1}{a^4} \int d\ln{k} \frac{k^3}{2\pi^2} \biggl\{ |g_{k}(\tau)|^2 + ( k^2 + {\mathcal H}^2) |f_{k}(\tau)|^2
- {\mathcal H}[ f_{k}^{*}(\tau)g_{k}(\tau) + f_{k}(\tau) g_{k}^{*}(\tau)] \biggr\}, 
\label{enmom1a}
\end{equation}
where the functions $f_{k}(\tau)$ and $g_{k}(\tau)$ are the mode functions obeying:
\begin{equation}
f_{k}' = g_{k},\qquad g_{k}' = - \biggl[ k^2 - \frac{a''}{a}\biggr].
\label{MF1a}
\end{equation}

If the energy momentum pseudo-tensor has the form derived in Eq. (\ref{PSsec1}), then the energy density 
will have, apparently, a slightly different form:
\begin{equation}
\rho_{\mathrm{gw}}^{(2)}(\tau) = \frac{1}{a^4} \int d\ln{k} \frac{k^3}{2\pi^2} \biggl\{ |g_{k}(\tau)|^2 + ( k^2 -7 {\mathcal H}^2) |f_{k}(\tau)|^2 + 3{\mathcal H}[ f_{k}^{*}(\tau)g_{k}(\tau) + f_{k}(\tau) g_{k}^{*}(\tau)] \biggr\}.
\label{enmom1b}
\end{equation}
From Eqs. (\ref{enmom1a}) and (\ref{enmom1b}) the corresponding critical fractions are
\begin{equation}
\Omega_{\mathrm{GW}}^{(1)}(k,\tau) = \frac{1}{\rho_{\mathrm{crit}}} \frac{d \rho^{(1)}_{\mathrm{GW}}}{d\ln{k}},\qquad
\Omega_{\mathrm{GW}}^{(2)}(k,\tau) = \frac{1}{\rho_{\mathrm{crit}}} \frac{d \rho^{(2)}_{\mathrm{GW}}}{d\ln{k}}.
\label{critfrac}
\end{equation}

The two relevant physical limits are for modes inside the Hubble radius (i.e. $k \tau >1$) and for modes 
outside the Hubble radius (i.e. $k \tau < 1$). If $ k /{\mathcal H} > 1$, then $f_{k}(\tau)$ will be, in the 
first approximation, plane waves and $ g_{k}(\tau) \simeq \pm i k f_{k}(\tau)$. In this limit:
\begin{eqnarray}
&&\Omega_{\mathrm{GW}}^{(1)}(k,\tau) = \frac{k^5 \, \ell_{\mathrm{P}}^2}{ 3 \pi^2  a^2 {\mathcal H}^2}
 |f_{k}(\tau)|^2 \biggl[1 + 
\frac{{\mathcal H}^2}{2 k^2} \biggr],
\label{cr1inside}\\
&&\Omega_{\mathrm{GW}}^{(2)}(k,\tau) = \frac{k^5 \, \ell_{\mathrm{P}}^2}{ 3 \pi^2  a^2 {\mathcal H}^2} |f_{k}(\tau)|^2 \biggl[1 -
\frac{7{\mathcal H}^2}{2 k^2} \biggr].
\label{cr2inside}
\end{eqnarray}
In the limit $k\tau >1$ we will have ${\mathcal H}^2 \ll k^2$. Thus Eqs. (\ref{cr1inside}) and (\ref{cr2inside}) 
coincide (up to corrections ${\mathcal O}({\mathcal H}^2/k^2)$). In this limit it is also possible to express
$\Omega_{\mathrm{GW}}(k,\tau)$ solely in terms of the power spectrum. Indeed, since 
\begin{equation}
|f_{k}(\tau)|^2 = \frac{\pi^2 a^2}{4 \ell_{\mathrm{P}}^2 k^3} {\mathcal P}_{\mathrm{T}}(k,\tau)
\label{psom}
\end{equation}
we will also have 
\begin{eqnarray}
&&\Omega_{\mathrm{GW}}^{(1)}(k,\tau) = \frac{k^2}{ 12  {\mathcal H}^2} {\mathcal P}_{\mathrm{T}}(k,\tau)\biggl[1 + 
\frac{{\mathcal H}^2}{2 k^2} \biggr],
\label{cr1insidea}\\
&&\Omega_{\mathrm{GW}}^{(2)}(k,\tau) = \frac{k^2 }{ 12 {\mathcal H}^2} {\mathcal P}_{\mathrm{T}}(k,\tau)\biggl[1 -
\frac{7{\mathcal H}^2}{2 k^2} \biggr].
\label{cr2insideb}
\end{eqnarray}
In the opposite limit, i.e. when the given wavelengths are smaller than the Hubble radius, the same 
analysis implies $\Omega_{\mathrm{GW}}^{(1)}(k,\tau) \neq \Omega_{\mathrm{GW}}^{(2)}(k,\tau)$. 
In short the idea is that, for $k \tau \ll 1$, $g_{k} = {\mathcal H} f_{k}$; indeed, for $k\tau \ll 1$, Eq. 
(\ref{EXPEC4}) implies 
\begin{equation}
f_{k}(\tau) = A_{k} a(\tau) + B_{k} a(\tau) \int^{\tau} \frac{d\tau'}{a^2(\tau')}, \qquad g_{k} \simeq {\mathcal H} f_{k},
\label{OUT}
\end{equation}
where the second term (going as $B_{k}/a(\tau)$) in Eq. (\ref{OUT}) 
is actually negligible for large times.
Therefore, in the limit $k \tau \ll 1$ it can be easily shown that  
$\Omega_{\mathrm{GW}}^{(1)}(k,\tau) \neq \Omega_{\mathrm{GW}}^{(2)}(k,\tau)$. It is curious to notice 
that, up to a factor of $2$, the first energy density (i.e. Eq. (\ref{enmom1a})) leads to the same 
value obtained for modes inside the Hubble radius.  
Therefore, for modes which are inside the Hubble radius:
\begin{equation}
\rho_{\mathrm{gw}}(\tau) = \frac{2}{a^4} \int d\ln{k} \frac{k^3}{2\pi^2} |f_{k}|^2 = \frac{1}{ 4 \ell_{\mathrm{P}}^2 a^4}
\int d\ln{k} k^2 {\mathcal P}_{\mathrm{T}}(k,\tau).
\end{equation}
Consequently, we can define the critical fraction of relic gravitons at a given time as
\begin{equation}
\Omega_{\mathrm{gw}}(k,\tau) = \frac{1}{\rho_{\mathrm{c}}} \frac{d \rho_{\mathrm{gw}}}{d \ln{k}} = 
\frac{k^2}{12 H^2 a^2} {\mathcal P}_{\mathrm{T}}(k,\tau)= \frac{k^2}{12 {\mathcal H}^2} 
{\mathcal P}_{\mathrm{T}}(k,\tau),
\label{DEF1}
\end{equation}
where, in the third equality it has been used that ${\mathcal H} = a H$.
Recalling Eq. (\ref{T11}) we can also express the critical fraction of relic gravitons in terms of the spectral amplitude. Indeed, according to Eq. (\ref{T11}) 
\begin{equation}
{\mathcal P}_{\mathrm{T}}(k,\tau) = 4 \nu {\mathcal P}_{\mathrm{T}}(\nu,\tau),
\end{equation}
and, therefore, 
\begin{equation}
\Omega_{\mathrm{gw}}(\nu,\tau) = \frac{4 \pi^2}{3 {\mathcal H}^2} \nu^3 {\mathcal S}_{h}(\nu,\tau).
\end{equation}
As long as the relevant wavelengths are shorter than the Hubble radius 
at a given time, different prescriptions for assigning the energy-momentum pseudo-tensor lead to the same result (see also the discussion in section \ref{sec6}).
In the opposite  limit different choices may exhibit quantitative 
differences. The limit of short wavelengths in comparison with the Hubble radius 
is the relevant one when discussing wide-band interferometers. Conversely, the initial 
conditions for the CMB anisotropies are set when the relevant wavelengths 
are larger than the Hubble radius before equality. 
\newpage
\renewcommand{\theequation}{6.\arabic{equation}}
\setcounter{equation}{0}
\section{Relic gravitons from the $\Lambda$CDM scenario}
\label{sec6}
 \subsection{Inflationary power spectra}
During the inflationary phase, the tensor power 
spectrum can be easily computed by solving Eq. (\ref{EXPEC4}):
\begin{equation}
F_{k}(\tau) = \frac{{\mathcal N}}{a(\tau) 
\sqrt{2 k}} \sqrt{ - k \tau} H_{\nu}^{(1)}( - k\tau),\qquad 
{\mathcal N} = \sqrt{\frac{\pi}{2}} e^{ i\pi(\nu + 1/2)/2},\qquad \nu = \frac{3- \epsilon}{2( 1 -\epsilon)}.
\label{TA1}
\end{equation}
where $H_{\nu}^{(1)}(z) = J_{\nu}(z) + i Y_{\nu}(z)$ is the Hankel function 
of first kind \cite{abr,tric} and where $\epsilon = - \dot{H}/{H^2}$. 
To pass from Eq. (\ref{EXPEC4}) to Eq. (\ref{TA1})  it is 
useful to bear in mind the following pair of identities 
\begin{equation}
{\mathcal H}^2 + {\mathcal H}^2 = a^2 H^2 ( 2 - \epsilon), \qquad  a H = - \frac{1}{\tau (1 - \epsilon)}.
\label{TA2aa}
\end{equation}
The second equality in Eq. (\ref{TA2aa}) follows (after integration 
by parts) from the relation between cosmic and conformal times, i.e. 
$a(\tau) d\tau = dt$.
 By substituting Eq. (\ref{TA1}) 
into Eq. (\ref{EXPEC7}) the standard expression of the tensor 
power spectrum  can be obtained. In particular, when the relevant modes exited the Hubble 
radius during inflation the power spectrum  becomes 
\begin{equation}
\overline{{\mathcal P}}_{\mathrm{T}}(k,\tau) = \ell_{\mathrm{P}}^2 H^2 \frac{2^{2 \nu}}{\pi^3} \Gamma^2(\nu) ( 1 - \epsilon)^{ 2 \nu -1} \biggl(\frac{k}{a H}\biggr)^{3 - 2 \nu}, \qquad 
\nu = \frac{3}{2} + \epsilon + {\mathcal O}(\epsilon^2),
\label{TA2}
\end{equation}
where the small argument limit of the Hankel functions has been taken.
In the slow-roll approximation, 
$\overline{M}_{\mathrm{P}}^2 H^2 \simeq V/3$; 
then Eq. (\ref{TA2}) implies that\footnote{Within the present notations, as already 
established in Eq. (\ref{defPL}), $\ell_{\mathrm{P}} = \sqrt{8\pi G} = 1/\overline{M}_{\mathrm{P}}= 
\sqrt{8\pi}{M_{\mathrm{P}}}$.} 
\begin{equation} 
\overline{{\mathcal P}}_{\mathrm{T}}(k) \simeq \frac{2}{3 \pi^2} \biggl(\frac{V}{\overline{M}_{\mathrm{P}}^4}\biggr)_{k \simeq a H} \simeq \frac{128}{3} 
\biggl(\frac{V}{M_{\mathrm{P}}^4}\biggr)_{k \simeq a H}.
\label{TA3}
\end{equation}
The spectral index defined from Eq. (\ref{TA3}) is, by definition, 
\begin{equation}
n_{\mathrm{T}} = \frac{d \ln{\overline{{\mathcal P}}_{\mathrm{T}}}}{d \ln{k}} = - \frac{2\epsilon}{1 - \epsilon} = - 2\epsilon + {\mathcal O}(\epsilon^2).
\label{TA4}
\end{equation}
 where the second equality follows from the identities obeyed by the slow-roll parameters \cite{maxbook}. 
 The tensor and scalar power spectra are customarily assigned at a reference scale (usually dubbed pivot scale):
\begin{equation}
\overline{{\mathcal P}}_{\mathrm{T}}(k) = {\mathcal A}_{\mathrm{T}} 
\biggl(\frac{k}{k_{\mathrm{p}}}\biggr)^{n_{\mathrm{T}}}, \qquad k_{\mathrm{p}} = 0.002 \,\, \mathrm{Mpc}^{-1},
\label{TA5}
\end{equation}
where, by definition, ${\mathcal A}_{\mathrm{T}}$ is the amplitude of the tensor power spectrum evaluated at the pivot scale $k_{\mathrm{p}}$. 
In the case of single-field inflationary models,  
the power spectrum computed at the pivot scale $k_{\mathrm{p}}$ (i.e.  ${\mathcal A}_{\mathrm{T}}$) and the 
spectral index $n_{\mathrm{T}}$ can be related.  Bearing in mind that the power spectrum of curvature perturbations is given, in single field inflationary models, as \cite{maxbook}
\begin{equation}
\overline{{\mathcal P}}_{{\mathcal R}}(k) = \frac{8}{3}\biggl(\frac{V}{\epsilon\, M_{\mathrm{P}}^4}\biggr)_{k \simeq a H} = 
{\mathcal A}_{{\mathcal R}} \biggl(\frac{k}{k_{\mathrm{p}}}\biggr)^{n_{\mathrm{s}} -1},
\label{TA6}
\end{equation}
the ratio between the tensor and the scalar power spectra is given by 
\begin{equation}
r_{\mathrm{T}} = \frac{\overline{{\mathcal P}}_{\mathrm{T}}(k)}{\overline{{\mathcal P}}_{{\mathcal R}}(k)}= \frac{{\mathcal A}_{\mathrm{T}}}{{\mathcal A}_{{\mathcal R}}} =  16 \epsilon, 
\label{TA6a}
\end{equation}
Equation (\ref{TA6a}) implies, recalling Eq. (\ref{TA4}), that  $r_{\mathrm{T}} = - 8 n_{\mathrm{T}}$. 
Since there is  a direct relation of the tensor spectral index to $r_{\mathrm{T}}$, the number of the parameters 
can be reduced from two to one.  In Tabs. \ref{TABLE1} and \ref{TABLE2} the values of  $r_{\mathrm{T}}$ have been reported as they can be estimated in few different analyses of the cosmological data sets. 

\subsection{Transfer functions for inflationary power spectra}

Equation (\ref{TA5}) correctly parametrizes the spectrum only when the relevant 
wavelengths are larger than the Hubble radius before matter-radiation equality (i.e. 
$k\tau\simeq k/{\mathcal H} < 1$ for $\tau <\tau_{\mathrm{eq}}$).
To transfer the spectrum inside the Hubble radius the procedure 
is to integrate numerically Eq. (\ref{EXPEC4}) (as well as Eqs. (\ref{FL1})--(\ref{FL3})) across the relevant transitions of the background geometry. While the geometry passes from inflation to radiation, Eq. (\ref{TA5}) implies that 
the tensor mode function is constant if the wavelength associated with the given Fourier mode is larger than the Hubble radius at the 
corresponding epoch:
\begin{equation}
F_{k}(\tau) = A_{k} + B_{k} \int \frac{d\tau'}{a^2(\tau')}, \qquad \frac{k}{a H} \ll 1, \qquad |A_{k}|^2 = \frac{\pi^2}{4 \ell_{\mathrm{P}}^2 k^3} \overline{P}_{\mathrm{T}}(k).
\label{TA7}
\end{equation}
The term proportional to $B_{k}$ in Eq. (\ref{TA7}) leads to a decaying mode and 
$F_{k}(\tau)$ is therefore determined, for $|k\tau|\ll 1$, by the first term whose squared modulus coincides with the spectrum computed in Eq. (\ref{TA3}) and parametrized as in Eq. (\ref{TA5}). The latter statement is true 
if the inflationary phase is suddenly followed 
by the radiation dominated epoch since, during radiation, $a(\tau) \simeq \tau$.
The situation can be different if, after inflation, a stiff age takes over. The
stiffer case still compatible with causality (i.e. $w_{\mathrm{t}} =1$) leads approximately 
to a scale factor $a(\tau) \simeq \sqrt{\tau}$. In the latter situation the second 
term in Eq. (\ref{TA7}) grows logarithmically and cannot be neglected 
in comparison with the constant contribution.
The evolution of the background (i.e. Eqs. (\ref{FL1})--(\ref{FL3})) and of the tensor 
mode functions (i.e. Eq. (\ref{EXPEC4})) should therefore be solved across the radiation matter transition and 
the usual approach is to compute the transfer function for the amplitude \cite{tur1} i.e. 
\begin{equation}
T_{h}(k) = \sqrt{\frac{\langle|F_{k}(\tau)|^2\rangle}{\langle|\overline{F}_{k}(\tau)|^2\rangle}}. 
\label{TA8}
\end{equation}
In Eq. (\ref{TA8}), $\overline{F}_{k}(\tau)$ denotes the approximate form of the mode function (holding during the 
matter-dominated phase); $F_{k}(\tau)$ denotes, instead, the solution obtained by fully numerical methods.  
The averages appearing in Eq. (\ref{TA8}) refer to the average over the oscillations: 
as the wavelengths are inside the Hubble radius, the solutions are all oscillating \footnote{The numerical average over the phases introduces some arbitrariness which 
can be cured by computing directly the transfer function for the spectral energy density.}. The calculation of $T_{h}(k)$ requires a careful matching over the phases between the numerical and the 
approximate (but analytic) solution. 
After matter-radiation equality, the scale factor is going, approximately, as $a(\tau) \simeq \tau^2$ and, therefore, the 
(approximate) solution of Eq. (\ref{EXPEC4})  is given by 
\begin{equation}
\overline{F}_{k}(\tau) = \frac{3 j_{1}(k\tau)}{k\tau} A_{k}, \qquad j_{1}(k\tau) = \frac{\sin{k\tau}}{(k\tau)^2} - 
\frac{\cos{k\tau}}{(k\tau)}.
\label{TA9}
\end{equation}
which is constant for $k\tau < 1$.
The result of the numerical integration for the amplitude transfer function 
$T_{h}^2(k/k_{\mathrm{eq}})$  has been recently revisited in \cite{mgn1,mgn2} and it turns out to be 
\begin{equation}
T_{h}(k/k_{\mathrm{eq}}) = \sqrt{1 + c_{1} \biggl(\frac{k}{k_{\mathrm{eq}}}\biggr) + b_{1} \biggl(\frac{k}{k_{\mathrm{eq}}}\biggr)^2}.
\label{TA10}
\end{equation}
where 
\begin{equation}
c_{1}= 1.260, \qquad b_{1}= 2.683.
\label{TAa10}
\end{equation}
The latter result agrees with the findings of  \cite{tur1} who obtain $ \overline{c}_{1}= 1.34$ and $\overline{b}_{1} = 2.50$. The value of $k_{\mathrm{eq}}$ can be 
obtained directly from the experimental data (see, for instance, last column of Tabs. \ref{TABLE1} and \ref{TABLE2} implying $k_{\mathrm{eq}} \simeq 
{\mathcal O}(0.009)\, \mathrm{Mpc}^{-1}$).  The WMAP  5-yr data combined with the supernova data 
and with the large-scale structure data would give $k_{\mathrm{eq}} = 0.00999^{+0.00028}_{-0.00027}\,\, \mathrm{Mpc}^{-1}$. A rather good analytic estimate 
for  $k_{\mathrm{eq}}$ is 
\begin{eqnarray}
k_{\mathrm{eq}} = 0.0082879 \,\, \biggl(\frac{h_{0}^2 \Omega_{\mathrm{M}0}}{0.1326}\biggr) \biggl(\frac{h_{0}^2 \Omega_{\mathrm{R}0}}{4.15 \times 10^{-5}}\biggr)^{-1/2}\,\, \mathrm{Mpc}^{-1},
\label{TA11}
\end{eqnarray}
where the typical value selected for 
$h_{0}^2 \Omega_{\mathrm{R}0}$  is given by the sum of the photon component (i.e. 
$h_{0}^2 \Omega_{\gamma0} = 2.47 \times 10^{-5}$) and of the neutrino component (i.e. $h_{0}^2 \Omega_{\gamma0} = 1.68 \times 10^{-5}$): the neutrinos, consistently with the $\Lambda$CDM paradigm, are taken to be massless and their (present) 
kinetic temperature is just a factor $(4/11)^{1/3}$ smaller than the (present) photon temperature. 

Equation (\ref{TA11})  stems from the observation that the exact solution of 
Eqs. (\ref{FL1})--(\ref{FL3}) for the matter-radiation transition can be given as $a(\tau) = a_{\mathrm{eq}}[ y^2 + 2 y]$ where $y= \tau/\tau_{1}$. The time-scale $\tau_{1} = \tau_{\mathrm{eq}}(\sqrt{2} +1)$ 
is related to the equality time $\tau_{\mathrm{eq}}$ which can be estimated as 
\begin{equation}
\tau_{\mathrm{eq}} = 
\frac{2 (\sqrt{2} -1)}{H_{0}} \frac{\sqrt{ \Omega_{\mathrm{R}0}}}{ \Omega_{\mathrm{M}0}} 
= 120.658 \biggl(\frac{h_{0}^2 \Omega_{\mathrm{M}0}}{0.1326}\biggr)^{-1} \biggl(\frac{h_{0}^2 \Omega_{\mathrm{R}0}}{4.15 \times 10^{-5}}\biggr)^{1/2}\,\,\mathrm{Mpc}.
\label{TA13}
\end{equation}
In the case of the WMAP 5-yr data combined with the supernova and large-scale structure data 
$h_{0}^2 \Omega_{\mathrm{M0}}= 0.1368^{0.0038}_{-0.0037}$.
Consequently, Eqs. (\ref{TA8}), (\ref{TA9}) and (\ref{TA10}) imply that  the spectrum of the tensor modes is given, at the present time, as 
\begin{equation}
{\mathcal P}_{\mathrm{T}}(k,\tau_{0}) = \frac{9 j_{1}^2(k\tau_{0})}{(k\tau_{0})^2} 
T^2_{h}(k/k_{\mathrm{eq}})\overline{{\mathcal P}}_{\mathrm{T}}(k).
\label{TA14}
\end{equation}
Within the standard approach, Eq. (\ref{TA14}) is customarily connected to the 
spectral energy density of the relic gravitons. 
In \cite{mgn1,mgn2} it has been observed that it is simpler and more accurate to compute directly the 
transfer function for the spectral energy density. In the following subsection 
this procedure will be illustrated in two different cases.

\subsection{Transfer function for the spectral energy}
\label{sec2b}
\begin{figure}[!ht]
\centering
\includegraphics[height=6.7cm]{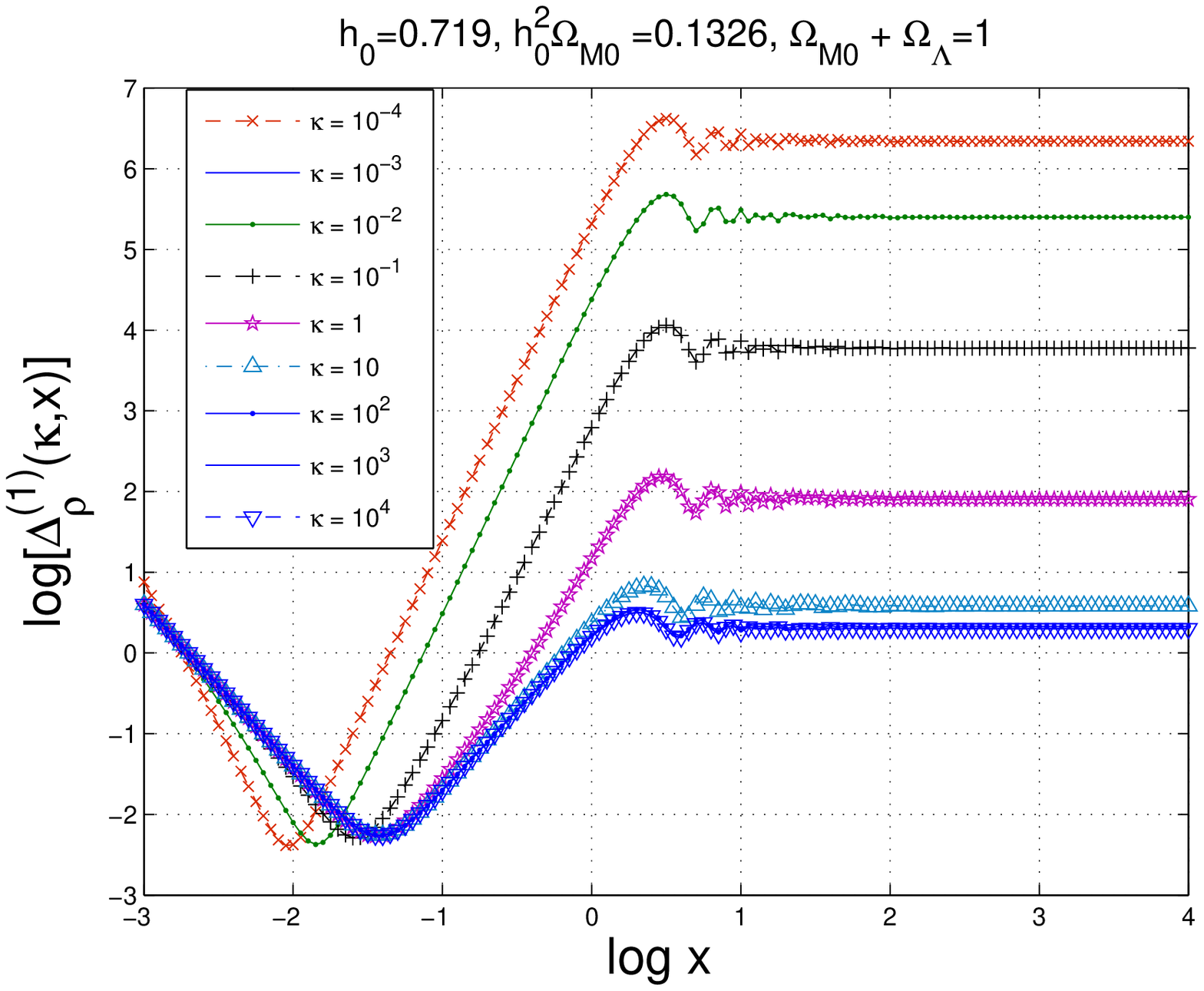}
\includegraphics[height=6.7cm]{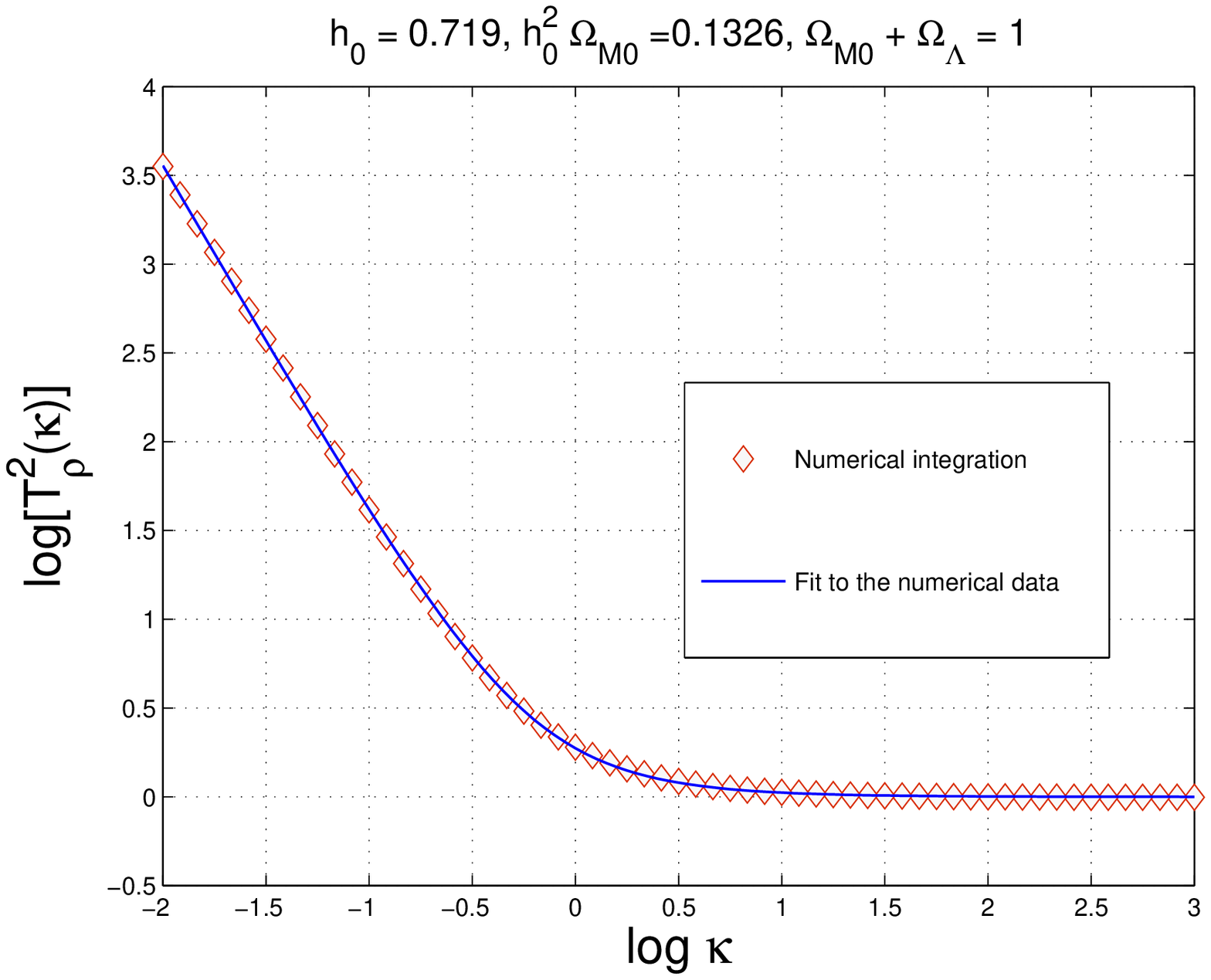}
\caption[a]{The functions given in Eqs. (\ref{DELTA1}) and (\ref{DELTA2}) are
numerically computed (plot at the left) for different 
values of $\kappa$ and in the case of the radiation-matter transition.
 In the plot at the right the transfer function for the spectral energy is illustrated.}
\label{Figure4}      
\end{figure}
Instead of computing the transfer function for the power spectrum it is more direct 
(and more accurate) to compute directly the spectral energy density and its 
related transfer function. 
As previously discussed there can be ambiguities in assigning the energy density because of possible 
different forms of the energy-momentum pseudo-tensor. In particular in Eqs. (\ref{enmom1a}) and (\ref{enmom1b}) 
two different expressions have been proposed on the basis of slightly different physical considerations.
The result of the numerical calculation are reported in Fig. \ref{Figure4} in terms 
of $\Delta^{(1)}_{\rho}(\kappa,x)$ (derived from Eq. (\ref{enmom1a})) and in terms 
of the transfer function of the spectral energy density
 (denoted by $T_{\rho}(\kappa)$).  The quantities $\Delta^{(1)}_{\rho}(\kappa, x)$ 
(and, analogously, $\Delta^{(2)}_{\rho}(\kappa, x)$) are defined as  
\begin{eqnarray}
\Delta^{(1)}_{\rho}(k,\tau) &=& \biggl\{ |g_{k}(\tau)|^2 + ( k^2 + {\mathcal H}^2) |f_{k}(\tau)|^2
- {\mathcal H}[ f_{k}^{*}(\tau)g_{k}(\tau) + f_{k}(\tau) g_{k}^{*}(\tau)] \biggr\},
\label{DELTA1}\\
\Delta^{(2)}_{\rho}(k,\tau) &=&\biggl\{ |g_{k}(\tau)|^2 + ( k^2 -7 {\mathcal H}^2) |f_{k}(\tau)|^2 + 3{\mathcal H}[ f_{k}^{*}(\tau)g_{k}(\tau) + f_{k}(\tau) g_{k}^{*}(\tau)] \biggr\}.
\label{DELTA2}
\end{eqnarray}
While Eq. (\ref{DELTA1}) follows from Eq. (\ref{enmom1a}), Eq. (\ref{DELTA2}) follows from Eq. (\ref{enmom1b});
Eqs. (\ref{DELTA1})--(\ref{DELTA2}) are related to the spectral energy densities in critical units, i.e.
\begin{equation}
\Omega^{(1)}_{\mathrm{GW}}(k,\tau) = \frac{k^3}{2\pi^2 a^4 \rho_{\mathrm{crit}}} \Delta^{(1)}_{\rho}(k,\tau), \qquad 
\Omega^{(2)}_{\mathrm{GW}}(k,\tau) = \frac{k^3}{2\pi^2 a^4 \rho_{\mathrm{crit}}} \Delta^{(2)}_{\rho}(k,\tau).
\label{DELTA3}
\end{equation}
As a function of $x = k\tau$ and $\kappa = k/k_{\mathrm{eq}}$, both 
$\Delta_{\rho}^{(1)}(\kappa,x)$  and $\Delta_{\rho}^{(2)}(\kappa,x)$ go to a constant value when the relevant modes are 
evaluated deep inside the Hubble radius. 
This occurrence allows to introduce the energy transfer function which is defined as: 
\begin{equation}
\lim_{x \gg 1} \Delta_{\rho}^{(1,2)}(\kappa,x) \equiv T^{2}_{\rho}(\kappa)  \Delta_{\rho}^{(1,2)}(\kappa,x_{\mathrm{i}}), \qquad x_{\mathrm{i}} \ll 1.
\label{DELTA4}
\end{equation}
\begin{figure}[!ht]
\centering
\includegraphics[height=6.7cm]{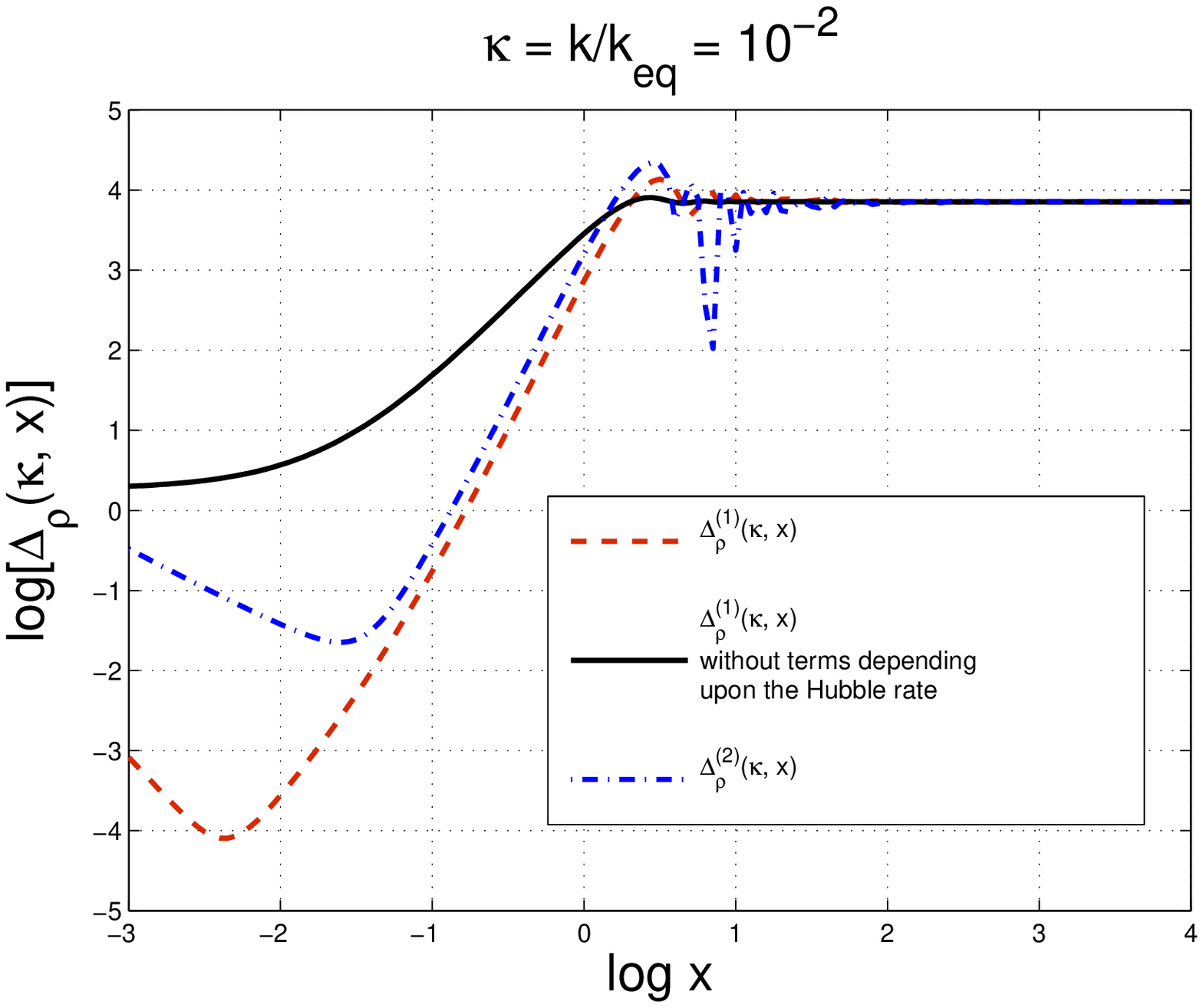}
\includegraphics[height=6.7cm]{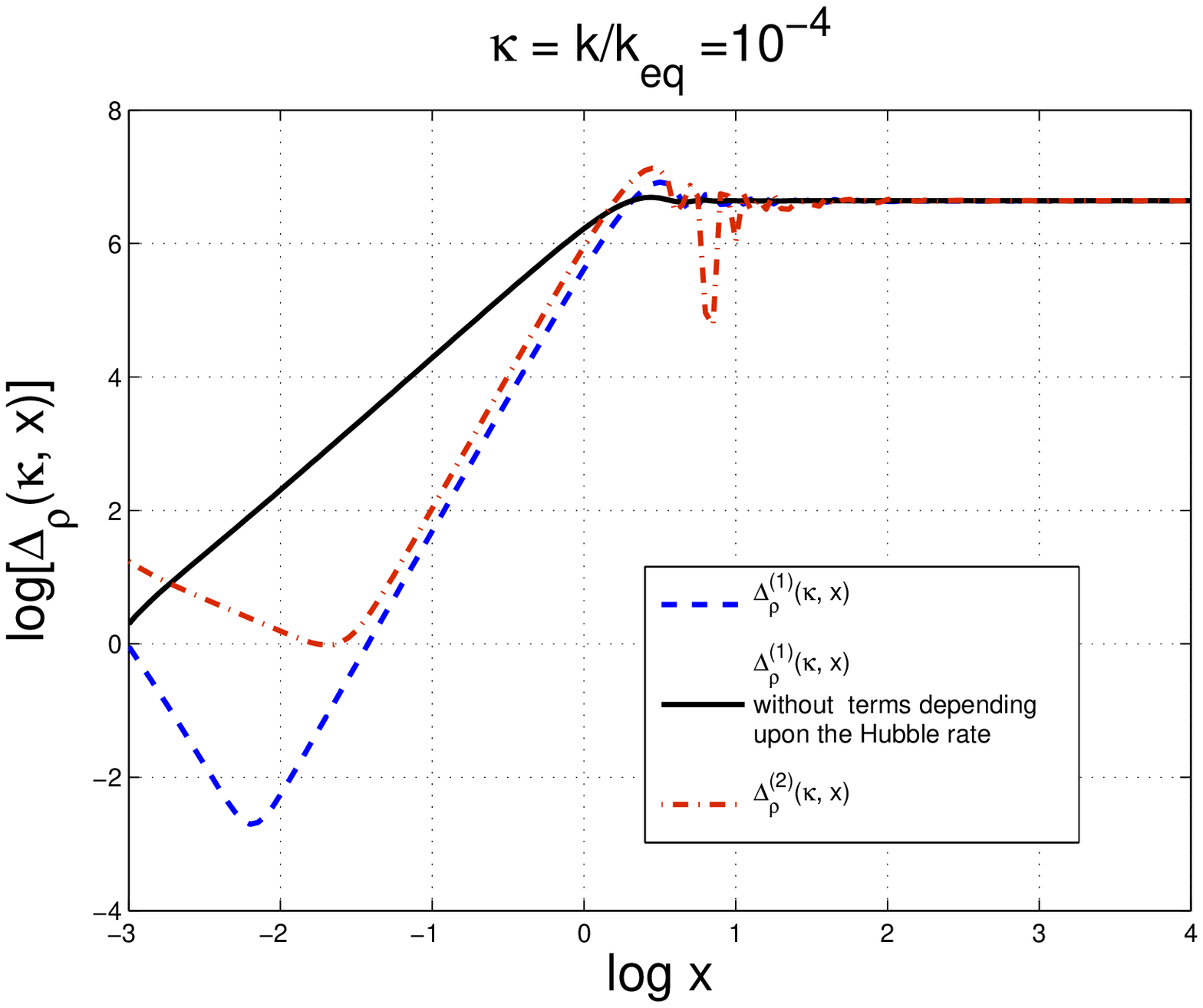}
\caption[a]{The different definitions of energy-momentum pseudo-tensor (i.e. 
Eqs. (\ref{DELTA1}) and (\ref{DELTA2})) are compared in the 
determination of the asymptotic value of the energy transfer function.}
\label{Figure5}      
\end{figure}
The specific form of the energy-momentum tensor is immaterial 
for the determination of $T_{\rho}^2(\kappa)$: different forms of the energy-momentum tensor 
of the relic gravitons will lead to the same result. This aspect can be appreciated 
by looking at Fig. \ref{Figure5} where $\Delta^{(1,2)}_{\rho}(k,\tau)$ has been 
reported for $\kappa = 10^{-2}$ (plot at the left) and for $\kappa = 10^{-4}$ (plot at the right). 
The dashed and the dot-dashed curves (in both plots) correspond, respectively, to $\Delta_{\rho}^{(1)}(\kappa, x)$ and 
to $\Delta_{\rho}^{(2)}(\kappa, x)$. The full line, in both plots, corresponds 
to the combination
\begin{equation}
 k^2 |f_{k}(\tau)|^2 +  |g_{k}(\tau)|^2= k( |c_{+}(k)|^2 + |c_{-}(k)|^2),
\label{comb1}
\end{equation}
where $c_{\pm}(k)$ are the mixing coefficients which
parametrize the solution for the tensor mode 
functions when the relevant wavelengths are, asymptotically,  inside the Hubble radius, i.e. 
\begin{equation}
\overline{f}_{k}(\tau) = \frac{1}{\sqrt{2 k}} \biggl[ c_{+}(k) e^{-i k \tau} + c_{-}(k) e^{i k\tau}\biggr],\qquad 
\overline{g}_{k}(\tau) = - i \sqrt{\frac{k}{2}} \biggl[ c_{+}(k) e^{-i k\tau} - c_{-}(k) e^{i k\tau}\biggr].
\label{BOG1}
\end{equation}
In Eq. (\ref{BOG1}) $\overline{f}_{k}(\tau)$ and $\overline{g}_{k}(\tau)$ are the solutions to leading order in the limit $k\tau\gg1 $.
From Eq. (\ref{BOG1}), $c_{\pm}(k)$ are given by
\begin{equation}
 c_{+}(k) = \frac{e^{i k\tau}}{\sqrt{2 k}}[ k \overline{f}_{k}(\tau) + i \overline{g}_{k}(\tau)],\qquad c_{-}(k) = \frac{e^{-i k\tau}}{\sqrt{2 k}}[ k \overline{f}_{k}(\tau) - i \overline{g}_{k}(\tau)],
\label{BOG2}
\end{equation}
Using Eqs. (\ref{BOG1})--(\ref{BOG2}), 
 Eqs. (\ref{DELTA1})--(\ref{DELTA2}) can be directly calculated in the limit 
  $x = k\tau \gg1$ with the result that 
\begin{equation}
 \Delta^{(1)}_{\rho}(\kappa,x_{\mathrm{f}}) = \Delta^{(2)}_{\rho}(\kappa,x_{\mathrm{f}}) = 
 \kappa ( |c_{+}(\kappa)|^2 + |c_{-}(\kappa)|^2)  + {\mathcal O}\biggl(\frac{1}{x_{\mathrm{f}}}\biggr),
 \label{BOG4}
 \end{equation}
 which proofs that the oscillating contributions are suppressed as $x_{\mathrm{f}}^{-1}$ for $x_{\mathrm{f}} \gg 1$.
 To get to the results illustrated in Figs. \ref{Figure4} and \ref{Figure5} the evolution equations of the mode functions have been integrated by setting 
initial conditions deep outside the Hubble radius (i.e. $x = k\tau \ll 1$), by following 
the corresponding quantities through the Hubble crossing (i.e. $ x \simeq 1$) and then, finally, deep inside 
the Hubble radius (i.e. $x \gg 1$). 
The integration of the mode functions is most easily performed in terms of $\tilde{f}_{\kappa}(x)= k f_{k}(\tau)$. Using  $\tilde{f}_{\kappa}(x)$, $\Delta_{\rho}(\kappa, x)\equiv \Delta^{(1)}_{\rho}(\kappa, x)$ can be written as 
\begin{equation}
\Delta_{\rho}(\kappa, x) = |\tilde{f}_{\kappa}(x)|^2 \biggl[ 1 + \biggl(\frac{d\ln{a}}{d x}\biggr)^2\biggr] + 
\left| \frac{d \tilde{f}_{\kappa}}{d x} \right|^2  - \biggl(\frac{d\ln{a}}{d x}\biggr)\biggl[  \biggl(\frac{d \tilde{f}_{\kappa}}{d x}\biggr)^{*} \tilde{f}_{\kappa} +  \biggl(\frac{d \tilde{f}_{\kappa}}{d x}\biggr)\tilde{f}_{\kappa}^{*}\biggl].
\label{INC1}
\end{equation}
 In the case of the conventional $\Lambda$CDM scenario, the Universe 
is suddenly dominated by radiation as soon as inflation ends. Equation (\ref{EXPEC4}) implies that $F_{k}$ 
is constant when $k\tau \ll 1$. The initial conditions are  fixed by requiring 
that, at the initial time of the numerical integration, 
\begin{equation}
\tilde{f}_{\kappa}(x_{\mathrm{i}}) = k f_{k}(\tau_{\mathrm{i}}) = k a(\tau) F_{k},\qquad x_{\mathrm{i}} = k \tau_{\mathrm{i}} \ll 1.
\label{INC2}
\end{equation}
In Figs. \ref{Figure4} and \ref{Figure4},  $x_{\mathrm{i}}=10^{-5}$ even if, 
for practical reasons, the scale on the horizontal axis has been narrowed.
Bearing in mind Eq. (\ref{TA13}), we can also write, for $x_{\mathrm{i}} \ll 1$,
\begin{equation}
\tilde{f}_{\kappa}(x_{\mathrm{i}}) = \frac{2 a_{\mathrm{eq}}}{\tau_{1}} F_{k} \, x_{\mathrm{i}}, \qquad 
\left|\frac{d \tilde{f}_{\kappa}}{d x} \right|_{x=x_{\mathrm{i}}} = \frac{2 a_{\mathrm{eq}}}{\tau_{1}} F_{k}. 
\label{INC3} 
\end{equation}
To avoid  unnecessary complications, the initial condition of the integrations 
illustrated in Figs. \ref{Figure4} and \ref{Figure5} have been set as 
$\tilde{f}_{\kappa}(x_{\mathrm{i}}) = x_{\mathrm{i}}$, i.e. the initial spectrum has been rescaled. 
The transfer function, by definition, must always depend only on the dynamics of the transition and not upon the features (e.g slope, amplitude) of the initial power spectrum.

In the plot at the right of Fig. \ref{Figure4}, the fit to the energy transfer function is reported with the full (thin) 
line on top of the diamonds defining the numerical points. The analytical form of the fit can then be 
written as:
\begin{equation}
T_{\rho}(k/k_{\mathrm{eq}}) = \sqrt{1 + c_{2}\biggl(\frac{k_{\mathrm{eq}}}{k}\biggr) + b_{2}\biggl(\frac{k_{\mathrm{eq}}}{k}\biggr)^2},\qquad c_{2}= 0.5238,\qquad
b_{2}=0.3537.
\label{ENTRANS}
\end{equation}
Equation (\ref{ENTRANS}) permits the accurate 
evaluation of the spectral energy density of relic gravitons, for instance, in the minimal version of the 
$\Lambda$CDM paradigm. 

Yet another relevant physical situation for the present considerations is the one where the background geometry, after inflation, 
transits from a stiff epoch to the ordinary radiation-dominated epoch. 
 In the primeval plasma, stiff phases can arise for various reasons. 
 Zeldovich \cite{ZEL1} (see also \cite{ZEL2}) suggested this possibility in connection with the entropy problem.  
In  \cite{mg1,mg2,mg3,mg4} it has been suggested that the stiff phase could take place after the inflationary phase 
with the main purpose of identifying a potential source of high-frequency gravitons.
 This possibility was  also prompted by a possible post-inflationary
dominance of a quintessence field.

The simplest consideration leading to the possibility of a post-inflationary phase stiffer than radiation 
is connected with our extreme ignorance of the thermal history of the Universe after inflation.
In a model-independent 
approach, it is plausible to think that the onset of the radiation-dominance could be 
delayed. This may happen, in concrete models, for various reasons. One possibility 
could be that the inflaton field does not decay but rather
changes its dynamical nature by acting as quintessence field \cite{PV} (see also 
\cite{Spok}).  In this kind of situations we are the geometry passes from a stiff phase where
\begin{eqnarray}
w_{\mathrm{t}}(\tau) &=& \frac{p_{\mathrm{t}}}{\rho_{\mathrm{t}}} > \frac{1}{3}, 
\label{ST1}\\
c^2_{\mathrm{st}}(\tau) &=& \frac{p_{\mathrm{t}}'}{\rho_{\mathrm{t}}'} = w_{\mathrm{t}} - \frac{w_{\mathrm{t}}'}{ 3 {\mathcal H} (w_{\mathrm{t}} + 1)} = w_{\mathrm{t}} - \frac{1}{3} \frac{d \ln{( w_{\mathrm{t}} + 1)}}{d \ln{a}}>  \frac{1}{3},
\label{ST2}
\end{eqnarray}
to a radiation-dominated phase where $c_{\mathrm{st}} = 1/\sqrt{3}$.
Note that, according to Eqs. (\ref{ST1}) and (\ref{ST2}), $c_{\mathrm{st}}^2 = w_{\mathrm{t}}$ iff the (total) barotropic 
index is constant in time. In the limiting case  $w_{\mathrm{t}} = 1 = c_{\mathrm{st}}^2$ 
and the speed of sound coincides with the speed of light. As argued 
in \cite{SpS}, barotropic indices $w_{\mathrm{t}} >1$ would not be 
compatible with causality (see, however, \cite{kessence1}). The presence of a suitable 
stiff phase has been also discussed recently as an effective way of suppressing entropic 
fluctuations \cite{sup} which are observationally constrained by the WMAP 5-yr data.

\begin{figure}[!ht]
\centering
\includegraphics[height=6.7cm]{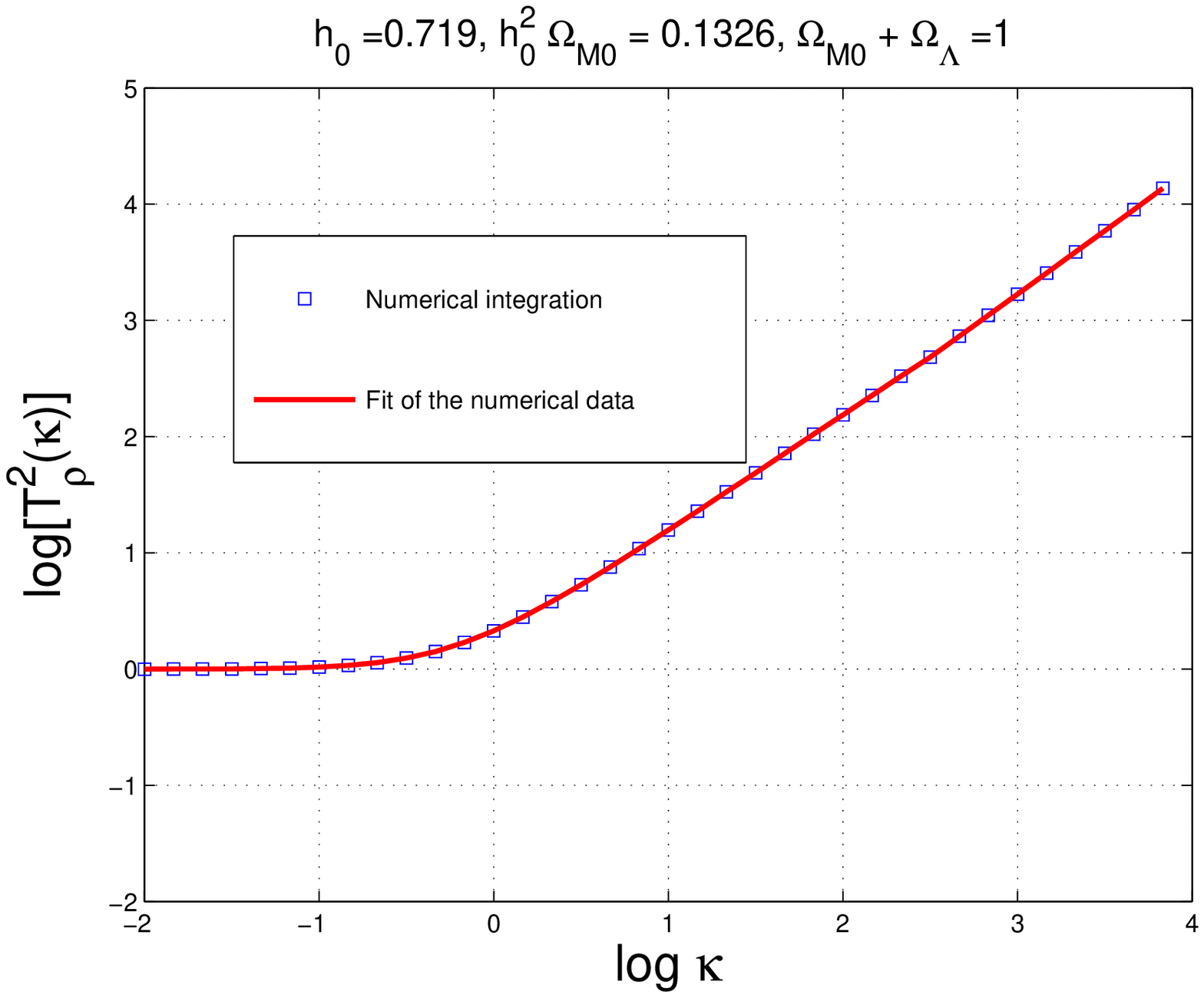}
\includegraphics[height=6.7cm]{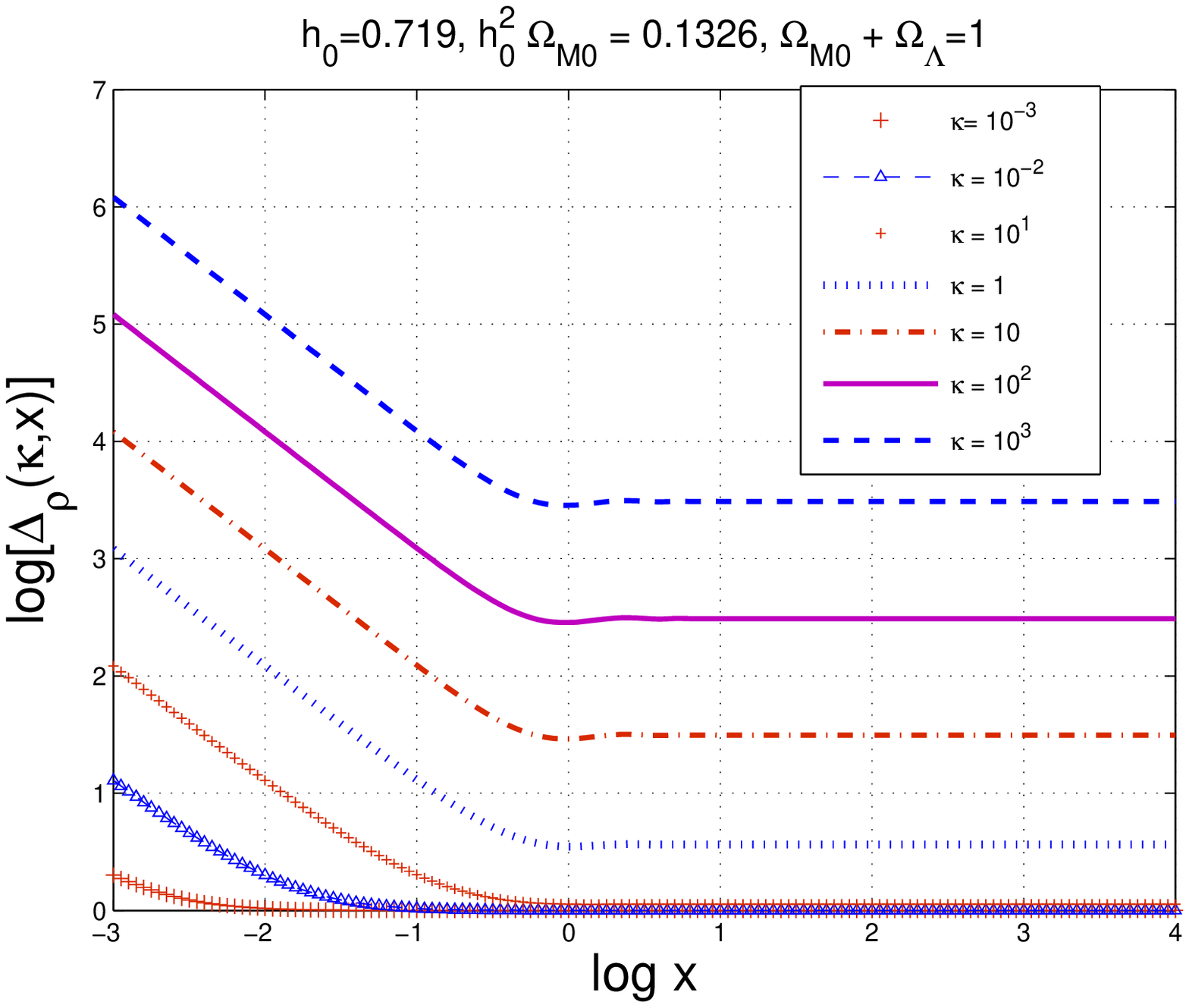}
\caption[a]{The transition between the stiff phase and the radiation phase is illustrated. 
The energy transfer function increases with the frequency while the opposite is true for the radiation-matter transition (see Fig. \ref{Figure4} where the analog results have been 
presented for the matter-radiation transition). }
\label{Figure6}      
\end{figure}

As in the case of the matter-radiation transition the transfer function only depends upon $\kappa$ which is defined, this time, 
as $\kappa = k/k_{\mathrm{s}}$, where $k_{\mathrm{s}} = \tau_{\mathrm{s}}^{-1}$ and $\tau_{\mathrm{s}}$ parametrizes 
the transition time. A simple analytical form of the transition regime is given by 
\begin{equation}
a(y) = a_{\mathrm{s}} \sqrt{y^2 + 2 y}, \qquad y = \frac{\tau}{\tau_{\mathrm{s}}},\qquad 
\tau_{\mathrm{s}} = \frac{1}{a_{\mathrm{i}} H_{\mathrm{i}}} \sqrt{\frac{\rho_{\mathrm{Si}}}{ \rho_\mathrm{Ri}}},
\label{ST3}
\end{equation}
where, by definition, $\rho_{\mathrm{si}} = \rho_{\mathrm{s}}(\tau_{\mathrm{i}})$ and $\rho_{\mathrm{Ri}} = 
\rho_{\mathrm{R}}(\tau_{\mathrm{i}})$. 
Equation (\ref{ST3}) is a solution of Eqs. (\ref{FL1})--(\ref{FL3}) when the radiation is present together 
with a stiff component which has, in the case of Eq. (\ref{ST3}) a sound speed which equals the speed of light. In the limit 
$y \to 0$ the scale factor expands as $a(y) = \sqrt{2 y}$ while,  in the opposite limit, 
$a(y) \simeq y$. 
In Fig. \ref{Figure6} (plot at the right) $\Delta_{\rho}(\kappa, x)$ is illustrated for different values of $\kappa$. We shall not dwell 
here (again) about the possible different forms of the energy momentum 
pseudo-tensor: provided the energy density is evaluated deep inside the Hubble radius the different approaches 
to the energy density of the relic gravitons give the same result.
The transfer function for the spectral energy density is numerically 
illustrated always in Fig. \ref{Figure6} (plot at the left).
The semi-analytical form of the transfer function becomes, this time, 
\begin{equation}
T_{\rho}^2(k/k_{\mathrm{s}}) = 1.0  + 0.204\,\biggl(\frac{k}{k_{\mathrm{s}}}\biggr)^{1/4} - 0.980 \,\biggl(\frac{k}{k_{\mathrm{s}}}\biggr)^{1/2}  + 3.389 \biggl(\frac{k}{k_{\mathrm{s}}}\biggr) -0.067\,\biggl(\frac{k}{k_{\mathrm{s}}}\biggr)\ln^2{(k/k_{\mathrm{s}})},
\label{ST10}
\end{equation}
where $k_{\mathrm{s}} = \tau_{\mathrm{s}}^{-1}$. The value of $k_{\mathrm{s}}$ can 
be computed in an explicit model\footnote{ In the context of quintessential 
inflation \cite{PV} (see also \cite{mg2,mg3}) $\rho_{\mathrm{Ri}} \simeq H_{\mathrm{i}}^4$ \cite{ford3}.} but it can also be left as a free parameter. 
Taking into account that the energy density of the inflaton will be exactly $\rho_{\mathrm{si}} \simeq H_{\mathrm{i}}^2 \overline{M}_{\mathrm{P}}^2$,  
the value of $k_{\mathrm{s}}$ (as well as the duration of the stiff phase) will be determined, grossly speaking, by 
$H_{\mathrm{i}}/\overline{M}_{\mathrm{P}}$.  
In Fig. (\ref{Figure6}) (plot at the left) the full line superimposed to the numerical points (illustrated by boxes) is the 
fit of Eq. (\ref{ST10}).

\subsection{Analytic results for the mixing coefficients}
The analytic results for the mixing coefficients are rather useful to 
obtain the final expression of the various transfer functions.
Indeed, defining as $k_{*}$ the typical wavenumber of the transition (e.g. $k_{*} = k_{\mathrm{eq}}$ in the case of the radiation-mattter transition), the slope of the
transfer function of the spectral energy density can analytically obtained in the limit $\kappa \gg 1$. This observation helps when we have, for instance, to fit 
the numerical data points with an analytical expression which will however reproduce 
the data not only for $\kappa > 1$ (as Figs. \ref{Figure4} and \ref{Figure6} clearly show).

For illustration of the method it is practical to consider the transition from a generic accelerated phase  to a decelerated stage of expansion. In this 
situation, by naming the transition point $-\tau_{1}$, the continuous 
and differentiable form of the scale factors can be written as:
\begin{eqnarray}
&& a_{\mathrm{i}}(\tau) = \biggl( - \frac{\tau}{\tau_{1}}\biggl)^{-\beta}, 
\qquad \tau < - \tau_{1},
\label{infsc}\\
&& a_{\mathrm{s}}(\tau) = \biggl[ \frac{\beta}{\alpha}\biggl(\frac{\tau}{\tau_{1}} + 1\biggr) + 1\biggr]^{\alpha}, \qquad 
\tau \geq -\tau_{1},
\label{stsc}
\end{eqnarray}
where the scale factors are continuous and differentiable at the transition 
point which has been generically indicated as $\tau_{1}$. 
The pump fields of the tensor mode functions turn out to be:
\begin{equation}
\frac{a_{\mathrm{i}}''}{a_{\mathrm{i}}} = \frac{\beta(\beta + 1)}{\tau^2},\qquad 
\frac{a_{\mathrm{s}}''}{a_{\mathrm{s}}} = \frac{\alpha (\alpha - 1)}{\biggl[
\tau + \biggl(\frac{\alpha}{\beta} + 1\biggr) \tau_{1}\biggr]^2}.
\end{equation}
The mode functions can then be written as:
\begin{eqnarray}
f_{\mathrm{i}}(\tau) &=& \frac{{\mathcal N}}{\sqrt{ 2 k}} \sqrt{- x}
 H_{\nu}^{(1)}( - x),\qquad \tau <  - \tau_{1},\qquad x = k \tau,
 \nonumber\\
\tilde{f}_{\mathrm{s}}(\tau) &=&  \frac{\sqrt{y}}{\sqrt{ 2 k}}[ {\mathcal M}
c_{+}(k)H_{\lambda}^{(2)}(y)  + 
{\mathcal M}^{*} c_{-}(k)H_{\lambda}^{(1)}(y)], \qquad \tau \geq -\tau_{1},
\end{eqnarray}
where 
\begin{eqnarray}
&&x = k \tau, \qquad y = k \tau + k\tau_{1} \biggl(1 + \frac{\alpha}{\beta}\biggr)
\nonumber\\
&& 
{\mathcal N} = \sqrt{\frac{\pi}{2}} e^{i(\nu + 1/2)\pi/2}, \qquad 
{\mathcal M} = \sqrt{\frac{\pi}{2}} e^{- i(\lambda + 1/2)\pi/2}.
 \end{eqnarray}
The continuity of the tensor mode functions at the transition point 
 \begin{equation}
 f_{\mathrm{i}}(-\tau_{1}) = \tilde{f}_{\mathrm{s}}(-\tau_{1}), \qquad 
 g_{\mathrm{i}}(-\tau_{1}) = \tilde{g}_{\mathrm{s}}(-\tau_{1}),
 \end{equation}
 implies that the mixing coefficients are given by: 
 \begin{eqnarray}
 c_{+}(k) &=& \frac{i \pi}{8 \sqrt{\alpha \beta}} e^{i \pi (\nu +\lambda)/2} 
 \{[\beta ( 2 \lambda + 1) + \alpha ( 2 \nu + 1)] H_{\nu}^{(1)}(x_1) H_{\lambda}^{(1)}(y_{1}) 
 \nonumber\\
 &-& 2 \alpha x_{1} [H_{\lambda}^{(1)}(y_{1}) H_{\nu+1}^{(1)}(x_1) + H_{\nu}^{(1)}(x_{1}) H_{\lambda + 1}^{(1)}(y_{1})]\}, 
 \nonumber\\
 c_{-}(k) &=&  \frac{i\pi}{8 \sqrt{\alpha \beta}} e^{i \pi (\nu - \lambda)/2} 
 \{[\beta ( 2 \lambda + 1) + \alpha ( 2 \nu + 1)] H_{\nu}^{(1)}(x_1) H_{\lambda}^{(2)}(y_{1}) 
 \nonumber\\
&-& 2 \alpha x_{1} [H_{\lambda}^{(2)}(y_{1}) H_{\nu+1}^{(1)}(x_1) + H_{\nu}^{(1)}(x_{1}) H_{\lambda + 1}^{(2)}(y_{1})]\},
\end{eqnarray} 
where, according to the notations previously established, $y_{1}
= y(-\tau_{1}) = (\alpha/\beta) x_{1}$.  The case $\alpha = \beta = 1$ corresponds 
to a transition from the inflationary phase 
to a radiation-dominated phase. In this case we do know which are the mixing 
coefficients. The previous expressions give us:
\begin{equation}
c_{-}(k) = \frac{e^{ 2 i x_{1}}}{2 x_{1}^2},\qquad 
c_{+}(k) = \biggl( 1 - \frac{1}{2 x_{1}^2} + \frac{i}{x_{1}}\biggr),
\label{pureDS}
\end{equation}
which clearly agree with previous results  \cite{flat1,flat6,flat8}. 
In the case of Eq. (\ref{pureDS}) $|c_{+}(k)|^2 - |c_{-}(k)|^2 =1$ and 
$ k^4 |c_{-}(k)|^2$ is exactly scale-invariant.
Another interesting situation is the one of the transition from inflation to stiff, 
 i.e. $\beta = 1$, $\alpha = 1/2$, $y_{1} = x_{1}/2$ which leads 
 to a logarithmic enhancement at small wavenumbers \cite{mg1,mg2}.
 In this situation the mixing coefficients can be written as:
 \begin{eqnarray}
 c_{-}(k) &=& \sqrt{\frac{\pi}{2}} \frac{( i - 1 )}{4 x_1^{3/2}} e^{i x_1}\biggl\{ \sqrt{2} e^{- i x_{1}/2} [ x_{1}^2 + 6 i x_{1} - 12] H_{0}^{(2)}(x_{1}/2) 
 \nonumber\\
 &+&  (i + x_{1}) [ i x_{1} H_{1}^{(2)}(x_{1}/2) - 3 i H_{0}^{(2)}(x_1/2)]\biggr\},
 \label{mix1}\\
 c_{+}(k) &=&\sqrt{\frac{\pi}{2}} \frac{( i + 1 )}{4 \sqrt{x_{1}}} e^{i x_1}\biggl\{x_{1} H_{0}^{(1)}(x_{1}/2) + i ( i + x_{1})  H_{1}^{(1)}(x_{1}/2)\biggr\}.
 \label{mix2}
 \end{eqnarray}
The above result can be expanded in for $x_{1}\ll 1$ and the result is:
\begin{eqnarray}
c_{+}(k) &=& \frac{-0.398( 1 -\, i)}{{x_1}^{\frac{3}{2}}}  + \sqrt{x_1}\,
   [ \left( 0.131 + 0.338\, i \right)  - 0.149\left( 1 - i  \right) \ln{x_{1}}]  + {\mathcal O}(x_{1}^{3/2}),
 \label{mix1a}\\
 c_{-}(k) &=& \frac{\left( 7.031 - 1.723\,i \right)  -  16.68 \left( 1+ \,i   \right) \,\ln{x_{1}}}{x_{1}^{3/2}} 
\nonumber\\
 &+& \sqrt{x_1}\,\left[ \left( -0.621 + 0.265 \,i  \right)  + 0.282\left(1 + i  \right) \, \ln{x_{1}} \right] + {\mathcal O}({x_{1}}^{3/2}).
\label{mix2a}
\end{eqnarray}
The logarithms arising in Eqs. (\ref{mix1a}) and (\ref{mix2a}) explain why, in Eq. (\ref{ST10}), the transfer 
function of the spectral energy density contains logarithms.
In spite of the fact that semi-analytical estimates can pin down the slope of the transfer functions in different intervals, they are insufficient for a faithful account of more realistic situations where the slow-roll corrections are relevant and when 
other dissipative effects (such as neutrino fee streaming) are taken into account.

\subsection{Exponential damping of the mixing coefficients}  
Consider the case of the  $\Lambda$CDM paradigm where the inflationary epoch 
is almost suddenly followed by the radiation-dominated phase. 
By denoting the transition time as $\tau_{\mathrm{i}}$, it is plausible to think that all the modes of the field such that 
$k > a_{\mathrm{i}} H_{\mathrm{i}} \simeq \tau_{\mathrm{i}}^{-1}$ are exponentially suppressed \cite{bir,gar}.  For the modes $k\tau_{\mathrm{i}} > 1$, the pumping action of the gravitational field is practically absent. 
There will be a given $k$, be it $k_{\mathrm{max}}$, for which $k \simeq \tau_{\mathrm{i}}^{-1}$. The latter wavenumber 
is, in practice, the maximal $k$ to be amplified and it can be estimated as :
\begin{eqnarray}
k_{\mathrm{max}} &=& 7.5959\times 10^{25}\,(\pi \epsilon {\mathcal A}_{{\mathcal R}})^{1/4} \biggl(\frac{h_{0}^2 \Omega_{\mathrm{R}0}}{4.15 \times 10^{-5}}\biggr)^{1/4} \,\, \mathrm{Mpc}^{-1},
\label{W1}\\
\nu_{\mathrm{max}}  &=& 1.1745\times 10^{11} \,(\pi \epsilon {\mathcal A}_{{\mathcal R}})^{1/4} \biggl(\frac{h_{0}^2 \Omega_{\mathrm{R}0}}{4.15 \times 10^{-5}}\biggr)^{1/4} \,\mathrm{Hz}.
\label{W2}
\end{eqnarray}
Equations (\ref{W1}) and (\ref{W2}) are derived by assuming that, right after inflation, the radiation-dominated 
phase takes over. Furthermore, recalling the slow-roll dynamics, $3 H^2_{\mathrm{i}} \overline{M}_{\mathrm{P}}^2 \simeq V$ 
and $V \propto \overline{M}_{\mathrm{P}}^4 \epsilon \overline{{\mathcal P}}_{\mathcal R}$. In Eqs. (\ref{W1}) and (\ref{W2}) 
${\mathcal A}_{{\mathcal R}}$ denotes, as already established, the amplitude of the curvature power spectrum  evaluated 
at the pivot scale. 

By taking as typical values of the curvature perturbations at the pivot scale the one 
endorsed, for instance, by the WMAP 5-yr data alone we will have we will have that Eqs. (\ref{W1}) and (\ref{W2}) can be 
written as: 
\begin{equation}
k_{\mathrm{max}} =  3.5661\times 10^{22} \,\biggl(\frac{\epsilon}{0.01}\biggr)^{1/4} 
\biggl(\frac{{\mathcal A}_{\mathcal R}}{2.41 \times 10^{-9}}\biggr)^{1/4}
\biggl(\frac{h_{0}^2 \Omega_{\mathrm{R}0}}{4.15 \times 10^{-5}}\biggr)^{1/4} \,\, \mathrm{Mpc}^{-1},
\label{W3}
\end{equation}
where the typical values of the slow-roll parameter have been derived by taking into account 
that, in the absence of running of the tensor spectral index, $r_{\mathrm{T}} = 16 \epsilon$; since, 
according to the WMAP 5-yr data alone, $r_{\mathrm{T}} < 0.43$,  $\epsilon \leq 0.01$.
\begin{figure}[!ht]
\centering
\includegraphics[height=6.7cm]{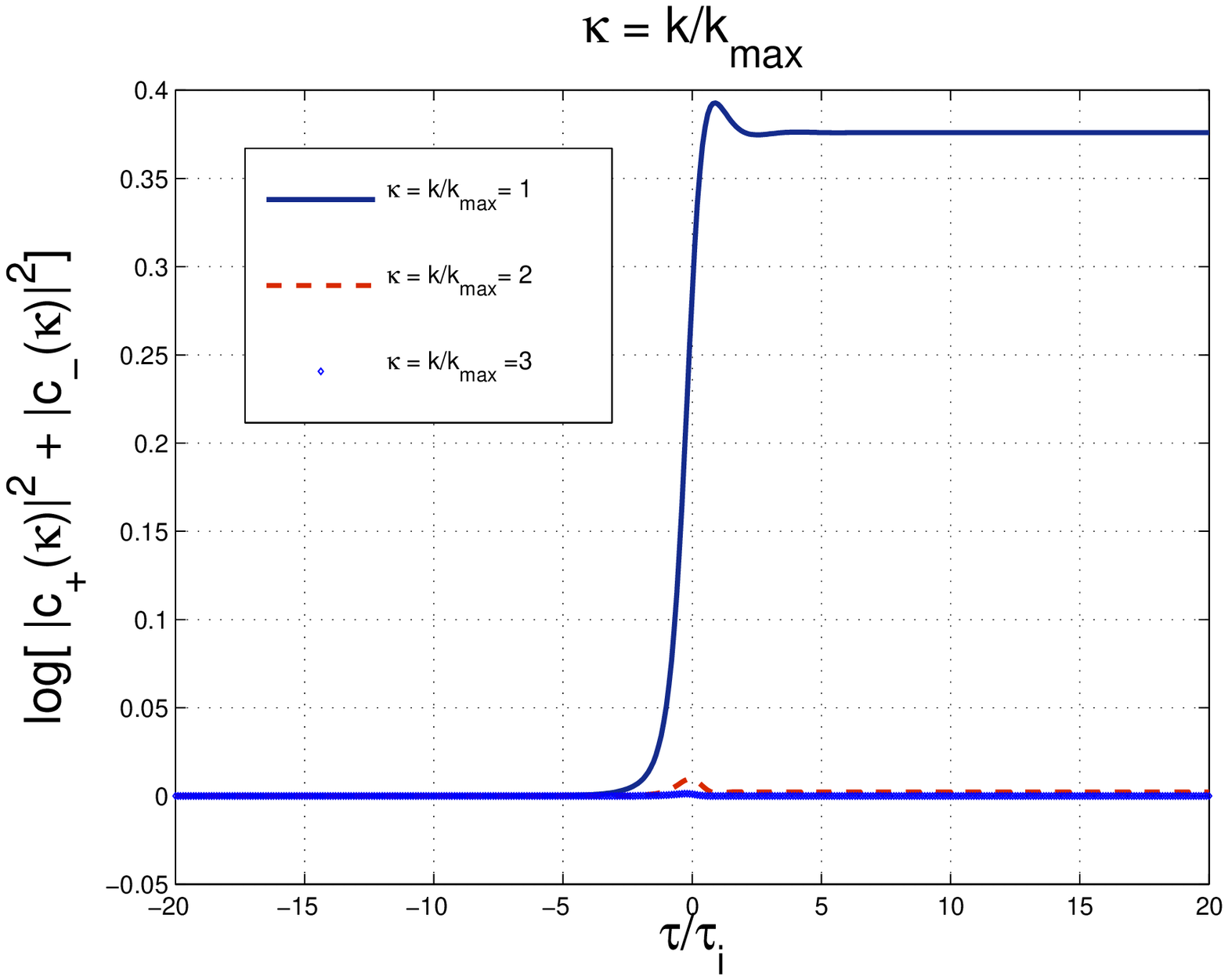}
\includegraphics[height=6.7cm]{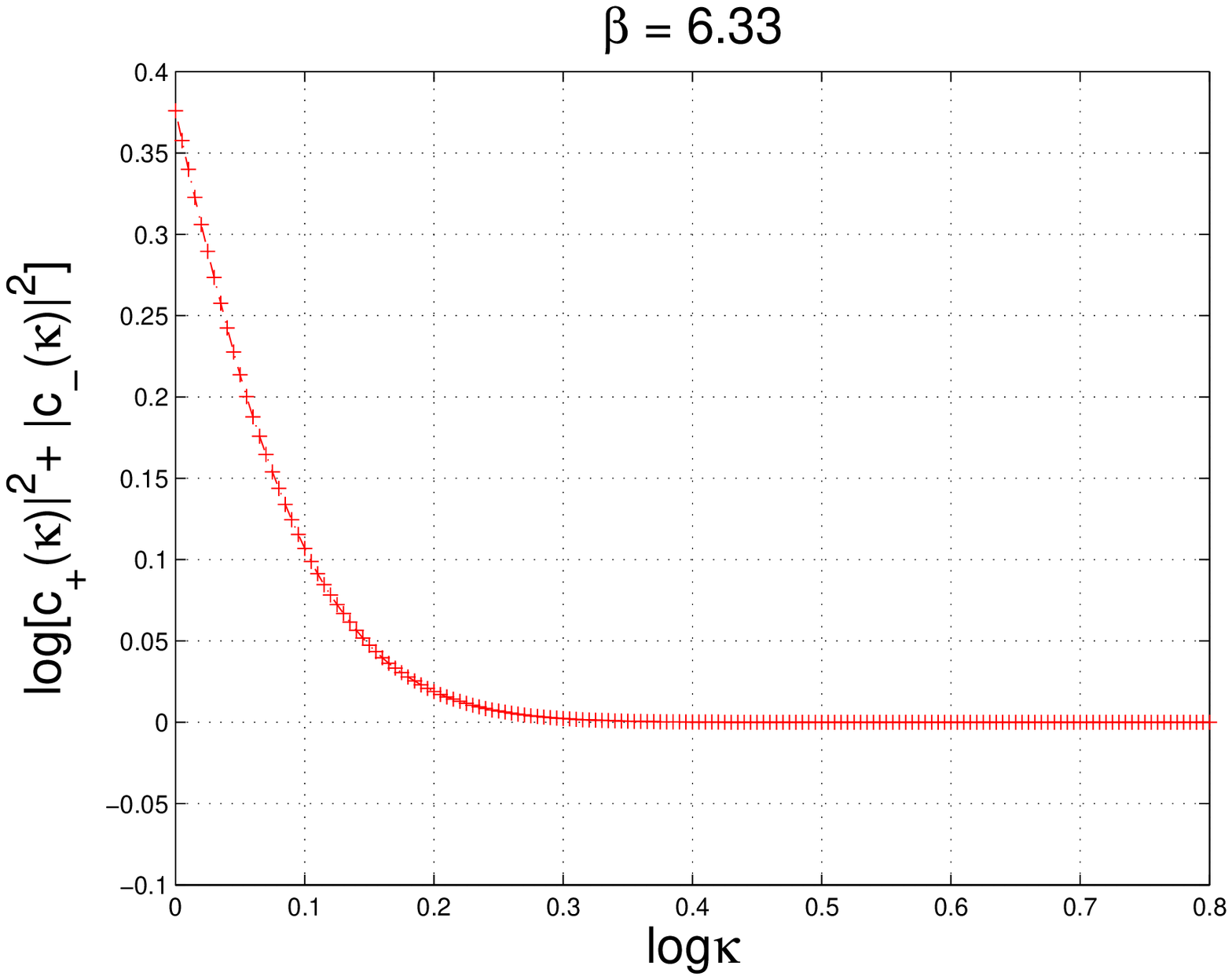}
\caption[a]{The time evolution of the mixing coefficients is reported at the left (on the horizontal axis the scale is linear). 
The exponential decay of the mixing coefficients is illustrated in the plot at the right in double 
logarithmic axes.}
\label{Figure7}
\end{figure}
For phenomenological purposes it can be also interesting to know what kind of exponential suppression 
we can expect. From the analysis of various transitions it emerges that the mixing coefficients for $k > k_{\mathrm{max}}$ (or $\nu > \nu_{\mathrm{max}}$) will satisfy 
\begin{equation}
|c_{+}(k)|^2 - |c_{-}|^2 =1 , \qquad |c_{+}(k)|^2 + |c_{-}(k)|^2 =  e^{- 2 \beta \frac{k}{k_{\mathrm{max}}}} + 1.
\label{W5}
\end{equation}
From Eq. (\ref{W5}) we can easily argue that, for $k > k_{\mathrm{max}}$, $|c_{+}(k)| \to 1$ and 
$|c_{-}(k)| \simeq 2^{-1/2} \exp{[- \beta k/k_{\mathrm{max}}]}$.  The point is then to estimate 
the value of $\beta$ which depends on the nature of the transition regime. Typically, however, $\beta > 2$ for sufficiently 
smooth transitions. To justify this statement it is interesting 
to consider the following toy model where the scale factor interpolates between a quasi-de Sitter phase and a radiation-dominated phase:
\begin{equation}
a(\tau) = a_{\mathrm{i}} [ \tau +  \sqrt{\tau^2 + \tau_{\mathrm{i}}^2}].
\label{W6}
\end{equation}
For  $\tau \to - \infty$ (i.e. $\tau \ll - \tau_{\mathrm{i}}$) , $a(\tau) \simeq - a_{\mathrm{i}}/\tau$ and the 
quasi de-Sitter dynamics is recovered. In the opposite 
limit (i. e. $\tau \gg + \tau_{\mathrm{i}}$), $a(\tau) \simeq a_{\mathrm{i}}\, \tau$ and the radiation dominance 
is recovered. 
In Fig. \ref{Figure5} (plot at the left) the exponential damping of the mixing coefficients is numerically illustrated. The 
curve at the top (full line) illustrates the case $\kappa = 1$. The cases $\kappa= 2$ and $\kappa = 3$ are barely 
distinguishable at the bottom of the plot. Notice, always in the right plot, the rather narrow 
range of times which are reported in a linear scale. In the plot at the right the asymptotic values of the mixing 
coefficients are reported for different values of $\kappa = k/k_{\mathrm{max}}$. By fitting the numerical data with 
with an equation of the form given in Eq. (\ref{W5}), the value of $\beta= 6.33$. 
Different examples can be presented on the same line of the one discussed in Fig. \ref{Figure5}. 
While it is pretty clear that the decay is indeed exponential, the value of $\beta$ may well vary. 
This can be summarized, for instance, in a rescaling of $k_{\mathrm{max}}$, i.e. by positing, for instance 
that $k_{\mathrm{max}} \to  \tilde{k}_{\mathrm{max}}/\beta$. Thus, the dynamics of the transition can slightly shift the 
numerical value of the upper cut-off by a numerical factor which depends upon 
the width of the transition regime.
\subsection{Nearly scale-invariant spectra}
\begin{figure}[!ht]
\centering
\includegraphics[height=6.7cm]{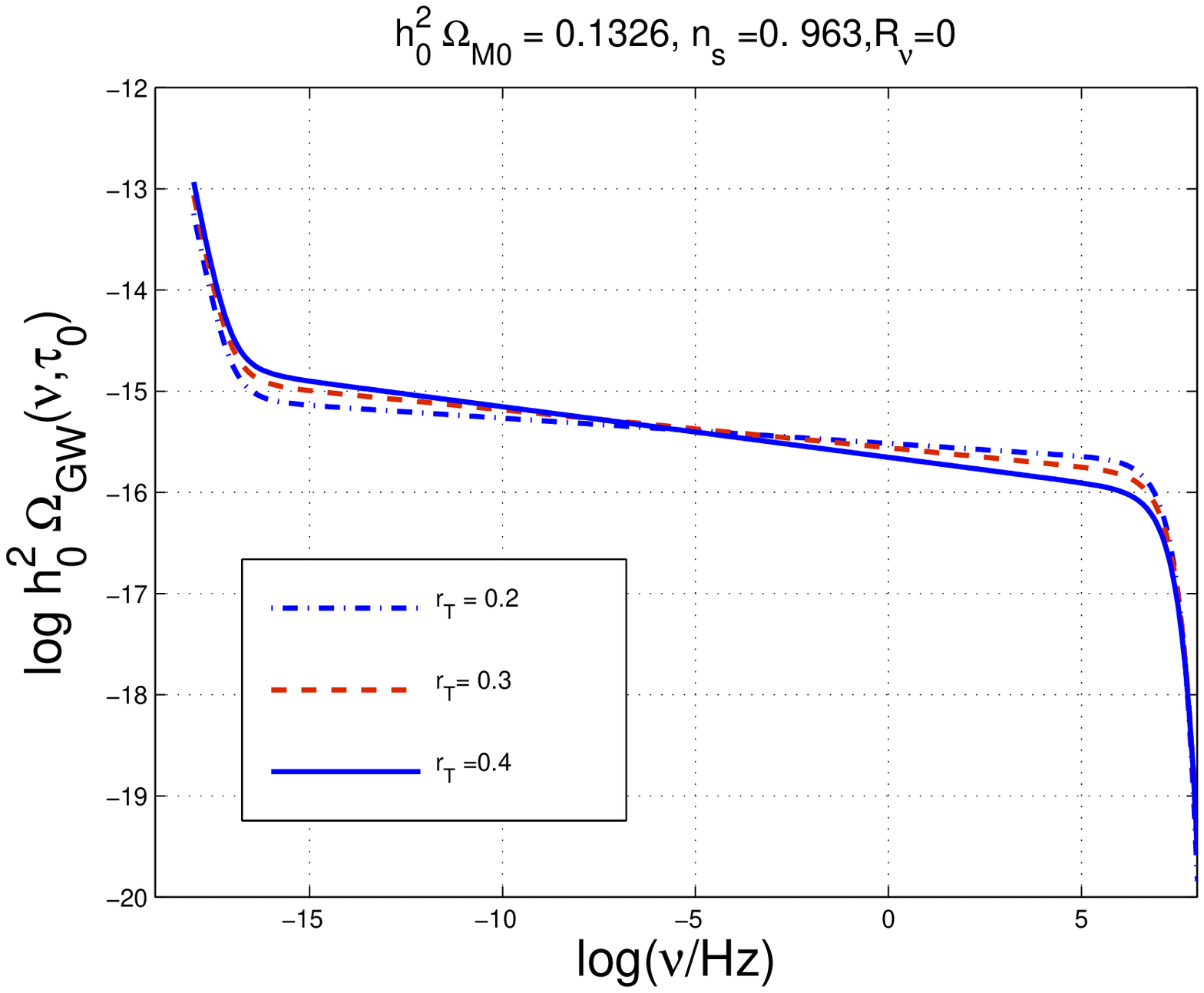}
\includegraphics[height=6.7cm]{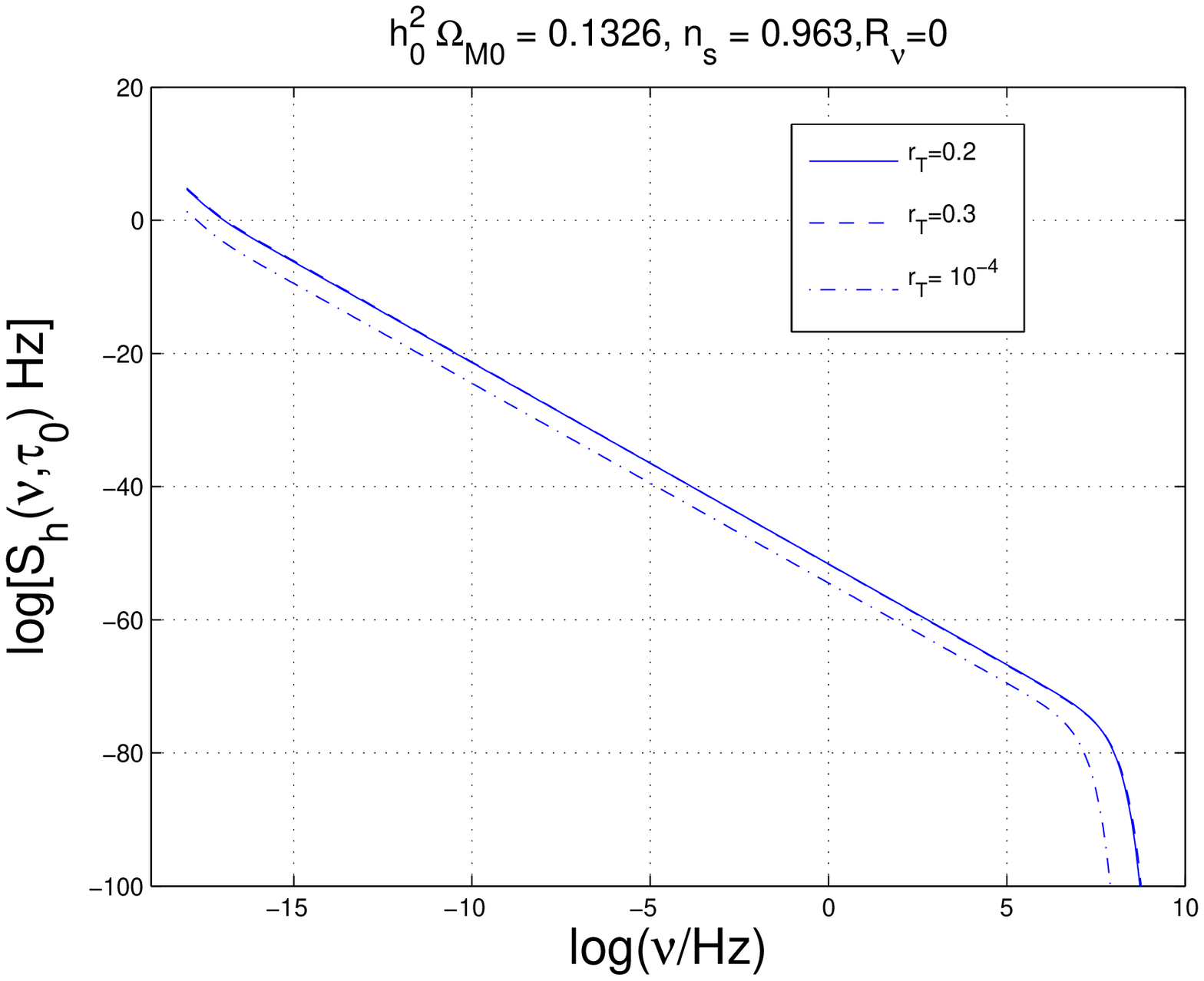}
\caption[a]{The spectral energy density of the relic gravitons (plot at the left) and the related $S_{h}(\nu,\tau_{0})$ (plot at the right) 
for different values of $r_{\mathrm{T}}$ and for the same set of fiducial parameters 
illustrated in Fig. \ref{Figure6}.}
\label{Figure9a}      
\end{figure}

By using the transfer function for the tensor amplitude, 
the spectral energy density for frequencies $\nu \gg \nu_{\mathrm{eq}}$ can be simply given by:
\begin{eqnarray}
h_{0}^2 \Omega_{\mathrm{GW}}(\nu,\tau_{0}) &=& {\mathcal N}_{h} \,\, r_{\mathrm{T}} \,\, \biggl(\frac{\nu}{\nu_{\mathrm{p}}}\biggr)^{n_{\mathrm{T}}} e^{- 2\beta \frac{\nu}{\nu_{\mathrm{max}}}},
\label{scaleinv1}\\
{\mathcal N}_{h} &=& 7.992 \times 10^{-15} \biggl(\frac{h_{0}^2 \Omega_{\mathrm{M}0}}{0.1326}\biggr)^{-2} 
\biggl(\frac{h_{0}^2 \Omega_{\mathrm{R}0}}{4.15\times 10^{-5}}\biggr) \biggl(\frac{d_{\mathrm{A}}}{1.4115 \times 10^{4}\, \mathrm{Mpc}}\biggr)^{-4},
\label{scaleinv2}
\end{eqnarray}
where $d_{\mathrm{A}}(z_{*})$ is the (comoving) angular diameter distance to decoupling.  In Eqs. (\ref{scaleinv1})--(\ref{scaleinv2}) (as well as in the program used for the 
numerical calculations) there are two complementary options. The first one is to use the angular diameter distance 
to decoupling which is directly inferred from the CMB data. For instance, in the case of the 5-yr WMAP data alone, 
$d_{A}(z_{*}) = 14115 \,\,\mathrm{Mpc}_{-191}^{+188}$. This is the strategy also adopted in other studies \cite{EF1} (in connection, obviously, with earlier 
releases of WMAP data). At the same time it is 
also possible to take the best fit value of the total matter fraction (i.e. $\Omega_{\mathrm{M}0}= 0.258$ for the case of the WMAP
5-yr data alone) and compute the comoving angular diameter distance according to the well know expression for spatially flat Universes: 
\begin{equation}
d_{\mathrm{A}}(z_{*}) =  \frac{1}{H_{0}} \int_{0}^{z_{*}} \frac{d z}{\sqrt{\Omega_{\mathrm{M}0} ( 1 + z)^3 + \Omega_{\Lambda} + \Omega_{\mathrm{R}0} (1 + z)^4}} = \frac{3.375}{H_{0}}= 14072\,\, \mathrm{Mpc},\qquad z_{*} = 1090.
\label{scaleinv3}
\end{equation}
The latter strategy has been used, for instance, in \cite{page}. The two strategies are  compatible and, moreover, this 
explains why, in Eq. (\ref{scaleinv2}) the dependence upon $\Omega_{\mathrm{M}0}$ does not cancel. 
In Eq. (\ref{scaleinv1}) $n_{\mathrm{T}}$ denotes, as usual, the tensor spectral index which can be also written as 
\begin{equation}
n_{\mathrm{T}} =  - 2 \epsilon + \frac{\alpha_{\mathrm{T}}}{2} \ln{(k/k_{\mathrm{p}})}, \qquad \alpha_{\mathrm{T}} = \frac{r_{\mathrm{T}}}{8}\biggl[(n_{\mathrm{s}} -1) + 
\frac{r_{\mathrm{T}}}{8}\biggr],
\label{int3a}
\end{equation}
If $\alpha_{\mathrm{T}} =0$, the standard case is recovered. It should be finally mentioned 
that, in the limit $\nu \gg \nu_{\mathrm{eq}}$, the oscillating terms have to be appropriately averaged: this is
 done by setting the terms going as $\cos^2{(2\pi \nu\tau_{0})}$ to $1/2$. The latter procedure has been employed, for instance, in the analyses of \cite{EF1,wnu3}.
In Fig. \ref{Figure9a} the transfer function for the spectral energy density (introduced 
in Eqs. (\ref{DELTA4}) and (\ref{ENTRANS})) has been 
consistently employed to estimate the spectral energy density itself and the spectral amplitude.
According to Eq. (\ref{T11}) the
spectral energy density can be written as  
\begin{equation}
\Omega_{\mathrm{GW}}(\nu,\tau) = \frac{4 \pi^2}{3 {\mathcal H}^2} \nu^3  S_{h}(\nu,\tau),
\label{DEF2}
\end{equation}
where we used that $k = 2 \pi \nu$ and that $H a = {\mathcal H}$.  
By making explicit the numerical factors in Eq. (\ref{DEF2}), $S_{h}(\nu,\tau_{0})$ can 
be expressed in terms of the spectral energy (in critical units)
\begin{equation}
S_{h}(\nu,\tau_{0}) = 7.981\times 10^{-43} \,\,\biggl(\frac{100\,\mathrm{Hz}}{\nu}\biggr)^3 \,\, h_{0}^2 \Omega_{\mathrm{GW}}(\nu,\tau_{0})\,\, \mathrm{Hz}^{-1},
\label{DEF3}
\end{equation}
where, we recall, $H_{0} = 3.24078\times 10^{-18}\,\,h_{0}\,\,\mathrm{Hz}$.
In Fig. \ref{Figure9a} the spectral energy of the relic gravitons as well as 
$S_{h}(\nu,\tau_{0})$ are reported for different values 
of $r_{\mathrm{T}}$.   
The spectra of Fig. \ref{Figure9a} have been obtained from the direct integration 
of the mode functions but can be parametrized, according to Eq. (\ref{ENTRANS}) as 
\begin{eqnarray}
h_{0}^2 \Omega_{\mathrm{GW}}(\nu,\tau_{0}) &=& {\mathcal N}_{\rho}  T^2_{\rho}(\nu/\nu_{\mathrm{eq}}) r_{\mathrm{T}} \biggl(\frac{\nu}{\nu_{\mathrm{p}}}\biggr)^{n_{\mathrm{T}}} e^{- 2 \beta \frac{\nu}{\nu_{\mathrm{max}}}}
\label{scaleinv4}\\
{\mathcal N}_{\rho} &=& 4.165 \times 10^{-15} \biggl(\frac{h_{0}^2 \Omega_{\mathrm{R}0}}{4.15\times 10^{-5}}\biggr).
\label{scaleinv5}
\end{eqnarray}
By comparing Eqs. (\ref{scaleinv1})--(\ref{scaleinv2}) to Eqs. (\ref{scaleinv4})--(\ref{scaleinv5}), the amplitude for $\nu\gg \nu_{\mathrm{eq}}$ differs, roughly,  
by a factor $2$. This coincidence is not surprising since  Eqs. (\ref{scaleinv1})--(\ref{scaleinv2}) have been obtained by averaging over the 
oscillations (i.e. by replacing cosine squared with $1/2$) and by imposing that $|g_{k}| = k |f_{k}|$.  These 
manipulations are certainly less accurate than the procedure used to derive the transfer function for the spectral energy density. 

So far the evolution of the tensor modes has been treated in the absence of anisotropic stress. 
This approximation is, strictly speaking, unrealistic. Indeed 
we do know that there are sources of anisotropic stress. After neutrino decoupling, the neutrinos free stream and the effective energy-momentum tensor acquires, to first-order in the amplitude 
of the plasma fluctuations, an anisotropic stress, i.e. 
\begin{equation}
\delta T_{i}^{j} = - \delta p  \delta_{i}^{j} + \Pi_{i}^{j},\qquad \partial_{i} \Pi_{j}^{i}= \Pi_{i}^{i} =0.
\label{ANIS1}
\end{equation}
The presence of the anisotropic stress clearly affects the evolution of the tensor 
modes. To obtain the wanted equation we perturb the Einstein equations to first-order and we get:
\begin{equation}
{h_{i}^{j}}'' + 2 {\mathcal H} {h_{i}^{j}}' - \nabla^2 h_{i}^{j} = - 16 \pi G a^2 \Pi_{i}^{j}.
\label{ANIS2}
\end{equation}
Equation (\ref{ANIS2}) reduces to an integro-differential equation which has been analyzed in 
\cite{wnu1} (see also \cite{wnu2,wnu3,wnu4}). The overall effect of collisionless particles is a reduction 
of the spectral energy density of the relic gravitons. Assuming that the only collisionless 
species in the thermal history of the Universe are the neutrinos, the amount 
of suppression can be parametrized by the function
\begin{equation}
{\mathcal F}(R_{\nu}) = 1 -0.539 R_{\nu} + 0.134 R_{\nu}^2 
\label{ANIS3}
\end{equation}
where $R_{\nu}$ is the fraction of neutrinos in the radiation plasma.
In the case $R_{\nu}=0$ (i.e. in the absence of collisionless patrticles) there is no suppression. If, on the contrary, 
$R_{\nu} \neq 0$ the suppression can even reach one order of magnitude. In the case $N_{\nu} = 3$, 
$R_{\nu} = 0.405$ and the suppression of the spectral energy density is proportional 
to ${\mathcal F}^2(0.405)= 0. 645$. This suppression will be effective for relatively 
small frequencies which are larger than $\nu_{\mathrm{eq}}$ and smaller than the
frequency corresponding to the Hubble radius at the time 
of big-bang nucleosynthesis, i.e. 
\begin{equation}
\nu_{\mathrm{bbn}} = 
2.252\times 10^{-11} \biggl(\frac{g_{\rho}}{10.75}\biggr)^{1/4} \biggl(\frac{T_{\mathrm{bbn}}}{\,\,\mathrm{MeV}}\biggr) 
\biggl(\frac{h_{0}^2 \Omega_{\mathrm{R}0}}{4.15 \times 10^{-5}}\biggr)^{1/4}\,\,\mathrm{Hz}.
\label{ANIS5}
\end{equation}
The effect of neutrino free-streaming (as well as all the other 
late time effects addressed in this subsection)  has been taken into account in Fig. \ref{Figure1} (see section \ref{sec1})
but it is absent in Fig. \ref{Figure9a}.
The second effect which has been taken into account in Fig. \ref{Figure1} is the damping associated with the (present) dominance of the dark energy component. Indeed 
the redshift of $\Lambda$-dominance is simply  defined by 
\begin{equation}
1 + z_{\Lambda} = \biggl(\frac{a_{0}}{a_{\Lambda}}\biggr) = 
 \biggl(\frac{\Omega_{\Lambda}}{\Omega_{\mathrm{M}0}}\biggr)^{1/3}.
\label{LAM1}
\end{equation}
Consider now the mode which will be denoted as $k_{\Lambda}$, i.e. the mode 
reentering the Hubble radius at $\tau_{\Lambda}$. By definition $k_{\Lambda} = H_{\Lambda} a_{\Lambda}$ must hold. But for $\tau > \tau_{\Lambda}$ 
is constant, i.e. $H_{\Lambda} \equiv H_{0}$ where $H_{0}$ is the present value of the Hubble rate.
Using now Eq. (\ref{LAM1}), it can be easily shown that $k_{\Lambda} = (\Omega_{M0}/\Omega_{\Lambda})^{1/3}k_{\mathrm{H}}$ 
where $k_{\mathrm{H}} = a_{0} H_{0}$. The frequency interval between $\nu_{\mathrm{H}}$ and $\nu_{\Lambda}$ is rather  tiny. 
Indeed, it turns out that 
\begin{eqnarray}
&& k_{\Lambda} = 1.686 \times 10^{-4} \, \biggl(\frac{h_{0}}{0.719}\biggr) \biggl(\frac{\Omega_{\mathrm{M}0}}{0.258}\biggr)^{1/3} \biggl(\frac{\Omega_{\Lambda}}{0.742}\biggr)^{1/3} \,\, \mathrm{Mpc}^{-1},
\label{LAM2}\\
&& \nu_{\Lambda} = 2.607 \times 10^{-19}  \biggl(\frac{h_{0}}{0.719}\biggr) \biggl(\frac{\Omega_{\mathrm{M}0}}{0.258}\biggr)^{1/3} \biggl(\frac{\Omega_{\Lambda}}{0.742}\biggr)^{1/3} \,\, \mathrm{Hz}.
\label{LAM3}
\end{eqnarray}
For the same choice of parameters of Eq. (\ref{LAM3}), $\nu_{\mathrm{H}} = 3.708 \times 10^{-19}$ Hz 
which is not so different than $\nu_{\Lambda} =  2.607 \times 10^{-19}$ Hz.  It would therefore seem that  this branch of the spectrum could be easily neglected. However, it turns 
out that the adiabatic damping of the mode function across $\tau_{\Lambda}$ reduces the amplitude of the spectral energy density by a factor 
$(\Omega_{\mathrm{M}0}/\Omega_{\Lambda})^2$. For the typical choice of parameters of Eqs. (\ref{LAM2}) and (\ref{LAM3}) we have that 
the suppression is of the order of $0.12$. Again this is a tiny number which is, anyway, comparable with the suppression due, for instance, to 
the neutrino free streaming. This class of effects has been repeatedly 
in a number of recent papers 
\cite{zh1} (see also \cite{zh2}).  The essence of the effect is captured by the following observation. Consider a mode $k$ which reenters before  $\tau_{\Lambda}$.
The present value of the amplitude $F_{k}(\tau) = f_{k}(\tau)/a(\tau)$ will be adiabatically suppressed since, as repeatedly stressed, in this 
regime $f_{k}(\tau)$ will simply be plane waves. Consequently, defining as $\tilde{F}_{k_{*}}$ the amplitude   at $k_{*} = H_{*} a_{*}$ when the given 
mode crosses the Hubble radius, we will also have that
 \begin{equation}
F_{k}(\tau_{0}) = \biggl(\frac{a_{k_{*}}}{a_{\Lambda}}\biggr)_{\mathrm{mat}} \biggl(\frac{a_{\Lambda}}{a_{0}}\biggr)_{\Lambda} \tilde{F}_{k_{*}} \equiv 
\biggl(\frac{k}{k_{\mathrm{H}}}\biggr)^{-2} \biggl(\frac{\Omega_{\mathrm{M}0}}{\Omega_{\Lambda}}\biggr)  \tilde{F}_{k_{*}},
\label{LAM4}
\end{equation}
where the subscripts (in the first equality) denote the time range over which the corresponding redshift is computed, i.e. either matter-dominated 
or $\Lambda$-dominated stages. The second equality follows from the first one by appreciating that $a(k_{*}) \simeq \tau_{*}^{2} \simeq k^{-2}$ and by using 
Eq. (\ref{LAM1}). Equation (\ref{LAM4}) implies, immediately, that the spectral energy density of relic gravitons is corrected in two different fashions. 
For $\nu < \nu_{\mathrm{H}}$ the frequency dependence will be different and will be proportional to 
$\Omega_{\mathrm{GW}}(\nu,\tau_{0}) \propto (\nu/\nu_{\mathrm{H}})^{n_{\mathrm{T}} -2} (\Omega_{\mathrm{M}0}/\Omega_{\Lambda})^{2}$.
Vice versa, in the range $\nu > \nu_{\mathrm{H}}$ the frequency dependence will be exactly the one already computed but, overall, the 
amplitude will be smaller by a factor $(\Omega_{\mathrm{M}0}/\Omega_{\Lambda})^2$. 
The modification of the frequency dependence is only effective between $0.36$ aHz and $0.26$ aHz: this effect is therefore unimportant and customarily ignored (see, for instance, \cite{EF1,zh1}) 
for phenomenological purposes. On the other hand, 
the overall suppression going as $(\Omega_{\mathrm{M}0}/\Omega_{\Lambda})^{2}$ must be taken properly into account on the same footing of other sources of suppression of the spectral energy density.
There is, in principle, a third effect which may arise and it has to do with the variation of the effective number of relativistic species. Recall, indeed, that 
the total energy and the total entropy densities of the plasma can be written as 
\begin{equation}
\rho_{\mathrm{t}} = g_{\rho}(T) \frac{\pi^2}{30} T^4,\qquad s_{\mathrm{t}} = g_{\mathrm{s}}(T) \frac{2 \pi^2}{45} T^3.
\label{EFF1}
\end{equation}
For temperatures much larger than the top quark mass, all the known species of the minimal standard model of particle interactions are in local thermal 
equilibrium, then $g_{\rho} = g_{\mathrm{s}} = 106.75$. Below, $T \simeq 175$ GeV the various species 
start decoupling, the notion of thermal equilibrium is replaced by the notion of kinetic equilibrium. The 
time evolution of the number of relativistic degrees of freedom effectively changes the evolution of the Hubble rate. 
In principle if a given mode $k$ reenters the Hubble radius at a temperature $T_{k}$ the spectral energy density 
of the relic gravitons is (kinematically) suppressed by a factor which can be estimated as (see, for instance, \cite{zh1})
 \begin{equation}
 \biggl(\frac{g_{\rho}(T_{k})}{g_{\rho0}}\biggr)\biggl(\frac{g_{\mathrm{s}}(T_{k})}{g_{\mathrm{s}0}}\biggr)^{-4/3}.
 \label{EFF2}
 \end{equation}
At the present time  $g_{\rho0}= 3.36$ and $g_{\mathrm{s}0}= 3.90$. In general terms the effect parametrized by 
Eq. (\ref{EFF2}) will cause a frequency-dependent suppression, i.e. a further modulation of the spectral 
energy density $\Omega_{\mathrm{GW}}(\nu,\tau_{0})$.  The maximal suppression one can expect 
can be obtained by inserting into Eq. (\ref{EFF2}) the largest $g_{\mathrm{s}}$ and $g_{\rho}$. 
So, in the case of the minimal standard model this would imply that the suppression (on $\Omega_{\mathrm{GW}}(\nu,\tau_{0})$)
 will be of the order of $0.38$. In popular supersymmetric extensions of the minimal standard models $g_{\rho}$ and $g_{s}$  can be as high as, approximately, $230$. This will bring down the figure given above to $0.29$.
 
All the three effects estimated in the last 
part of the present section (i.e. free streaming, dark energy, evolution of relativistic degrees of freedom) have common features.
Both in the case of the neutrinos and in the case of the evolution of the relativistic degrees of freedom the potential impact of the effect
could be more pronounced. 
For instance, suppose that, in the early Universe, the particle model has many more degrees of freedom and many more 
particles which can free stream, at some epoch. At the same time 
we can say that all the aforementioned effects {\em decrease} rather than {\em increasing} the spectral energy density.
Taken singularly, each of the effects will decrease $\Omega_{\mathrm{GW}}$ by less than  one 
order of magnitude. The net result of the combined effects will then be, roughly,  a suppression of $\Omega_{\mathrm{GW}}(\nu,\tau_{0})$
which is of the order of $3 \times 10^{-2}$ (for $10^{-16}\,\,\mathrm{Hz} < \nu <10^{-11} \, \mathrm{Hz}$) and of the order of $4\times 10^{-2}$ for $\nu > 10^{-11}$Hz.
These figures are comparable with the possible inaccuracies stemming from the calculation of the transfer function and, therefore, this is a further motivation, to use the transfer function of the spectral energy density.
Finally the late time effects reduce a quantity which is already pretty small, i.e., as computed, $h_{0}^2 \Omega_{\mathrm{GW}}(\nu,\tau_{0}) \simeq 10^{-15}$ 
for $\nu \gg \nu_{\mathrm{eq}}$.
\newpage
\renewcommand{\theequation}{7.\arabic{equation}}
\setcounter{equation}{0}
\section{B-modes induced by long wavelength gravitons}
\label{sec7}
In the minimal realization of the $\Lambda$CDM scenario the scalar fluctuations of the geometry induce an E-mode polarization which has been observed and which is now subjected to closer scrutiny \cite{WMAP51,WMAP52,WMAP53,WMAP54,WMAP55}
The tensor modes of the geometry not only induce an E-mode polarization but also a B-mode polarization.
The detected angular power spectra due to the presence of a putative 
(adiabatic) curvature perturbation are the temperature autocorrelation 
(TT angular power spectrum) the E-mode autocorrelation (EE angular power spectrum) and their cross correlation (i.e. the TE angular power spectrum). The various angular power spectra of the temperature 
and polarization observables have been already defined in section 
\ref{sec2} (see, in particular, Eqs. (\ref{intN5})--(\ref{intN6}) and discussions 
therein). Long wavelength gravitons  contribute not only to the TT, EE and TE angular power spectra but also to the B-mode autocorrelations, i.e. 
the BB angular power spectra. 
The effect of long wavelength gravitons on the temperature and polarization 
observables can be studied by deriving the evolution equations 
of the brightness perturbations which are related, in loose terms, to the 
fluctuations of the Stokes parameters.  The tensor nature of the fluctuation defined 
in Eq. (\ref{SI2}) plays, in this respect, a decisive role. In particular the following 
two points should be borne in mind: 
\begin{itemize}
\item{} in the case of the scalar modes of the geometry the heat 
transfer equations have an azimuthal symmetry; 
\item{} in the case of the tensor modes the fluctuations of the brightness 
do depend, both, upon $\mu = \cos{\vartheta}$ as well as upon $\varphi$; this is ultimately, the rationale for the existence 
of a B-mode polarization.
\end{itemize}
The heat transfer equations can be schematically written as  
\begin{equation}
\biggl(\frac{d f}{d \tau}\biggr) = - \epsilon' f + \frac{3 \epsilon'}{16 \pi } \int {\mathcal M}(\Omega,\Omega') f(\Omega') d \Omega',
\label{Boltz1}
\end{equation}
where $\epsilon' = x_{\mathrm{e}} \tilde{n}_{\mathrm{e}} \sigma_{\mathrm{T}} a$ is the differential optical depth
\footnote{In the differential optical depth enters not only the cross section but also the electron 
concentration $\tilde{n}_{\mathrm{e}}$ and the ionization fraction $x_{\mathrm{e}}$.
The notation for the differential optical depth varies: some authors prefer $\kappa'$ some 
other $\dot{\tau}$. Given the notations used for the conformal time coordinate we will stick to the choice made in 
Eq. (\ref{Boltz1}).}. In the expression for the differential optical depth  $\sigma_{\mathrm{T}}= (8/3) \pi r_{0}^2$ 
where $r_{0}= e^2/m_{\mathrm{e}}$ is the classical radius of the electron.
The (differential) cross section for Compton scattering of polarized photons can be written in terms 
of $\sigma_{\mathrm{T}}$ and it takes the usual Klein-Nishina form:
\begin{equation}
\frac{d \sigma}{d \Omega} = \frac{3 \sigma_{\mathrm{T}}}{32 \pi} \biggl(\frac{E_{\mathrm{f}}}{E_{\mathrm{i}}}\biggr)^2 \biggl[ \frac{E_{\mathrm{f}}}{E_{\mathrm{i}}} + \frac{E_{\mathrm{i}}}{E_{\mathrm{f}}} -2  + 4 (\hat{e}\,\cdot\,\hat{e}')^2 \biggr],
\label{Boltz1a}
\end{equation}
where $\hat{e}$ and $\hat{e}'$ are, respectively,  the outgoing and ingoing polarizations 
of the photons; $E_{\mathrm{f}}$ and $E_{\mathrm{i}}$ are, respectively, 
the energies of the outgoing and of the ingoing photons. In the limit $ E_{\mathrm{f}} \simeq E_{\mathrm{i}}$  the differential cross section becomes, as anticipated
\begin{equation}
\frac{d \sigma}{d \Omega} = \frac{3 \sigma_{\mathrm{T}}}{8 \pi} (\hat{e}\,\cdot\,\hat{e}')^2,
\label{Boltz1b}
\end{equation}
The right left side of Eq. (\ref{Boltz1}) constitutes the collisionless 
term while the right hand side is the collisional contribution. 
At the right-hand side 
of Eq. (\ref{Boltz1b}) ${\mathcal M}(\Omega,\Omega')$ is, in general, a matrix whose dimensionality depends upon the specific problem. As it will be shown ${\mathcal M}(\Omega,\Omega')$ can be easily computed from Eq. (\ref{Boltz1b}). In similar terms $f(\Omega)$ should be understood as a column matrix whose components are the various Stokes parameters.

\subsection{Collisionless Boltzmann equation for the tensor modes} 
The collisionless part of the Boltzmann equation can be written as:
\begin{equation}
\frac{d f}{ d \tau} = \frac{\partial f}{\partial \tau} + \frac{d x^{i}}{d \tau} \frac{\partial f}{\partial x^{i}} +  
\frac{d q^{i}}{d\tau} \frac{\partial f}{\partial q^{i}},
\label{Boltz2}
\end{equation}
where $q^{i}$ is the comoving three-momentum of the photon.  
If the tensor fluctuation of the geometry is parametrized as in Eq. (\ref{SI2})
\begin{equation}
\delta_{\mathrm{t}}^{(1)} g_{ij} = - a^2(\tau) h_{ij}(\vec{x},\tau),
\label{Boltz3}
\end{equation}
then, the right hand side of Eq. (\ref{Boltz2}) can be written as 
\begin{equation} 
\frac{\partial f}{\partial \tau} + \hat{n}^{i}\frac{\partial f}{\partial x^{i}}  - \frac{1}{2} \hat{n}^{i} \hat{n}^{j} h_{ij}' 
\frac{\partial f}{\partial q},
\label{Boltz4}
\end{equation}
where the prime denotes a derivation with respect to $\tau$ and where the following identities have been used 
\begin{equation}
\frac{\partial f}{\partial q^{i}} = \hat{n}^{i}   \frac{\partial f}{\partial q}, \qquad 
\hat{n}^{i} = \frac{d x^{i}}{d\tau} = \frac{q^{i}}{|\vec{q}|}, \qquad 
\frac{d q^{i}}{d \tau} = - \frac{1}{2} h_{ij}' q^{j}. 
\label{Boltz5}
\end{equation}
In Eq. (\ref{Boltz5}) $\hat{n}^{i}$ denotes the direction of the photon 
momentum. The third identity appearing in Eq. (\ref{Boltz5}) stems 
directly from the definition of comoving three-momentum, i.e. 
\begin{equation}
- \delta_{i j} p^{i} p^{j} = g_{ij} P^{i} P^{j}, \qquad q^{i} = a p^{i}
\label{Boltz6}
\end{equation}
where $P^{i}$ is a generic spatial component of the canonical momentum 
obeying 
\begin{equation}
g_{\alpha\beta} P^{\alpha} P^{\beta} =0, \qquad P^{\alpha} = \frac{d x^{\alpha}}{d\lambda}. 
\label{Boltz7}
\end{equation}
Using Eq. (\ref{SI2}) the full metric $g_{ij}$ appearing in Eq. (\ref{Boltz6}) becomes 
\begin{equation}
g_{ij} = - a^2(\tau) \delta_{ij}  + \delta_{\mathrm{t}}^{(1)} g_{ij} \equiv -a^2(\tau) ( \delta_{ij} + h_{ij}).
\label{Boltz8}
\end{equation}
Consequently the comoving three-momentum $q^{i}$ and its conformal time derivative 
can be expressed as:
\begin{eqnarray}
&& q^{i} = a^2 \biggl( \delta^{i}_{j} + \frac{1}{2} h_{j}^{i}\biggr) \, P^{j}, 
\label{Boltz9}\\
&& \frac{d q^{i}}{d \tau} = \biggl[ 2 a^2 {\mathcal H} \biggl( \delta_{j}^{i} + \frac{1}{2} h_{j}^{i} \biggr) P^{j}  + 
\frac{a^2}{2} {h_{j}^{i}}' P^{j} + a^2 \biggl( \delta_{j}^{i} + \frac{1}{2} h_{j}^{i} \biggr) \frac{d P^{j}}{d \tau} \biggr].
\label{Boltz10}
\end{eqnarray}
To obtain an explicit expression for $d q^{i}/d\tau$ the conformal time derivative of the canonical momentum
should be made explicit by using the geodesic equation, i.e.
\begin{equation}
\frac{d P^{i}}{d\tau} + \Gamma^{i}_{\mu\nu} \frac{P^{\mu} P^{\nu}}{P^{0}} =0, \qquad P^{\mu} = \frac{d x^{\mu}}{ d \lambda} 
= P^{0} \frac{d x^{i}}{d\tau}.
\label{Boltz11}
\end{equation}
To first order in the tensor fluctuations of the geometry the Christoffel connection will then lead to the following expression: 
\begin{equation}
\Gamma_{j0}^{i} = \biggl( {\mathcal H} \delta_{j}^{i} + \frac{1}{2} {h_{j}^{i}}' \biggr).
\label{Boltz12}
\end{equation}
Using Eq. (\ref{Boltz12}) into Eqs. (\ref{Boltz10}) and (\ref{Boltz11}), the third identity reported in 
Eq. (\ref{Boltz5}) is quickly recovered.  
Depending upon the problem at hand the collisionless part of the Boltzmann equation 
can be written in different (but equivalent) forms. Equation (\ref{SI4}) allows to write $h_{ij}'$ in terms of the 
two polarizations: 
\begin{equation}
h_{ij}' = \sum_{\lambda} h_{(\lambda)}' \epsilon^{(\lambda)}_{ij}  \equiv h'_{\oplus} \epsilon_{ij}^{\oplus} + 
h_{\otimes}' \epsilon_{ij}^{\otimes}.
\label{Boltz13}
\end{equation}
Using Eqs. (\ref{Boltz5}) and (\ref{Boltz13}), Eq. (\ref{Boltz4}) can be written,  in Fourier space, as 
\begin{equation}
\frac{\partial f}{\partial \tau} + i k \mu f - \frac{1}{2} \frac{\partial f}{\partial \ln{q}} \biggl\{ [(\hat{n}\cdot \hat{a})^2  - (\hat{n}\cdot \hat{b})^2] h_{\oplus}' + 2 (\hat{n} \cdot \hat{a}) (\hat{n} \cdot \hat{b})  h_{\otimes}'\biggr\} = {\mathcal C}_{\mathrm{coll}},
\label{Boltz14}
\end{equation}
where ${\mathcal C}_{\mathrm{coll}}$ denotes, for completeness, the collisional part of the Boltzmann equation; 
$\mu = \hat{k} \cdot \hat{n}$ is the projection of the photon momentum on the direction of the Fourier mode. If the direction of propagation of the gravitational wave 
coincides with the third (Cartesian) component, i.e. $\hat{k} = \hat{z}$. It is then easy to see that 
 the unit vectors $\hat{a}$ and $\hat{b}$ must be directed, respectively, along the $\hat{x}$ and $\hat{y}$ 
 Cartesian directions. This particular choice of the coordinate system implies then 
 \begin{eqnarray}
&& (\hat{n}\cdot \hat{a})^2  - (\hat{n}\cdot \hat{b})^2 = n_{x}^2 - n_{y}^2 = (1 - \mu^2) \cos{2 \varphi},
\label{Boltz15}\\
&& 2 (\hat{n} \cdot \hat{a}) (\hat{n} \cdot \hat{b}) = 2 n_{x} n_{y} = (1 - \mu^2) \sin{2 \varphi}.
\label{Boltz16}
\end{eqnarray}
Equations (\ref{Boltz15}) and (\ref{Boltz16}) are not symmetric for $\varphi \to - \varphi$: while Eq. (\ref{Boltz15}) 
is left unchanged, Eq. (\ref{Boltz16}) acquires a minus sign. 
Conversely, the evolution equations of the scalar modes of the geometry are symmetric for $\varphi \to - \varphi$.  

\subsection{Azimuthal structure of the collisional contribution} 
The differential cross section of Eq. (\ref{Boltz1b}) depends upon $\hat{e}$ and $\hat{e}'$ which are, 
respectively, the polarizations of the outgoing and of the ingoing photons. 
If $\hat{k}$ coincides with the radial direction $\hat{e}$ and $\hat{e}'$ can be written, respectively, as 
\begin{eqnarray}
&& \hat{k} = (\sin{\vartheta}\,\cos{\varphi},\sin{\vartheta}\,\sin{\varphi},\, \cos{\vartheta}), 
\nonumber\\
&& \hat{e}_{x} = (\cos{\vartheta}\,\cos{\varphi},\, \cos{\vartheta}\,\sin{\varphi},\, -\sin{\vartheta}), \qquad 
\hat{e}_{y} = (\sin{\varphi},\, - \cos{\varphi},\, 0),
\label{Boltz18}\\
&&\hat{k}' = (\sin{\vartheta'}\,\cos{\varphi'},\, \sin{\vartheta'}\,\sin{\varphi'},\, \cos{\vartheta'}), 
\nonumber\\
&& \hat{e}_{x}' = (\cos{\vartheta'}\,\cos{\varphi'},\, \cos{\vartheta'}\,\sin{\varphi'},\, -\sin{\vartheta'}), \qquad 
\hat{e}_{y}' = (\sin{\varphi'},\, - \cos{\varphi'},\, 0).
\label{Boltz19}
\end{eqnarray}
In Eqs. (\ref{Boltz18}) and (\ref{Boltz19}) the components of $\hat{e}$ and $\hat{e}'$  have been 
identified with the $\hat{x}$ and $\hat{y}$ directions. 
This is just to guide the intuition since the components of  $\hat{e}$ and $\hat{e}'$ should only 
be orthogonal to each others and orthogonal to the direction of propagation. Furthermore 
notice that, in the case $\vartheta' = \varphi' =0$,  
\begin{equation}
\hat{k}'=(0,\,0,\,1),\qquad \hat{e}_{x}'=(1,\,0,\,0),\qquad \hat{e}_{y}'=(0,\,-1,\,0). 
\label{Boltz19a}
\end{equation}
Under the transformation 
 $\varphi \to - \varphi$ and $\varphi' \to - \varphi'$,  Eqs. (\ref{Boltz18}) and (\ref{Boltz19}) lead to another valid choice of orthogonal frame.  The intensity and the polarization of the (outgoing) radiation field will be given, respectively, as 
\begin{equation}
I(\tau,\vartheta,\varphi)= {\mathcal I}_{x}(\tau,\vartheta,\varphi) + 
{\mathcal I}_{y}(\tau,\vartheta,\varphi), \qquad  Q(\tau,\vartheta,\varphi) 
= {\mathcal I}_{x}(\tau, \vartheta,\varphi) - {\mathcal I}_{y}(\tau,\vartheta,\varphi),
\label{Boltz20}
\end{equation}
Since $Q$ and $U$ are not invariant for a rotation in the polarization plane and as soon as 
 $Q$ is generated also $U$ follows (see, e.g., Eqs. (\ref{Tr9}) and discussion therein). 
 For pedagogical purposes,  it is useful to consider, as a warm-up the simplification provided by Eq. (\ref{Boltz20}). 
 Consider the case where the ingoing polarization is fixed (e. g. 
$\vartheta' = \varphi' =0$); suppose also, for sake of simplicity, that the dependence of the Stokes parameters upon the conformal time coordinate is trivial. Using  Eq. (\ref{Boltz1b}) the outgoing Stokes parameters are given by 
\begin{eqnarray}
 && {\mathcal I}_{x}(\vartheta,\varphi) 
 = \frac{3 \epsilon'}{8\pi} \biggl[(\hat{e}_{x} \,\hat{e}_{x}')^2 {\mathcal I}_{x}'+ 
 (\hat{e}_{x}\, \hat{e}_{y}')^2 {\mathcal I}_{y}' \biggr],
 \label{Boltz21a}\\
 && {\mathcal I}_{y}(\tau, \vartheta,\varphi)= \frac{3 \epsilon'}{8\pi} \biggl[(\hat{e}_{y}\,\hat{e}_{x}')^2 {\mathcal I}_{x}' + 
 (\hat{e}_{y}\,\hat{e}_{y}')^2 {\mathcal I}_{y}'\biggr].
\label{Boltz22a}
\end{eqnarray}
In the generic situation the ingoing Stokes parameters do depend upon 
$\varphi'$ and $\vartheta'$ and, therefore, the outgoing Stokes parameters are
\begin{eqnarray}
 && {\mathcal I}_{x}(\vartheta,\varphi) = \frac{3 \epsilon'}{8\pi} \,\int d\Omega'  \biggl[(\hat{e}_{x} \,\hat{e}_{x}')^2 {\mathcal I}_{x}'(\vartheta',\varphi')+ 
 (\hat{e}_{x}\, \hat{e}_{y}')^2 {\mathcal I}_{y}'(\vartheta',\varphi') \biggr],
 \label{Boltz21}\\
 && {\mathcal I}_{y}(\vartheta,\varphi)= \frac{3 \epsilon'}{8\pi} \,\int d\Omega'  \biggl[(\hat{e}_{y}\,\hat{e}_{x}')^2 {\mathcal I}_{x}'(\vartheta',\varphi') + 
 (\hat{e}_{y}\,\hat{e}_{y}')^2 {\mathcal I}_{y}'(\vartheta',\varphi')\biggr].
\label{Boltz22}
\end{eqnarray}
 From  the definitions of $\hat{e}$ and $\hat{e}'$, simple trigonometric identities lead to the following expressions for the four products appearing in Eqs. (\ref{Boltz21}) and (\ref{Boltz22}):
\begin{eqnarray}
&& (\hat{e}_{x}\, \hat{e}_{x}') = \mu \, \mu' \cos{(\varphi' - \varphi)} + \sqrt{1 -\mu^2} \, \sqrt{1 - {\mu'}^2}, \qquad 
(\hat{e}_{x}\, \hat{e}_{y}')  = \mu \, \sin{(\varphi' - \varphi)},
\label{Boltz23}\\
&& (\hat{e}_{y}\, \hat{e}_{x}')  =- \mu' \, \sin{(\varphi' - \varphi)},\qquad 
(\hat{e}_{y}\, \hat{e}_{y}')  = \cos{(\varphi' - \varphi)}.
\label{Boltz24}
\end{eqnarray}
As already mentioned, in Eqs. (\ref{Boltz23}) and (\ref{Boltz24}) the notations 
\begin{equation}
\mu = \cos{\vartheta},\qquad \mu' = \cos{\vartheta'},\qquad \sin{\vartheta} =\sqrt{1 -\mu^2},\qquad 
\sin{\vartheta'} = \sqrt{1 -{\mu'}^2},
\label{Boltz24a}
\end{equation}
have been used. The results of Eqs. (\ref{Boltz21})--(\ref{Boltz22}) and of
 Eqs. (\ref{Boltz23})--(\ref{Boltz24})  allow for an explicit evaluation of the the collisional part of the Boltzmann equation which can be written, component by component, in the simplified  case where $f(\tau, \vartheta,\varphi)$ has only two components which can be 
 identified with the intensities of the radiation field along the $x$ and $y$ axes
 \begin{equation}
 f(\tau, \vartheta,\varphi) = ({\mathcal I}_{x}(\tau,\vartheta,\varphi),\, {\mathcal I}_{y}(\tau,\vartheta,\varphi)).
\label{Boltz24b} 
\end{equation}
The result is:
\begin{eqnarray}
&& \frac{d {\mathcal I}_{x}}{ d \tau} = - \epsilon' {\mathcal I}_{x} 
\nonumber\\
&&+ \frac{3 \epsilon'}{16 \pi}
\int_{-1}^{1} d \mu' \int_{0}^{2\pi} d\varphi' [ {\mathcal M}_{11}(\mu,\mu',\varphi,\varphi') {\mathcal I}_{x}'(\tau,\mu',\varphi') + {\mathcal M}_{12}(\mu,\mu',\varphi,\varphi') {\mathcal I}_{y}'(\tau,\mu',\varphi')]
\label{Boltz25}\\
&&  \frac{d {\mathcal I}_{y}}{ d \tau} = - \epsilon' {\mathcal I}_{y} 
\nonumber\\
&&+ \frac{3 \epsilon'}{16 \pi}
\int_{-1}^{1} d \mu' \int_{0}^{2\pi} d\varphi' [ {\mathcal M}_{21}(\mu,\mu',\varphi,\varphi') {\mathcal I}_{x}'(\tau,\mu',\varphi') + {\mathcal M}_{22}(\mu,\mu',\varphi,\varphi') {\mathcal I}_{y}'(\tau,\mu',\varphi')],
\label{Boltz26}
\end{eqnarray}
where, by definition, 
\begin{eqnarray}
 {\mathcal M}_{11}(\mu,\mu',\varphi,\varphi') &=& 2 (\hat{e}_{x} \, \hat{e}_{x}')^2 = 
\mu^2 \mu'^2 + 2 ( 1 - \mu^2) (1 - \mu'^2) + \mu^2 \mu'^2 \cos{2 (\varphi' - \varphi)}
\nonumber\\
&+& 4 \mu\, \mu' \sqrt{1 -\mu^2} \sqrt{1 - {\mu'}^2} \cos{(\varphi' - \varphi)}, 
\label{Boltz27}\\
  {\mathcal M}_{12}(\mu,\mu',\varphi,\varphi')&=& 2 (\hat{e}_{x} \, \hat{e}_{y}')^2 =\mu^2 [ 1 - \cos{2 (\varphi' - \varphi)}], 
\label{Boltz28}\\
{\mathcal M}_{21}(\mu,\mu',\varphi,\varphi') &=& 2 (\hat{e}_{y} \, \hat{e}_{x}')^2 = \mu'^2 - \mu'^2 
\cos{2 (\varphi' - \varphi)}, 
\label{Boltz29}\\
 {\mathcal M}_{22}(\mu,\mu',\varphi,\varphi') &=& 2 (\hat{e}_{y} \, \hat{e}_{y}')^2 = 1+ 
\cos{2 (\varphi' - \varphi)}.
\label{Boltz30}
\end{eqnarray}
The second equality appearing in Eqs. (\ref{Boltz27})--(\ref{Boltz30}) follows immediately from 
Eqs. (\ref{Boltz23}) and (\ref{Boltz24}). If  ${\mathcal I}_{x}'$ and ${\mathcal I}_{y}'$ do not depend on $\varphi'$, Eqs. (\ref{Boltz25}) and (\ref{Boltz26}) 
 can be simplified by integrating explicitly upon $\varphi'$: 
 \begin{eqnarray}
 && \frac{d {\mathcal I}_{x}}{d \tau} = - \epsilon' {\mathcal I}_{x} +
 \frac{3 \epsilon'}{8} \int_{-1}^{1} d\mu' \{ 
 [ \mu^2 \mu'^2 + 2 ( 1- \mu^2) ( 1 - \mu'^2)] {\mathcal I}_{x}'(\tau,\mu') + \mu^2 
 {\mathcal I}_{y}'(\tau,\mu')\},
 \label{Boltz31}\\
 &&  \frac{d {\mathcal I}_{y}}{d \tau} = - \epsilon' {\mathcal I}_{y} + 
 \frac{3 \epsilon'}{8} \int_{-1}^{1} d\mu' \{ 
 [ \mu'^2 {\mathcal I}_{x}'(\tau,\mu') + {\mathcal I}_{y}'(\tau,\mu')\}.
\label{Boltz32}
\end{eqnarray}
Equations (\ref{Boltz31}) and (\ref{Boltz32}) is just a useful warm-up in view of the realistic situation where:
\begin{itemize}
\item{} the components of $f(\vartheta,\varphi)$ are not $2$ but $3$, i.e. 
${\mathcal I}_{x}(\vartheta,\varphi)$ and  ${\mathcal I}_{y}(\vartheta,\varphi)$ are supplemented by
$U(\vartheta,\varphi)$;
\item{} all the $3$ components of $f(\vartheta,\varphi)$ do depend upon $\varphi$; in the analog 
of Eqs. (\ref{Boltz31}) and (\ref{Boltz32}) on top of the integration over $\mu'$ an integration 
over $\varphi'$ will appear.
\end{itemize}
Since $f(\vartheta,\varphi)$ has now three components 
\begin{equation}
 f(\tau,\vartheta,\varphi) = ({\mathcal I}_{x}(\tau,\vartheta,\varphi),\, {\mathcal I}_{y}(\tau,\vartheta,\varphi),\, U(\tau,\vartheta,\varphi)).
\label{Boltz32a}
\end{equation}
${\mathcal M}(\Omega,\Omega')$ will be a $3\times3$  matrix whose entries can be computed in full analogy with 
 what has been already done in Eqs. (\ref{Boltz27})--(\ref{Boltz30}). 
 In general terms, the ingoing electric field can be expressed as
 \begin{equation}
 \vec{E}'(\tau,\vartheta',\varphi')= E_{x}'(\tau,\vartheta',\varphi') \,\hat{e}_{x}' + 
 E_{y}'(\tau,\vartheta',\varphi') \,\hat{e}_{y}',
 \label{Boltz33}
\end{equation}
implying that  the outgoing Stokes parameters can be written as 
\begin{eqnarray}
&& {\mathcal I}_{x}(\tau,\vartheta,\varphi) = 
\frac{3 \epsilon'}{8\pi}\int d\Omega' (\hat{e}_{x}\,\cdot\vec{E}\,')^2,
\label{Boltz34}\\
&& {\mathcal I}_{y}(\tau,\vartheta,\varphi) =  
\frac{3 \epsilon'}{8\pi}\int d\Omega' (\hat{e}_{y}\,\cdot\vec{E}\,')^2, 
\label{Boltz35}\\
&& U(\tau,\vartheta,\varphi) = \frac{ 3 \epsilon'}{ 8\pi}\int d\Omega'  [ 2 (\hat{e}_{x} \cdot \vec{E}\,')  (\hat{e}_{y}\cdot \vec{E}\,')].
\label{Boltz36}
\end{eqnarray}
where, for the moment, the collisionless part of the Boltzmann equation has been neglected.
Inserting Eq. (\ref{Boltz33}) into Eqs. (\ref{Boltz34})--(\ref{Boltz36}) 
the explicit form of ${\mathcal M}(\Omega,\Omega')$ since 
the Stokes parameters of the ingoing radiation field are nothing but 
\begin{equation}
{\mathcal I}_{x}' = E_{x}'^2,\qquad {\mathcal I}_{y}' = E_{y}'^2,\qquad 
U' = 2 \, E_{x}' E_{y}'. 
\label{Boltz37}
\end{equation}
The terms 
${\mathcal M}_{11}$, ${\mathcal M}_{12}$, ${\mathcal M}_{21}$, ${\mathcal M}_{22}$ 
will be exactly the ones already evaluated in Eqs. (\ref{Boltz27})--(\ref{Boltz30}). 
The remaining entries are found to be 
\begin{eqnarray}
{\mathcal M}_{13}(\mu,\mu',\varphi, \varphi') &=& 2\,(\hat{e}_{x}' \,\hat{e}_{x})(\hat{e}_{x}\,\hat{e}_{y}')=
 \mu^2 \mu' \sin{2 (\varphi' - \varphi)} 
\nonumber\\ 
&+& 
2 \mu \sqrt{1 -\mu^2} \sqrt{1 - \mu'^2} \sin{(\varphi' - \varphi)},
\label{Boltz38}\\
 {\mathcal M}_{23}(\mu,\mu',\varphi, \varphi') &=& 2\,(\hat{e}_{x}' \,\hat{e}_{y})(\hat{e}_{y}\,\hat{e}_{y}')= - \mu' \sin{2 (\varphi' - \varphi)},
\label{Boltz39}\\
 {\mathcal M}_{31}(\mu,\mu',\varphi, \varphi') &=& 4  (\hat{e}_{x}' \,\hat{e}_{x})(\hat{e}_{y}\,\hat{e}_{x}')=
- 2\mu'^2 \mu \sin{2(\varphi' - \varphi)}
\nonumber\\
&-& 4 \mu' \sqrt{1 - \mu^2} \sqrt{1 - \mu'^2} 
\sin{(\varphi' - \varphi)},
\label{Boltz40}\\
 {\mathcal M}_{32}(\mu,\mu',\varphi, \varphi') &=& 4(\hat{e}_{x} \,\hat{e}_{y}')(\hat{e}_{y}\,\hat{e}_{y}')= 2 \mu \sin{2 (\varphi' - \varphi)},
\label{Boltz41}\\
 {\mathcal M}_{33}(\mu,\mu',\varphi, \varphi') &=&  2(\hat{e}_{x} \,\hat{e}_{x}')(\hat{e}_{y}\,\hat{e}_{y}') + 2 (\hat{e}_{x} \,\hat{e}_{y}')(\hat{e}_{y}\,\hat{e}_{x}') = 2 \mu \mu' \cos{2 (\varphi' - \varphi)} 
\nonumber\\
&+& 
2 \sqrt{1 - \mu^2} \sqrt{1 - \mu'^2} \cos{(\varphi' - \varphi)}.
\label{Boltz42}
\end{eqnarray}
As in Eqs. (\ref{Boltz27})--(\ref{Boltz30}), the second equality in each of Eqs. (\ref{Boltz38})--(\ref{Boltz42}) 
 follows immediately from Eqs. (\ref{Boltz23}) and (\ref{Boltz24}).
A final remark on the symmetry properties of the various entries 
of ${\mathcal M}(\Omega,\Omega')$ is in order:
\begin{itemize}
\item{} ${\mathcal M}_{11}$, ${\mathcal M}_{12}$, ${\mathcal M}_{21}$, ${\mathcal M}_{22}$ and ${\mathcal M}_{33}$ are all 
symmetric under the simultaneous transformation $\varphi \to - \varphi$  and $\varphi' \to - \varphi'$;
\item{} for the same transformation, the remaining entries flip their respective sign.
\end{itemize}

\subsection{Different parametrizations of the full equation} 
The phase 
space distribution obeying the collisionless part of the Boltzmann equation (see, e. g., Eq. (\ref{Boltz14})) does 
depend upon $k=|\vec{k}|$ (i.e. the Fourier mode). In the previous subsection the dependence 
upon $k$ has been suppressed (for sake of simplicity) since all the 
aforementioned considerations 
could be separately repeated for each Fourier mode. Similarly, from Eq. (\ref{Boltz14}), it is clear 
that the phase-space distribution does depend not only upon $\hat{n}$ (i.e. the direction 
of the photon) but also upon its comoving three-momentum. 
The phase space distribution consists of $3$ independent functions
 whose dependence can be written, in Fourier space, as:
\begin{equation}
f(k,\tau,\mu,\varphi, q) = [ {\mathcal I}_{x}(k,\tau,\mu,\varphi, q), {\mathcal I}_{y}(k,\tau,\mu,\varphi,q), U(k,\tau,\mu,\varphi,q)].
\label{Boltz43}
\end{equation}
As  already discussed it is more handy to use directly $\mu= \cos{\vartheta}$ as independent 
variable.
The full Boltzmann equation can then be solved by iteration around the equilibrium configuration 
provided by the Bose-Einstein distribution $f_{0}(q)$; this means that the $3$ components 
of $f(\tau,\mu,\varphi) $ can be written as:
\begin{eqnarray}
&& {\mathcal I}_{x}(k,\tau,\mu,\varphi,q) = f_{0}(q) [ 1 + \overline{{\mathcal I}}_{x}(k, \tau,\mu,\varphi)],
\label{Boltz44}\\
&& {\mathcal I}_{y}(k, \tau,\mu,\varphi,q) = f_{0}(q) [ 1 + \overline{{\mathcal I}}_{y}(k, \tau,\mu,\varphi)],
\label{Boltz45}\\
&& U(k, \tau,\mu,\varphi,q) = f_{0}(q) \overline{U}(k, \tau,\mu,\varphi).
\label{Boltz46}
\end{eqnarray}
The solution of the problem will then require the determination of 
$\overline{{\mathcal I}}_{x}(k, \tau,\mu,\varphi)$, $\overline{{\mathcal I}}_{y}(k, \tau,\mu,\varphi)$ and 
$\overline{U}(k, \tau,\mu,\varphi)$.
The form of the final solution for $\overline{{\mathcal I}}_{x}(k, \tau,\mu,\varphi)$, $\overline{{\mathcal I}}_{y}(k, \tau,\mu,\varphi)$ and $\overline{U}(k, \tau,\mu,\varphi)$ will depend upon the polarization 
of the relic graviton.  This aspect cal be appreciated by looking at the collisionless part 
of the full Boltzmann equation, i.e. Eq. (\ref{Boltz14}).
Consider, for sake of concreteness, the case of the  
$\oplus$ polarization.  Inserting Eq.  (\ref{Boltz43}) into  Eq. (\ref{Boltz14}) and using the perturbative 
scheme of Eqs. (\ref{Boltz44})--(\ref{Boltz46}), 
 the three independent components of the Boltzmann equation become:
 \begin{eqnarray}
&&  \frac{\partial \overline{{\mathcal I}}_{x}}{\partial \tau}+ ( i k \mu + \epsilon')  \overline{{\mathcal I}}_{x} - 
\frac{1}{2} \frac{\partial h_{\oplus}}{\partial\tau} \cos{2 \varphi} ( 1 - \mu^2) \frac{ \partial \ln{f_{0}}}{\partial \ln{q}}=
{\mathcal C}_{x}(k,\tau,\mu,\varphi),
\label{Boltz47}\\
&& \frac{\partial \overline{{\mathcal I}}_{y}}{\partial \tau} + ( i k \mu + \epsilon')  \overline{{\mathcal I}}_{y} - 
\frac{1}{2} \frac{\partial h_{\oplus}}{\partial\tau} \cos{2 \varphi} ( 1 - \mu^2) \frac{ \partial \ln{f_{0}}}{\partial \ln{q}}=
{\mathcal C}_{y}(k,\tau, \mu,\varphi),
\label{Boltz48}\\
&& \frac{\partial \overline{U}}{\partial \tau} + ( i k \mu  + \epsilon') \overline{U} = {\mathcal C}_{U}(k,\tau,\mu,\varphi),
\label{Boltz49}
\end{eqnarray}
where, for convenience, the collision terms 
${\mathcal C}_{x}(k,\tau,\mu,\varphi)$, $ {\mathcal C}_{y}(k,\tau,\mu,\varphi)$ and  
${\mathcal C}_{U}(k,\tau,\mu,\varphi)$ have been expressed as:
\begin{eqnarray}
&&{\mathcal C}_{x}(k,\tau,\mu,\varphi)= \frac{3 \epsilon'}{16\pi} \int d\mu' d\varphi'
[{\mathcal M}_{11}(\mu,\mu',\varphi,\varphi') \overline{{\mathcal I}}_{x}(k,\tau,\mu',\varphi')
+ {\mathcal M}_{12}(\mu,\mu',\varphi,\varphi') \overline{{\mathcal I}}_{y}(k,\tau,\mu',\varphi')
\nonumber\\
&&+ {\mathcal M}_{13}(\mu,\mu',\varphi,\varphi') \overline{U}(k,\tau,\mu',\varphi')],
\label{Boltz50}\\
&&  {\mathcal C}_{y}(k,\tau,\mu,\varphi)=\frac{3 \epsilon'}{16\pi} \int d\mu' d\varphi'
[{\mathcal M}_{21}(\mu,\mu',\varphi,\varphi') \overline{{\mathcal I}}_{x}(k,\tau,\mu',\varphi')
+ {\mathcal M}_{22}(\mu,\mu',\varphi,\varphi') \overline{{\mathcal I}}_{y}(k,\tau,\mu',\varphi')
\nonumber\\
&&+ 
{\mathcal M}_{23}(\mu,\mu',\varphi,\varphi') \overline{U}(k,\tau,\mu',\varphi')],
\label{Boltz51}\\
&& {\mathcal C}_{U}(k,\tau,\mu,\varphi)= \frac{3 \epsilon'}{16\pi} \int d\mu' d\varphi'
[{\mathcal M}_{31}(\mu,\mu',\varphi,\varphi') \overline{{\mathcal I}}_{x}(k,\tau,\mu',\varphi')
+ {\mathcal M}_{32}(\mu,\mu',\varphi,\varphi') \overline{{\mathcal I}}_{y}(k,\tau,\mu',\varphi')
\nonumber\\
&& + 
{\mathcal M}_{33}(\mu,\mu',\varphi,\varphi') \overline{U}(k,\tau,\mu',\varphi')].
\label{Boltz52}
\end{eqnarray}
In the case of the $\otimes$ polarization Eqs. (\ref{Boltz47})--(\ref{Boltz49}) 
lead to a similar result where, however, the relevant part of the collisionless 
contribution is different and it is given by the replacement
\begin{equation}
\frac{1}{2} \frac{\partial h_{\oplus}}{\partial \tau} \cos{2 \varphi} ( 1 - \mu^2) \frac{ \partial \ln{f_{0}}}{\partial \ln{q}}
\to \frac{1}{2} \frac{\partial h_{\otimes}}{\partial \tau} 
\sin{2 \varphi} ( 1 - \mu^2) \frac{ \partial \ln{f_{0}}}{\partial \ln{q}},
\label{Boltz53}
\end{equation}
as it can be argued directly from Eq. (\ref{Boltz14}).
By inspecting Eqs. (\ref{Boltz47}), (\ref{Boltz48}) 
and (\ref{Boltz49}) it is possible to argue that the azimuthal structure of the equations, appearing 
in connection with the polarization of the gravitational wave, can be 
decoupled from the radial structure by appropriately writing the various Stokes parameters.
In the case of the $\oplus$ polarization, symmetry considerations demand that a sound ansatz 
for the full solution can be written in terms of two independent functions, i.e. 
$\beta(k,\tau,\mu,\varphi)$ and $\zeta(k,\tau,\mu,\varphi)$:
\begin{eqnarray}
&& \overline{{\mathcal I}}_{x}(k,\tau,\mu,\varphi) = 
\frac{\zeta(k,\tau,\mu)}{2}( 1 - \mu^2) \cos{2\varphi} 
+  \frac{\beta(k,\tau,\mu)}{2}( 1 + \mu^2) \cos{2\varphi},
\label{Boltz54}\\
&&\overline{{\mathcal I}}_{y}(k,\tau,\mu,\varphi) = 
\frac{\zeta(k,\tau,\mu)}{2}( 1 - \mu^2) \cos{2\varphi} 
-  \frac{\beta(k,\tau,\mu)}{2}( 1 + \mu^2) \cos{2\varphi},
\label{Boltz55}\\
&& \overline{U}(k,\tau,\mu,\varphi) = 2 \mu \beta(k,\tau,\mu) \sin{2 \varphi}.
\label{Boltz56}
\end{eqnarray}
In the case 
of the $\otimes$ polarization the azimuthal structure of the collisionless Boltzmann equation 
changes because of the replacement of Eq. (\ref{Boltz53}). The analog 
of Eqs. (\ref{Boltz54})--(\ref{Boltz56}) become, in the case of the $\otimes$ polarization, 
\begin{eqnarray}
&& \overline{{\mathcal I}}_{x}(k,\tau,\mu,\varphi) = 
\frac{\zeta(k,\tau,\mu)}{2}( 1 - \mu^2) \sin{2\varphi} 
+  \frac{\beta(k,\tau,\mu)}{2}( 1 + \mu^2) \sin{2\varphi},
\label{Boltz57}\\
&&\overline{{\mathcal I}}_{y}(k,\tau,\mu,\varphi) = 
\frac{\zeta(k,\tau,\mu)}{2}( 1 - \mu^2) \sin{2\varphi} 
-  \frac{\beta(k,\tau,\mu)}{2}( 1 + \mu^2) \sin{2\varphi},
\label{Boltz58}\\
&& \overline{U}(k,\tau,\mu,\varphi) = - 2 \mu \beta(k,\tau,\mu) \cos{2 \varphi}.
\label{Boltz59}
\end{eqnarray}
Using, alternatively, either Eqs. (\ref{Boltz54})--(\ref{Boltz56}) or Eqs. (\ref{Boltz57})--(\ref{Boltz59})
the explicit evolution equations obeyed by $\beta(k,\tau,\mu,\varphi)$ and $\zeta(k,\tau,\mu,\varphi)$
can be derived. 
For sake of concreteness consider again the case of the $\oplus$ polarization and 
insert Eqs. (\ref{Boltz54}), (\ref{Boltz55}) and (\ref{Boltz56}) into Eqs. (\ref{Boltz50}), (\ref{Boltz51})
and (\ref{Boltz52}). In the collision terms the integrations over $\varphi'$ can be performed by recalling that 
\begin{eqnarray} 
&& \int_{0}^{2\pi} \cos{2(\varphi' - \varphi)} \, \cos{2\varphi'}\, d\varphi' = 
\int_{0}^{2\pi} \sin{2(\varphi' - \varphi)} \, \sin{2\varphi'}\, d\varphi' = \pi \cos{2\varphi},
\label{simpint1}\\
&& \int_{0}^{2\pi} \sin{2(\varphi' - \varphi)} \, \cos{2\varphi'}\, d\varphi' = -\pi \sin{2 \varphi},
\label{simpint2}\\
&& \int_{0}^{2\pi} \cos{2(\varphi' - \varphi)} \, \sin{2\varphi'}\, d\varphi' = \pi \sin{2 \varphi}.
\label{simpint3}
\end{eqnarray}
Consequently, Eqs. (\ref{Boltz50}), (\ref{Boltz51}) and (\ref{Boltz52}) become 
\begin{eqnarray}
{\mathcal C}_{x}(k,\mu,\varphi) &=& - \epsilon'\, \mu^2 \cos{2\varphi} \int_{-1}^{1} [ \zeta ( 1 - x^2)^2 - \beta (1 + x^2)^2 - 4 \beta x^2] dx,
\label{expC1}\\
 {\mathcal C}_{y}(k,\mu,\varphi) &=&  \epsilon'\,  \cos{2\varphi} \int_{-1}^{1} [ \zeta ( 1 - x^2)^2 - \beta (1 + x^2)^2 - 4 \beta x^2] dx,
\label{expC2}\\
{\mathcal C}_{U}(k,\mu,\varphi) &=& - 2 \mu  \epsilon' \,  \sin{2\varphi} \int_{-1}^{1} [ \zeta ( 1 - x^2)^2 - \beta (1 + x^2)^2 - 4 \beta x^2] dx.
\label{expC3}
\end{eqnarray}
Now the essential steps of the derivation are the following:
\begin{itemize}
\item{} Eqs. (\ref{Boltz54})--(\ref{Boltz56}) must be inserted, respectively, 
 at left hand side of Eqs. (\ref{Boltz47})--(\ref{Boltz49}); 
\item{} Eqs. (\ref{expC1})--(\ref{expC3}) must be inserted, respectively, at the right 
hand side of Eqs. (\ref{Boltz47})--(\ref{Boltz49}).
\end{itemize}
The final result of the previous pair of manipulations can be explicitly written as \footnote{As in the previous sections the $h_{\oplus}'$ and $h_{\otimes}'$ denote the time derivatives of the polarizations 
with respect to the conformal time coordinate $\tau$. This notation has been avoided in the previous 
equations of the present section since it could have been confused with the angular 
variables describing the polarizations of the outgoing photons. From now on this 
possible clash of notations does not arise.}
\begin{eqnarray}
&& ( 1 -\mu^2)\biggl[ \frac{\partial\zeta}{\partial \tau} + (i k \mu + \epsilon') \zeta - 
\frac{\partial \ln{f_{0}}}{\partial \ln{q}} h_{\oplus}'\biggr] + 
( 1 + \mu^2) \biggl[ \frac{\partial\beta}{\partial \tau} + (i k \mu + \epsilon') \beta\biggr] = - 2 \epsilon' \mu^2  \Psi 
\label{Boltz60}\\
&&  ( 1 -\mu^2)\biggl[ \frac{\partial\zeta}{\partial \tau} + (i k \mu + \epsilon') \zeta - 
\frac{\partial \ln{f_{0}}}{\partial \ln{q}} h_{\oplus}'\biggr] -
( 1 + \mu^2) \biggl[ \frac{\partial\beta}{\partial \tau} + (i k \mu + \epsilon') \beta\biggr] =   2 \epsilon'  \Psi 
\label{Boltz61}\\
&&  \frac{\partial\beta}{\partial \tau} + (i k \mu + \epsilon') \beta = -\epsilon' \Psi 
\label{Boltz62}
\end{eqnarray}
where $\Psi(k,\tau)$ is given by
\begin{equation}
\Psi(k,\tau) = \frac{3}{32}\int_{-1}^{1} d x  [ -(1 + x^2)^2 \beta + (1  - x^2)^2 \zeta - 4 x^2 \beta].
\label{Boltz63}
\end{equation}
As expected, one of the three equations appearing in Eqs. (\ref{Boltz60})--(\ref{Boltz61}) is redundant.
Inserting Eq. (\ref{Boltz62}) into Eq. (\ref{Boltz61}) the two independent 
equations turn out to be 
\begin{eqnarray}
&& \frac{\partial\beta}{\partial \tau} + (i k \mu + \epsilon') \beta =  -\epsilon' \Psi, 
\label{Boltz64}\\
&&  \frac{\partial\zeta}{\partial \tau} + (i k \mu + \epsilon') \zeta - 
\frac{\partial \ln{f_{0}}}{\partial \ln{q}} h_{\lambda}' = \epsilon' \Psi, 
\label{Boltz65}
\end{eqnarray}
Concerning the result obtained in Eqs. (\ref{Boltz64}) and (\ref{Boltz65}) few comments 
are in order:
\begin{itemize}
\item{} in Eqs. (\ref{Boltz64})--(\ref{Boltz65}) $h_{\lambda}'$ denotes indifferently either
$h_{\oplus}'$ or $h_{\otimes}'$: the derivation 
reported in the case of $h_{\oplus}$ can be 
repeated in the case of $h_{\otimes}$  bearing in mind the differences in the angular structure
(i.e.  Eqs. (\ref{Boltz57})--(\ref{Boltz59}) should be used instead of Eqs. (\ref{Boltz54})--(\ref{Boltz56}));
\item{} by changing $\varphi \to - \varphi$ and $\varphi' \to - \varphi'$, Eqs. (\ref{Boltz54})--(\ref{Boltz56}) and 
Eqs. (\ref{Boltz57})--(\ref{Boltz59}) will be different because if the various sines appearing the 
various expressions; therefore the angular structure of the Stokes parameters will change but Eqs. 
(\ref{Boltz64}) and (\ref{Boltz65}) will keep their form.
\end{itemize}
The rationale for the latter statement is that while $\varphi \to - \varphi$ and $\varphi' \to - \varphi'$
changes the angular structure of the Stokes parameters, also some of the entries of ${\mathcal M}(\Omega,\Omega')$ will change. The net result, as already mentioned, will be that Eqs. 
(\ref{Boltz64}) and (\ref{Boltz65}) will still be valid. 
The result expressed by Eqs. (\ref{Boltz64}) and (\ref{Boltz65}) has been firstly obtained by Polnarev \cite{P1} and then exploited for different 
semi-analytical discussions of the problem (see, e. g. \cite{P2,P3,P4}).
The polarization decomposition leading to Eqs. (\ref{Boltz64}) and (\ref{Boltz65}) can be related to slightly different treatments (see, for instance, \cite{B1}) using the 
brightness perturbations rather than $\beta$ and $\zeta$. The connection 
between the different formalisms will be explored later in this section.
By summing up Eqs. (\ref{Boltz64}) and (\ref{Boltz65}) the collision terms cancel: 
\begin{equation} 
\frac{\partial(\zeta + \beta)}{\partial \tau} + (i k \mu + \epsilon') (\zeta + \beta) - 
\frac{\partial \ln{f_{0}}}{\partial \ln{q}} h_{\lambda}' =0.
\label{Boltz66}
\end{equation}
Defining $ \xi = \beta + \zeta$ the collision term of Eq. (\ref{Boltz63}) 
can also be written as:
\begin{equation}
\Psi(k,\tau) = \frac{3}{32} \int_{-1}^{1} dx [ ( 1 - x^2)^2 \xi(k,\tau) - 2 ( 1 + x^2)^2 \, \beta(k,\tau)].
\label{Boltz67}
\end{equation}
The collision term can be further simplified by expanding in multipoles 
the unknowns $\beta(k,\mu,\tau)$ and $\zeta(k,\mu,\tau)$
\begin{eqnarray}
\beta(k,\mu,\tau) = \sum_{\ell} (-i)^{\ell} ( 2 \ell + 1) \beta_{\ell}(k,\tau) P_{\ell}(\mu), \qquad
\zeta(k,\mu,\tau) = \sum_{\ell} (-i)^{\ell} ( 2 \ell + 1) \zeta_{\ell}(k,\tau) P_{\ell}(\mu),
\label{Boltz69}
\end{eqnarray}
where $P_{\ell}(\mu)$ denote, as usual, the Legendre polynomials. 
By recalling that the first three Legendre polynomials of even order are 
\begin{equation}
P_{0}(x) = 1,\qquad P_{2}(x) = \frac{3 x^2 -1}{2}, \qquad P_{4}(x) = \frac{35 x^4 - 30 x^2 + 3}{8},
\label{Boltz70}
\end{equation}
the orthogonality of the Legendre polynomials imply that the integral over $x$ in Eq. (\ref{Boltz67}) leads to 
\begin{equation}
\Psi(k,\tau) = \frac{\zeta_{0}}{10} - \frac{3}{5} \beta_{0} + \frac{\zeta_{2}}{7} + \frac{6}{7} \beta_{2} + \frac{3}{70}(\zeta_{4} - \beta_{4}). 
\label{Boltz71}
\end{equation}
So far the discussion has been conducted in terms of one single polarization 
at the time. For both polarizations the evolution equations of $\zeta$ and $\beta$ turn out 
to be, of course, the same. However the azimuthal structure of the relevant Stokes parameters will be different for different poolarizations. In view of specific applications it is useful to introduce a 
unified notations allowing for the simultaneous treatment of the two different cases:
\begin{eqnarray}
 \overline{{\mathcal I}}(\vec{k},\tau,\mu,\varphi) &=& \overline{{\mathcal I}}_{x}(\vec{k},\tau,\mu,\varphi) + 
\overline{{\mathcal I}}_{y}(\vec{k},\tau,\mu,\varphi) 
\nonumber\\
&\equiv&
 (1- \mu^2)\biggl[ e^{2 i \varphi} e_{1}(\vec{k}) + e^{ - 2 i \varphi}
e_{2}(\vec{k})\biggr] \frac{\zeta(k,\tau,\mu)}{\sqrt{2}}, 
\label{Boltz72}\\
\overline{Q}(\vec{k},\tau,\mu,\varphi) &=& \overline{{\mathcal I}}_{x}(\vec{k},\tau,\mu,\varphi) -
\overline{{\mathcal I}}_{y}(\vec{k},\tau,\mu,\varphi) 
\nonumber\\
&\equiv&( 1 + \mu^2) \biggl[ e^{2 i \varphi} e_{1}(\vec{k}) + e^{ - 2 i \varphi}
e_{2}(\vec{k})\biggr] \frac{\beta(k,\tau,\mu)}{\sqrt{2}}, 
\label{Boltz73}\\
\overline{U}(\vec{k},\tau,\mu,\varphi) &=& - 2 i \mu \biggl[ e^{2 i \varphi} e_{1}(\vec{k}) - e^{ - 2 i \varphi}
e_{2}(\vec{k})\biggr] \frac{\beta(k,\tau,\mu)}{\sqrt{2}},
\label{Boltz74}
\end{eqnarray}
where, following Eqs. (\ref{STa1})--(\ref{STa3}) (see also \cite{B1}) the stochastic variables
\begin{equation}
e_{1}(\vec{k}) = \frac{e_{\oplus}(\vec{k}) - i e_{\otimes}(\vec{k})}{\sqrt{2}}, \qquad 
e_{2}(\vec{k}) = \frac{e_{\oplus}(\vec{k}) + i e_{\otimes}(\vec{k})}{\sqrt{2}}.
\label{STa1rep}
\end{equation}
have been introduced.  Note that, Eqs. (\ref{Boltz73}) and (\ref{Boltz74}) reproduce 
exactly Eqs. (\ref{Boltz54})--(\ref{Boltz56}) (when $e_{\otimes}(\vec{k})=0$). Similarly, when 
$e_{\oplus}(\vec{k})=0$, Eqs. (\ref{Boltz57})--(\ref{Boltz59}) are correctly recovered.
The variables of Eq. (\ref{STa1rep})  assume (see also Eqs. (\ref{STa2})--(\ref{STa3})) that the time dependence can be factorized by in terms of an appropriate transfer function for the amplitude.
By linearly combining Eqs. (\ref{Boltz73}) and (\ref{Boltz74}) it is also 
easy to obtain
\begin{eqnarray}
&& \overline{Q}(\vec{k},\tau,\mu,\varphi) + i \overline{U}(\vec{k},\tau,\mu,\varphi) = \biggl[(1 + \mu)^2 e^{2 i \varphi} e_{1}(\vec{k}) + ( 1- \mu)^2 e^{- 2 i \varphi}e_{2}(\vec{k})\biggr] 
\frac{\beta(k,\mu,\tau)}{\sqrt{2}},
\label{Boltz76}\\
&& \overline{Q}(\vec{k},\tau,\mu,\varphi) - i \overline{U}(\vec{k},\tau,\mu,\varphi) = \biggl[(1 - \mu)^2 e^{2 i \varphi} e_{1}(\vec{k}) + ( 1+ \mu)^2 e^{- 2 i \varphi}e_{2}(\vec{k})\biggr] 
\frac{\beta(k,\mu,\tau)}{\sqrt{2}}.
\label{Boltz77}
\end{eqnarray}
The dependence of Eqs. (\ref{Boltz65}) and (\ref{Boltz66}) 
on the derivative of $f_{0}(q)$ with respect to the comoving three-momentum $q$ can be further 
simplified by rephrasing the Boltzmann equations 
in terms of the appropriate brightness perturbations. Recalling that, by definition 
of brightness perturbations, 
\begin{equation}
f_{0}\biggl(\frac{q}{ 1 + \Delta_{\mathrm{I}}}\biggr) = f_{0}(q) \biggl[ 1 - \frac{\partial \ln{f_{0}}}{\partial \ln{q}} \Delta_{\mathrm{I}}\biggr],
\label{Boltz78}
\end{equation}
the intensity $\overline{{\mathcal I}}$ introduced in Eq. (\ref{Boltz72}) 
can be written as ${\mathcal I} = - \Delta_{\mathrm{I}} (\partial \ln{f_{0}}/\partial\ln{q})$. 
Consequently, Eqs. (\ref{Boltz72}) and (\ref{Boltz76})--(\ref{Boltz77}) 
can also be written as:
\begin{eqnarray}
&& \Delta_{\mathrm{I}}(\vec{k},\tau,\mu,\varphi) = \biggl[ (1-\mu^2) e^{2 i \varphi} e_{1}(\vec{k}) + 
( 1 - \mu^2) e^{- 2 i\varphi}e_{2}(\vec{k})\biggr] \Delta_{\mathrm{T}}(k,\tau,\mu),
\label{Boltz79}\\
&& \Delta_{+}(\vec{k},\tau,\mu,\varphi) = \biggl[ (1+\mu)^2 e^{2 i \varphi} e_{1}(\vec{k}) + 
( 1 - \mu)^2  e^{- 2 i\varphi}e_{2}(\vec{k})\biggr] \Delta_{\mathrm{P}}(k,\tau,\mu),
\label{Boltz80}\\
&& \Delta_{-}(k,\tau,\mu,\varphi) = \biggl[ (1-\mu)^2 e^{2 i \varphi} e_{1}(\vec{k}) + 
( 1 + \mu)^2  e^{- 2 i\varphi}e_{2}(\vec{k})\biggr] \Delta_{\mathrm{P}}(k,\tau,\mu),
\label{Boltz81}
\end{eqnarray}
where the relation of $\Delta_{\mathrm{T}}(k,\tau,\mu)$ and $\Delta_{\mathrm{P}}(k,\tau,\mu)$ to 
$\beta(k,\tau,\mu)$ and $\zeta(k,\tau,\mu)$ is
\begin{equation}
\zeta(k,\tau,\mu) = - \sqrt{2} \frac{\partial\ln{f_{0}}}{\partial\ln{q}} \Delta_{\mathrm{T}}(k,\tau,\mu)
,\qquad \beta(k,\tau,\mu) = - \sqrt{2} \frac{\partial\ln{f_{0}}}{\partial\ln{q}} \Delta_{\mathrm{P}}(k,\tau,\mu).
\label{Boltz82}
\end{equation}
Using Eq. (\ref{Boltz82}) inside Eqs. (\ref{Boltz65})--(\ref{Boltz66}),
the evolution equations obeyed by $\Delta_{\mathrm{P}}(k,\tau,\mu)$ and $\Delta_{\mathrm{T}}(k,\tau,\mu)$  can be obtained and it is:
\begin{eqnarray}
&& \frac{\partial \Delta_{\mathrm{P}}}{\partial \tau} + ( i k \mu + \epsilon') \Delta_{\mathrm{P}} = 
- \epsilon' S,
\label{Boltz83}\\
&&  \frac{\partial \Delta_{\mathrm{T}}}{\partial \tau} + ( i k \mu + \epsilon') \Delta_{\mathrm{T}} = 
- h' + \epsilon' S,
\label{Boltz84}\\
&& S(k,\tau) = \biggl(\frac{\Delta_{\mathrm{T}0}}{10} + \frac{\Delta_{\mathrm{T}2}}{7} + 
\frac{3}{70} \Delta_{\mathrm{T}4}\biggr) - \biggl(\frac{3}{5} \Delta_{\mathrm{P}0} - \frac{6}{7}
\Delta_{\mathrm{P}2} + \frac{3}{70} \Delta_{\mathrm{P}4}\biggr).
\label{Boltz85}
\end{eqnarray}
In Eq. (\ref{Boltz84}) $h$ is the canonical amplitude of the graviton. Recalling the results of section \ref{sec3} 
it has been defined that $h_{\oplus} = \sqrt{2} \ell_{\mathrm{P}} h$ and $h_{\otimes} = 
\sqrt{2} \ell_{\mathrm{P}} h$. The $\sqrt{2}$ factor simplifies once 
the brightness perturbations of Eqs. (\ref{Boltz82}) are used. Using the common practice Eqs. (\ref{Boltz83})--(\ref{Boltz85}) have been written in units $\ell_{\mathrm{P}} =1$.

After Eqs. (\ref{Boltz18}) and (\ref{Boltz19}) it has been stressed that by flipping the 
sign of $\varphi$ and $\varphi'$ the outgoing and ingoing 
polarizations change but the choice of orthnormal frame is still valid. The discussion 
conducted so far can be repeated by assuming the polarizations of Eqs. (\ref{Boltz18}) and (\ref{Boltz19})
but with $\varphi\to - \varphi $ and $\varphi' \to -\varphi'$. The same sign flip can be 
followed throughout the derivation and it can be appreciated that the results of Eqs. (\ref{Boltz83}), (\ref{Boltz84}) 
and (\ref{Boltz85}) are left unchanged (see also discussion after Eqs. (\ref{Boltz41})--(\ref{Boltz42})).
The flip in the sign of $\varphi$ and $\varphi'$ affects, however, the brightness perturbations.
The intensity of the radiation field, as expected,  is left invariant  when $\varphi\to -\varphi$
(see Eq. (\ref{Boltz79})). On the contrary, by repeating all the steps  of the derivation, Eqs. (\ref{Boltz80}) and (\ref{Boltz81}) are modified. Consequently, by flipping the sign of $\varphi$,  Eqs. (\ref{Boltz79}), (\ref{Boltz80}) and (\ref{Boltz81}) become
\begin{eqnarray}
&& \Delta_{\mathrm{I}}(\vec{k},\tau,\mu,\varphi) = \biggl[ (1-\mu^2) e^{2 i \varphi} e_{1}(\vec{k}) + 
( 1 - \mu^2) e^{- 2 i\varphi}e_{2}(\vec{k})\biggr] \Delta_{\mathrm{T}}(k,\tau,\mu),
\label{Boltz79a}\\
&& \Delta_{+}(\vec{k},\tau,\mu,\varphi) = \biggl[ (1-\mu)^2 e^{2 i \varphi} e_{1}(\vec{k}) + 
( 1 + \mu)^2  e^{- 2 i\varphi}e_{2}(\vec{k})\biggr] \Delta_{\mathrm{P}}(k,\tau,\mu),
\label{Boltz86}\\
&& \Delta_{-}(\vec{k},\tau,\mu,\varphi) = \biggl[ (1+\mu)^2 e^{2 i \varphi} e_{1}(\vec{k}) + 
( 1 - \mu)^2  e^{- 2 i\varphi}e_{2}(\vec{k})\biggr] \Delta_{\mathrm{P}}(k,\tau,\mu).
\label{Boltz87}
\end{eqnarray}
In \cite{B1} the brightness perturbations for the poolarization have been assigned as in Eqs. (\ref{Boltz86})--
(\ref{Boltz87}) and not as Eqs. (\ref{Boltz80}) and (\ref{Boltz81}). 
The ladder operators of Eqs. (\ref{Kplus}) and (\ref{Kminus}) are not invariant 
under the transformation $\varphi \to - \varphi$.  More specifically the spin weight 
of a given function changes the sign whenever $\varphi \to - \varphi$.
Equations  (\ref{Kplus}) and (\ref{Kminus}) 
are consistent with the brightness perturbations written as in Eqs. (\ref{Boltz86}) and 
(\ref{Boltz87}).  To be consistent with the customary notations we will therefore 
adhere to the conventions stipulated in \cite{B1}.

\subsection{B-modes from relic gravitons}
The analytic and numerical solutions of the evolution equations 
for the brightness perturbations allow for an explicit evaluation 
of the temperature and polarization observables. 
The results obtained so far imply that long wavelength gravitons will produce both 
temperature and polarization anisotropies. More specifically, following the 
terminology of section \ref{sec2} (see, in particular, Eqs. (\ref{intN5})--(\ref{intN7})) 
the relevant angular power spectra 
induced by the relic gravitons will be the  TT, EE, TE and BB angular power spectra. 

The explicit form of $a^{(\mathrm{B})}_{\ell\,m}$ and $a^{(\mathrm{E})}_{\ell\,m}$
can be derived from the general expressions already encountered, respectively, 
in Eqs. (\ref{EB8}) and (\ref{EB9}).
The integrands of Eqs. (\ref{EB8}) and (\ref{EB9}) can be 
computed from the action of the the differential operators of Eqs. 
(\ref{Kplus})--(\ref{Kminus}), i.e., more specifically, 
\begin{eqnarray}
&& K_{-}^{(1)}(\hat{n})[K_{-}^{(2)}(\hat{n}) \Delta_{+}(\hat{n})]= ( 1 - \mu^2) \Delta_{+}'' - 4 \mu \Delta_{+}' - 2 
\Delta_{+} - \frac{\partial_{\varphi}^2 \Delta_{+}}{1 - \mu^2} 
\nonumber\\
&+& 
2i \biggl[ \partial_{\varphi} \Delta_{+}' - \frac{\mu}{1 - \mu^2} \partial_{\varphi} \Delta_{+}\biggr],
\label{Bex3}\\
&& K_{+}^{(-1)}(\hat{n})[K_{+}^{(-2)}(\hat{n}) \Delta_{-}(\hat{n})]  = ( 1 - \mu^2) \Delta_{-}'' - 4 \mu \Delta_{-}' - 2 
\Delta_{-} - \frac{\partial_{\varphi}^2 \Delta_{-}}{1 - \mu^2} 
\nonumber\\
&-& 
2i \biggl[ \partial_{\varphi} \Delta_{-}' - \frac{\mu}{1 - \mu^2} \partial_{\varphi} \Delta_{-}\biggr].
\label{Bex4}
\end{eqnarray}
Equations (\ref{EB8}) and (\ref{EB9}) become then 
\begin{eqnarray}
a^{(\mathrm{B})}_{\ell\,m} &=& - \overline{N}_{\ell}\int d\hat{n} Y_{\ell\,m}^{*}(\hat{n}) \biggl\{
( 1 -\mu^2) \Delta_{\mathrm{U}}'' - 4 \mu \Delta_{\mathrm{U}}' - 2 \Delta_{\mathrm{U}} - 
\frac{\partial_{\varphi}^2 \Delta_{\mathrm{U}}}{1-\mu^2} 
\nonumber\\
&+& 2 \biggl[ \partial_{\varphi} \Delta_{\mathrm{Q}}' - \frac{\mu}{1 - \mu^2} \partial_{\varphi} \Delta_{\mathrm{Q}}\biggr]\biggr\},
\label{Bex5}\\
a^{(\mathrm{E})}_{\ell\,m} &=& - \overline{N}_{\ell}
\int d\hat{n} Y_{\ell\,m}^{*}(\hat{n}) \biggl\{
( 1 -\mu^2) \Delta_{\mathrm{Q}}'' - 4 \mu \Delta_{\mathrm{Q}}' - 2 \Delta_{\mathrm{Q}} - 
\frac{\partial_{\varphi}^2 \Delta_{\mathrm{Q}}}{1-\mu^2} 
\nonumber\\
&-& 2 \biggl[ \partial_{\varphi} \Delta_{\mathrm{U}}' - \frac{\mu}{1 - \mu^2} \partial_{\varphi} \Delta_{\mathrm{U}}\biggr]\biggr\}.
\label{Bex6}
\end{eqnarray}
In Eqs. (\ref{Bex5})--(\ref{Bex6}) the prime denotes a derivation with respect to $\mu = \cos{\vartheta}$.
This notation will be consistently followed, for sake of simplicity,  but only in this subsection.
Equations (\ref{Bex5}) and (\ref{Bex6}) can be easily written in even more explicit terms and, 
as before, it is easier to focus on a single polarization, e. g. $\oplus$:
\begin{equation}
\Delta_{\mathrm{Q}}(k,\hat{n},\tau) = \sqrt{2}\,( 1 + \mu^2) \, \cos{2 \varphi}\Delta_{\mathrm{P}}(k,\mu,\tau), \qquad 
\Delta_{\mathrm{U}}(k,\hat{n},\tau) = - 2 \sqrt{2}\,\mu\, \sin{2 \varphi} 
\Delta_{\mathrm{P}}(k,\mu,\tau).
\label{Bex7}
\end{equation}
Inserting Eq. (\ref{Bex7}) into Eqs. (\ref{Bex5}) and (\ref{Bex6}) the resulting 
expressions will be 
\begin{eqnarray}
&& a^{(\mathrm{B})}_{\ell\,m} = 2 \sqrt{2} \, \overline{N}_{\ell} \int d\hat{n} Y_{\ell\,m}^{*}(\hat{n})\, \sin{ 2 \varphi} \,( 1 - \mu^2)[\mu \Delta_{\mathrm{P}}'' + 4 \Delta_{\mathrm{P}}'],
\label{Bex8}\\
&& a^{(\mathrm{E})}_{\ell\,m} = - \sqrt{2}\, \overline{N}_{\ell}  \int d\hat{n} \cos{2\varphi} Y_{\ell\,m}^{*} ( 1 - \mu^2) [ ( 1 + \mu^2) \Delta_{\mathrm{P}}'' + 8 \mu \Delta_{\mathrm{P}}' + 12 \Delta_{\mathrm{P}}]. 
\label{Bex9}
\end{eqnarray}
In Eqs. (\ref{Bex8}) 
and (\ref{Bex9}) the explicit integration over the angular variables can be performed in explicit terms.  
Consider, in particular,  Eq. (\ref{Bex8}), i.e. 
\begin{equation}
a^{(\mathrm{B})}_{\ell\,m} = 2 \sqrt{2} \, \overline{N}_{\ell} \int_{-1}^{1} d\mu \int_{0}^{2\pi} d\varphi 
 Y_{\ell\,m}^{*}(\mu,\varphi)\, \sin{ 2 \varphi} \,( 1 - \mu^2)[\mu \Delta_{\mathrm{P}}'' + 4 \Delta_{\mathrm{P}}'].
\label{Bex10}
\end{equation}
Bearing in mind the connection between spherical harmonics and the associated 
Legendre functions, the $Y_{\ell\,m}(\hat{n})$ will be essentially given by the product 
of $P_{\ell}^{m}(\mu)$ and of $e^{i m\varphi}$, up to a well know 
normalization coefficient. In Eq. (\ref{Bex10}) 
the integration over $\varphi$ can be carried on in explicit terms and 
since there is a $\sin{2 \varphi}$ in the integrand the whole integral will be proportional 
to $(\delta_{m,\,2} - \delta_{m\,-2})$; in more precise terms the integration 
over $\varphi$ will bring Eq. (\ref{Bex10}) in the form:
\begin{equation}
a_{\ell\, m}^{(\mathrm{B})} = - 2\, i\,\pi\, \sqrt{2} \frac{(\ell -2)!}{(\ell + 2)!} \sqrt{\frac{2 \ell + 1}{4\pi}} (\delta_{m,\,2} - \delta_{m\,-2})
\int_{-1}^{1} P_{\ell}^{(2)}(\mu) [ ( 1 - \mu^2) \mu \Delta_{\mathrm{P}}'' + 4 ( 1 - \mu^2) 
\Delta_{\mathrm{P}}'].
\label{Bex11}
\end{equation}
Equation (\ref{Bex11}) can be further simplified by using the following three relations:
\begin{eqnarray}
&&\Delta_{\mathrm{P}}(k,\mu,\tau) = \sum_{j} 
( 2 j+ 1) (-i)^{j} \Delta_{\mathrm{P}\,j}(k,\tau) \, P_{j}(\mu),
\label{Bex12}\\
&& \mu P_{j}^{m}(\mu) = \frac{( j -m + 1) P_{j + 1}^{m}(\mu) + ( j + m) P_{j-1}^{m}(\mu)}{(2 j + 1)},
\label{Bex13}\\
&& \sqrt{1 - \mu^2} P_{j}^{m-1}(\mu) = \frac{P_{j - 1}^{m}(\mu)  - P_{j +1}^{m}(\mu)}{2 j + 1}.
\label{Bex14}
\end{eqnarray}
Equations (\ref{Bex13}) and (\ref{Bex14}) are well known recurrence relations for the 
for the associated Legendre functions (see p. 24 of Ref. \cite{edmonds} for a derivation).
Equation (\ref{Bex12}) can be written in slightly different ways. 
Some authors do not include, at the right hand side of Eq. (\ref{Bex12}), the factor $(-i)^{j}$ inside the sum (see 
for instance \cite{B2}). 
The analog of Eq. (\ref{Bex11}) can also be obtained within the approach of Ref. \cite{B2} whose conventions 
(for the $\oplus$ polarization) are
\begin{equation}
    P^{ab}_\oplus(k,\hat{n}) ={T_0\over8} \sum_j (2j+1)\,
     \Delta_{Qj}^+(k)\, M_j^{ab}(\hat{n}),
\label{Bex15}
\end{equation}
with
\begin{equation}
     M_{(j)}^{ab}(\hat{n}) = P_j(\mu)\left( 
     \begin{array}{cc}
     (1+\mu^2)\cos{2\varphi} & -2  \frac{\mu}{\sqrt{ 1 - \mu^2}}\sin 2\varphi\\
     \noalign{\vskip6pt}
     -2   \frac{\mu}{\sqrt{ 1 - \mu^2}}\sin 2\varphi &
     -\frac{(1+\mu^2)}{1 -\mu^2}\cos2\varphi\\
     \end{array} \right).
\label{Bex16}
\end{equation}
For $\times$ polarization, the angular structure of Eq. (\ref{Bex16}) will be different and can 
be obtained from the results for the $\oplus$ polarization by replacing
$\cos2\varphi \rightarrow \sin2\varphi$ and $\sin2\varphi \rightarrow -\cos 2\varphi$.
Eqs. (\ref{Bex15}) and (\ref{Bex16}) are nothing but the explicit expression of the polarization 
tensor on the sphere introduced in Eq. (\ref{SP2}) with the difference 
that the indices (in Eq. (\ref{SP2})) are both covariant while in Eq. (\ref{Bex15}) they are 
both contravariant. The conventions of \cite{B2} imply also a difference of a factor 
of $1/8$ in comparison with the conventions adopted here. 
Equations (\ref{Bex13}) and (\ref{Bex14}) can  be used inside Eq. (\ref{Bex11}). As an example 
the term $(1 - \mu^2)\mu \Delta_{\mathrm{P}}''$, after using Eq. (\ref{Bex12}) will produce a factor 
$\mu ( 1 - \mu^2) P_{j}''$, i.e. 
\begin{equation}
\mu ( 1 - \mu^2) \frac{d^2 P_{j}}{d\mu^2} = \mu P_{j}^{(2)}(\mu) \equiv 
\frac{(j - 1) P_{j + 1}^{(2)}(\mu) + ( j + 2) P_{j -1}^{(2)}(\mu)}{2 j + 1},
\label{Bex17}
\end{equation}
where the first equality follows from the definition of the associated Legendre functions 
(i.e. Eq. (\ref{Ydef2})) while the second follows from Eq. (\ref{Bex13}) with $m = \pm 2$. 
The second term in the integrand of Eq. (\ref{Bex11}), i.e. $( 1 - \mu^2) 
\Delta_{\mathrm{P}}'$ will produce a factor $( 1 - \mu^2) P_{j}'$, i.e. 
\begin{equation}
(1 - \mu^2) \frac{d P_{j}}{d\mu} = \sqrt{1 - \mu^2} P_{j}^{(1)}(\mu) \equiv \frac{P_{j - 1}^{(2)}(\mu) - 
P_{j +1}^{(2)}(\mu)}{(2 j + 1)},
\label{Bex18}
\end{equation}
where, again, the first and second equalities follow, respectively, 
 from Eqs. (\ref{Ydef2}) and Eq. (\ref{Bex13}) both in the case $m = \pm 2$.
Inserting Eqs. (\ref{Bex17}) and (\ref{Bex18}) inside Eq. (\ref{Bex11}) the integration over $\mu$ 
can be performed by simply using the orthogonality of the associated Legendre functions, i.e. 
\begin{equation}
\int_{-1}^{1} d\mu \,\, P_{j}^{m}(\mu) P_{j'}^{m}(\mu) = \frac{2}{2 j + 1} \, \frac{(j + m)!}{(j-m)!}\, \delta_{j\,j'}.
\label{Bex19}
\end{equation}
The final result for the B-mode will then be:
\begin{equation}
a^{(\mathrm{B})}_{\ell\,m} = 2 \sqrt{\frac{\sqrt{2}\, \pi}{2 \ell + 1}} \, (-i)^{\ell} (\delta_{m,\,2} - 
\delta_{m,\,-2}) [ \Delta_{\mathrm{P}\,\ell -1} - (\ell -1)  \Delta_{\mathrm{P}\,\ell +1}].
\label{Bex20}
\end{equation}
If the factor $(-i)^{j}$ would be absent from the expansion of Eq.  (\ref{Bex12}) there would 
not be $(-i)^{\ell}$ at the right-hand side of Eq. (\ref{Bex20}); for the same reason,
 the relative sign of the two terms 
inside the squared bracket
would be plus instead of minus. Finally, the factor $(1/8)$ appearing at the right-hand side 
of Eq. (\ref{Bex15}) will also modify slightly the prefactor  of Eq. (\ref{Bex20}) which will be 
\begin{equation}
     a^{({\rm C}),\oplus}(\vec{k})_{lm} = -{i\over 4}(\delta_{m\,2}-\delta_{m\,-2})
\sqrt{2\pi\over 2\ell+1} \left[ (\ell+2)\tilde\Delta^{\oplus}_{\mathrm{Q}\,\ell-1}(k) +
(\ell-1)\tilde\Delta^{\oplus}_{\mathrm{Q}\,\ell +1}(k)\right],
\label{Bex21}
\end{equation}
 where Eq. (\ref{SP26}) has been used to relate $a_{\ell\,m}^{(\mathrm{B})}$ and $a_{\ell\,m}^{(\mathrm{C})}$.
 In the case of the E-mode, always working with the $\oplus$ polarization, Eqs. (\ref{Bex6})  and (\ref{Bex9}) will give, after 
integration over $\varphi$, 
\begin{equation}
a_{\ell\,m}^{(\mathrm{E})}= - \sqrt{2}\, \pi \frac{(\ell - 2)!}{(\ell +2)!} \, \sqrt{\frac{2 \ell + 1}{4\pi}} (\delta_{m,\,2} + \delta_{m\,-2})  \int_{-1}^{1} d\mu ( 1 - \mu^2) [ ( 1 + \mu^2) \Delta_{\mathrm{P}}'' + 8 \mu \Delta_{\mathrm{P}}' + 12 \Delta_{\mathrm{P}}],
\label{Bex22}
\end{equation}
where Eqs. (\ref{Ydef1}) and (\ref{Ydef2}) have been used.  To complete the calculation it is necessary to perform the integration over 
$\mu$. This can be done, as previously shown, by using wisely the recursion relations of the associate Legendre 
functions.  From Eq. (\ref{Bex22}) we have that, using the expansion of Eq. (\ref{Bex12}), 
\begin{eqnarray}
&&\int_{-1}^{1} d\mu ( 1 - \mu^2) [ ( 1 + \mu^2) \Delta_{\mathrm{P}}'' + 8 \mu \Delta_{\mathrm{P}}' + 12 \Delta_{\mathrm{P}}]=
\nonumber\\
&& \sum_{j} \Delta_{\mathrm{P}j} (-i)^{j} ( 2 j + 1) \int_{-1}^{1} P_{\ell}^{(2)}(\mu) ( 1 -\mu^2)[ (1 + \mu^2) P_{j}'' + 2 \mu P_{j}' + 12 P_{j}] \, d\mu.
\label{Bex23}
\end{eqnarray}
As in the case of the B-mode, the main idea of the calculation is to express the relevant part of the integrand appearing at 
the right hand side of Eq. (\ref{Bex23}) in terms of products of associated Legendre functions 
with the same $m$ but different values of $j$ so that the orthogonality relation (\ref{Bex19}) can be 
exploited. Using the definition of the associated Legendre functions of Eq. (\ref{Ydef2}) it is easy to show that 
\begin{equation}
( 1 -\mu^2)[ (1 + \mu^2) P_{j}'' + 2 \mu P_{j}' + 12 P_{j}]  = ( 1 + \mu^2) P_{j}^{(2)} - 8\mu \sqrt{1 -\mu^2} P_{j}^{(1)} + 
12 (1 -\mu^2) P_{j},
\label{Bex24}
\end{equation}
where the minus sign in the second term at the right hand side of Eq. (\ref{Bex24}) arises since 
the Condon-Shortley phase has been included in the definition of the associated Legendre functions 
(see Eqs. (\ref{Ydef1})--(\ref{Ydef2}) and discussion therein). The right hand side 
of Eq. (\ref{Bex24}) now be simplified by using the following steps:
\begin{itemize}
\item{} Eq. (\ref{Bex13}) can be used to simplify the term $\mu^2 P_{j}^{(2)}(\mu)$;
\item{} Eq. (\ref{Bex14}) can be used to simplify the term $8\mu \sqrt{1 -\mu^2} P_{j}^{(1)}(\mu)$;
\item{}  the term $(1- \mu^2)P_{j}(\mu)$ can be simplified by using the equation of the Legendre 
polynomials and the recursion relations (\ref{Bex13}) and (\ref{Bex14}).
\end{itemize}
Using these three steps, a rather lengthy but straightforward algebra leads to the following 
expression: 
\begin{eqnarray}
( 1 -\mu^2)[ (1 + \mu^2) P_{j}'' + 2 \mu P_{j}' + 12 P_{j}]  &=& \frac{ 6 (j -1) (j + 2)}{( 2 j -1) (2 j + 3)} P_{j}^{(2)}(\mu) + 
\frac{(j + 3)( j+ 4)}{(2 j + 1)( 2 j + 3)} P_{j + 2}^{(2)}(\mu) 
\nonumber\\
&+& \frac{(j - 3) ( j -2)}{( 2 j + 1) ( 2 j -1)} P_{j -2}^{(2)}(\mu).
\label{Bex27}
\end{eqnarray}
To get to Eq. (\ref{Bex27}) it is useful to recall, on a side, that 
\begin{equation}
( 1 -\mu^2) P_{j}(\mu) = \frac{P_{j + 2}^{(2)}(\mu)}{(2 j + 1) ( 2 j + 3)} - \frac{2 P_{j}^{(2)}(\mu)}{(2 j - 1) ( 2 j + 3)} + 
\frac{P_{j -2}^{(2)}(\mu)}{( 2 j + 1) (2 j -1)},
\label{Bex28}
\end{equation}
which can be derived by recalling the equation obeyed by the Legendre polynomials $P_{j}(\mu)$ as well as 
the recursion relations reported in Eqs. (\ref{Bex13})--(\ref{Bex14}).
Using Eq. (\ref{Bex28}) inside Eqs. (\ref{Bex22})--(\ref{Bex23}) it is easy to obtain, after simple algebra, that 
\begin{eqnarray}
a_{\ell\, m}^{(\mathrm{E})} &=& - \sqrt{2\pi} \sqrt{ 2\ell +1} (-i)^{\ell} ( \delta_{m\,2} + \delta_{m\,-2}) \biggl[ 
\frac{6 (\ell -1) (\ell + 2)}{(2\ell -1) ( 2 \ell + 3)} \Delta_{\mathrm{P}\,\ell}
\nonumber\\
&-& \frac{(\ell + 1) (\ell + 2)}{(2 \ell + 1) ( 2 \ell -1)} \Delta_{\mathrm{P}\, \ell-2} - \frac{\ell (\ell -1)}{(2 \ell + 1) (2 \ell + 3)} \Delta_{\mathrm{P}\,\ell +2}\biggr]
\label{Bex29}
\end{eqnarray}
As in the case of the B-mode the presence of the factor $(-i)^{j}$ in Eq. (\ref{Bex12}) leads 
to a sign difference in the terms  $\Delta_{\mathrm{P}\, \ell\pm2}$ appearing in Eq. (\ref{Bex29}).   
The absolute values of the coefficients of the three terms appearing  (inside the the square bracket) in 
 Eq. (\ref{Bex29}) are independent on the conventions related to Eq. (\ref{Bex12}). The coefficient 
 of the term $\Delta_{\mathrm{P}\,\ell}$ appears to be different from the 
 homologue term reported in Eq. (4.41) of \cite{B2}. After many checks and cross-checks 
 of the derivations leading to Eq. (\ref{Bex29}) it has been concluded 
 that the difference (i.e. $6\ell(\ell +1)$ instead of $6 (\ell -1) (\ell + 2)$ in the numerator 
 of the coefficient of $\Delta_{\mathrm{P}\ell}$) is just the result of a typo and the 
 correct expression is $6 (\ell -1) (\ell + 2)$ as in Eq. (\ref{Bex29}).

\subsection{Angular power spectra induced by long wavelength gravitons}
\begin{figure}[!ht]
\centering
\includegraphics[height=6.7cm]{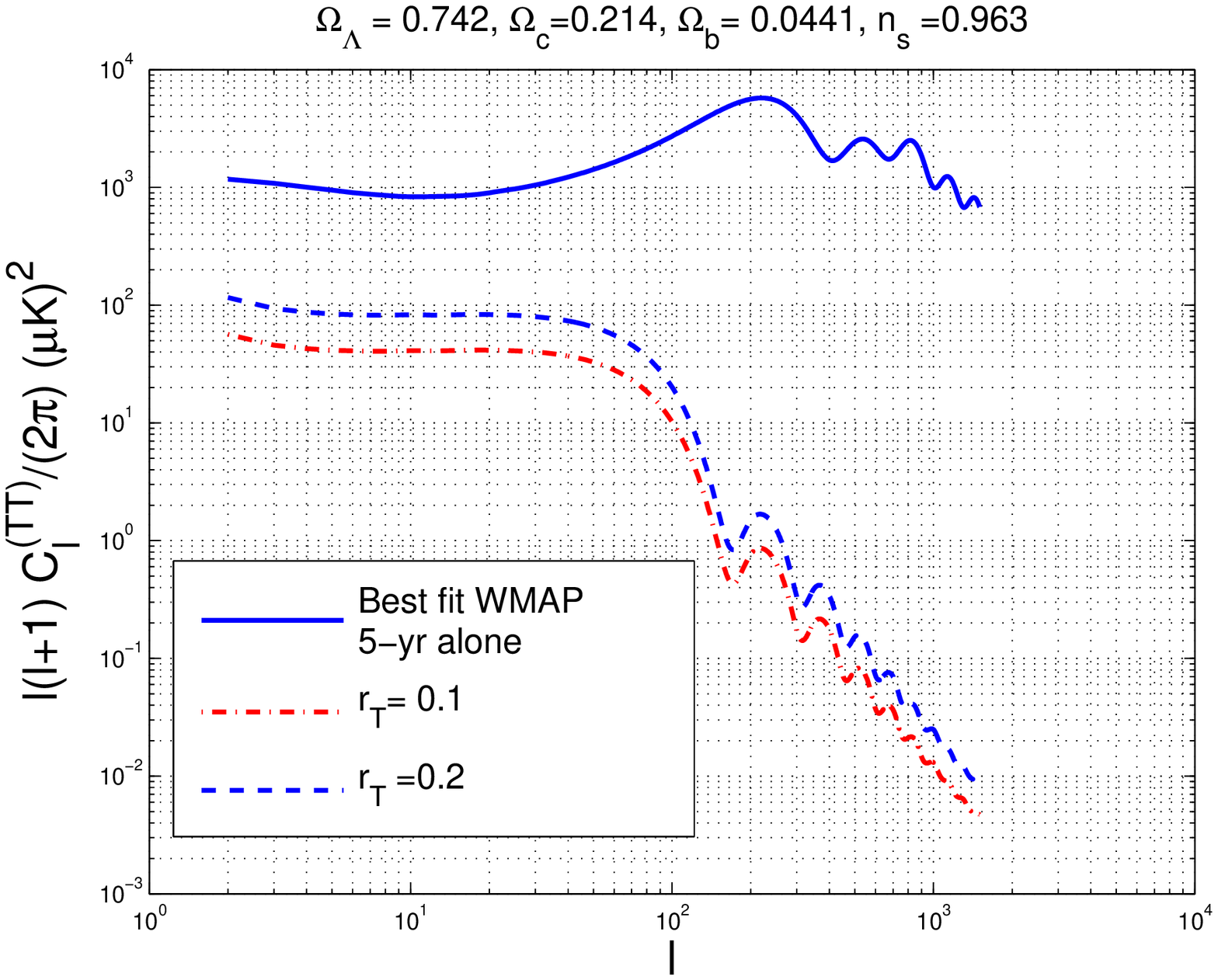}
\includegraphics[height=6.7cm]{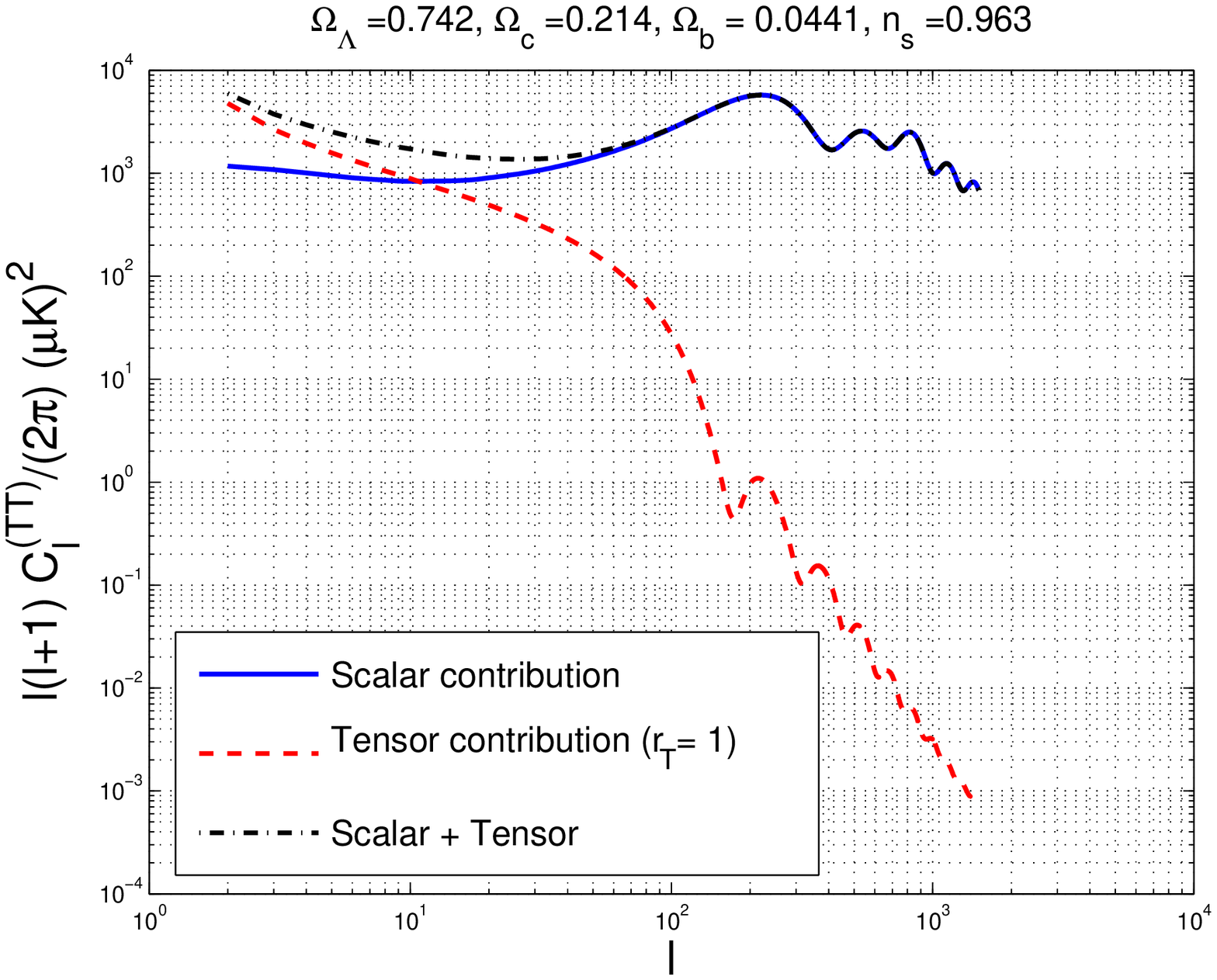}
\caption[a]{The temperature autocorrelations of the standard adiabatic 
scenario (full line) are compared with the TT angular power spectra induced by the tensor modes.}
\label{APS1}      
\end{figure}

The tensor modes of the geometry produce both, temperature and polarization anisotropies.
Consider, for simplicity, the $\Lambda$CDM parameters determined from the best 
fit of the WMAP 5-yr data alone. These parameters have been reported in Eq. (\ref{bestfit}). 
In Fig. \ref{APS1} the TT angular power spectra stemming from the standard 
adiabatic mode and the temperature autocorrelations induced by the tensor modes are 
illustrated.
In both plots of Fig. \ref{APS1} the full line denotes the TT angular power spectra 
corresponding to the best fit of the WMAP 5-yr data alone. In the plot at the 
left the dashed and dot-dashed lines illustrate, respectively, the cases $r_{\mathrm{T}}=0.1$ and 
$r_{\mathrm{T}} =0.2$. The tensor spectral index is fixed according to Eqs. (\ref{int1}) and (\ref{int4})  (see also Eqs. (\ref{TA3}) and (\ref{TA4})). In both plots of Fig. \ref{APS1} the spectral index does not run, i.e. 
$\alpha_{\mathrm{T}}=0$ in Eq. (\ref{int4}).
Always in Fig. \ref{APS1} (plot at the right) the dashed line denotes the tensor contribution 
in the case $r_{\mathrm{T}} =1$ while the dot-dashed line denotes the 
total TT angular power spectrum.
The temperature and polarization observables can be obtained numerically (see, e.g. \cite{B1,B2}) 
but useful analytic results  do exist in the literature \cite{P1,P2,P3,P4,zh1}. An essential step 
for both approaches is to derive the angular power spectra in more explicit terms and as a function 
of the solution of the heat transfer equations. The solution of Eqs. (\ref{Boltz83}), (\ref{Boltz84}) 
and (\ref{Boltz85}) can be formally written by using the integration along the line of sight which is also 
common to the scalar case (see, e.g. \cite{maxbook}, for an introduction):
\begin{eqnarray}
&&\Delta_{\mathrm{P}}(k,\mu,\tau_{0}) =\int_{0}^{\tau_{0}} d \tau S_{\mathrm{P}}(k,\tau) e^{- i \mu x},
\label{LS1}\\
&& \Delta_{\mathrm{T}}(k,\mu,\tau_{0}) =\int_{0}^{\tau_{0}} d \tau S_{\mathrm{T}}(k,\tau) e^{- i \mu x},
\label{LS2}
\end{eqnarray}
where $x = k(\tau_{0} - \tau)$. Note that $S_{\mathrm{T}}(k,\tau)$ and $S_{\mathrm{P}}(k,\tau)$ are related 
to the source terms of, respectively, Eqs. (\ref{Boltz83}) and (\ref{Boltz84}), i.e. 
\begin{equation}
S_{\mathrm{P}}(k,\tau)=  - {\mathcal K}(\tau) \, S(k,\tau),\qquad 
S_{\mathrm{T}}(k,\tau)=  - h' e^{- \epsilon(\tau,\tau_{0})} + {\mathcal K}(\tau) \, S(k,\tau),
\label{LS3}
\end{equation}
where $S(k,\tau)$ is given by Eq. (\ref{Boltz85}) and where 
\begin{equation}
{\mathcal K}(\tau) = \epsilon' e^{-\epsilon(\tau,\tau_{0})},\qquad \epsilon(\tau,\tau_{0}) 
= \int_{\tau}^{\tau_{0}} a \tilde{n}_{\mathrm{e}}\, x_{\mathrm{e}} \sigma_{\mathrm{T}}\, d\tau, \qquad 
\epsilon' =  a \tilde{n}_{\mathrm{e}}\, x_{\mathrm{e}} \sigma_{\mathrm{T}}.
\label{LS4}
\end{equation}
is the visibility function expressed in terms of the differential optical depth $\epsilon'$ and in terms 
of the optical depth $\epsilon(\tau,\tau_{0})$. The visibility function ${\mathcal K}(\tau)$ gives 
that probability that a photon is last scattered between $\tau$ and $\tau + d\tau$. In loose 
terms the visibility function is peaked around the redshift of recombination. 
In analytic discussions the visibility function is often approximated 
by means of a Gaussian profile (see, e. g., \cite{tur1,zh1,maxbook} and also \cite{zeld1,wyse,pav1,pav2}):
\begin{equation}
{\mathcal K}(\tau) = {\mathcal N}(\sigma_{\mathrm{rec}}) e^{- \frac{(\tau - \tau_{\mathrm{rec}})^2}{2 \sigma_{\mathrm{rec}}^2}},\qquad \int_{0}^{\tau_{0}} {\mathcal K}(\tau) d\tau = 1.
\label{LS5}
\end{equation}
As indicated in Eq. (\ref{LS5}),  ${\mathcal N}(\sigma_{\mathrm{rec}})$ is determined by requiring that
the integral of ${\mathcal K}(\tau)$ over $\tau$ is normalized to $1$. The WMAP data suggest a thickness 
(in redshift space) $\Delta z_{\mathrm{rec}} \simeq 195 \pm 2$ \cite{WMAPfirst1} which would imply 
that $\sigma_{\mathrm{rec}}$, in units of the (comoving) angular diameter distance to recombination, 
can be estimated as  $\sigma_{\mathrm{rec}}/\tau_{0} \simeq 1.43 \times 10^{-3}$. 
When $\tau_{0} \gg \tau_{\mathrm{rec}}$ and 
 $ \tau_{0} \gg \sigma_{\mathrm{rec}}$ the normalization appearing in Eq. (\ref{LS5}) can be estimated as ${\mathcal N}(\sigma_{\mathrm{rec}}) \to\sigma_{\mathrm{rec}}^{-1}\, \sqrt{2/\pi}$. In \cite{zh1} it has been suggested that a better approximation, for the case of the tensor modes,  is to assume that ${\mathcal K}(\tau)$ 
 has two different widths, respectively, for $\tau< \tau_{\mathrm{rec}}$ and for $\tau> \tau_{\mathrm{rec}}$.
Recalling now Eq. (\ref{Tr12}) and (\ref{intN6}) the $a_{\ell\, m}^{(\mathrm{T})}$ can be written as
\begin{eqnarray}
&& a_{\ell\, m}^{(\mathrm{T})} = \int d\hat{n} \, Y_{\ell m}^{*} \, \Delta_{\mathrm{I}}(\hat{n})
\nonumber\\
&& =\frac{1}{(2\pi)^{3/2}} \int d^{3} k \, \int_{0}^{2\pi}  d\varphi \int_{-1}^{1}\, d\mu  ( 1 - \mu^2) [ e^{2i\varphi} e_{1}(\vec{k}) + e^{ - 2 i \varphi} e_{2}(\vec{k}) ]
Y_{\ell\,m}^{*}(\mu, \varphi) \Delta_{\mathrm{T}}(k,\mu,\tau_{0}),
\label{LS6}
\end{eqnarray}
where the second equality follows from the  Fourier transform of $\Delta_{\mathrm{I}}(\hat{n})$ and by using 
Eq. (\ref{Boltz79a}) in the obtained expression. The angular integrations can now be performed with 
the approach already exploited in previous subsection for the polarization observables. 
It is however useful to solve Eq. (\ref{LS6}) with a slightly different method (see, e. g. \cite{B1}) 
where explicit use is made of the solutions (\ref{LS1}) and (\ref{LS2}).
\begin{figure}[!ht]
\centering
\includegraphics[height=6.7cm]{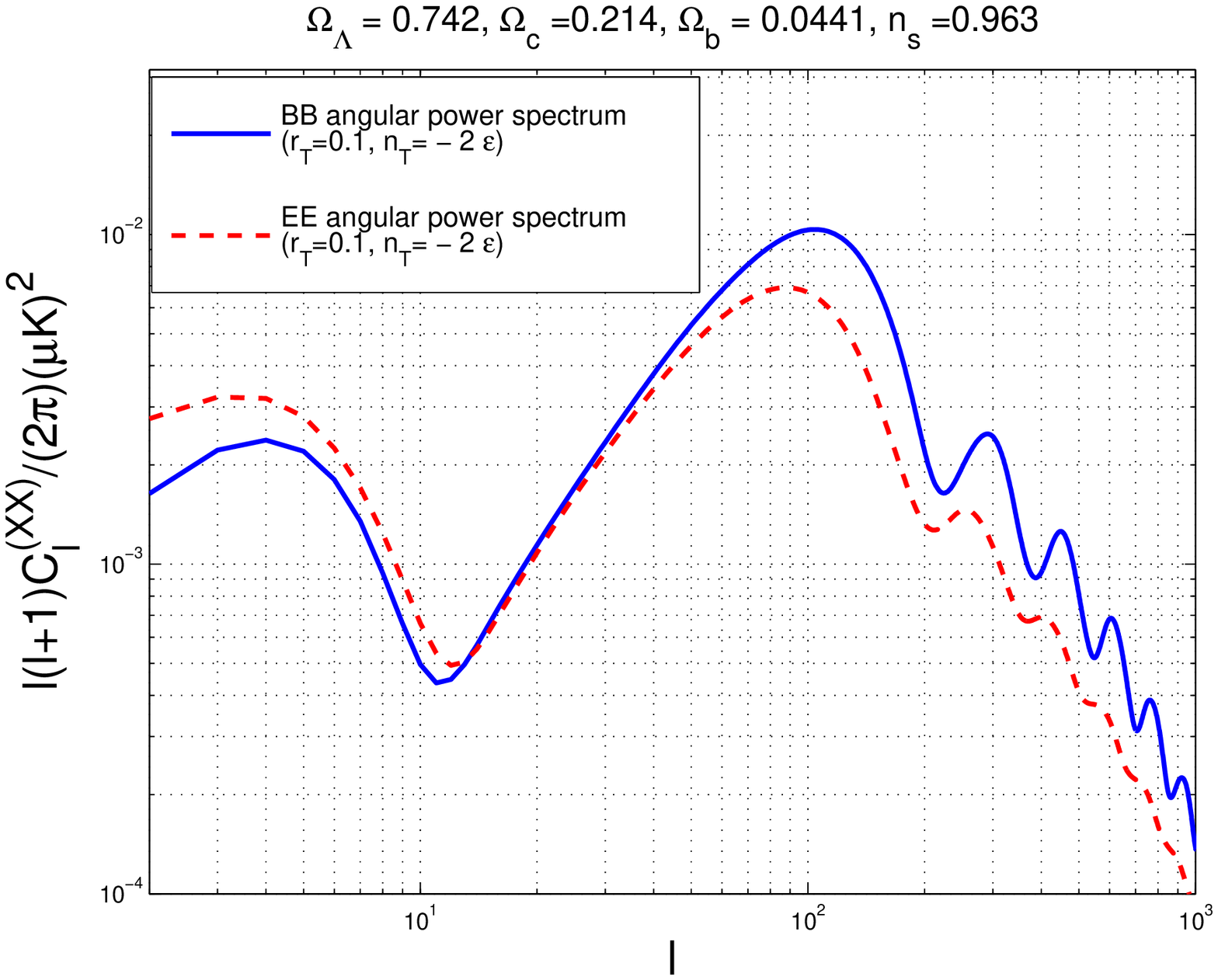}
\includegraphics[height=6.7cm]{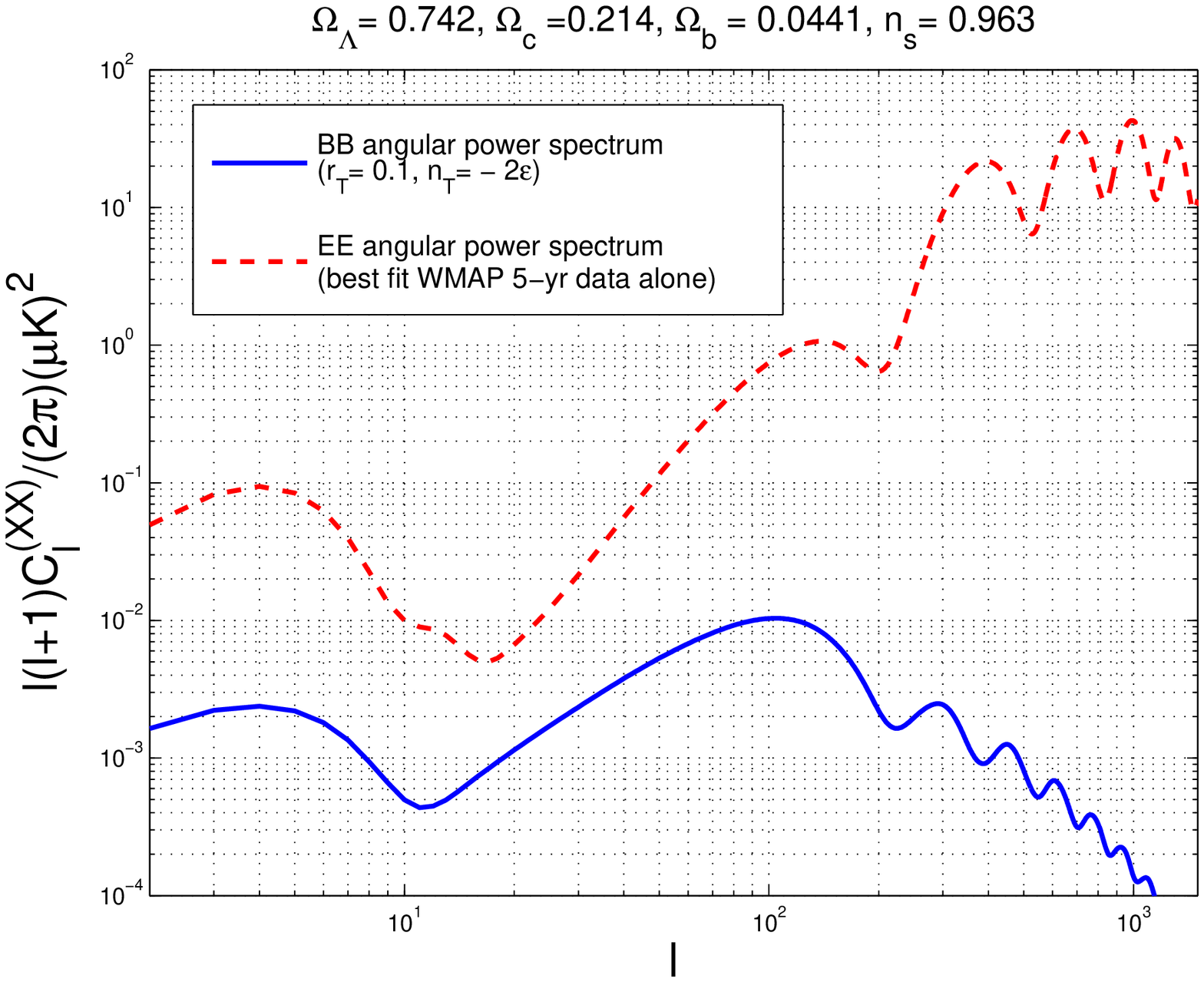}
\caption[a]{The polarization autocorrelations induced by the tensor modes 
are compared among themselves and with the EE angular power spectra produced by 
the adiabatic mode.}
\label{APS2}      
\end{figure}
Indeed, inserting Eq. (\ref{LS1}) into Eq. (\ref{LS6}) $a^{(\mathrm{T})}_{\ell\, m}$ can be written as 
\begin{eqnarray}
a_{\ell\, m}^{(\mathrm{T})} &=& \frac{1}{(2\pi)^{3/2}} \int d^{3} k\, \int_{0}^{\tau_{0}} d\tau \, S_{\mathrm{T}}(k,\tau) [ 
{\mathcal I}^{(\mathrm{T})}_{+}(\ell, m, x) e_{1}(\vec{k}) + {\mathcal I}^{(\mathrm{T})}_{-}(\ell, m, x) e_{2}(\vec{k}) ],
\label{LS7}\\
{\mathcal I}^{(\mathrm{T})}_{\pm}(\ell, m, x) &=& \int_{0}^{2\pi} d\varphi Y_{\ell\, m}^{*}(\mu,\varphi) (1- \mu^2) 
e^{\pm 2 i \varphi} e^{- i \mu x}.
\label{LS8}
\end{eqnarray}
In ${\mathcal I}^{(\mathrm{T})}_{\pm}$ the integration over $\varphi$ is trivial since the (ordinary) spherical harmonics 
depend upon $\varphi$ as $e^{i m \varphi}$ (see, e. g., Eqs. (\ref{Ydef1}) and (\ref{Ydef2})).
Thus, as expected on the basis of the results of the previous subsection, 
 ${\mathcal I}^{(\mathrm{T})}_{\pm} \propto 2\pi \delta_{m,\, \pm 2}$. The latter results allows for a 
 simplification so that, for instance, 
\begin{eqnarray}
 {\mathcal I}^{(\mathrm{T})}_{+}(\ell, m, x) &=& 2\pi \delta_{m,\,2} \, \sqrt{\frac{2\ell + 1}{4\pi}} \, 
 \sqrt{\frac{(\ell -2)!}{(\ell +2)!}} \int_{-1}^{1} d\mu  ( 1 - \mu^2)^2 \, \frac{d^2 P_{\ell}}{d\mu^2} e^{- i \mu x}
 \nonumber\\
 &=& - 4 \pi (-i)^{\ell} \delta_{m,\,2} \, \sqrt{\frac{2\ell + 1}{4\pi}} \, 
 \sqrt{\frac{(\ell -2)!}{(\ell +2)!}} ( 1 + \partial_{x}^2)^2 [ x^2 j_{\ell}(x)],
 \label{LS9}
 \end{eqnarray}
 where the second equality follows by first integrating by parts and by then expanding $e^{- i\mu x}$ 
 in series of Legendre polynomials. Indeed, in Eq. (\ref{LS9}), $j_{\ell}(x)$ are the spherical Bessel functions. 
 Note, finally that, inside the integral of Eq. (\ref{LS9}) expressions like $(1 -\mu^2)^2 e^{- i\mu x}$ can be 
 traded for $ ( 1 + \partial_{x}^2)^2 \, e^{- i\mu x}$. Recalling the properties of the associated Legendre 
 (see second relation of Eq. (\ref{Ydef2})) the other integral, i.e.  ${\mathcal I}^{(\mathrm{T})}_{-}(\ell, m, x)$ 
 can be performed exactly in the same way. By making explicit the ensemble 
 averages and by using Eq. (\ref{intN6}), the angular power spectrum illustrated in Fig. \ref{APS1} 
 can be obtained and it is given by:
 \begin{eqnarray}
C_{\ell}^{(\mathrm{TT})} &=& 4\pi \int \frac{d k}{k} {\mathcal P}_{\mathrm{T}}(k) |\Delta_{\mathrm{T}\ell}(k, \tau_{0})|^2,
\nonumber\\
\Delta_{\mathrm{T}\ell}(k, \tau_{0}) &=& \overline{N}_{\ell} \int_{0}^{\tau_{0}} S_{\mathrm{T}}(k,\tau) \frac{j_{\ell}(x)}{x^2},
\label{LS10}
\end{eqnarray}
where, as in Eq. (\ref{intN4}), $\overline{N}_{\ell} = \sqrt{(\ell -2)!/(\ell +2)!}$. 
Having discussed the temperature autocorrelation induced by the long wavelength gravitons 
it is now the moment of discussing the polarization observables.
In Fig. \ref{APS2} (plot at the left) the EE angular power spectra (full line) 
are compared with the B-mode autocorrelations  (dashed line) induced by the tensor 
modes in the case $r_{\mathrm{T}}=0.1$ while the other cosmological parameters 
are fixed to the best-fit values of the WMAp 5-yr data alone. 
Always in Fig. \ref{APS2} (plot at the right) the full line illustrates the B-mode autocorrelation 
in the case $r_{\mathrm{T}}=0.1$ while the dashed line 
illustrated the EE angular power spectrum stemming from the standard adiabatic mode.
In Fig. \ref{APS3} (plot at the left) the BB angular power spectra are illustrated for different values 
of $r_{\mathrm{T}}$ compatible with current bounds on $r_{\mathrm{T}}$ (see Tabs. \ref{TABLE1} and 
\ref{TABLE2}).

The E-mode and the B-mode angular power spectra can be obtained from Eqs. (\ref{Boltz86}) and (\ref{Boltz87}) 
by carefully following  all the steps leading to the temperature autocorrelation of Eq. (\ref{LS10}). The crucial 
difference will be, of course, that $a_{\ell\, m}^{(\mathrm{E})}$ and $a_{\ell\, m}^{(\mathrm{B})}$ 
arise, respectively, in the expansion of $\Delta_{\mathrm{E}}(\hat{n})$ and $\Delta_{\mathrm{B}}(\hat{n})$ 
as reported in Eq. (\ref{intN4}). To compute $a_{\ell\, m}^{(\mathrm{E})}$ and $a_{\ell\, m}^{(\mathrm{B})}$
in terms of $S_{\mathrm{P}}(k,\tau)$ the steps are, in short, the following:
\begin{itemize}
\item{} $a_{\ell\, m}^{(\mathrm{E})}$ and $a_{\ell\, m}^{(\mathrm{B})}$ should be first 
expressed in terms of $\Delta_{\mathrm{E}}(\hat{n})$ and $\Delta_{\mathrm{B}}(\hat{n})$ as in Eq. 
(\ref{intN4});
\item{} then Eqs. (\ref{Boltz86}) and (\ref{Boltz87}) should be inserted into Eqs. (\ref{EX1}) and (\ref{EX2});
\item{} the evolution for $\Delta_{\mathrm{P}}(k,\mu,\tau)$ can then be solved in Fourier space 
as in Eq. (\ref{LS1}). 
\end{itemize}
The first and the second step has been already (partially) performed in the previous subsection and the results 
have been given in Eqs. (\ref{Bex5})--(\ref{Bex6})  with the difference that, now, the two polarizations can be 
treated simultaneously so that, in Fourier space,
\begin{eqnarray}
\Delta_{\mathrm{Q}}(\vec{k}, \tau, \mu,\varphi) &=& ( 1 + \mu^2) [ e^{ 2 i \varphi} e_{1} (\vec{k}) + e^{-2 i \varphi} e_{2} (\vec{k})] \Delta_{\mathrm{P}}(k,\tau,\mu),
\label{LS11}\\
\Delta_{\mathrm{U}}(\vec{k}, \tau, \mu,\varphi) &=& 2 \mu i [ e^{ 2 i \varphi} e_{1} (\vec{k}) - e^{-2 i \varphi} e_{2} (\vec{k})] \Delta_{\mathrm{P}}(k,\tau,\mu).
\label{LS12}
\end{eqnarray}
Note that Eqs. (\ref{LS11}) and (\ref{LS12}) follow directly from Eqs. (\ref{Boltz86}) and (\ref{Boltz87}) by 
definition of $\Delta_{\pm}(\vec{k}, \tau, \mu,\varphi)$. 
It should be appreciated that one derivation with respect to $\varphi$ changes the azimuthal structure 
of $\Delta_{\mathrm{Q}}(\vec{k}, \tau, \mu,\varphi)$ and $\Delta_{\mathrm{U}}(\vec{k}, \tau, \mu,\varphi)$, i.e. 
\begin{eqnarray}
&& \partial_{\varphi} \Delta_{\mathrm{U}} = - 4 \mu [ e^{ 2 i \varphi} e_{1}(\vec{k}) + e^{ - 2 i \varphi} e_{2}(\vec{k})]
\Delta_{\mathrm{P}}(k, \tau, \mu),
\label{LS13}\\
&& \partial_{\varphi} \Delta_{\mathrm{Q}} = 2 i (1 + \mu^2)  [ e^{ 2 i \varphi} e_{1}(\vec{k}) - e^{ - 2 i \varphi} e_{2}(\vec{k})]\Delta_{\mathrm{P}}(k, \tau, \mu). 
\label{LS14}
\end{eqnarray}
Using now Eqs. (\ref{LS11})--(\ref{LS12}) and (\ref{LS13})--(\ref{LS14}) inside Eqs. (\ref{Bex5}) and (\ref{Bex6}) 
we do get, after Fourier transform,
\begin{eqnarray}
 a^{(\mathrm{B})}_{\ell m} &=& - 2 i \frac{\overline{N}_{\ell}}{(2\pi)^{3/2}} \int d^{3} k \, \int_{0}^{2\pi}  d\varphi \int_{-1}^{1}\, d\mu  [ e^{2i\varphi} e_{1}(\vec{k}) - e^{ - 2 i \varphi} e_{2}(\vec{k}) ]
Y_{\ell\,m}^{*}(\mu, \varphi)
\nonumber\\
&\times& ( 1 - \mu^2) [\mu \partial_{\mu}^2 + 4 \partial_{\mu}]\Delta_{\mathrm{P}}(k,\mu,\tau_{0}),
\label{LS15}\\
a^{(\mathrm{E})}_{\ell m} &=& -  \frac{\overline{N}_{\ell}}{(2\pi)^{3/2}} \int d^{3} k \, \int_{0}^{2\pi}  d\varphi \int_{-1}^{1}\, d\mu   [ e^{2i\varphi} e_{1}(\vec{k}) + e^{ - 2 i \varphi} e_{2}(\vec{k}) ]
Y_{\ell\,m}^{*}(\mu, \varphi) 
\nonumber\\
&\times&( 1 - \mu^2)[( 1 +\mu^2) \partial_{\mu}^2 + 8 \mu \partial_{\mu} + 12 ]\Delta_{\mathrm{P}}(k,\mu,\tau_{0}).
\label{LS16}
\end{eqnarray}
According to Eq. (\ref{LS1})  $\Delta_{\mathrm{P}}(k,\mu,\tau_{0})$ can be related to $S_{\mathrm{P}}(k,\tau)$ and the resulting coefficients are:
\begin{eqnarray}
a_{\ell\, m}^{(\mathrm{B})} &=& - \frac{\overline{N}_{\ell}}{(2\pi)^{3/2}}  \int_{0}^{2\pi} d\varphi \int_{-1}^{1} d\mu 
 Y_{\ell\, m}^{*} (\mu,\varphi) \int 
d^{3} k  [ e^{2i\varphi} e_{1}(\vec{k}) - e^{ - 2 i \varphi} e_{2}(\vec{k}) ] 
\nonumber\\
&\times& ( 1 - \mu^2) {\mathcal O}_{\mathrm{B}}(x) \int_{0}^{\tau_{0}} S_{\mathrm{P}}(k,\tau) \, e^{- i\mu x}  d\tau 
\label{LS17}\\
a_{\ell\, m}^{(\mathrm{E})} &=& - \frac{\overline{N}_{\ell}}{(2\pi)^{3/2}}  \int d \hat{n} Y_{\ell\, m}^{*} (\mu,\varphi) \int 
d^{3} k  [ e^{2i\varphi} e_{1}(\vec{k}) + e^{ - 2 i \varphi} e_{2}(\vec{k}) ]
\nonumber\\
&\times& ( 1 - \mu^2) {\mathcal O}_{\mathrm{E}}(x) \int_{0}^{\tau_{0}} S_{\mathrm{P}}(k,\tau) \, e^{- i\mu x}  d\tau, 
\label{LS18}
\end{eqnarray}
where ${\mathcal O}_{\mathrm{B}}(x)$ and ${\mathcal O}_{\mathrm{E}}(x)$ are two differential operators, i.e. 
\begin{equation} 
{\mathcal O}_{\mathrm{B}}(x) = 2 x^2 \partial_{x}^2 + 8 x,\qquad {\mathcal O}_{\mathrm{E}}(x) = x^2 ( 1 - \partial_{x}^2) - 8 x \partial_{x} -12.
\label{LS19}
\end{equation}
Now Eqs. (\ref{LS17}) and (\ref{LS18}) can be used into Eqs. (\ref{intN5}) and the angular power 
spectra become:
\begin{eqnarray}
&& C_{\ell}^{(\mathrm{BB})} = 4\pi \int \frac{d k}{k} {\mathcal P}_{\mathrm{T}}(k) |\Delta_{\mathrm{B}\ell}(k,\tau_{0})|^2,
\label{LS20}\\
&& C_{\ell}^{(\mathrm{EE})} = 4\pi \int \frac{d k}{k} {\mathcal P}_{\mathrm{T}}(k) |\Delta_{\mathrm{E}\ell}(k,\tau_{0})|^2, 
\label{LS21}
\end{eqnarray}
where 
\begin{figure}[!ht]
\centering
\includegraphics[height=6.7cm]{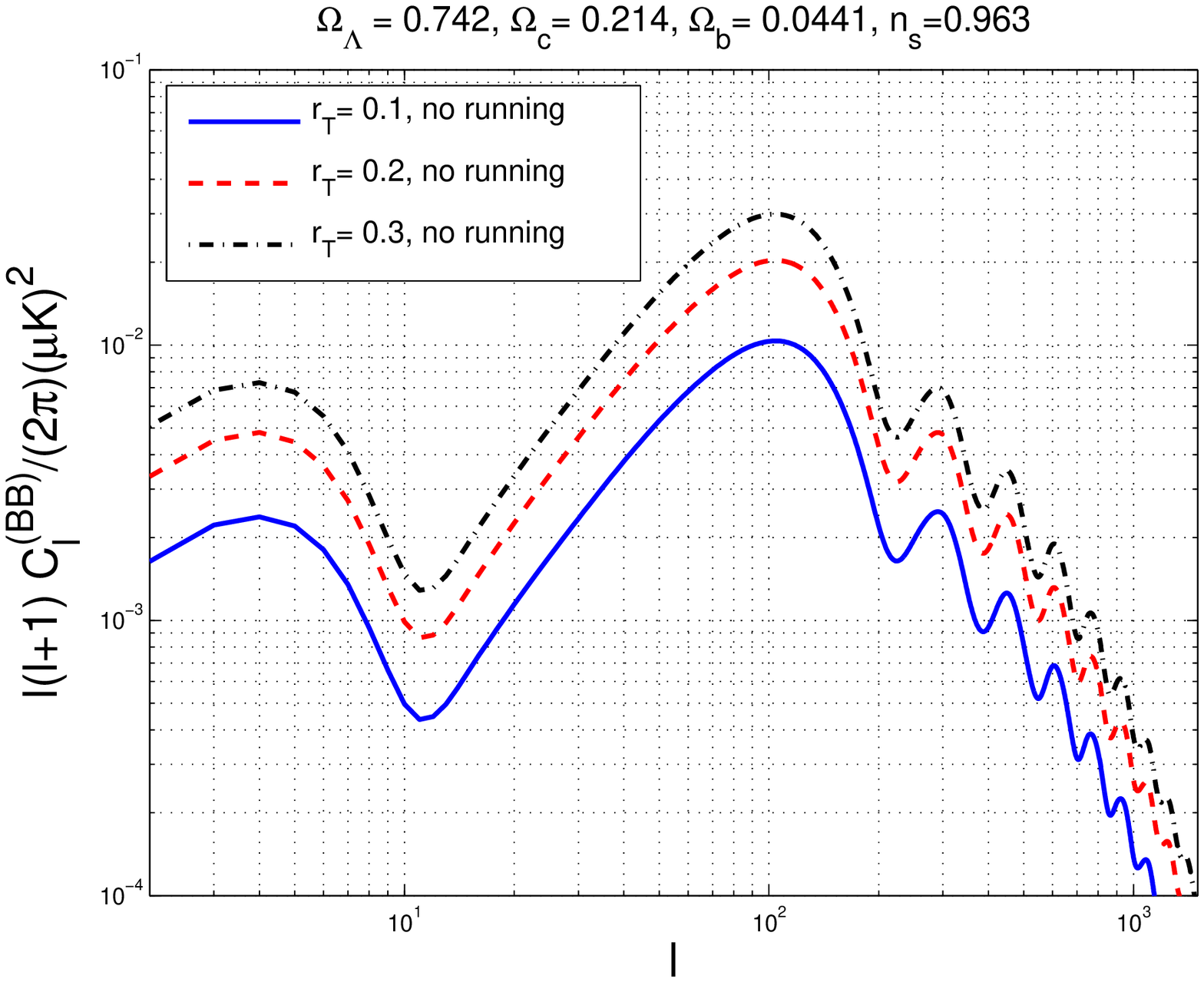}
\includegraphics[height=6.7cm]{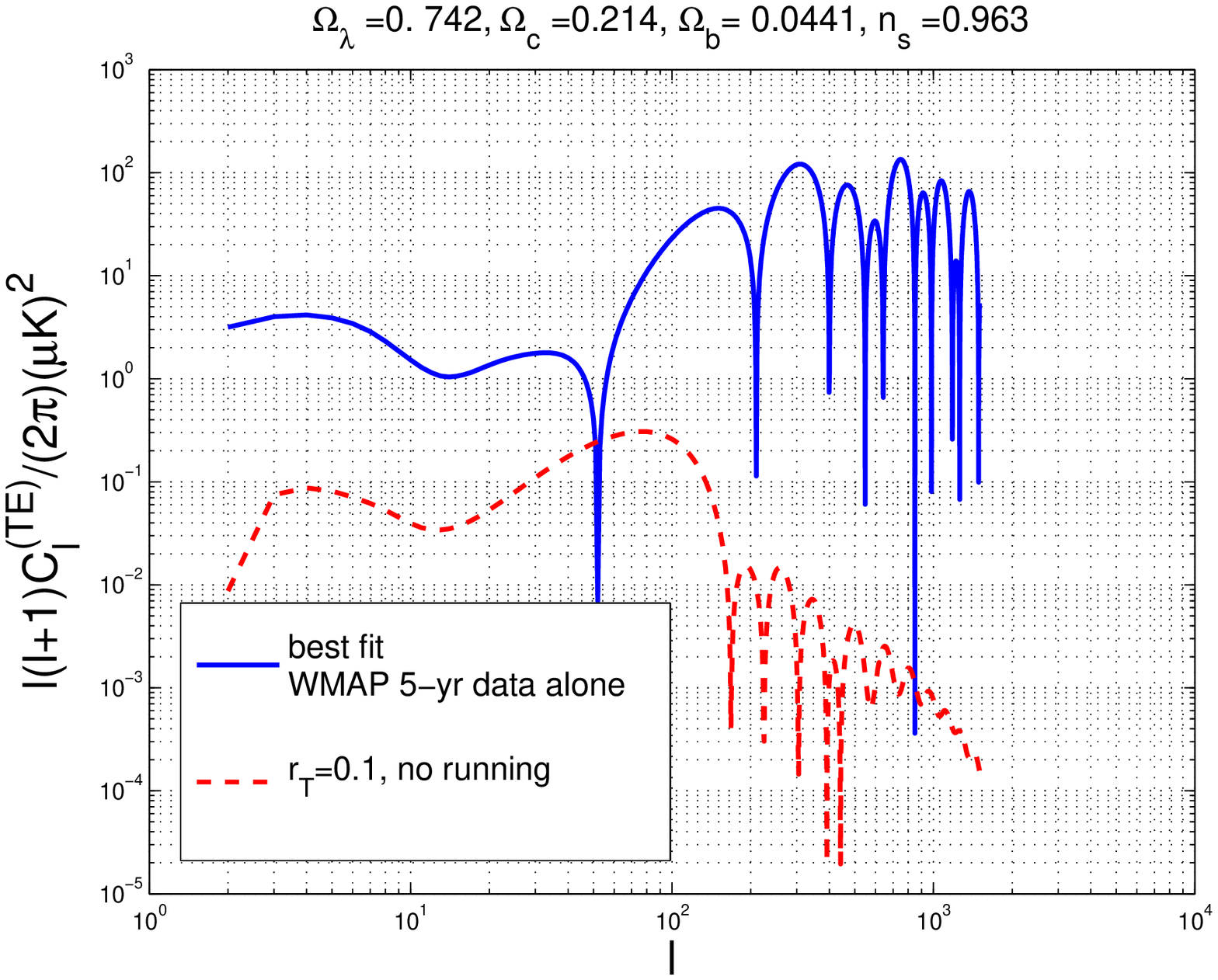}
\caption[a]{The B-mode autocorrelation for different values of $r_{\mathrm{T}}$ (plot at the left). The TE cross
correlation induced by the tensor modes is compared with the corresponding angular power spectrum induced by the standard adiabatic mode.}
\label{APS3}      
\end{figure}
\begin{eqnarray}
&& \Delta_{\mathrm{E}\ell}(k,\tau_{0}) = \biggl[ \frac{4}{x} \partial_{x}j_{\ell}(x) + \frac{2}{x^2} j_{\ell}(x) + \partial^2_{x} j_{\ell}(x) - j_{\ell}(x)\biggr] \int_{0}^{\tau_{0}} d\tau S_{\mathrm{P}}(k,\tau),
\label{LS22}\\
&& \Delta_{\mathrm{B}\ell}(k,\tau_{0}) = \biggl[ 2 \partial_{x}j_{\ell}(x) + \frac{4}{x^2} j_{\ell}(x)\biggr] \int_{0}^{\tau_{0}} d\tau S_{\mathrm{P}}(k,\tau).
\label{LS23}
\end{eqnarray}
The tensor modes also induce a TE angular power spectrum (i.e. a cross-correlation between the temperature and 
the E-mode polarization). In Fig. \ref{APS3} (plot at the right) the absolute value of the TE (tensor) angular power 
spectrum is compared with the corresponding angular power spectrum induced by the standard adiabatic mode.
The cross-correlation between temperature and polarization can be obtained inserting 
Eqs. (\ref{LS8})  and (\ref{LS16}) into Eq. (\ref{intN6}) and the result is 
\begin{equation}
C_{\ell}^{(\mathrm{TE})} = 4\pi \int \frac{d k}{k} {\mathcal P}_{\mathrm{T}}(k) \Delta_{\mathrm{E}\ell}(k,\tau_{0}) \Delta_{\mathrm{T}\ell}(k,\tau_{0}),
\label{LS24}
\end{equation}
where, $\Delta_{\mathrm{T}\ell}(k,\tau_{0})$ and  $\Delta_{\mathrm{E}\ell}(k,\tau_{0})$ 
are given, respectively, by Eqs. (\ref{LS10}) and (\ref{LS22}). 
It should finally be mentioned that the derivatives with respect to $x$ appearing in Eqs. 
(\ref{LS22}) and (\ref{LS23}) can be transformed into derivatives with respect to $\tau$ 
by making use of integration by parts \cite{B1}. This step is carried on in full analogy with 
what happens with the scalar modes of the geometry \cite{maxbook}.

Various experiments 
provided, so far, direct limits on the B-mode polarization. 
\begin{table}[!ht]
\begin{center}
\begin{tabular}{|c|c|c|c|}
\hline
Experiments & Years  & Pivot frequencies &Upper limits on BB \\
\hline
Dasi \cite{dasi1,dasi2,dasi3} & 2000-2003 & 
$26$--$36$ GHz & $\ell (\ell +1) C_{\ell}^{(\mathrm{BB})}(2\pi) \leq 50 \,(\mu\,\mathrm{K})^2$  
\\
Boomerang \cite{boom1}&2003  & $145$ GHz& $\ell(\ell +1)C_{\ell}^{(\mathrm{BB})}/(2\pi)\leq 8.6 \,(\mu\,\mathrm{K})^2$ 
 \\
Maxipol \cite{maxip1,maxip2} & 2003 &  $140$ GHz & 
$\ell(\ell+1)C_{\ell}^{(\mathrm{BB})}/(2\pi)\leq 112.3 \,(\mu\,\mathrm{K})^2$ 
\\
Quad \cite{quad1,quad2,quad3} & 2007 & $100$/ $150$ GHz&
$\ell (\ell +1)C_{\ell}^{(\mathrm{BB})}/(2\pi) \leq  10 \,(\mu\,\mathrm{K})^2$
\\
Cbi \cite{cbi1,cbi2} & 2002-2005&  $26$--$36$ GHz& 
$\ell (\ell +1)C_{\ell}^{(\mathrm{BB})}/(2\pi) \leq  3.76\,(\mu\,\mathrm{K})^2$
\\
Capmap  \cite{capmap1,capmap2} & 2004-2005& $35$--$46$/$84$--$100$ GHz& 
$\ell(\ell+1) C_{\ell}^{(\mathrm{BB})}/(2\pi)\leq  4.8\,(\mu\,\mathrm{K})^2$
\\
Wmap \cite{WMAP51,WMAP52,WMAP54} & 2001-2006& $23$--$94$ GHz& 
$\ell (\ell+1)C_{\ell}^{(\mathrm{BB})}/(2\pi) \leq  0.15\,(\mu\,\mathrm{K})^2$
\\
\hline
\end{tabular}
\caption{The main polarization experiments, their typical observational frequencies, and the upper limits on the B-mode polarization are illustrated.}
\label{TABLE3}
\end{center}
\end{table}
The limits on $r_{\mathrm{T}}$ reported in Tabs. \ref{TABLE1} and \ref{TABLE2} follow 
from a combined analysis of the TT, EE and TE angular power spectra which allows 
for  a tensor contribution. As Tab. \ref{TABLE3} indicates the upper limits on the B-mode polarization 
are still rather loose and, often, derived on the basis of a limited range of harmonics. 
\newpage
\renewcommand{\theequation}{8.\arabic{equation}}
\setcounter{equation}{0}
\section{High-frequency spikes in the relic graviton background}
\label{sec8}

In the $\Lambda$CDM paradigm long wavelength gravitons can affect the CMB polarization. 
As the frequency increases 
towards the region accessible to wide band interferometers, the $\Lambda$CDM signal 
can only decrease for number of independent reasons:
\begin{itemize}
\item{}  in the $\Lambda$CDM paradigm the spectral energy density is only approximately scale-invariant 
(see, e.g. Fig. \ref{Figure1}) but the scaling violations always tend to make the spectral energy density 
smaller at high frequencies;
\item{} there are various secondary effects associated, for instance, with the variation of the effective number of relativistic degrees of freedom, with the neutrino anisotropic stress and with the transition 
to a phase dominated by the dark-energy contribution: all these features reduce 
the spectral energy density in different frequency regions;
\item{} the addition of supplementary scalar fields driving the inflationary phase 
does not change the two previous statements.
\end{itemize}
The previous conclusions are all based (either directly or indirectly) on the assumption 
that the thermal history of the Universe is minimal in the sense that, after inflation, the Universe 
soon becomes dominated by relativistic particles so that the sound speed 
of the plasma soon reaches values $c_{\mathrm{st}} = 1/\sqrt{3}$.
The post-inflationary thermal history might not be minimal. For instance 
it could happen that the transition to the radiation-dominated regime 
is not instantaneous \cite{mg1}. More specifically it can happen that, after inflation, the sound 
speed of the plasma is such that $c_{\mathrm{st}} > 1/\sqrt{3}$.  When the sound 
speed is larger than $1/\sqrt{3}$, the fluid is said to be stiff.
In the system of units used 
in the present paper the speed of light is  such that $c= 1$. A natural 
 upper limit for the sound speed is exactly $1$ which is 
the maximally stiff fluid compatible with causality \cite{SpS} (see, however, also \cite{kessence1}). 
If the thermal history of the Universe contemplates a post-inflationary phase stiffer 
than radiation, a spike in the relic graviton spectrum is expected at high frequencies \cite{mg1} 
(see also \cite{mg2,mg3} as well as Eqs. (\ref{ST1})--(\ref{ST2}) and discussion therein).
The possibility of having a post-inflationary phase stiffer than radiation has been 
also investigated in different contexts such as in \cite{REP1,REP2,REP3}.

In the early Universe, the dominant energy condition might be violated and this observation will also 
produce scaling violations in the spectral energy density  \cite{DOC1}. 
If we assume the validity of the $\Lambda$CDM paradigm, a violation 
of the dominant energy condition implies that, during an early stage of the life of the Universe, the effective enthalpy density of the sources driving the geometry was negative and this may happen in the presence 
of bulk viscous stresses \cite{DOC1} (see also \cite{DOC2,DOC3} for interesting reprises of this idea). In what follows the focus will be on the more mundane possibility that the thermal history of the plasma includes a phase where the  speed of sound was close to the speed of light. 

Absent any indirect tests on the thermal history of the Universe prior to the formation of light nuclear elements, 
it is legitimate to investigate situations where, before nucleosyntheis, 
the sound speed of the plasma was larger than  $1/\sqrt{3}$, at most equalling the speed of light. 
In this plausible extension of the current cosmological paradigm, 
hereby dubbed Tensor-$\Lambda$CDM (i.e. T$\Lambda$CDM) scenario, high-frequency gravitons are copiously 
produced \cite{mgn1,mgn2}. Without conflicting with the bounds on the tensor to scalar ratio stemming from the combined analysis of the three standard cosmological 
data sets (i.e. cosmic microwave background anisotropies, large-scale structure data and observations of type Ia
supenovae),  the spectral energy density of the relic gravitons in the T$\Lambda$CDM scenario can be potentially observable by wide-band interferometers (in their advanced version) operating  in a frequency window 
which ranges between few Hz and few kHz. 

The presence  of a stiff phase increases the spectral energy density  for frequencies larger than a pivotal frequency $\nu_{\mathrm{s}}$ which is related to the total duration of the stiff phase.
If the stiff phase takes place before  BBN, then $\nu_{\mathrm{s}} > 10^{-2}$ nHz. If the 
stiff phase takes place for equivalent temperatures larger than $100$ GeV, then $\nu_{\mathrm{s}} \geq \mu\mathrm{Hz}$. If the stiff phase takes place for $T \geq 100$ TeV,  then $\nu_{\mathrm{s}} > \mathrm{mHz}$.  
The frequency $\nu_{\mathrm{s}}$ marks the beginning of a branch of the spectrum where  the tilt of the spectral energy density is blue (i.e. increasing in slope) rather than nearly scale invariant or slightly red (as it is the case in the conventional scenario). 

\subsection{Scaling violations}
In the $\Lambda$CDM paradigm, for frequencies larger than $\nu_{\mathrm{eq}}$ the spectral energy density of the relic gravitons is, approximately, scale invariant (see Fig. \ref{Figure1}).  In the context 
of the T$\Lambda$CDM scenario the approximate scale-invariance of the flat plateau is violated.
This situation is illustrated, for instance, in Fig. \ref{Figure2a}. In the present subsection 
a more detailed account of the typical frequencies of the problem will be presented.
If there is some delay between the end of inflation and the onset of radiation 
the maximal wavenumber of the spectrum will be given by: 
\begin{equation}
k_{\mathrm{max}} = M_{\mathrm{P}} \biggl(\frac{H}{M_{\mathrm{P}}}\biggr)^{1 - \alpha}
 \biggr(\frac{H_{\mathrm{r}}}{M_{\mathrm{P}}}\biggl)^{\alpha - 1/2} 
 \biggl(\frac{H_{\mathrm{eq}}}{M_{\mathrm{P}}}\biggr)^{1/2} \biggl(\frac{a_{\mathrm{eq}}}{a_{0}}\biggr)
 \label{STFR1}
 \end{equation}
 where $\alpha = 2/[3(w_{\mathrm{t}} +1)]$ (with $w_{\mathrm{t}} > 1/3$) is related to the specific kind of stiff dynamics. Equation (\ref{STFR1}) can also be written as 
 \begin{equation}
k_{\mathrm{max}} = M_{\mathrm{P}} \Sigma^{-1}
 \biggl(\frac{H_{\mathrm{eq}}}{M_{\mathrm{P}}}\biggr)^{1/2} \biggl(\frac{a_{\mathrm{eq}}}{a_{0}}\biggr).
 \label{STFR2}
 \end{equation}
 where 
 \begin{equation}
 \Sigma = \biggl(\frac{H}{M_{\mathrm{P}}}\biggr)^{\alpha-1 }
 \biggr(\frac{H_{\mathrm{r}}}{M_{\mathrm{P}}}\biggl)^{1/2-\alpha }.
 \label{STFR3}
\end{equation}
In the case $\Sigma = {\mathcal O}(1)$ (as it happens in the case $\alpha = 1/3$ if the initial radiation is in the 
form of quantum fluctuations) $\nu_{\mathrm{max}} = k_{\mathrm{max}}/(2\pi) \simeq 100\,\,\mathrm{GHz}$, more precisely:
\begin{equation}
\nu_{\mathrm{max}} = 1.177 \times 10^{11} \Sigma^{-1} 
\biggl(\frac{h_{0}^2 \Omega_{\mathrm{R}0}}{4.15 \times 10^{-5}}\biggr)^{1/4}\,\,\mathrm{Hz}.
\label{STFR4}
\end{equation}
On top of the standard parameters of the $\Lambda$CDM scenario (see, e. g., Eq. (\ref{bestfit})) 
the minimal T$\Lambda$CDM scenario demands two supplementary parameters
\begin{itemize}
\item{} the frequency $\nu_{\mathrm{s}}$ defining the region of the spectrum at which the scaling violations take place;
\item{} the slope of the spectrum arising during the stiff phase.
\end{itemize}
The frequency $\nu_{\mathrm{s}}$ can be dynamically related to the frequency of the maximum and, consequently, the first parameter can be trated for $\Sigma$. As it will be shown in a moment, typical values 
for $\Sigma$ range between $0.01$ and $0.5$.
The slope of the spectrum during the stiff phase depends 
upon the total barotropic index and can therefore be traded for $w_{\mathrm{t}}$.
The curvature scale $H_{\mathrm{r}}$ determines $k_{\mathrm{s}}$ (or $\nu_{\mathrm{s}}$), i.e. the frequency at which the spectral energy density starts increasing.
Supposing that from the end of inflation there is a 
single stiff phase (as it is natural to assume in a minimalistic 
persepective) the value of $k_{\mathrm{s}}$ is 
\begin{equation}
k_{\mathrm{s}} =  M_{\mathrm{P}} \biggl(\frac{H_{\mathrm{eq}}}{M_{\mathrm{P}}}\biggr)^{1/2} \biggl(\frac{a_{\mathrm{eq}}}{a_{0}}\biggr) \sqrt{\frac{H_{\mathrm{r}}}{M_{\mathrm{P}}}}.
\label{STFR5}
\end{equation}
Using the relation of $H_{\mathrm{r}}$ to $\Sigma$, Eq. (\ref{STFR5}) can also be written as 
\begin{equation}
k_{\mathrm{s}} =  M_{\mathrm{P}} \biggl(\frac{H_{\mathrm{eq}}}{M_{\mathrm{P}}}\biggr)^{1/2} 
\biggl(\frac{a_{\mathrm{eq}}}{a_{0}}\biggr)  \Sigma^{ 1/(1 - 2\alpha)}  \biggl(\frac{H}{M_{\mathrm{P}}}\biggr)^{(\alpha -1)/(2 \alpha -1)}.
\label{STFR6}
\end{equation}
and as 
\begin{equation}
\nu_{\mathrm{s}} = 1.173 \times 10^{11} \Sigma^{ 1/(1-2\alpha)} \,\,(\pi \epsilon {\mathcal A}_{{\mathcal R}})^{\frac{\alpha -1}{2(2\alpha -1)}}\,\,\biggl(\frac{h_{0}^2 \Omega_{\mathrm{R}0}}{4.15 \times 10^{-5}}\biggr)^{1/4}\,\,\mathrm{Hz}.
\label{STFR7}
\end{equation}

The quantity $\Sigma$ which parametrizes the location, in the spectrum, where the scaling violations appear must be  smaller than $1$ or, at most, of order $1$.  This is what happens within specific models. For instance, 
if the radiation present at the end of inflation comes from amplified quantum fluctuations (i.e. Gibbons-Hawking radiation), 
quite generically, at the end of inflation $\rho_{\mathrm{r}} \simeq H^4$. More specifically 
\begin{equation}
\rho_{\mathrm{r}}  = \frac{\pi^2}{30} N_{\mathrm{eff}} T_{H}^4 = \frac{N_{\mathrm{eff}} H^4}{480 \pi^2}.
\label{STFR13}
\end{equation}
In Eq. (\ref{STFR13}) $N_{\mathrm{eff}}$ is the number of species contributing 
to the quantum fluctuations during the quasi-de Sitter stage of expansion. 
In \cite{ford3} (see also \cite{mg2,mg3,PV}) it has been argued that this quantity could be evaluated using a perturbative expansion valid in the limit of quasi-conformal 
coupling. It should be clear that $N_{\mathrm{eff}}$ is conceptually 
different from the number of relativistic degrees of freedom $g_{\rho}$. 
Given $H$ and $N_{\mathrm{eff}}$ the length of the stiff phase is fixed, in this case,  by \cite{PV}
\begin{equation}
\lambda H^4 \biggl(\frac{a_{\mathrm{i}}}{a_{\mathrm{r}}}\biggr)^{4} =  H^2 M_{\mathrm{P}}^2  \biggl(\frac{a_{\mathrm{i}}}{a_{\mathrm{r}}}\biggr)^{3(w +1)} 
= H^2 M_{\mathrm{P}}^2  \biggl(\frac{a_{\mathrm{i}}}{a_{\mathrm{r}}}\biggr)^{2/\alpha}, 
\label{STFR14}
\end{equation}
where we used the fact that $\alpha = 2/[3(w +1)]$ and where we defined  $\lambda= N_{\mathrm{eff}}/(480\pi^2)$.
Equation (\ref{STFR14}) implies that
\begin{equation}
\biggl(\frac{a_{\mathrm{i}}}{a_{\mathrm{r}}}\biggr) = \lambda^{\frac{\alpha}{2 - 4\alpha}} \biggl(\frac{H}{M_{\mathrm{P}}}\biggr)^{\frac{\alpha}{1 - 2 \alpha}}.
\label{STFR15}
\end{equation}
\begin{figure}[!ht]
\centering
\includegraphics[height=6.7cm]{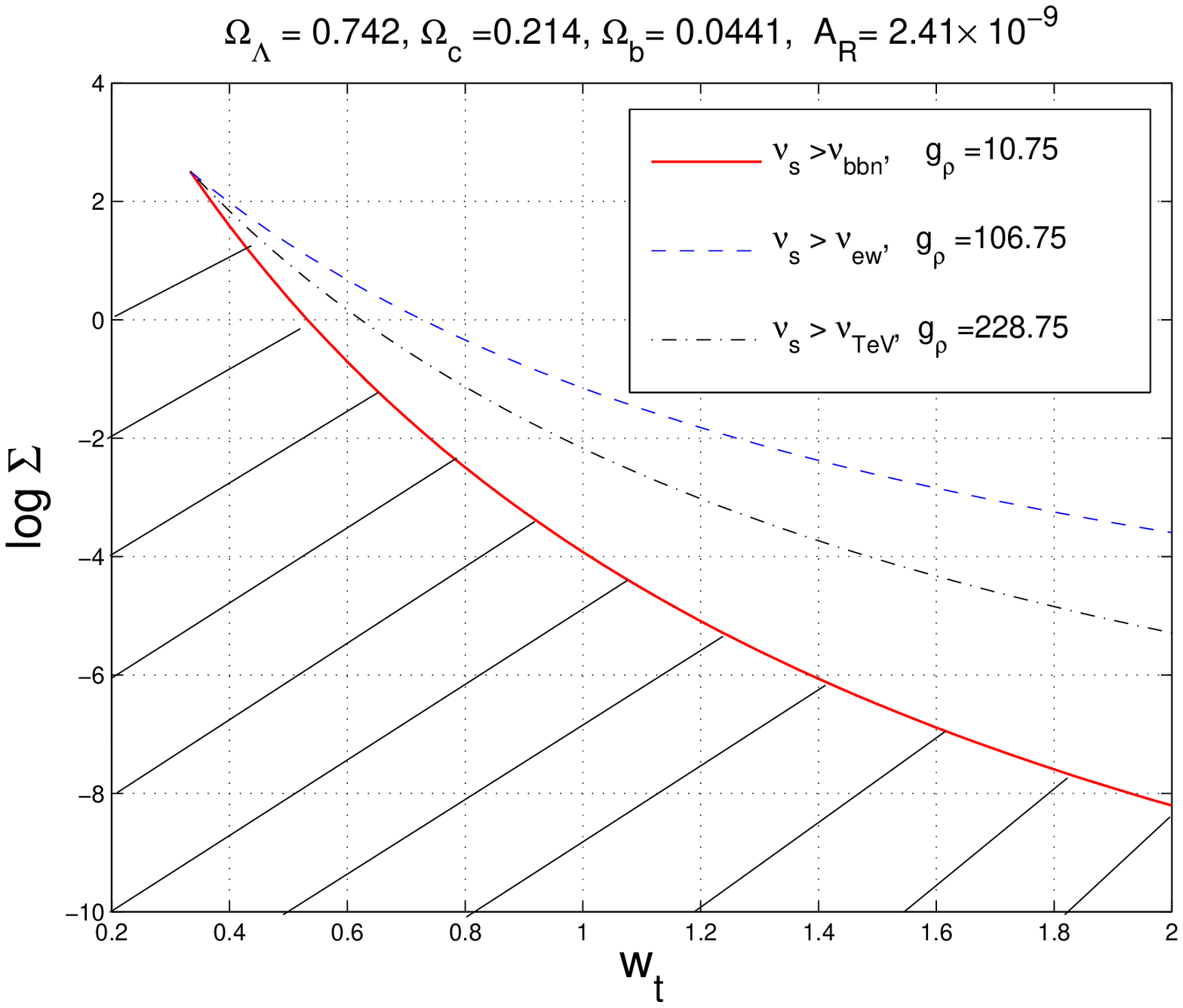}
\includegraphics[height=6.7cm]{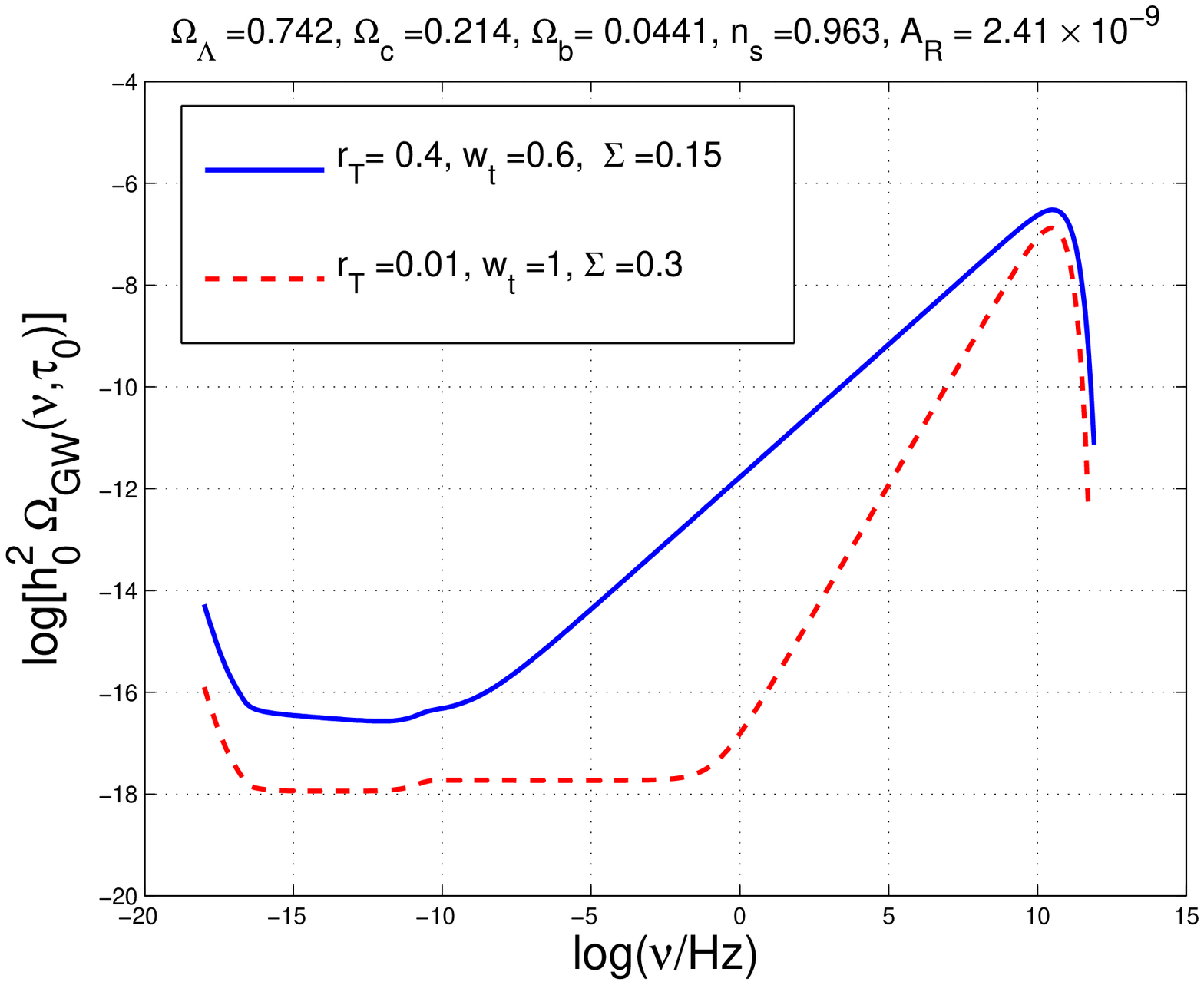}
\caption[a]{The bounds on $\nu_{\mathrm{s}}$ in the $(\Sigma, w_{\mathrm{t}})$ plane; the spectral energy density of the relic gravitons in the T$\Lambda$CDM scenario.}
\label{FigEX}      
\end{figure}
Recalling that $H_{\mathrm{r}} = H (a_{\mathrm{i}}/a_{\mathrm{r}})^{1/\alpha}$, we also have that Eq. (\ref{STFR15}) implies 
\begin{equation}
\biggl(\frac{H_{\mathrm{r}}}{M_{\mathrm{P}}}\biggr) = \lambda^{\frac{1}{2 ( 1 - 2 \alpha)}} \biggl(\frac{H}{M_{\mathrm{P}}}\biggr)^{\frac{2 ( 1 - \alpha)}{(1 -2 \alpha)}}.
\label{STFR16}
\end{equation}  
Using Eq. (\ref{STFR16})  and the definition of  $\Sigma$ (see Eq. (\ref{STFR3})) it turns out that  $\Sigma = \lambda^{1/4}$, which is always smaller than $1$ and, at most, ${\mathcal O}(1)$. 

Instead of endorsing an explicit model by pretending to know 
the whole thermal history of the Universe in reasonable detail, it is more 
productive to keep $\Sigma$ as a free parameter  and to require that the  scaling violations in the spectral energy density will take place before BBN.  
Consequently the variation of $\Sigma$, $w$ and $r_{\mathrm{T}}$
can be simultaneously bounded \cite{mgn1,mgn2}.
If the stiff dynamics takes place before big-bang nucleosynthesis, then $\nu_{\mathrm{s}}>\nu_{\mathrm{bbn}}$
 (see also Eq. (\ref{ANIS5})). This requirement guarantees that the stiff dynamics will be over by the time light nuclei start being formed. In a complementary approach one might also require that $\nu_{\mathrm{s}} > \nu_{\mathrm{ew}}$ where $\nu_{\mathrm{ew}}$ corresponds to the value of the Hubble rate at the electroweak 
epoch, i.e. 
\begin{eqnarray}
k_{\mathrm{ew}} &=& 2.5764\times10^{9} \biggl(\frac{g_{\rho}}{106.75}\biggr)^{1/4} \biggl(\frac{T_{*}}{100\,\,\mathrm{GeV}}\biggr) 
\biggl(\frac{h_{0}^2 \Omega_{\mathrm{R}0}}{4.15 \times 10^{-5}}\biggr)^{1/4}\,\,\mathrm{Mpc}^{-1}
\label{EW1}\\
\nu_{\mathrm{ew}} &=& 3.998\times 10^{-6} \biggl(\frac{g_{\rho}}{106.75}\biggr)^{1/4} \biggl(\frac{T_{*}}{100\,\,\mathrm{GeV}}\biggr) 
\biggl(\frac{h_{0}^2 \Omega_{\mathrm{R}0}}{4.15 \times 10^{-5}}\biggr)^{1/4}\,\,\mathrm{Hz}.
\label{EW2}
\end{eqnarray}
Finally, yet a different requirement could be to impose that $\nu > \nu_{\mathrm{Tev}}$ where $\nu_{\mathrm{TeV}}$ is defined as
\begin{equation}
\nu_{\mathrm{TeV}} = 4.819\times 10^{-3} \biggl(\frac{g_{\rho}}{228.75}\biggr)^{1/4} \biggl(\frac{T_{*}}{100\,\,\mathrm{TeV}}\biggr) 
\biggl(\frac{h_{0}^2 \Omega_{\mathrm{R}0}}{4.15 \times 10^{-5}}\biggr)^{1/4}\,\,\mathrm{Hz}.
\label{EW3}
\end{equation}
The latter requirement would imply that the stiff age is already finished by the time 
the Universe has a temperature of the order of $100$ TeV when, presumably, the number of relativistic degrees of freedom was much larger than in the minimal standard model (in Eq. (\ref{EW3}) the typical value of $g_{\rho}$ is 
the one arising in the minimal supersymmetric extension of the standard model). 

In Fig. \ref{FigEX} (plot at the left) the constraints on $\Sigma$, $w_{\mathrm{t}}$ are illustrated.
The value of $\Sigma$ controls the position of the frequency at which 
the nearly scale-invariant slope of the spectrum will be violated. The barotropic index $w_{\mathrm{t}}$ 
is taken to be always larger than $1/3$ and with a maximal value of $1$. The shaded region corresponds 
to the region excluded in the most constraining case, i.e. the one demanding $ \nu_{\mathrm{s}} > \nu_{\mathrm{bbn}}$. Also the values of $r_{\mathrm{T}}$ are bounded by the same kind of considerations. Indeed, 
$\nu_{\mathrm{s}}$ depends upon $\epsilon$ which is related to $r_{\mathrm{T}}$ (see, e.g., Eq. (\ref{STFR7})).
Therefore, a lower bound on $\nu_{\mathrm{s}}$ also implies a bound in the 
($r_{\mathrm{T}},\,w_{\mathrm{t}}$) plane.

\subsection{Spectral energy density in the minimal T$\Lambda$CDM  scenario}

In Fig. \ref{FigEX} (plot at the right) two examples of the scaling violations on the spectral energy density 
are illustrated. Both examples are compatible with the bounds illustrated in the plot at the left. 
Similar examples have been already illustrated in Fig. \ref{Figure2a}. These examples will now 
be discussed in greater detail. The spectral energy density has been computed by using the 
numerical approach presented in section \ref{sec6}.

In the two examples of  Fig. \ref{FigEX}  (plot at the right) the $\Lambda$CDM  parameters are fixed to the 
values reported in Eq. (\ref{bestfit}) (see \cite{WMAP51,WMAP52,WMAP53,WMAP54,WMAP55}). The spectral index has been allowed to run, i.e. 
$\alpha_{\mathrm{T}} \neq 0$ (see Eqs. (\ref{int4}) and (\ref{int3a})).
The two supplementary parameters should be identified 
with the sound speed during the stiff phase (i.e. $c_{\mathrm{st}}$) and
the threshold frequency (i.e. $\nu_{\mathrm{s}}$). Besides $c_{\mathrm{st}}$ 
and $\nu_{\mathrm{s}}$, there will also be $r_{\mathrm{T}}$ 
which controls, at once, the normalization and the slope of the 
low-frequency  branch of the spectral energy density.
At the moment wide band interferometers 
have sensitivities which are insufficient for cutting through the 
phenomenologically interesting region \cite{LIGOS1,LIGOS2,LIGOS3}. In the 
near future, however, there is the hope of a dramatic improvement 
of the sensitivity: even 5 or 6 orders of magnitude 
at least heeding the original design (see e.g. \cite{LIGO}) 
together with the recent proposals for an advanced Ligo program. 

As specifically discussed in Eqs. (\ref{STFR5})--(\ref{STFR6}) the frequency of the elbow, i.e. $\nu_{\mathrm{s}}$,  is fully determined by $\Sigma$ (see Eq. (\ref{STFR3}) and discussion therein). 
The two supplementary parameters $\nu_{\mathrm{s}}$ and $c_{\mathrm{st}}$ can be traded 
for $\Sigma$ and $w_{\mathrm{t}}$ as already done in Fig. \ref{FigEX} (plot at the left).
In doing so there is also a potential advantage since, according to Eq. (\ref{STFR4}), 
$\Sigma$ shifts the maximal frequency of the spectrum. 

As soon as the frequency increases from the aHz up to the nHz (and even larger) the spectral energy density 
increases sharply in comparison with the nearly scale-invariant case where the spectral energy density was, 
for $\nu > \mathrm{nHz}$, at most ${\mathcal O}(10^{-16})$.
In the case of Fig. \ref{FigEX} the spectral energy density is clearly much larger. 
The accuracy in the determination of the infra-red branch of the spectrum is a condition 
for the correctness of the estimate of the spectral energy density of the high-frequency branch. The plots of 
Fig. \ref{FigEX} (see also Fig. \ref{Figure2a}) demonstrate that the low-frequency bounds on $r_{\mathrm{T}}$ do not forbid a larger signal at higher frequencies. 

A decrease of  $r_{\mathrm{T}}$ implies a suppression of the nearly scale-invariant plateau  in the region 
$\nu_{\mathrm{eq}} < \nu <\nu_{\mathrm{s}}$. At the same time the amplitude of the spectral energy density still 
increases for frequencies larger than the frequency of the elbow (i.e. $\nu_{\mathrm{s}}$). 
The latter trend can be simply understood since, at high frequency,  
the transfer function for the spectral energy density grows faster than the power spectrum of inflationary origin. 
For instance, in the case $w_{\mathrm{t}} =1$ and neglecting logarithmic corrections,
$\Omega_{\mathrm{GW}}(\nu,\tau_{0}) \propto \nu^{n_{\mathrm{T}}+1}$  for $\nu\gg \nu_{\mathrm{s}}$. Now, recall that 
$n_{\mathrm{T}}$ is given by Eq. (\ref{int3}). If $r_{\mathrm{T}}\to 0$, the combination 
 $(n_{\mathrm{T}} +1)$ will be much closer to $1$ than in the case when, say, $r_{\mathrm{T}} \simeq 0.3$.
 This aspect can be observed in Fig. \ref{Figure2a}
 where different values of $r_{\mathrm{T}}$ have been reported.  By decreasing the $w_{\mathrm{t}}$ from $1$ to, say, $0.6$ the extension of the nearly flat plateau gets narrower. 
 This is also a general effect which is particularly evident by comparing the two curves of Fig. \ref{FigEX} (plot 
 at the right).  
\begin{figure}[!ht]
\centering
\includegraphics[height=6.7cm]{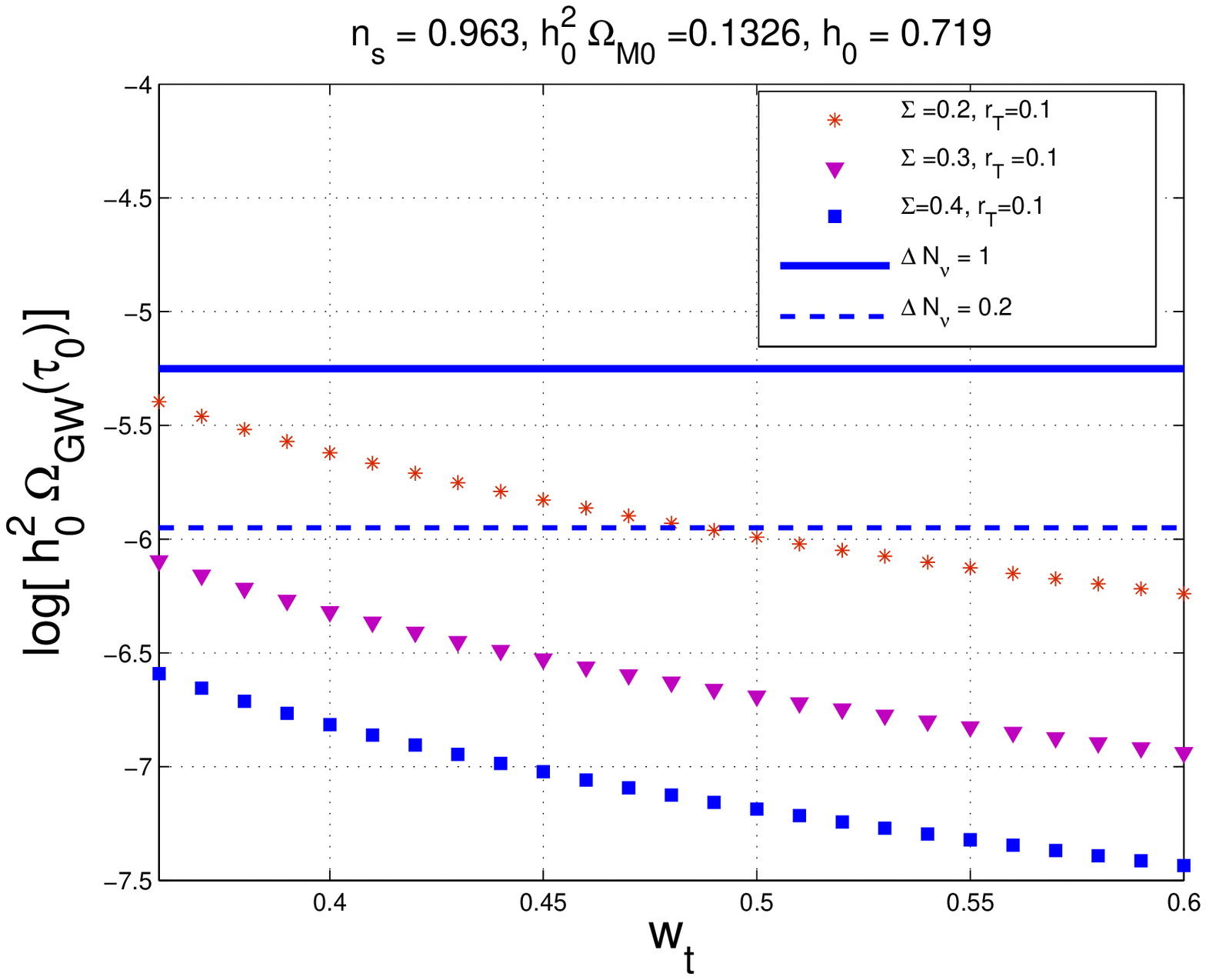}
\includegraphics[height=6.7cm]{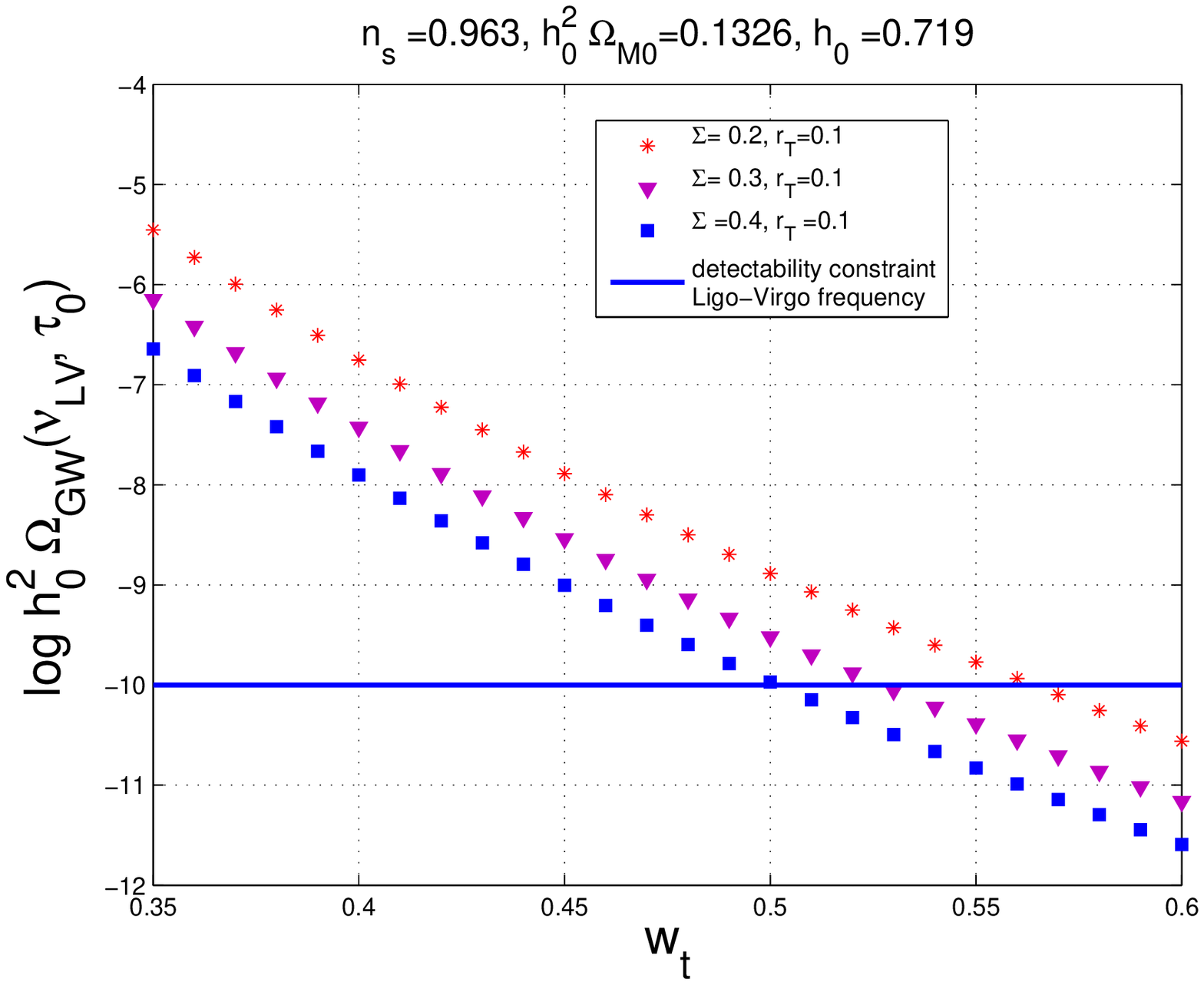}
\caption[a]{The bounds stemming from the amount of extra-relativistic species at the epoch 
of the synthesis of light nuclei are applied to the relic graviton spectra from the stiff epoch (plot at the left). The detectability constraint (full line in the plot at the right) stemming from the putative sensitivities of wide-band interferometers in their 
advanced version. The points corresponding to the spectral energy density should lie above the full lines to be potentially interesting for those instruments.}
\label{FigEX2}      
\end{figure}
 The slope of the high-frequency branch of the graviton energy spectrum can be easily deduced with 
 analytic methods and it turns out to to be 
\begin{equation} 
\frac{d\ln{\Omega_{\mathrm{GW}}}}{d\ln{\nu}}= 
 \frac{6 w_{\mathrm{t}} -2}{3 w_{\mathrm{t}} +1},\qquad \nu> \nu_{\mathrm{s}},
 \label{estimate}
 \end{equation}
  up to logarithmic corrections. 
 The  result of Eq. (\ref{estimate}) stems  from the simultaneous integration of the  background evolution equations and of the tensor mode functions according to the techniques described in section \ref{sec3}. 
 The semi-analytic estimate of the 
 slope (see \cite{mg1}) agrees with the results obtained by means 
 of the transfer function of the spectral energy density. 

To conclude the discussion it is appropriate to elaborate on the interplay between 
the stiff spectra and the phenomenological bounds mentioned in subsection \ref{sec16}.
Let us start with millisecond pulsar bound of Eq. (\ref{PUL}).

Assuming the maximal growth of the spectral energy density and the minimal 
 value of $\nu_{\mathrm{s}}$, i.e. $\nu_{\mathrm{bbn}}$ we will have 
\begin{equation}
h_{0}^2 \Omega_{\mathrm{GW}}(\nu,\tau_{0}) \propto \nu, \qquad \nu 
\geq \nu_{\mathrm{s}} \simeq \nu_{\mathrm{bbn}}.
\label{PUL2}
\end{equation}
Since $\nu_{\mathrm{pulsar}} \simeq 10^{3} \nu_{\mathrm{bbn}}$,
Eq. (\ref{PUL2}) implies  that 
$h_{0}^2\Omega_{\mathrm{GW}}(\nu_{\mathrm{pulsar}},\tau_{0}) \simeq 10^{-13}$  or even $10^{-14}$ depending upon 
$r_{\mathrm{T}}$.  But this value is always much smaller than the constraint stemming from pulsar timing measurements (see Eq. (\ref{PUL})). If either $\nu_{\mathrm{s}} \gg \nu_{\mathrm{bbn}}$ 
or $c_{\mathrm{st}} < 1$ the value of 
$h_{0}^2\Omega_{\mathrm{GW}}(\nu_{\mathrm{pulsar}},\tau_{0})$ 
will be even smaller\footnote{This conclusion follows immediately from the hierarchy between $\nu_{\mathrm{pulsar}}$ and 
$\nu_{\mathrm{bbn}}$. If either $\nu_{\mathrm{s}} \gg \nu_{\mathrm{bbn}}$, $h_{0}^2 \Omega_{\mathrm{GW}}$ can only grow very little and certainly much less than 
required to violate the bound of Eq. (\ref{PUL}).}. Consequently, even in the extreme cases when the 
frequency of the elbow  is 
close to $\nu_{\mathrm{bbn}}$, the spectral energy density is always much smaller than the requirement of Eq. (\ref{PUL}). 

As noticed in the past \cite{mg0,mg1}, the most significant constraint on the stiff spectra stems from BBN.
The models illustrated in Fig. \ref{FigEX} (plot at the right)  are on the verge 
of saturating the bounds of Eqs. (\ref{BBN1})--(\ref{BBN2}).  This conclusion 
stems directly from the form of spectral energy density: the broad spike  dominates
the (total) energy density of relic gravitons which are inside the Hubble radius 
at the time of big bang nucleosynthesis. 
For consistency with the low-frequency 
 determinations of the tensor power spectrum $r_{\mathrm{T}}$ must be bounded 
 from above according to the values reported, for instance, in 
 Tabs. \ref{TABLE1} and \ref{TABLE2}.
Once the the value of $r_{\mathrm{T}}$ has been selected, 
the constraints of Eqs. (\ref{BBN1})--(\ref{BBN2}) can be imposed.
From Fig. \ref{FigEX} a large signal is expected for $\nu_{\mathrm{LV}} \simeq 100$ Hz for $0.35 < w_{\mathrm{t}} < 0.6$.  This range turns out to be compatible with the bounds 
of Eqs. (\ref{BBN1})--(\ref{BBN2}). 
In the opposite limit (e. g. $w_{\mathrm{t}} \simeq 1$) 
the spike becomes narrower, the elbow frequency augments and the signal at the interferometer scale diminishes. 

In Fig. \ref{FigEX2}  the energy density of the relic gravitons inside the Hubble 
radius at the nucleosynthesis epoch is reported in the case $r_{\mathrm{T}} =0.1$ and for different 
values of $\Sigma$. In the plot at the left $n_{\mathrm{s}} =0.963$ as implied by the WMAP 5-yr data 
alone. The acceptable region of the parameter space must stay below the horizontal lines which 
illustrate different values of $\Delta N_{\nu}$ (see Eqs. (\ref{BBN1})--(\ref{BBN2})).
As the scalar spectral index diminishes, the constraints are better satisfied since 
$n_{\mathrm{s}}$ controls $\alpha_{\mathrm{T}}$ and, consequently,  the frequency dependence of the tensor spectral index $n_{\mathrm{T}}$ (see Eq. (\ref{int3})) in the case $\alpha_{\mathrm{T}} \neq 0$.
\begin{figure}[!ht]
\centering
\includegraphics[height=6.7cm]{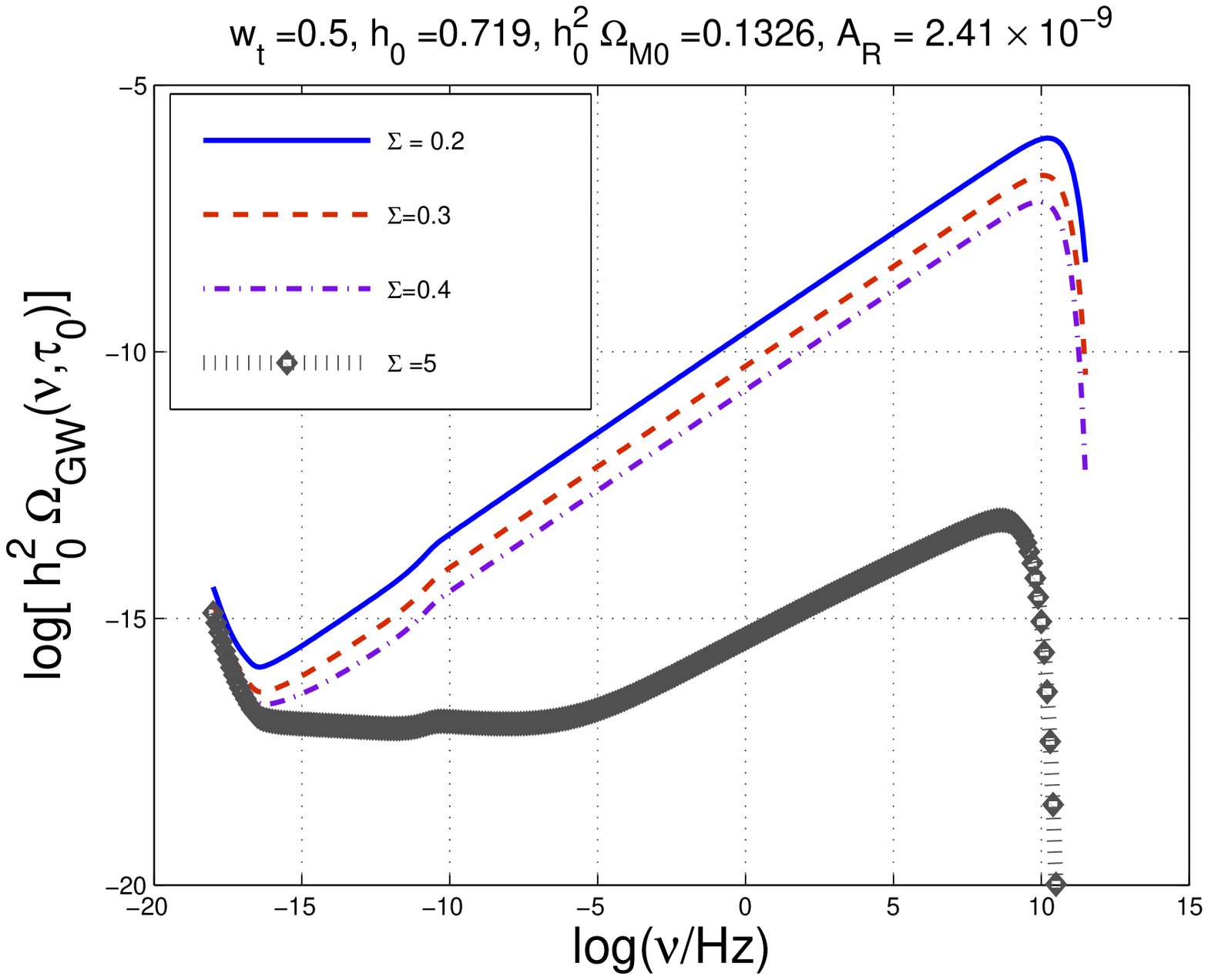}
\includegraphics[height=6.7cm]{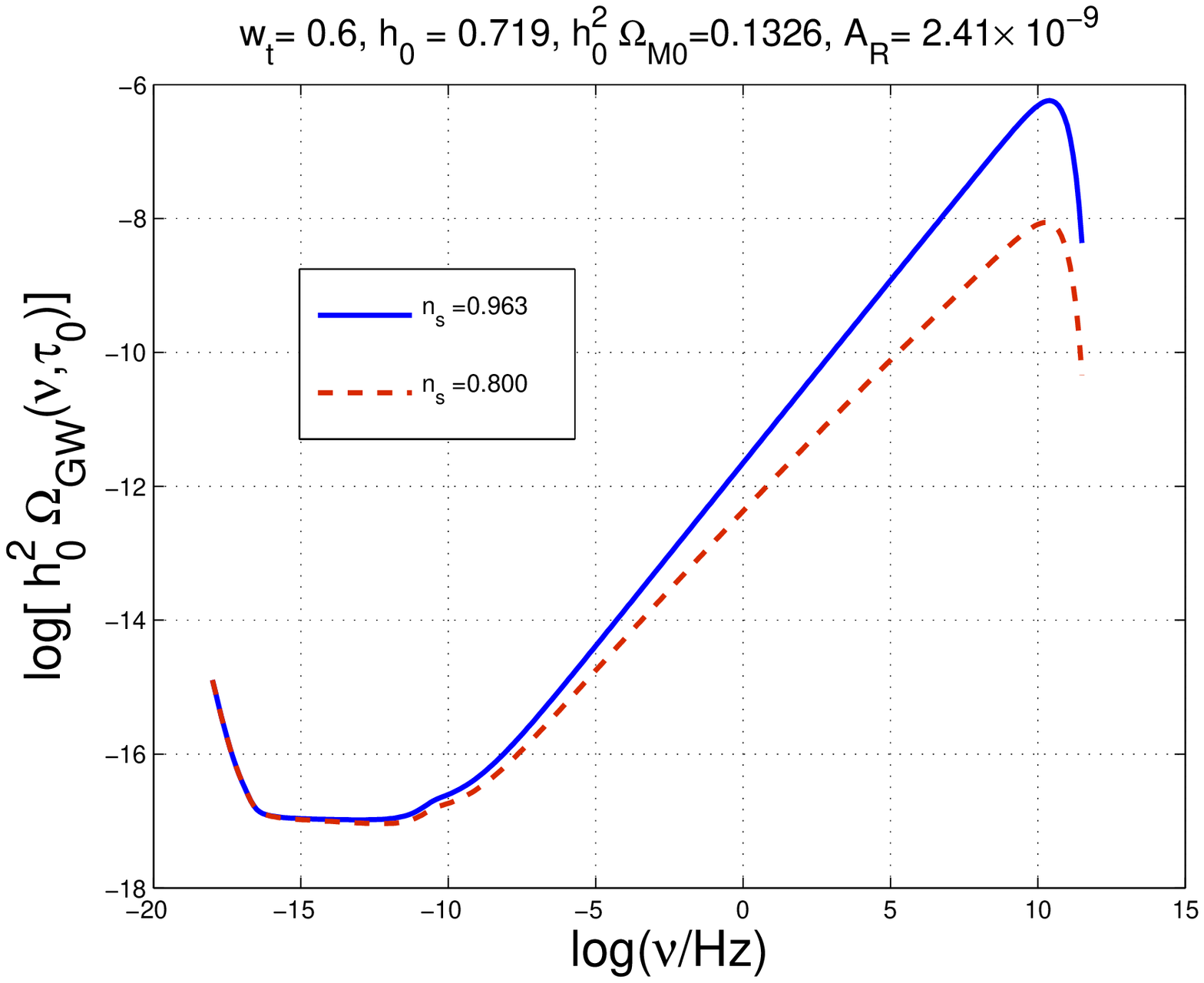}
\caption[a]{The spectral energy density is illustrated for small values of $w_{\mathrm{t}}$
and different values of $\Sigma$ (plot at the left).
In the plot at the right $\Sigma =0.2$ and $w_{\mathrm{t}} =0.6$.}
\label{FigEX3}      
\end{figure}
\subsection{Detectability prospects}
By lowering $w_{\mathrm{t}}$, $h_{0}^2 \Omega_{\mathrm{GW}}(\nu,\tau_{0})$ 
increases for $\nu = \nu_{\mathrm{LV}}\simeq 0.1\, \mathrm{kHz}$.
This  trend can be inferred from Fig. \ref{FigEX2} (plot at the right) 
where the spectral energy density is evaluated exactly
for $\nu = \nu_{\mathrm{LV}}$.  To be detectable by wide band interferometers 
the parameters of the T$\Lambda$CDM must lie above the full lines.
The region of low barotropic indices emerging neatly from Fig. \ref{FigEX2},
 leads to spectral energy densities which are progressively flattening as $w_{\mathrm{t}}$ diminishes towards $1/3$.  
Low values of $w_{\mathrm{t}}$ bring the frequency of the elbow, i.e. $\nu_{\mathrm{s}}$ below 
$10^{-10}$ Hz which is unacceptable since it would mean that, during nucleosynthesis, the Universe was dominated by the stiff fluid. 
In Fig. \ref{FigEX} (plot at the left) the region above the full line corresponds to a range of parameters  for 
which $\nu_{\mathrm{s}} > \nu_{\mathrm{bbn}}$: in such a range a decrease of $w_{\mathrm{t}}$ demands an increase of $\Sigma$. The latter occurrence  is illustrated in Fig. \ref{FigEX3} 
where, at the left, $w_{\mathrm{t}} =0.5$.
The full, dashed and dot-dashed curves illustrated in Fig. \ref{FigEX3} (plot at 
the left) are incompatible with the phenomenological constraints 
since the frequency of the elbow is systematically smaller than $\nu_{\mathrm{bbn}}$.
Once more, this choice of parameters would contradict the bounds of  Fig. \ref{FigEX} and would imply that the stiff phase is not yet finished at the BBN time. In the left plot of Fig. \ref{FigEX3} the diamonds denote a model which is compatible with BBN considerations but whose signal at the frequency of interferometers 
is rather small (always three orders of magnitude larger than in the case of conventional inflationary models).

The compatibility  with the phenomenological constraints demands that the parameters
of the T$\Lambda$CDM paradigm must lie above the full lines of Fig. \ref{FigEX} (plot at the left).
The requirements of Fig. \ref{FigEX} suggest, therefore, that 
$\Sigma$ should be raised a bit. In this case the frequency of the elbow gets 
shifted to the right but, at the same time, the overall amplitude of the spike 
diminishes.  The putative amplitude remains still much larger than 
the conventional inflationary signal reported, for instance, in Fig. \ref{Figure2}.
\begin{figure}[!ht]
\centering
\includegraphics[height=6.7cm]{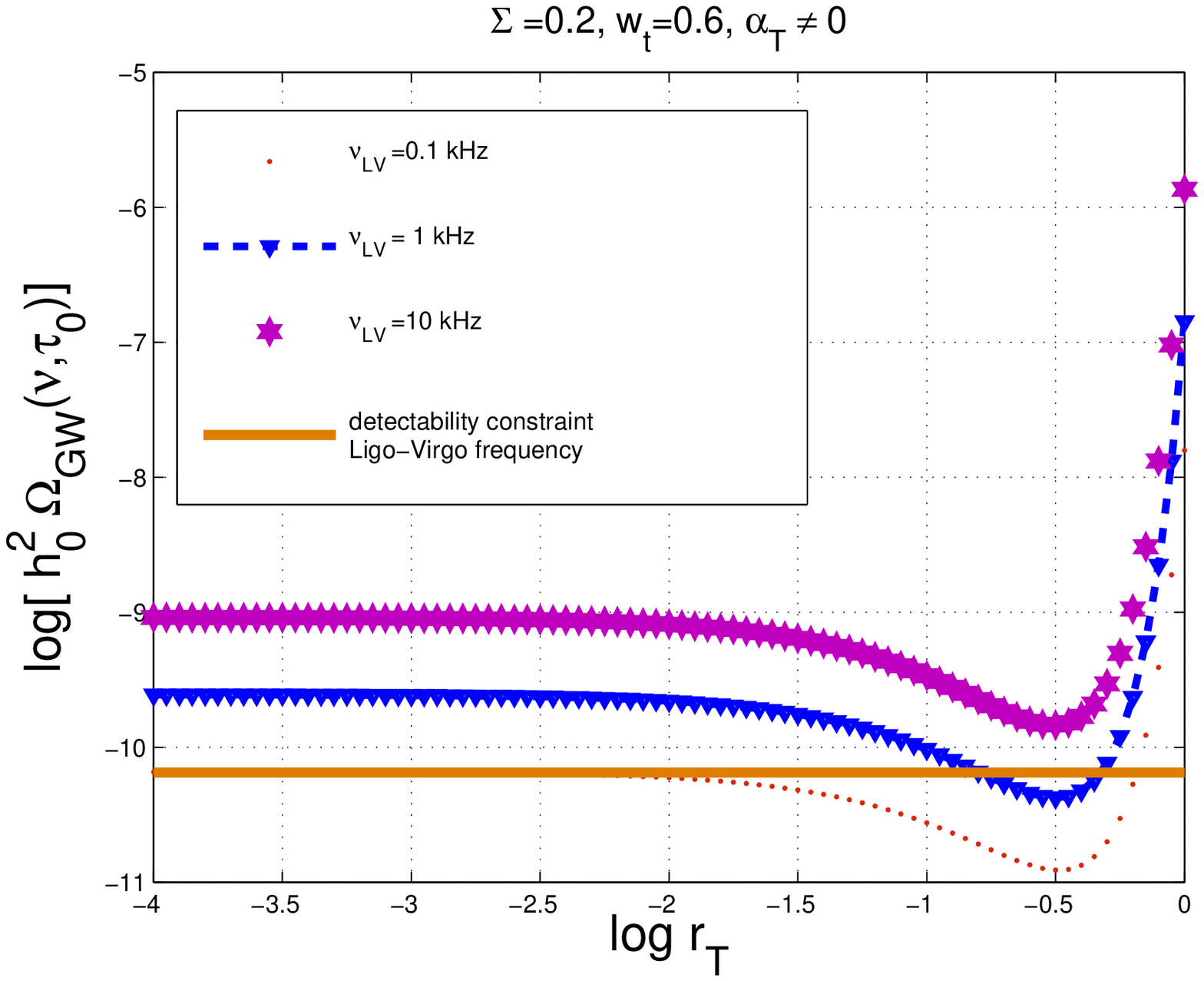}
\includegraphics[height=6.7cm]{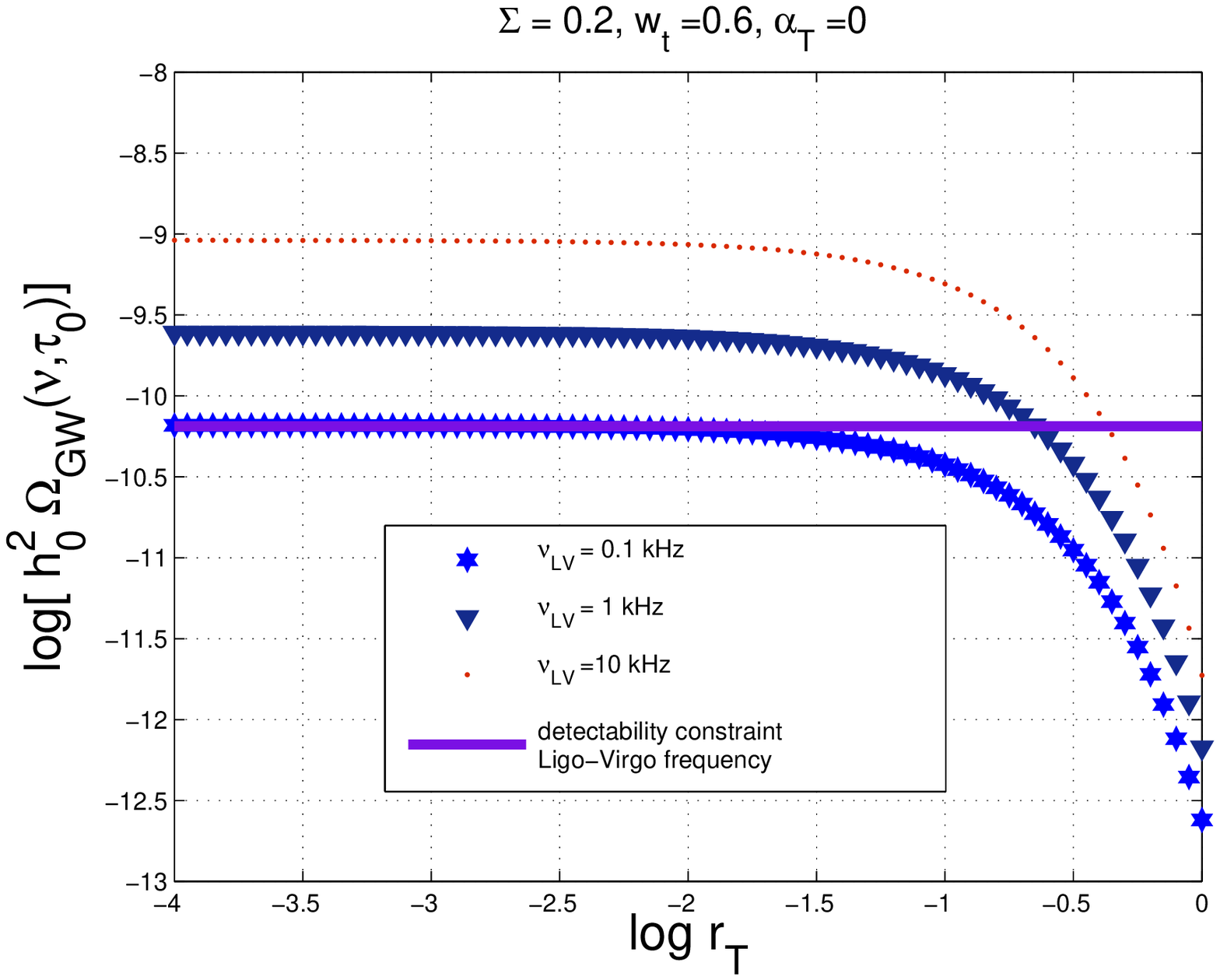}
\caption[a]{The graviton energy spectrum is illustrated, in the T$\Lambda$CDM scenario,  for $\nu= \nu_{\mathrm{LV}}$  and as a function of $r_{\mathrm{T}}$ (plot at the left) in the case $\alpha_{\mathrm{T}} \neq 0$. In the plot at the right $\alpha_{\mathrm{T}} =0$.}
\label{FigEX4}      
\end{figure}
In Fig. \ref{FigEX4} the spectral energy density of the relic gravitons is illustrated as a function of $r_{\mathrm{T}}$ 
for a choice of parameters which is compatible with all the bounds applicable to the stochastic backgrounds of the 
relic gravitons. The three curves refer to three different frequencies, i.e. $0.1$ kHz, $1$ kHz and $10$ kHz. Indeed, if the spectrum is 
nearly scale-invariant (as in the case o Fig. \ref{Figure2}) we can compare the potential signal with the central frequency of the 
window. If the signal increases with frequency it is interesting to plot the same curve for some significant 
frequencies inside the window of wide-band interferometers. Even if the frequency window extends from 
few Hz to $10$ kHz the maximal sensitivity is in the central region and depends upon various important factors which will 
now be briefly discussed. 

To illustrate more quantitatively this point we remind the 
expression of the signal-to-noise ratio (SNR)  in the 
context of optimal processing  required for the detection of stochastic backgrounds: 
\begin{equation}
{\rm SNR}^2 \,=\,\frac{3 H_0^2}{2 \sqrt{2}\,\pi^2}\,F\,\sqrt{T}\,
\left\{\,\int_0^{\infty}\,{\rm d} \nu\,\frac{\gamma^2 (\nu)\,\Omega^2_{{\rm GW}}(\nu,\tau_{0})}
{\nu^6\,S_n^{\,(1)} (\nu)\,S_n^{\,(2)} (\nu)}\,\right\}^{1/2}\; ,
\label{SNR1}
\end{equation}
 ($F$ depends upon 
the geometry of the two detectors and in the case of the correlation between 
two interferometers $F=2/5$; $T$ is the observation time). 
In Eq. (\ref{SNR1}), $S_n^{\,(k)} (f)$ is the (one-sided) noise power 
spectrum (NPS) of the $k$-th 
$(k = 1,2)$ detector. The NPS contains the important informations concerning the 
noise sources (in broad terms seismic, thermal and shot noises)
 while $\gamma(\nu)$ is the overlap reduction function 
which is determined by the relative locations and orientations 
of the two detectors. In \cite{mg4} Eq. (\ref{SNR1}) has been used to assess the 
detectability prospects of gravitons coming from a specific model of stiff evolution with $w_{\mathrm{t}} =1$.
At that time the various suppressions of the low-frequency amplitude as well as the free-streaming effects 
were not taken into account. Furthermore, the evaluation of the energy transfer function was obtained, in 
\cite{mg6}, not numerically but by matching of the relevant solutions. We do know, by direct comparison, that 
such a procedure is justified but intrinsically less accurate than the one proposed here. 
It would be interesting to apply Eq. (\ref{SNR1})  for the (more accurate) assessment of the sensitivities 
of different instruments to a potential signal stemming from the stiff age \footnote{
For intermediate frequencies the integral of Eq. (\ref{SNR1}) is sensitive to the form of the overlap reduction 
function which depends upon the mutual position and relative orientations of the interferometers. The 
function $\gamma(\nu)$ effectively cuts-off the integral which defines the signal to noise ratio for a typical 
frequency $\nu\simeq 1/(2 d)$ where $d$ is the separation between the two detectors.
Since $\Omega_{\mathrm{GW}}$ increases 
with frequency (at least in the case of relic gravitons from stiff ages) at most as $\nu$ and since there is a $\nu^{-6}$ in the 
denominator, the main contribution to the integral should occur for $\nu < 0.1$ kHz.  This argument can be 
explicit verified in the case of the calculations carried on in \cite{mg4} and it would be interesting 
to check it also in our improved framework.}.

Equation (\ref{SNR1}) assumes that the intrinsic noises of the detectors are stationary, Gaussian, 
uncorrelated, much larger in amplitude than the gravitational strain, and 
statistically independent on the strain itself \cite{int1,int2,int3,int4}.  The integral appearing in Eq. (\ref{SNR1})
extends over all the frequencies. However, the noise power spectra of the detectors are defined 
in a frequency interval ranging from few Hz to $10$ kHz. In the latter window, for very small frequencies 
the seismic disturbances are the dominant source of noise. For intermediate and high frequencies 
the dominant sources of noise are, respectively,  thermal and electronic (i.e. shot) noises.  
The wideness of the band is very important when cross-correlating two detectors: typically 
the minimal detectable $h_{0}^2\Omega_{\mathrm{GW}}$
 will become smaller (i.e. the sensitivity will increase) by a factor $1/\sqrt{\Delta \nu T}$ where 
 $\Delta \nu$ is the bandwidth and $T$, as already mentioned, is the observation time.
 Naively, if the minimal detectable signal (by one detector ) is 
 $h_{0}^2\Omega_{\mathrm{GW}} \simeq 10^{-5}$, then the cross-correlation of two 
 identical detector with overlap reduction $\gamma(\nu) =1$ will detect  
 $h_{0}^2\Omega_{\mathrm{GW}} \simeq 10^{-10}$ provided $\Delta \nu \simeq 100$ Hz and 
 $T\simeq {\mathcal O}(1\mathrm{yr})$ (recall that $1\mathrm{yr} = 3.15 \times 10^{7} \mathrm{Hz}^{-1}$). 
The achievable sensitivity of a pair of wide band interferometers crucially 
depends upon the spectral slope of the theoretical energy spectrum in the 
operating window of the detectors. So, a flat spectrum will lead 
to an experimental sensitivity which might not be similar to the 
sensitivity achievable in the case of a blue or violet spectra.   
Previous calculations \cite{mg4,mg5,mg6} showed that, however, 
to get a reasonable idea of the potential signal it is sufficient to compare 
the signal with the sensitivity to flat spectrum which has been 
reported in Eq. (\ref{SENS}). Of course any experimental 
improvement in comparison with the values of Eq. (\ref{SENS}) 
will widen the detectability region by making the prospects
of the whole discussion more rosy. 

In the T$\Lambda$CDM paradigm the maximal signal 
occurs in a frequency region between the MHz and the GHz. 
This intriguing aspect led to the suggestion \cite{mg4,mg5} that 
microwave cavities  \cite{HF1a} can be used as GW detectors precisely in the 
mentioned frequency range. Prototypes of these detectors \cite{HF1b} have been 
described  and the possibility of further improvements in their sensitivity received 
recently attention \cite{HF1c,HF1d,HF2,HF3,HF4,HF5}. 
Different groups are now concerned with high-frequency 
gravitons. In \cite{HF1d} the ideas put forward in \cite{HF1a,HF1b,HF1c} 
have been developed by using electromagnetic cavities (i.e. static electromagnetic fields).
In  \cite{HF2,HF3,HF4} dynamical electromagnetic fields (i.e. wave guides) have been studied always 
for the purpose of detecting relic gravitons. Yet a different approach to the problem has been 
described in \cite{HF5}. In \cite{HF4} an interesting prototype 
detector was described with frequency of operation of the order of $100$ MHz (see also \cite{HF6}). 
It is not clear if, in the near future, the improvements in the terrestrial technologies will allow 
the detection of relic gravitons for frequencies, say, larger than the MHz. It is, however, a rather 
intriguing possibility. 

\newpage

\section{Final remarks}
The incoming score year might witness direct experimental 
evidences of relic gravitons either from small frequency experiments 
or from high-frequency experiments.  By low-frequency experiments 
we mean, as in the bulk of this review, the CMB experiments 
possibly analyzed together with the two remaining cosmological data sets 
(i.e. large-scale structure determinations of the matter power spectrum and 
type Ia supernova observations). By high-frequency experiments 
we mean the appropriately advanced versions 
of wide band interferometers such as Ligo, Virgo, Geo and Tama. 

The main observables related to relic gravitons have been reviewed in a self-contained manner and in the framework of the $\Lambda$CDM paradigm 
with specific attention to the complementarity between the low-frequency 
and the high-frequency branches of the relic graviton spectrum.
Any model claiming a signal 
coming from high-frequency gravitons should  be compatible with the 
$\Lambda$CDM paradigm in the low-frequency branch of the 
relic graviton spectrum.

 It is instructive to go back to the comparison drawn, in Fig. \ref{FigureSP}, between the electromagnetic spectrum and the relic 
graviton spectrum. The gap between the graviton frequencies explored by CMB experiments and the graviton frequencies probed by wide-band interferometers is of the order of the frequency gap between radio waves and $\gamma$-rays. 

A detection of long wavelength gravitons in CMB experiments can be direct 
(i.e. from the B-mode polarization) or indirect (i.e. from some global fit 
of CMB observables including the tensor contribution). In the context 
of the $\Lambda$CDM paradigm the  CMB detection of long wavelength gravitons 
will fix the overall normalization of the spectral energy density. Even in the 
absence of such a direct detection, the current upper limits 
on the contribution of long wavelength gravitons to CMB observables 
implies a minute signal at higher frequencies. The (hoped) 
sensitivities achievable by the advanced wide-band interferometers 
are still insufficient for a direct detection of high-frequency gravitons. 

The latter statement summarizes the standard lore 
of the problem which may well be realized.  At the same time 
it might be unwise to assume (or presume) that 
the current success of the $\Lambda$CDM paradigm 
also fixes the whole thermal history of the Universe for 
temperatures larger than the MeV.  The general ideas conveyed in the present 
review suggest that the high-frequency branch of the 
relic graviton spectrum is rather sensitive to the whole 
post-inflationary thermal history of the Universe.  
If the post-inflationary evolution is dominated by stiff sources, for 
instance, it is not impossible, as explicitly shown, to have 
positive detection of relic gravitons both at small and high 
frequencies even enforcing the current bounds 
on the tensor contribution to CMB observables. 

If we assume (or strongly believe) the standard lore (i.e. that 
relic gravitons will probably not be seen by wide-band detectors)
it is useless to demand theoretical accuracy. For instance it is useless 
to ask what would be the effect of changing $\Omega_{\Lambda}$ 
on a signal which is anyway $6$ or even $7$ orders of magnitude 
smaller than the most optimistic sensitivities.
A positive detection of relic graviton backgrounds at high-frequencies 
would demand, however, more accurate estimates of the theoretical 
signal in different models. Absent a direct detection of relic gravitons by wide-band 
inteferometers, accurate theoretical calculations can be used to set 
bounds on possible deviations of the post-inflationary thermal history.

\end{document}